\documentclass[12pt]{iopart}
\usepackage{amsfonts}
\usepackage{graphicx}
\usepackage{url}
\usepackage{cite}
\usepackage{caption}
\usepackage{calrsfs}
\DeclareMathAlphabet{\pazocal}{OMS}{zplm}{m}{n}
\usepackage{geometry}
\begin{document}
\title[The inherent community structure of hyperbolic networks]{The inherent community structure of hyperbolic networks}

\author{Bianka Kovács$^1$ and Gergely Palla$^{1,2,3}$}

\address{$^1$ Dept. of Biological Physics, Eötvös Loránd University, H-1117 Budapest, Pázmány P. stny. 1/A, Hungary}
\address{$^2$ MTA-ELTE Statistical and Biological Physics Research Group, H-1117 Budapest, Pázmány P. stny. 1/A, Hungary}
\address{$^3$ %Digital Health and Data Utilisation Team, Health Services Management Training Centre, Faculty of Health and Public Administration, Semmelweis University, Budapest, Hungary
Health Services Management Training Centre, Semmelweis University,  H-1125, Kútvölgyi út 2, Budapest, Hungary}
\ead{pallag@hal.elte.hu}

Keywords: hyperbolic networks; communities; PSO model; $\mathbb{S}^1/\mathbb{H}^2$ model%When you submit an article, you will be asked to supply some keywords relevant to your work. If your article is accepted for publication, we will display these keywords on the published article, and they will be used to index your article, helping to make it more discoverable. When choosing keywords, think about the kinds of terms you would use when searching online for related articles.

\begin{abstract}
    A remarkable approach for grasping the relevant statistical features of real networks with the help of random graphs is offered by hyperbolic models, centred around the idea of placing nodes in a low-dimensional hyperbolic space, and connecting node pairs with a probability depending on the hyperbolic distance. It is widely appreciated that these models can generate random graphs that are small-world, highly clustered and scale-free at the same time; thus, reproducing the most fundamental common features of real networks. In the present work, we focus on a less well-known property of the popularity-similarity optimisation (PSO) model and the $\mathbb{S}^1/\mathbb{H}^2$ model from this model family, namely that the networks generated by these approaches also contain communities for a wide range of the parameters, which was certainly not an intention at the design of the models. We extracted the communities from the studied networks using well-established community finding methods such as Louvain, Infomap and label propagation. The observed high modularity values indicate that the community structure can become very pronounced under certain conditions. In addition, the modules found by the different algorithms show good consistency, implying that these are indeed relevant and apparent structural units. Since the appearance of communities is rather common in networks representing real systems as well, this feature of hyperbolic models makes them even more suitable for describing real networks than thought before.
\end{abstract}

\captionsetup[figure]{font=small,justification=justified,labelsep=period,labelfont=bf}

\section{Introduction}
\label{sect:Intro}

Complex network theory is a rapidly expanding interdisciplinary field, strongly interwoven with statistical physics, concentrating on the interesting non-trivial statistical features of the graphs representing the connections/interactions between entities of complex systems \cite{Laci_revmod,Dorog_book,Newman_Barabasi_Watts,Jari_Holme_Phys_Rep,Vespignani_book}. Over the last two decades, the vast number of studies of real networks have shown that some of these features seem to be almost universal, such as the small-world property \cite{Milgram_small_world,Kochen_book}, the relatively high clustering coefficient \cite{Watts-Strogatz}, the inhomogeneous degree distribution \cite{Faloutsos,Laci_science}, and the presence of communities \cite{Fortunato_coms,Fortunato_Hric_coms,Cherifi_coms}. Grasping these properties in a unified modelling framework is a non-trivial problem; however, a very notable approach pointing in this direction is given by hyperbolic network models \cite{PSO,EPSO_HyperMap,GPA_PSOsoftComms,nPSO,nPSO_2,S1,S1softComms,von_Looz_generate}, centred around the idea of placing nodes on a hyperbolic plane, and drawing links with a probability depending on the metric distance. 

%\cite{Milgram_small_world,Kochen_book}, the high clustering coefficient \cite{Watts-Strogatz} and the scale-free degree distribution \cite{Faloutsos,Laci_science}.

Probably the most well-known model from this family is the popularity-similarity optimisation (PSO) model \cite{PSO}, working in the native disk representation of the two-dimensional hyperbolic space. Here the nodes are introduced one by one with logarithmically increasing radial coordinates and uniformly random angular coordinates, and the newly appearing nodes connect to the previous ones with a probability decreasing with the hyperbolic distance. This model is known to be capable of generating networks that are small-world, highly clustered and scale-free at the same time. Roughly speaking, the degree of the nodes is determined by their radial coordinate -- with the inner nodes becoming eventually hubs -- and due to a parameter-controlled outward shift of the nodes (corresponding to popularity fading), the decay exponent of the degree distribution is also tuneable in the model. By changing the cutoff of the connection probability as a function of the hyperbolic distance with another parameter called the temperature, the clustering coefficient of the resulting random graphs can be adjusted as well. 

Another remarkable hyperbolic network model, capable of generating small-world, highly clustered and scale-free random graphs is given by the $\mathbb{S}^1/\mathbb{H}^2$ model \cite{S1,S1H2_Mercator}. In the $\mathbb{S}^1$ model nodes are placed on a circle and are given a hidden variable drawn from a power-law distribution. Here the connection probability depends on the angular distance between the nodes and the hidden variables. By converting the hidden variables to radial coordinates in the native disk representation of the hyperbolic plane, we arrive to the equivalent $\mathbb{H}^2$ model, where the connection probabilities depend on the hyperbolic distance between the nodes in a similar way as in the PSO model.

In parallel with the success of hyperbolic models, there have also been several studies carried out focusing on possible hidden metric spaces behind real networks, starting with the examination of the self-similarity of scale-free networks \cite{S1}, followed by reports on the hyperbolicity of 
protein interaction networks \cite{Higham_geom_protein_2008,Kuchaiev_geom_protein_2009}, the Internet \cite{Boguna_2009_nat_phys, Boguna_Krioukov_Internet_2010,Jonckhere_Internet_2011,Bianconi_internet_2015,Chepoi_Internet_2017}, brain networks \cite{Cannistraci_brain_2013,Tadic_brain_2018}, or the world trade network \cite{Boguna_trade_net_2016}. Furthermore, a connection between the navigability of networks and hyperbolic spaces was shown \cite{Boguna_2009_nat_phys,Gulyas_natcoms}, the geometric nature of weights \cite{Boguna_geometric_weights_ncoms} and clustering \cite{Candellero_clust_hyp_geom,Krioukov_clust_hyp_geom} was demonstrated,
methods for measuring the hyperbolicity of networks were introduced \cite{Kennedy_measure_hyper,Borassi_measury_hyper}, and practical fast algorithms for generating hyperbolic networks were proposed \cite{von_Looz_generate}. Hyperbolic networks are also closely related to network models based on simplicial complexes \cite{Bianconi_simplex_emergent_hyp,Mulder_geom_complexity}, where the emergent geometry of the generated random graphs was shown to be hyperbolic. In addition, significant achievements were obtained related to the problem of hyperbolic embedding as well \cite{Boguna_Krioukov_Internet_2010,EPSO_HyperMap,Alanis-Lobato_LE_embedding,linkWeights_coalescentEmbedding,Alanis-Lobat_liekly_LE_emb,S1H2_Mercator,our_embedding}, where the task is to find the most suitable node coordinates in a hyperbolic space given an input network topology. 

Returning to hyperbolic network models, in the recent years there have also been efforts devoted to the development of generative methods capable of producing hyperbolic random graphs with an apparent community structure \cite{nPSO,nPSO_2,GPA_PSOsoftComms,S1softComms}. Clusters or communities in hyperbolic networks usually correspond to separated angular regions \cite{commSector_hypEmbBasedOnComms_2016,commSector_commDetMethod,commSector_linkPred,commSector_hypEmbBasedOnComms_2019,Boguna_hyp_embed_coms,Cannistraci_ASI}. In accordance with this, in Refs.\cite{nPSO,nPSO_2} the uniform angular distribution of the nodes was replaced by a multimodal distribution, where communities arise naturally at the peaks. The appearance of communities in Refs.\cite{GPA_PSOsoftComms,S1softComms} was achieved by applying a geometric preferential attachment process, also inducing the formation of denser angular regions corresponding to communities.

Although the above-mentioned ideas do provide very interesting models with 'built-in' community formation, in the present paper we would like to draw the attention to the lesser-known but somewhat surprising fact that angular inhomogeneity is not a necessary condition for the presence of communities in hyperbolic network models, and that communities can appear in networks generated by the 'plain' PSO model or the $\mathbb{S}^1/\mathbb{H}^2$ model as well. This was first shown for the E-PSO model (a generalisation of the PSO model \cite{EPSO_HyperMap}) in Refs.\cite{commSector_hypEmbBasedOnComms_2016,commSector_linkPred} and for the $\mathbb{S}^1/\mathbb{H}^2$ model in Ref.\cite{commSector_commDetMethod}, along with the proposition of the "Community-Sector hypothesis", supposing that most members of a community gather in the same angular sector on the hyperbolic plane. In the closely related study of Ref.\cite{commSector_hypEmbBasedOnComms_2019}, the dependence of the modularity (a commonly used quality score for communities introduced in Ref.\cite{Newman_modularity_original}) on the temperature parameter $T\in[0,1)$ of the E-PSO model (controlling the clustering coefficient) for communities found by the Louvain method \cite{Louvain} was also studied to some extent. According to the results, the modularity can be even above $0.7$ when $T$ is low, and gradually decreases when $T$ is increased; however, can still stay above $0.3$ when $T$ approaches $1$. In parallel with these studies, in Ref.\cite{analogyBetweenHypEmbAndComms} the analogy between the hyperbolic embedding and the community structure was studied mostly for real networks and partly for synthetic graphs generated by the PSO model, where again, the PSO networks was observed to have a notable community structure, just like the real networks.

Even though the above results already provide important signs related to the presence of communities in hyperbolic networks with homogeneous angular node distribution, here we revisit this phenomenon in a detailed in-depth study, motivated by the following. First of all, in spite that a modularity value above $0.3$ can be a good community indicator in practice \cite{Clauset_Newman_Moore_coms}, it is important to note that a high modularity value alone is not always accompanied by a true modular structure, as e.g. Erd{\H o}s--R{\'e}nyi random graphs \cite{ER_model} or scale-free networks obtained with the Barab{\'a}si--Albert model \cite{BA_model} can also yield modularity values above $0.8$ under certain circumstances \cite{Guimera_ER_modul,Good_modul_max}. Thus, in order to have a truly solid claim about the presence of communities in random graph models without any explicit community formation mechanism, it is best to back up the large modularity values with further analysis of the supposed modular structure from multiple aspects.

Another task of high importance is the more detailed exploration of the parameter space. Apart from simple parameters such as the network size and the average degree, both the PSO model and the $\mathbb{S}^1/\mathbb{H}^2$ model have basically two parameters: one controlling the decay exponent $\gamma$ of the scale-free degree distribution and the other controlling the clustering coefficient. By analysing the effect of these parameters on the communities, we can gain a clear picture about what sort of modular structure can be expected when the aim is to generate a hyperbolic random graph with specified $\gamma$ and clustering coefficient values. 

Along this line, here we generate random graphs according to the PSO and the $\mathbb{S}^1/\mathbb{H}^2$ models in a wide range of parameter settings and examine their community structure with the help of three well-established community finding algorithms given by the Louvain method \cite{Louvain}, the Infomap algorithm \cite{Infomap} and asynchronous label propagation \cite{alabprop}. The Louvain approach is known to be a very efficient modularity maximising method, while the other two algorithms included do not build on the modularity and extract the modular structure of the studied networks based on different concepts. By applying independent community finding methods, the comparison between the found modules can reveal whether they correspond to strong, significant structures that can be located consistently in several different ways or not. In order to gain a quantitative comparison between the communities found by the different methods, we rely on the concept of the adjusted mutual information (AMI) \cite{AMI}, a well-known information-theoretic similarity measure. Besides the modularity, we also examine the angular separation index (ASI) of the communities \cite{Cannistraci_ASI} corresponding to a measure developed specifically for hyperbolic networks, characterising the angular mixing of the groups of nodes (communities) on the native disk. %developed specifically for hyperbolic networks where groups of nodes tend to gather in well-defined angular regions of the native disk. 

%The high modularity values of the obtained results indicate that the hyperbolic networks under study come with an inherent community structure for a considerable range of the model parameters.

The paper is organised as follows. In section~\ref{sect:Methods} we describe the PSO and $\mathbb{S}^1/\mathbb{H}^2$ models used for network generation, together with a short summary of the applied community finding methods and the quality measures used for evaluating the detected community structures. This is followed by the results in section~\ref{sect:Results}, whereas we discuss the implications of our findings in section~\ref{sect:Discuss}.

\section{Methods and preliminaries}
\label{sect:Methods}
We begin the description of the used methods with a brief introduction to hyperbolic network models in section~\ref{sect:Hyp_models}, including both the PSO model in section~\ref{sect:PSO} and the $\mathbb{S}^1/\mathbb{H}^2$ model in section~\ref{sect:S1H1}. The community related measures and algorithms are summarised in section~\ref{sect:coms}, starting with the concept of modularity in section~\ref{sect:modularity}, the angular separation index in section~\ref{sect:ASI} and the adjusted mutual information in section~\ref{sect:AMI}, followed by the description of the used community finding algorithms in sections~\ref{sect:alabelprop}–\ref{sect:Infomap}.

\subsection{Hyperbolic network models}
\label{sect:Hyp_models}

When studying the underlying hyperbolic geometry of complex networks, commonly the native representation  of the two-dimensional hyperbolic space is used \cite{hyperGeomBasics}, in which the hyperbolic plane of constant curvature $K<0$ is represented by a disk of radius $R$ in the Euclidean plane (for which $K=0$). In this representation the Euclidean angles between hyperbolic lines are equal to their hyperbolic values, and the radial coordinate $r$ of a point (defined as its Euclidean distance from the disk centre) is equal to its hyperbolic distance from the disk centre. The hyperbolic distance between two points is measured along their connecting hyperbolic line, which is either the arc of the Euclidean circle going through the given points and intersecting the disk's boundary perpendicularly or -- if the disk centre falls on the Euclidean line connecting the two points in question -- the corresponding diameter of the disk. The hyperbolic distance $x$ between two points at polar coordinates $(r,\theta)$ and $(r',\theta')$ fulfills the hyperbolic law of cosines written as
\begin{equation}
    \mathrm{cosh}(\zeta x)=\mathrm{cosh}(\zeta r)\,\mathrm{cosh}(\zeta r')-\mathrm{sinh}(\zeta r)\,\mathrm{sinh}(\zeta r')\,\mathrm{cos}(\Delta\theta),
    \label{eq:hypDist}
\end{equation}
where $\zeta=\sqrt{-K}$ and $\Delta\theta=\pi-|\pi-|\theta-\theta'||$ is the angle between the examined points. According to Ref. \cite{hyperGeomBasics}, for $2\cdot\sqrt{e^{-2\zeta r}+e^{-2\zeta r'}}<\Delta\theta$ and sufficiently large $\zeta r$ and $\zeta r'$, the hyperbolic distance can be approximated as
\begin{equation}
    x\approx r+r'+\frac{2}{\zeta}\cdot\ln\left(\frac{\Delta\theta}{2}\right).
    \label{eq:hypDistApprox}
\end{equation}

\subsubsection{The PSO model for network generation}
\label{sect:PSO}
In the popularity-similarity optimisation model, nodes are placed one by one in the above described native disk representation of the hyperbolic plane and connected with probabilities depending on the hyperbolic distance. The parameters of the model can be listed as follows:
\begin{itemize}
	\item The curvature $K<0$ of the hyperbolic plane, controlled by $\zeta=\sqrt{-K}>0$. Changing the value of $\zeta$ corresponds to a simple rescaling of the hyperbolic distances; the usual custom is to set the value of $\zeta$ to $1$ (i.e. set $K$ to $-1$).
	\item The final number of nodes $N\in\mathbb{Z}^+$ in the network.
	\item The number of connections $m\in\mathbb{Z}^+$ established by the newly appearing nodes, corresponding to the half of the average degree $\left< k\right>$. (The first $m$ nodes of the network form a complete graph).
    \item The popularity fading parameter $\beta\in(0,1]$, controlling the outward drift of the nodes on the native disk. %The probability density of the node radial coordinate at time $1\leq t\leq N$ is $\rho_{\mathrm{radial}}(r)=\frac{\zeta}{2\beta}e^{\frac{\zeta}{2\beta}\big(r-\frac{2}{\zeta}\mathrm{ln}(t)\big)}$ \cite{EPSO_HyperMap}.
    The exponent $\gamma$ of the power-law decaying tail of the degree distribution is related to the popularity fading parameter as $\gamma=1+1/\beta$.
    \item The temperature $T\in[0,1)$, controlling the average clustering of the network, where lower temperature results in a higher average clustering coefficient.
    %\item $\rho_{\mathrm{angular}}(\theta)$, the probability density, according to which the angular coordinates of the nodes are sampled. In the original (uniform) PSO model $\rho_{\mathrm{angular}}(\theta)\equiv1/2\pi$.
\end{itemize}
During the random graph generation process, initially the network is empty, and at each time step $i=1,2,...,N$ a new node joins the network as follows:
\begin{enumerate}
	\item The new node $i$ appears at polar coordinates $(r_{ii},\theta_i)$, where the radial coordinate $r_{ii}$ is set to $\frac{2}{\zeta}\mathrm{ln}(i)$ and the angular coordinate $\theta_i$ is sampled from $[0,2\pi)$ uniformly at random.
    \item The radial coordinate of each previously (at time $j<i$) appeared node $j$ is increased according to the formula $r_{ji}=\beta r_{jj}+(1-\beta)r_{ii}$ in order to simulate popularity fading.
    \item The new node $i$ establishes connections with previously appeared nodes. Only single links are permitted.
    \begin{enumerate}
        \item If the number of previously appeared nodes is not larger than $m$, node $i$ connects to all of them.
        \item Otherwise, the new node $i$ connects to $m$ of the previously appeared nodes, where the connection probabilities are determined by the hyperbolic distances between the node pairs, which can be calculated based on equation~(\ref{eq:hypDist}). If $T=0$, node $i$ simply connects to the $m$ hyperbolically closest nodes, whereas at temperatures $T>0$, any previous node $j=1,2,...,i-1$ gets connected to node $i$ with probability
            \begin{equation}
            p(x_{ij})=\frac{1}{1+e^{\frac{\zeta}{2T}(x_{ij}-R_i)}},
            \label{eq:PSO_con_prob}
            \end{equation}
            where the cutoff distance $R_i$ is set to
            \begin{equation}
             R_i =\left\lbrace \begin{array}{ll}
             r_{ii}-\frac{2}{\zeta}\mathrm{ln}\left(\frac{2T}{\mathrm{sin}(T\pi)}\cdot\frac{1-e^{-\frac{\zeta}{2}(1-\beta)r_{ii}}}{m(1-\beta)}\right) & \mbox{if}\; \beta<1, \\
             r_{ii}-\frac{2}{\zeta}\mathrm{ln}\left(\frac{T}{\mathrm{sin}(T\pi)}\cdot\frac{\zeta r_{ii}}{m}\right) & \mbox{if}\; \beta=1,
             \end{array} \right.    
            \end{equation}
            ensuring that the expected number of nodes connecting to the new node $i$ at its arrival is equal to $m$.
    \end{enumerate}
\end{enumerate}

\subsubsection{The $\mathbb{S}^1/\mathbb{H}^2$ model for network generation}
\label{sect:S1H1}
In the $\mathbb{S}^1$ model \cite{S1}, first the $N$ number of nodes are placed on a one-dimensional sphere (i.e. a circle) and each is given a hidden variable $\kappa_i\in[\kappa_0,\infty),\,i=1,2,...,N$. Then, each pair of nodes becomes connected with a probability taking into account both the angular distance and the hidden variables. In the below described algorithm \cite{S1H2_Mercator,S1_code}, $\kappa_i$ corresponds to the expected degree $\bar{k}_i$ of node $i$ in the thermodynamic limit. Thus, the connection rule can be phrased in a simple, intuitive way, namely the nodes that are closer in the hidden metric space underlying the network are more likely to be connected, but in the meantime nodes with higher degree obtain farther-reaching connections as well. In the equivalent $\mathbb{H}^2$ model \cite{S1H2_Mercator}, the hidden variable $\kappa_i$ is converted into the radial coordinate $r_i$ in the native representation of the hyperbolic plane, and the connection probability depends on the hyperbolic distance between the nodes, that expresses the effect of both the similarity and the node degrees (the popularity). 

The parameters of these models can be listed as follows:
\begin{itemize}
\item The total number of nodes $N$. 
\item The average degree $\left< k\right>$.
\item The exponent $2<\gamma$ of the tail of the degree distribution following a power law of the form $P(k)\sim k^{-\gamma}$. 
\item The parameter $1<\alpha$, controlling the average clustering coefficient $\left< c\right>$ of the generated network ($\lim\limits_{\alpha\to1}\left< c\right> =0$).
\end{itemize}
%Note that in the E-PSO model the same exponent $\gamma$ can be expressed with the popularity fading parameter $\beta$ as $\gamma=1+1/\beta$, and the temperature $T$ plays the same role as $1/\alpha$.

In the $\mathbb{S}^1$ model, a network of $N$ number of nodes -- each of them indexed by $i\in[1,N]$ -- is generated through the following steps:
\begin{enumerate}
    \item For each node $i$ an angular coordinate $\theta_i$ is sampled from the interval $[0,2\pi)$ uniformly at random.
    \item For each node $i$ a hidden variable $\kappa_i$ is sampled from the interval $[\kappa_0,\infty)$ according to the distribution $\rho(\kappa)=(\gamma-1)\cdot\frac{\kappa^{-\gamma}}{\kappa_0^{1-\gamma}}$, where $\kappa_0=\frac{\gamma-2}{\gamma-1}\cdot\left< k\right>$.
    \item Each pair of nodes $i-j$ is connected with probability
    \begin{equation}
    p_{ij}=\frac{1}{1+\left(\frac{N\cdot\Delta\theta_{ij}}{2\pi\cdot\mu\cdot\kappa_i\cdot\kappa_j}\right)^{\alpha}},
    \label{eq:S1_con_prob}
    \end{equation}
    where $\Delta\theta_{ij}=\pi-|\pi-|\theta_i-\theta_j||$ is the angular distance between the nodes, and $\mu=\frac{\alpha}{2\pi\left< k\right>}\cdot\sin\left(\frac{\pi}{\alpha}\right)$.
\end{enumerate}

To facilitate a straightforward comparison with the PSO model, we converted the hidden variable associated to the nodes into a radial coordinate in the native representation of the hyperbolic plane (at $K=-1$ curvature) as 
\begin{equation}
r_i=\hat{R}-2\ln\left(\frac{\kappa_i}{\kappa_0}\right),
\end{equation}
where $\hat{R}=2\ln\left(\frac{N}{\mu\pi\kappa_0^2}\right)$. Note that using this hyperbolic representation (i.e. the $\mathbb{H}^2$ model) the connection probability (\ref{eq:S1_con_prob}) becomes $p_{ij}=\left[1+e^{\frac{\alpha}{2}\cdot(x_{ij}-\hat{R})}\right]^{-1}$, depending on the hyperbolic distance $x_{ij}$ in the same way as the connection probability in equation~(\ref{eq:PSO_con_prob}).

\subsection{Finding and evaluating communities}
\label{sect:coms}

Communities (also referred to as modules, cohesive groups, clusters) are frequently occurring structural units in complex networks having usually a larger internal and a smaller external link density, lacking however a widely accepted unique definition. Finding, evaluating and comparing communities are all non-trivial problems, with a vast number of different solutions suggested in the literature \cite{Fortunato_coms,Fortunato_Hric_coms,Cherifi_coms}. Here we first describe the concept of modularity in section~\ref{sect:modularity}, corresponding to the most widely used measure for quantifying the quality of communities. This is followed by the angular separation index in section~\ref{sect:ASI}, providing a score specific for hyperbolic networks, measuring the angular intermixing between communities in the hyperbolic disk, and the adjusted mutual information in section~\ref{sect:AMI}, allowing the quantitative comparison between community partitions found by different methods. In our studies, we have picked three well-grounded, commonly used methods for detecting communities, namely the asynchronous label propagation, detailed in section~\ref{sect:alabelprop}, the Louvain algorithm, described in section~\ref{sect:Louvain}, and Infomap, summarised in section~\ref{sect:Infomap}.

\subsubsection{Modularity}
\label{sect:modularity}
Probably the most well-known quality measure for communities is given by the modularity \cite{Newman_modularity_original}, comparing the observed density of links between the members of the same community with the expected link density based on some random null model, written in general as
\begin{equation}
    Q = \frac{1}{2L}\sum_{i=1}^N\sum_{j=1}^N\left[A_{ij} -P_{ij}\right]\delta_{c_i,c_j},
\end{equation}
where $N$ is the number of nodes in the network, $A_{ij}$ denotes an element the adjacency matrix ($A_{ij}\equiv A_{ji}=1$ if $i$ is connected to $j$ and otherwise $A_{ij}\equiv A_{ji}=0$), $P_{ij}$ gives the connection probability between nodes $i$ and $j$ in the null model, $L$ stands for the total number of links in the network, $c_i$ is the community to which node $i$ belongs and the Kronecker delta $\delta_{c_i,c_j}$ ensures that non-zero contribution can come only from node pairs in the same community. This quality measure can take values in the $Q\in[-1/2,1]$ interval, where larger values of $Q$ indicate stronger communities that have a significantly larger internal link density compared to the random expectation.

In practice, a natural choice for the null model is provided by the configuration model, where the connection probability between nodes $i$ and $j$ can be given with the node degrees $k_i$ and $k_j$ simply as $P_{ij}=\frac{k_ik_j}{2L}$. This form has also been extended to weighted networks \cite{weightedModularityDef}, where the number of links $L$ is replaced by $M=\frac{1}{2}\cdot\sum\limits_{i=1}^N\sum\limits_{j=1}^N w_{ij}$ (with $w_{ij}$ denoting the link weight between nodes $i$ and $j$), and the node degrees are replaced by the node strengths defined e.g. for node $i$ as $s_i = \sum_{\ell=1}^N w_{i\ell}$, resulting in
\begin{equation}
    Q=\frac{1}{2M}\cdot\sum\limits_{i=1}^N\sum\limits_{j=1}^N \left[w_{ij}-\frac{s_is_j}{2M}\right]\delta_{c_i,c_j}.
    \label{eq:modularityDef}
\end{equation}

In order to take into account the hyperbolic distances along the links, we adopted the practice suggested in Ref.\cite{linkWeights_coalescentEmbedding}, and used in our community analysis a link weight defined as 
\begin{equation}
    w_{ij}\equiv w_{ji}=\frac{1}{1+x_{ij}}
    \label{eq:linkWeights}
\end{equation}
for adjacent nodes $i$ and $j$, where the hyperbolic distance $x_{ij}$ was calculated based on equation~(\ref{eq:hypDist}) using $\zeta=1$. In our analysis, we used the code available from Ref.\cite{modularity_code} for calculating the weighted modularity of the detected community structures.

%We measured the quality/strength of the detected community structures by the modularity, which compares the observed fraction of links within the communities with its expected value after the/a degree-preserving randomization/randomisation of the network (kéne hivatkozni az eredeti modularitás definíciót, aminek még nem súlyozott hálókra volt?). For undirected, weighted networks(,) the modularity $Q\in[-1,1]$ is defined by equation~(\ref{eq:modularityDef}) \cite{weightedModularityDef}, where $w_{ij}$ denotes the weight of the edge between node $i$ and node $j$, $k_i^{\mathrm{w}}=\sum\limits_{\ell=1}^N w_{i\ell}$ is the weighted degree of node $i$,  $c_i$ stands for the community to which node $i$ belongs and $\delta_{c_i,c_j}$ is the Kronecker delta. Larger values of $Q$ indicates stronger community structures. We used the link weights given in equation~(\ref{eq:linkWeights}) \cite{linkWeights_coalescentEmbedding}, where the hyperbolic distance $x_{ij}$ between the network nodes $i$ and $j$ was calculated based on equation~(\ref{eq:hypDist}) using $\zeta=1$. (Ugyanezzel az élsúlyozással segítettük mindhárom vizsgált csoportkereső algoritmus munkáját, de ez mind3 csoportkeresőnél le van írva.) A modularitásszámolásra használt program: \cite{modularity_code}.

\subsubsection{Angular separation index}
\label{sect:ASI}
In networks embedded into the hyperbolic disk, communities usually occupy well-defined angular regions, having little or no overlap with the region of the other communities \cite{commSector_hypEmbBasedOnComms_2016,commSector_commDetMethod,commSector_linkPred,commSector_hypEmbBasedOnComms_2019,Boguna_hyp_embed_coms,Cannistraci_ASI}. A quantitative score characterising this tendency is given by the angular separation index (ASI) \cite{Cannistraci_ASI}. Its basic idea is to compare the number of "mistakes" in the angular arrangement -- i.e. the number $o_i$ of nodes belonging to other communities falling between the boundaries of the given module $i$ -- summed over all the $C$ communities of the network with the highest total number of mistakes obtained with the same clustering of the nodes when the angular coordinates are shuffled at random. Formally, the ASI can be expressed as
\begin{equation}
    \mathrm{ASI} = 1-\frac{\sum\limits_{i=1}^{C}o_i}{\max\limits_{r}\left(\sum\limits_{i=1}^{C}o_i^{(r)}\right) },
\end{equation}
where the maximisation in the denominator is over a fixed number of random shuffles (we used 1000 shuffles, i.e. $r=1,2,...,1000$, as suggested in Ref.\cite{Cannistraci_ASI}). Accordingly, an ASI value close to $1$ indicates well-separated clusters with a low intermixing in the angular coordinates of the members, and an ASI value close to $0$ is obtained when the angular arrangement of the members of different clusters is random.

\subsubsection{Adjusted mutual information}
\label{sect:AMI}
In the field of community detection, together with the rapid increase in the number of different algorithms proposed, came the need for well-grounded methods for comparing the results of the different approaches. Since e.g. the number of found communities and the sizes of the modules can show large variations across the different methods, judging the extent of similarity between two community partitions is non-trivial. Given two sets of communities $A$ and $B$ over the same network, hosting $C_A$ and $C_B$ number of communities each, a well-known information theoretic similarity measure is offered by the normalised mutual information (NMI) \cite{Danon_mutinfo,Lancichinetti_mutinfo}, that can be defined based on the mutual information 
\begin{equation}
    \mathrm{MI}(A,B) = -\sum_{i=1}^{C_A}\sum_{j=1}^{C_B}\frac{N_{ij}}{N}\ln\left(\frac{N_{ij}N}{N_iN_j}\right)
\end{equation}
and the entropies 
\begin{equation}
    H(A) = -\sum_{i=1}^{C_A}\frac{N_i}{N}\ln\left(\frac{N_i}{N}\right), \;\;\; H(B) = -\sum_{j=1}^{C_B}\frac{N_j}{N}\ln\left(\frac{N_j}{N}\right),
\end{equation}
where $N_{ij}$ denotes the number of shared members of communities $i$ and $j$, $N_i$ and $N_j$ stand for the number of nodes in the individual communities, and the total number of nodes in the network is given by $N$. There are several different possibilities for normalising the mutual information $\mathrm{MI}(A,B)$, e.g. we can divide it by the maximum, the arithmetic mean or the geometric mean of the entropies $H(A)$ and $H(B)$ \cite{AMI}. In the present study we used the maximum of the entropies; thus, throughout the paper 
\begin{equation}
    \mathrm{NMI}(A,B)\equiv \frac{ \mathrm{MI}(A,B)}{\max\left[H(A),H(B)\right]}.
\end{equation}
%where $I(A,B)$ is corresponding to the mutual information between $A$ and $B$, whereas $H(A)$ and $H(B)$ denote the information theoretic entropy of $A$ and $B$, respectively. In terms of the number of communities $C_A$ and $C_B$, the number of shared members $N_{ij}$ and the number of nodes in the individual communities given by $N_i$ and $N_j$, the NMI can be given as
%\begin{equation}
% \mathrm{NMI}(A,B) = \frac{-2\sum_{i=1}^{C_A}\sum_{j=1}^{C_B}N_{ij}\ln\left(\frac{N_{ij}N}{N_iN_j}\right)}{\sum_{i=1}^{C_A}N_i\ln\left(\frac{N_i}{N}\right)+\sum_{j=1}N_j\ln\left(\frac{N_j}{N}\right)}.   
% \label{eq:NMI}
%\end{equation}
%where $N_i$ and $N_j$ stand for the number of nodes in communities $i$ and $j$ respectively, $N$is the total number of nodes in the network and $N_{ij}$ denotes the number of common nodes between $i$ and $j$. 
This quantity becomes 1 if and only if the partitions $A$ and $B$ are identical, otherwise its value is lower than 1. 

The concept of adjusted mutual information (AMI) supplements this consistent upper bound with a consistent zero expectation corresponding to the similarity we can expect by random chance \cite{AMI,McCarthy_AMI}. To achieve this, the average mutual information of random partitions $A'$ and $B'$ is subtracted from the nominator, and the average maximum entropy of random partitions is subtracted from the denominator yielding
\begin{equation}
    \mathrm{AMI}(A,B)=\frac{\mathrm{MI}(A,B)-\left<\mathrm{MI}(A',B')\right>_{\rm rand}}{\max\left[H(A),H(B)\right]-\left< \max\left[H(A'),H(B')\right]\right>_{\rm rand}}.
    \label{eq:AMI}
\end{equation}
In our analysis, we used the code available from Ref.\cite{AMI_code} for calculating the AMI between the found community partitions.

\subsubsection{Asynchronous label propagation algorithm for community detection}
\label{sect:alabelprop}

The asynchronous label propagation algorithm \cite{alabprop} simulates the diffusion of labels along the links in the examined network, where the nodes are labelled by the identifier of the community to which they belong, and these labels are regularly updated based on the labels of the neighbouring nodes using a majority rule. The idea behind this method is that as the labels propagate, the densely connected groups of nodes will reach a consensus on a unique label. This approach is %uses solely the network structure to guide the community detection and does 
not aimed at optimising any predefined measure or function.

Initially, a unique community label is assigned to each node in the network. Afterwards, the following asynchronous update process is repeated until every node in the network has at least as many neighbours within its own community as it has in any other communities:
%Then the algorithm repeats the following asynchronous label updating process until every node in the network has at least as many neighbours within its own community as it has in any other community:/.
\begin{enumerate}
    \item Nodes are arranged in a random order.
    %Arrange the network nodes in a random order. (Minden lépésben újrasorsolva, ezért külön pont!)
    \item According to this order, we iterate over the nodes and update their label one by one based on their neighbours: each node joins the community to which most of its neighbours currently belong. Note that the label of the neighbours may have already been updated in the given iteration. The neighbouring labels are weighted based on the strength of their link connected to the current node, and 
%    a could have been updated azt jelentené, hogy frissítve lehettek volna, de nem lettek
%    Iterate over the nodes in the given order and update their label one by one based on the current -- either already updated in the current iteration or not -- labels of their neighbours: a node always joins the community to which most of its neighbours currently belong. 
    ties in the weighted number of neighbours are broken at random. 
\end{enumerate}
Due to the random propagation of the labels, in this approach it is possible that distinct communities may eventually settle to the same label. Therefore, after the termination of the above algorithm (where we used the code available from Ref.\cite{alabprop_code} with link weights calculated from equation~(\ref{eq:linkWeights})), we also applied a breadth-first search on the subgraphs of each individual community to separate the disconnected (i.e. connected only via nodes of different communities in the original network) groups of nodes having the same label, as suggested in Ref.\cite{alabprop}.

%In our investigations, we weighted the examined synthetic networks based on equation~(\ref{eq:linkWeights}) and complemented/completed the asynchronous label propagation algorithm \cite{alabprop_code} with a breadth-first search on the subnetworks of each individual community to separate the disconnected (i.e. connected only via nodes of different communities in the original/whole network) groups of nodes having the same label \cite{alabprop}. We performed the community detection with asynchronous label propagation once for each network.

\subsubsection{Louvain algorithm for community detection}

\label{sect:Louvain}
Though finding the exact maximum of modularity is a computationally hard problem \cite{npmodularity}, over the years several heuristic modularity optimisation methods were proposed \cite{Fortunato_coms,Fortunato_Hric_coms}, and one of the most popular among these is the Louvain algorithm \cite{Louvain}. This approach is capable of unfolding a complete hierarchical community structure (where modules can be composed of submodules) within a relatively short time even for extremely large networks. The algorithm is repeating two phases iteratively until the modularity stops improving: %(a használt programnál: ha újModularitás - eddigiModularitás $<$ MINérték, akkor befejezzük a két fázis ismételgetését; MINérték=0.0000001):
\begin{enumerate}
    \item Searching for a local maximum in the modularity at the given organisation level of the network.
    \begin{itemize}
        \item First, a unique community is assigned to each node of the current network.
        \item This is followed by a repeated iteration over the nodes until the modularity does not increase any further (or, in our case, until the gain in the modularity does not decrease below a threshold of $\Delta Q_{\rm min}=10^{-7}$). %(a használt programnál: ha újModularitás - eddigiModularitás $<$ MINérték, akkor kilépés a pontokon iterálásból; MINérték=0.0000001)).
        \begin{itemize}
            \item We evaluate the changes in the modularity that would take place if the current node $i$ was transferred to the community of each of its neighbours. %Evaluate the modularity change that would take place by transferring the given node $i$ from its current community to the community of each of its neighbouring nodes.
            \item If all the calculated modularity changes are negative, node $i$ stays in its current community. %Otherwise, node $i$ is transferred to the community with the largest improvement in the modularity.
            Otherwise, we carry out the transfer of node $i$  where the improvement in the modularity is the largest.
        \end{itemize}
    \end{itemize}
    \item Moving up to the next organisation level of the system represented by the network between the just found communities:
%    Community aggregation: Move(?) to the upper next organisation level by creating a network of the communities found in phase 1.
    \begin{itemize}
        \item Each community is considered as a single node.
        \item A self-loop is created for each new node, weighted by twice the sum of the link weights within the corresponding community.
        \item The new nodes are connected by links weighted by the sum of the link weights between the corresponding community members on the previous organisation level. %(The links between members of the same community appear as self-connections on the nodes on the next level). 
        %Connect the new nodes with links weighted by the sum of the weight of the links connecting to each other the members of the corresponding two communities at/on(?) the previous organisation level.
    \end{itemize}
\end{enumerate}
In our investigations, we weighted the links in the examined hyperbolic networks according to equation~(\ref{eq:linkWeights}) and considered only the final partition (i.e. the top-level community structure, having the highest modularity among the different organisation levels) found by the implementation of the algorithm available from Ref. \cite{Louvain_code}.  %corresponding to the top level of the network's hierarchical community structure, 
%characterised by the highest modularity among the different organisation levels. %We performed the community detection with Louvain once for each network.

\subsubsection{Infomap algorithm for community detection}
\label{sect:Infomap}
The Infomap algorithm, as suggested by its name, provides an information-theoretic approach for finding communities in networks \cite{Infomap} based on a correspondence between the optimal community structure and the most parsimonious description of an infinitely long random walk trajectory on the network. The random walk can be considered as a proxy for the flow in the network (travelling passengers, spreading ideas, etc.), making its components interdependent to varying extents. It is intuitive to assume that communities correspond to localized regions of the network where random walkers spend a lot of time. We can take advantage of this property of communities when aiming for the most compact description of a random walker trajectory as follows. 
%Thus, communities play a distinguished role in determining(?) the flow on the network, which can be utilized for compressing the description of a random walker trajectory.

In a simple approach, the trajectory is corresponding to the sequence of the visited nodes, each labelled with a unique codeword. However, trajectories can be defined more concisely by using a map-like description following the principle of geographic maps, where e.g. the same street names appear in multiple cities. In a similar manner, after naming the communities, the code words of the nodes can be recycled among the different communities, and only the members of the same community have to be given unique names.
%After naming the communities, "vertex codewords can be recycled among the different communities" and used for multiple vertices ("vertices with identical name can be distinguished by specifying the community they belong to"), and only the vertices of the same community have to be given unique names. 
By limiting the number of different code words used to denote the nodes, the length of these code words can be reduced, leading to a considerable saving in the length of the trajectory description. Naturally, the recycling of the code words also comes at a cost, namely one has to indicate when the random walker leaves a given community to enter a new one by specifying the code word of the new community. Nevertheless, if communities are well separated from each other, then the transition between communities is not frequent, and we gain in the length of the trajectory description even with this extra cost taken into account.
%("once the random walker enters a community, it tends to stay there for a long time") and the "codewords of the communities will not be repeated many times" in the description of a random walker trajectory.

For a map-like trajectory description based on a given community structure, the efficiency can be evaluated by the so-called map equation \cite{Infomap}, expressing the optimal code length (i.e. the theoretical lower bound of the code length) for an average movement of an infinitely long random walk. The Infomap algorithm itself searches for the multi-level, hierarchical network partition minimising the map equation in a heuristic manner, splitting modules into submodules, subsubmodules and so on in order to reduce the description length. If the splitting of a given leaf in the community hierarchy does not decrease the description length anymore, the downward growth of the given branch in the hierarchy is stopped. 
%If adding submodules does not decrease the description length anymore, the given branch of the hierarchy does not grow further downwards.
%method searches for that multi-level, hierarchical network partition which yields the shortest description length, i.e. (which) minimises the hierarchical map equation. This algorithm splits modules (in)to submodules, subsubmodules and so on in order to reduce the description length. If adding submodules does not decrease the description length anymore, the given branch of the hierarchy does not grow further downwards.
In our community analysis, we used the code available from Ref.\cite{Infomap_code} with link weights calculated according to equation~(\ref{eq:linkWeights}) and queried from the output of the algorithm the communities corresponding to the leaves of the community hierarchy.

%we inputted the generated networks to/into the Infomap algorithm \cite{Infomap_code} after the link weighting given in equation~(\ref{eq:linkWeights}) and used/queried the leaf modules of the resulted hierarchy (...mert itt ilyen kód van: https://mapequation.github.io/infomap/python/). We performed the community detection with Infomap once for each network.

%[Leaf modulokat kérdeztünk le (1 leaf=1 csoport). Feltehetően az algoritmus addig szabdal szét csoportokat, amíg ezáltal a kódhossz rövidíthető, tehát a leaf modulok az optimális kódhosszt adják? - Emiatt értelmes éppen ezeket a modulokat használni?
%NEM, mert a leaf modulokig tartó TELJES elérési utak adják meg az optimális leírást... (ha csak a leaf modul indexeket és melléjük pontindexeket használunk, azt nem tudni milyen leírást ad, ilyet nem is néz a program, csak hogy az az optimális, ha a leaf modulokig minden szintű csoportindexet és a pontok indexeit használjuk)]

\section{Results} %lehetnek egyedi fejezet és alfejezetcímek is, nem kötelező ezeket használni
\label{sect:Results}

We generated random graphs using the PSO and the $\mathbb{S}^1/\mathbb{H}^2$ models in a wide range of parameter settings, and used the obtained networks as inputs for the community finding methods given by the asynchronous label propagation, the Louvain and the Infomap algorithms. According to the results, the hyperbolic random graphs seemed to possess a strong community structure for quite a few combinations of the network generation parameters.

\begin{figure}
    \centering
    \captionsetup{width=\textwidth}
    \includegraphics[width=\textwidth]{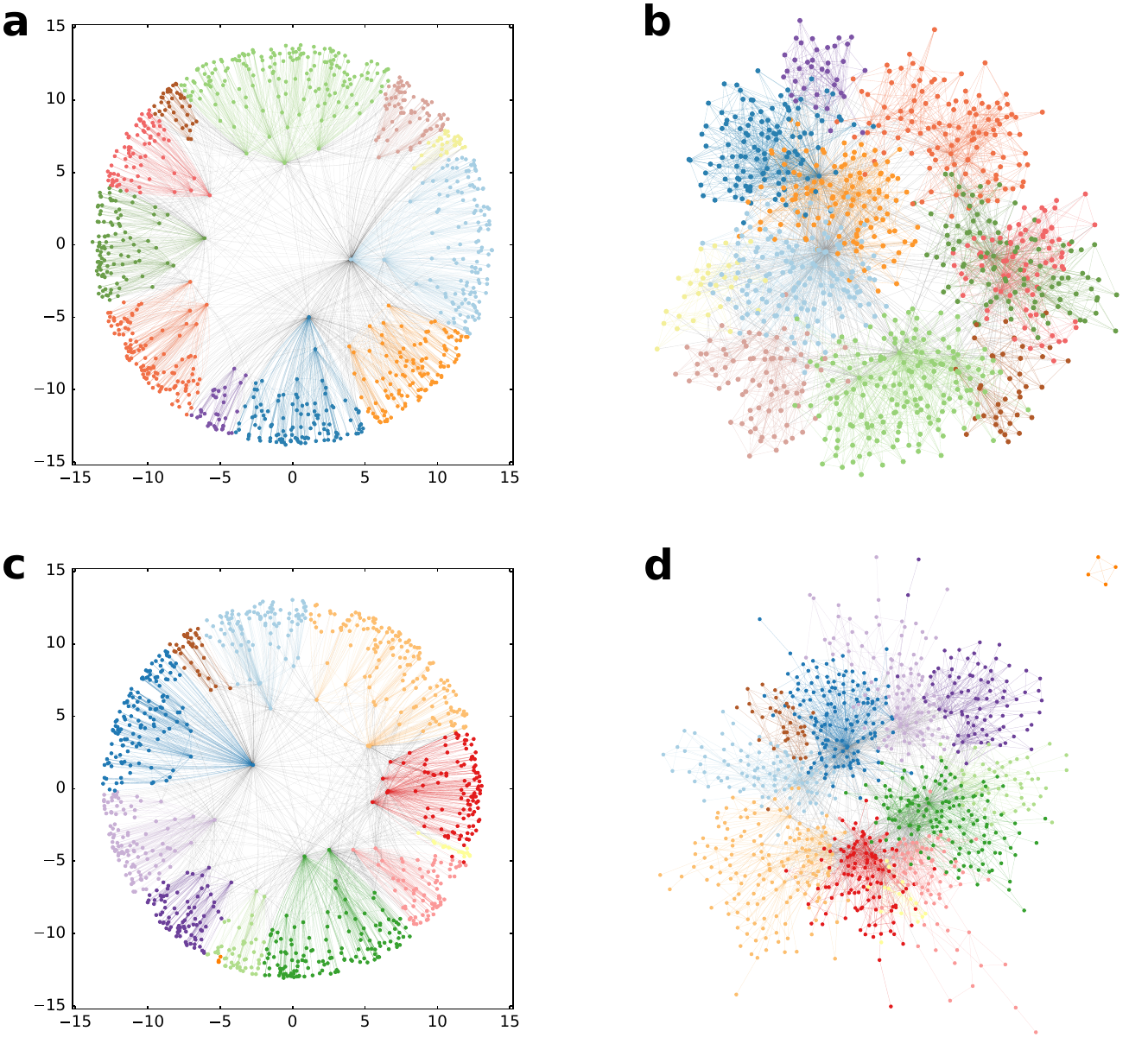}
    \caption{{\bf Communities found by the Louvain algorithm in hyperbolic networks.} (a) The obtained communities (colour coded) in a network with $N=1000$ number of nodes, generated by the PSO model with parameters $m=5$ (corresponding to $\left<  k\right> = 10$), $\beta = 0.7$ (corresponding to $\gamma=2.43$) and $T=0.2$ (resulting in an average clustering coefficient of 0.58). The layout shows the network in the native disk representation of the two-dimensional hyperbolic space of curvature $K=-1$, with the nodes arranged according to their coordinates assigned during the network generation process. The weighted modularity for the found partition is $Q=0.75$ and the angular separation index is $\mathrm{ASI}=1.0$. (b) Layout of the network shown in panel (a) on the Euclidean plane. (c) The detected communities in a network generated by the $\mathbb{S}^1/\mathbb{H}^2$ model with parameters $N=1000$, $\left< k\right>=10$, $\gamma=2.43$ and $\alpha=5$ (resulting in an average clustering coefficient of 0.71), shown in the native disk representation of the hyperbolic plane of curvature $K=-1$. The weighted modularity of the shown partition is $Q=0.74$, the angular separation index is $\mathrm{ASI}=0.998$. (d) Layout of the same network as in panel (c) on the Euclidean plane.
    %A különböző csoportokba tartozó pontok közti élek szürkék. Mindkét hálónál a várt átlagfokszám 10 és a várt gamma=2.43. A PSO-nál T=0.2, az S1-nél alpha=5(=1/0.2), amik elvileg nagy klaszterezettséghez vezetnek -> az ábrázolt PSO-s háló átlagos klaszterezettsége 0.5834698, az S1-esé 0.70923284. A súlyozott modularitás a PSO-s hálónál 0.747286, az S1-esnél 0.737985.
    }
    \label{fig:coms_illustrate}
\end{figure}

As an illustration, in figure~\ref{fig:coms_illustrate} we show the partition found by the Louvain algorithm in networks of size $N=1000$ both according to the layout in the native disk representation of the two-dimensional hyperbolic space and according to a standard layout in the Euclidean plane. In figures~\ref{fig:coms_illustrate}(a) and \ref{fig:coms_illustrate}(c), the sets of nodes grouped together by Louvain occupy well-defined angular regions in the hyperbolic disk with barely any overlap with the region of the neighbouring communities. However, according to figures~\ref{fig:coms_illustrate}(b) and \ref{fig:coms_illustrate}(d), the detected communities are clearly outlined even in such layouts which do not build on the hyperbolic origin of the networks.

\begin{figure}
    \centering
    \captionsetup{width=0.85\textwidth}
    \includegraphics[width=0.85\textwidth]{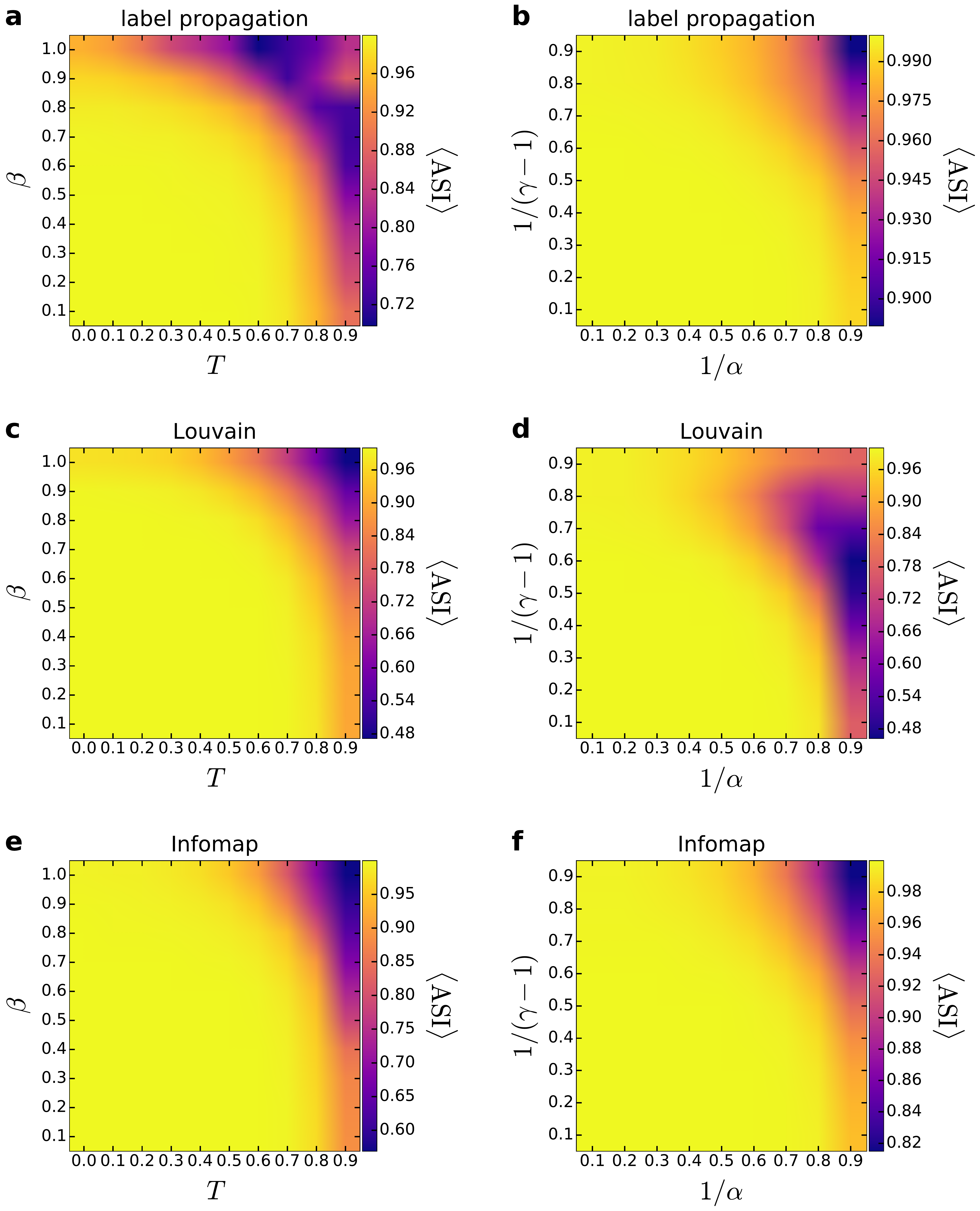}
    \caption[width=0.85\textwidth]{ {\bf Angular separation index in the PSO and the $\mathbb{S}^1/\mathbb{H}^2$ models.} The results for the PSO model are given in the left column (panels (a), (c) and (e)), whereas the ASI obtained for the $\mathbb{S}^1/\mathbb{H}^2$ model appears in the right column (panels (b), (d) and (f)). The ASI for the communities detected by asynchronous label propagation is given in the top row (panels (a) and (b)), the ASI regarding the results of Louvain is shown in the middle row (panels (c) and (d)) and the ASI for the partitions found by Infomap is presented in the bottom row (panels (e) and (f)). We show the measured ASI (indicated by the color, averaged over 100 samples) as a function of the model parameters $T$ and $\beta$, or $1/\alpha$ and $1/(\gamma -1)$ for networks of size $N=10,000$ and expected average degree $\left< k\right> =10$.}
    \label{fig:ASI}
\end{figure}

%annak érdekében, hogy megmutassuk, hogy az 1. ábra nem egy speciális eset, hanem tipkus, plottoljuk az ASI-t a paraméterk függvényében a 2.ábrán
We found that the angular separation of the detected modules exemplified by figures~\ref{fig:coms_illustrate}(a) and \ref{fig:coms_illustrate}(c) is quite general in the hyperbolic disk. Using the angular separation index (ASI) described in section~\ref{sect:ASI}, we evaluated quantitatively the angular separation of the modules obtained with the asynchronous label propagation, the Louvain and the Infomap algorithms for a large variety of the network generation parameters. In the case of the PSO model, for both the temperature $T$ and the popularity fading parameter $\beta$ we took 10 equidistant data points between 0 and 1 (altogether 100 parameter combinations in the $T-\beta$ parameter plane) and generated 100 networks with each parameter setting. In the case of the $\mathbb{S}^1/\mathbb{H}^2$ model, to allow a straightforward comparison with the results seen for the PSO model, instead of the original model parameters $\alpha$ and $\gamma$ we changed to $1/\alpha$ (analogous to the temperature $T$ in the PSO model) and $1/(\gamma-1)$ (equivalent to the popularity fading parameter $\beta$ in the PSO model). Similarly to the studies of the PSO model, we considered a $9\times9$ grid in the $1/\alpha-1/(\gamma-1)$ parameter plane (in the $\mathbb{S}^1/\mathbb{H}^2$ model $2<\gamma$ and $\alpha$ is finite; hence, this model is not defined for $\beta=1$ and $T=0$), and generated 100 networks for each parameter combination. As it is shown in figure~\ref{fig:ASI}, for PSO and $\mathbb{S}^1/\mathbb{H}^2$ networks of size $N=10,000$ and average degree $\left< k\right>=10$ a considerably high ASI can be obtained with all three community finding methods for most of the $T-\beta$ and $\alpha-\gamma$ parameter settings. 
%The widespread appearance of high ASI values reinforces that the clear angular separation of the detected communities is typical for these hyperbolic network models.
%Using the angular separation index (ASI), we evaluated quantitatively the angular separation of the modules in figures~\ref{fig:coms_illustrate}(a) and \ref{fig:coms_illustrate}(c): for the PSO network $\mathrm{ASI}=...$ and for the $\mathbb{S}^1/\mathbb{H}^2$ network $\mathrm{ASI}=...$. As it is shown in figure~\ref{fig:ASI}, for PSO and $\mathbb{S}^1/\mathbb{H}^2$ networks of size $N=10,000$ and average degree $\left< k\right>=10$ a considerably high ASI can be obtained with all three community finding methods at most of the $\beta-T$ and $\gamma-\alpha$ parameter settings. Based on this, the clear angular separation of the detected modules seems to be typical for these hyperbolic network models.
%Szokás azt gondolni, hogy a hiperb.térben a csoportok szög szerint szétválnak - jellemzően azt tapasztaltuk, hogy a csoportkeresők valamiféle szög szerinti struktúrát találnak meg. Ennek igazolására ASI-t mértünk.

\begin{figure}
    \centering
    \captionsetup{width=0.85\textwidth}
    \includegraphics[width=0.85\textwidth]{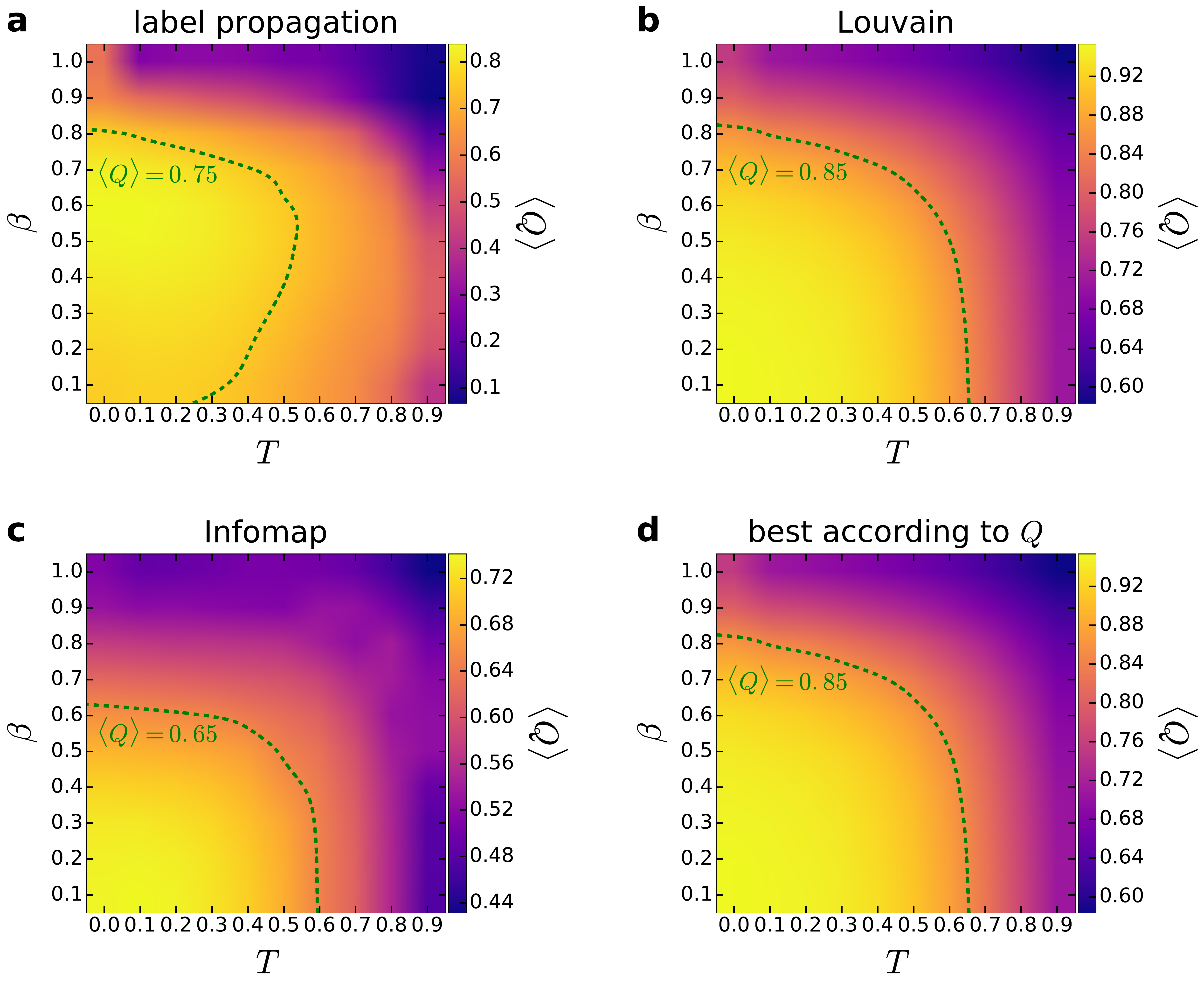}
    \caption{ {\bf Modularity in the PSO model}. We show the weighted modularity $Q$ (indicated by the color, averaged over 100 samples) as a function of the model parameters $T$ and $\beta$ for networks of size $N=10,000$ and expected average degree $\left< k\right> =10$. The panels correspond to the results obtained with asynchronous label propagation (a), Louvain (b), Infomap (c), and when the best community partition is taken from the three methods according to $Q$ (d).}
    \label{fig:PSO_heatmaps}
\end{figure}

\begin{figure}
    \centering
    \captionsetup{width=0.85\textwidth}
    \includegraphics[width=0.85\textwidth]{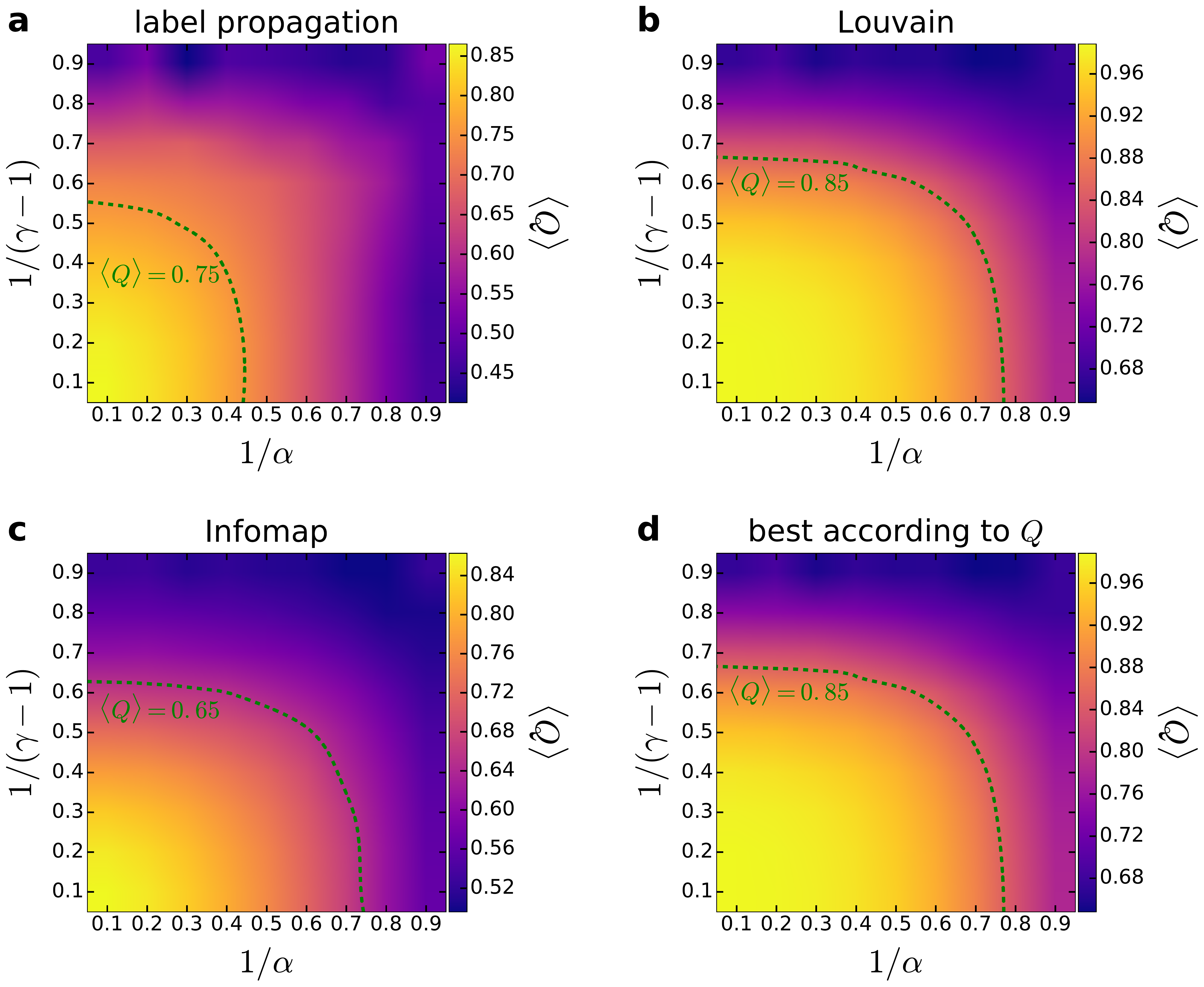}
    \caption{ {\bf Modularity in the $\mathbb{S}^1/\mathbb{H}^2$ model}. We show the weighted modularity $Q$ (indicated by the color, averaged over 100 samples) as a function of the model parameters $1/\alpha$ and $1/(\gamma -1)$ for networks of size $N=10,000$ and expected average degree $\left< k\right> =10$. The panels correspond to the results obtained with asynchronous label propagation (a), Louvain (b), Infomap (c), and when the best community partition is taken from the three methods according to $Q$ (d).}
    \label{fig:S1_heatmaps}
\end{figure}

%Ezek a szög szerinti struktúrák amúgy csoportok
In order to verify that the angularly separated modules detected by the asynchronous label propagation, the Louvain and the Infomap algorithms are indeed relevant structural units of the networks, we measured the quality of the extracted community partitions by the weighted modularity $Q$ given in equation~(\ref{eq:modularityDef}). In figures~\ref{fig:PSO_heatmaps} and \ref{fig:S1_heatmaps} we show the corresponding results for networks of size $N=10,000$ and expected average degree $\left< k\right>=10$, where the weighted modularity is plotted as a function of the model parameters with the help of heat maps. According to figure~\ref{fig:PSO_heatmaps}, for a considerably large region in the parameter plane the modularity averaged over 100 networks is larger than 0.65 for the communities found by Infomap (figure~\ref{fig:PSO_heatmaps}(c)), larger than 0.75 for the communities extracted by asynchronous label propagation (figure~\ref{fig:PSO_heatmaps}(a)) and larger than 0.85 for the communities located by Louvain (figure~\ref{fig:PSO_heatmaps}(b)).
For Louvain and Infomap, the highest scores in the modularity are achieved at low $T$ and $\beta$ parameters, corresponding to networks with a high average clustering coefficient and a rather homogeneous degree distribution. The modularity is high in this region also for the asynchronous label propagation; however, in this case the highest modularity values occur for mid-range $\beta$ values. 
%(Az alabpropnál közepes bétákra lett a legnagyobb az érték! Az S1-nél nincs ilyen baj. Louvain és Infomap: ahogy béta nő, Q átlaga csökken, szórása nő. alabprop: béta=0.1-től 0.6-ig nő a Qátlag, utána végig csökken, a Q szórása végig nő béta növekedésével kis T-kre; az alabprop szórása jellemzően ~1 nagyságrenddel nagyobb, mint a másik 2 módszeré (lásd a Suppl1-et szórásos színtérképekért)) 
When $\beta$ approaches 1, the observed $Q$ seems to decrease for all community finding methods. Nevertheless, $Q$ can still take relatively high values at e.g. $\beta=0.6$, where the generated network is expected to be scale-free with a degree decay exponent of $\gamma\simeq 2.67$. According to the results displayed in figure~\ref{fig:S1_heatmaps}, the maximum of $Q$ for the $\mathbb{S}^1/\mathbb{H}^2$ model is in the low-value regime of the $1/\alpha-1/(\gamma-1)$ parameter plane for all three community finding methods, where the modularity values seem to be higher by a small margin compared to the case of the PSO model, e.g. reaching up to $\left< Q\right>=0.99$ for the communities found by Louvain.

As mentioned in the Introduction, a large modularity value alone does not always indicate a true modular structure as e.g. both Erd{\H o}s--R{\'e}nyi random graphs and Barab{\'a}si--Albert random graphs have been shown to display relatively high modularity values under certain circumstances \cite{Guimera_ER_modul,Good_modul_max}. However, for random graphs generated by the aforementioned two classical models with the same size and link density as in figures~\ref{fig:PSO_heatmaps} and \ref{fig:S1_heatmaps}, the modularity can reach up to only about $0.28$, which is significantly smaller compared to the $Q$ values we observed in the studied hyperbolic networks. Furthermore, in the present study 2 out of the 3 community finding methods applied are not based on modularity maximisation, and they still find communities that yield high $Q$ values. 

%consistent when comparing the modules obtained from different methods. Furthermore,  Based on the above, the communities appearing in the PSO and $\mathbb{S}^1/\mathbb{H}^2$ models seem to be truly significant structural units that arise due to the interesting properties of the underlying hyperbolic geometry, as shall be discussed later.

\begin{figure}
    \centering
    \captionsetup{width=0.85\textwidth}
    \includegraphics[width=0.85\textwidth]{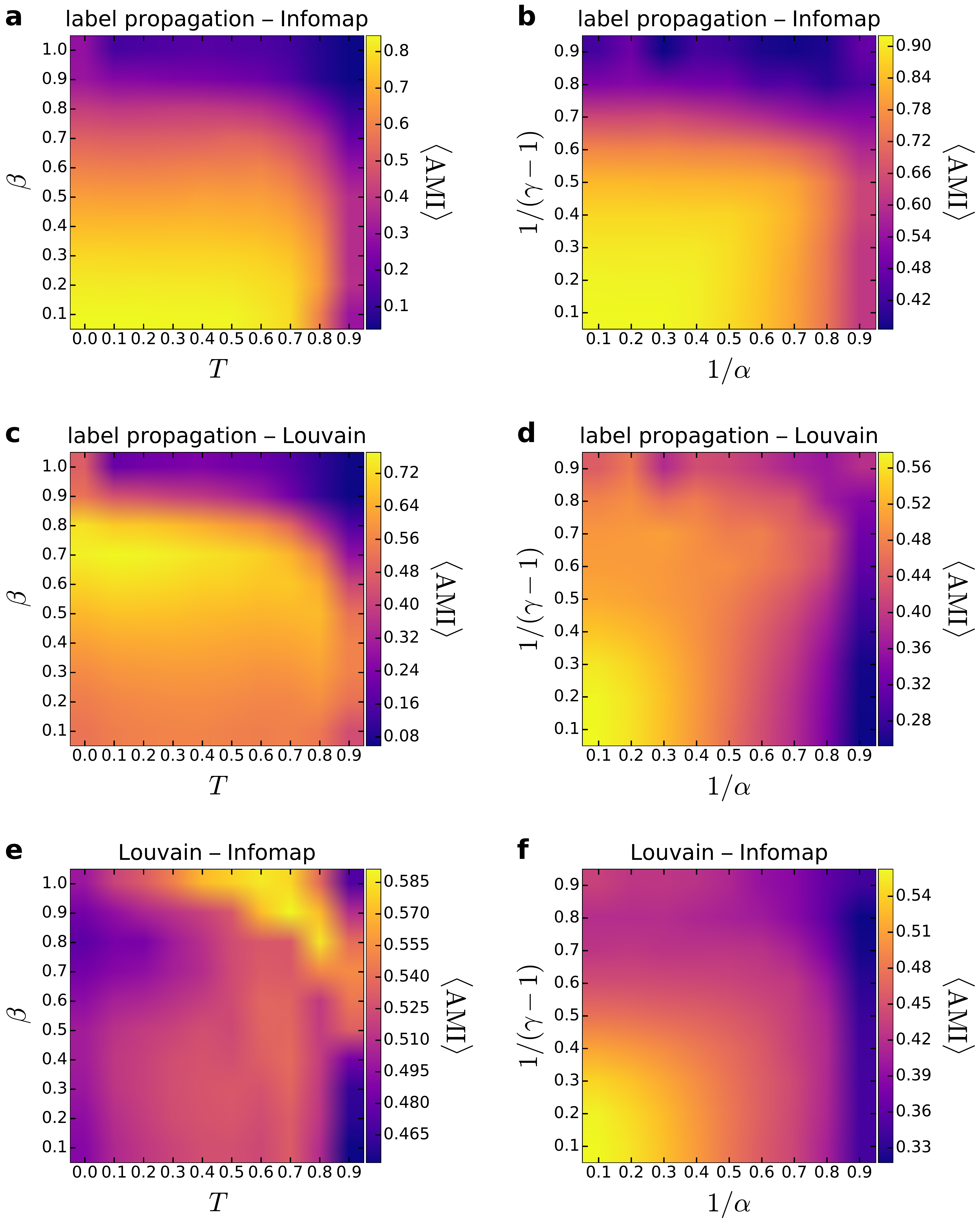}
    \caption{ {\bf Adjusted mutual information between the different community partitions.} The results for the PSO model are given in the left column (panels (a), (c) and (e)), whereas the AMI obtained for the $\mathbb{S}^1/\mathbb{H}^2$ model appears in the right column (panels (b), (d) and (f)). The AMI between the communities detected by asynchronous label propagation and Infomap is given in the top row (panels (a) and (b)), the AMI regarding the results of asynchronous label propagation and Louvain is shown in the middle row (panels (c) and (d)) and the AMI between the partitions found by Louvain and Infomap is presented in the bottom row (panels (e) and (f)). We show the measured AMI (indicated by the color, averaged over 100 samples) as a function of the model parameters $T$ and $\beta$, or $1/\alpha$ and $1/(\gamma -1)$ for networks of size $N=10,000$ and expected average degree $\left< k\right> =10$.}
    \label{fig:AMI}
\end{figure}

In order to examine the significance of the found communities from another aspect, we also compared the community partitions obtained with the different methods using the adjusted mutual information described in section~\ref{sect:AMI}. The results are displayed in figure~\ref{fig:AMI} with the help of heat maps, showing the AMI averaged over 100 networks as a function of the model parameters in the studied parameter planes.
According to the figure, the highest similarity values occur between the communities found by asynchronous label propagation and Infomap (figures~\ref{fig:AMI}(a) and \ref{fig:AMI}(b)). These can reach up to even $\left< \mathrm{AMI}\right>=0.9$, indicating an almost one-to-one correspondence between the modules of the different partitions. On the other hand, the lowest similarity values can be observed for Louvain and Infomap (figures~\ref{fig:AMI}(e) and \ref{fig:AMI}(f)), where the typical value of the AMI is about 0.5. However, this is still in the range of acceptable consistency between the different partitions and is definitely way higher than what we would expect e.g. for random partitions. Therefore, based on figure~\ref{fig:AMI} we can say that in those parameter regions where the communities are characterised by relatively high modularity scores, the partitions obtained with the different community detection methods also show significant consistency with each other. This fact reassures that the modules we observe in the studied hyperbolic networks are indeed relevant and apparent structural units that can be detected based on multiple approaches in a consistent way.

\begin{figure}
    \centering
    \captionsetup{width=0.85\textwidth}
    \includegraphics[width=0.85\textwidth]{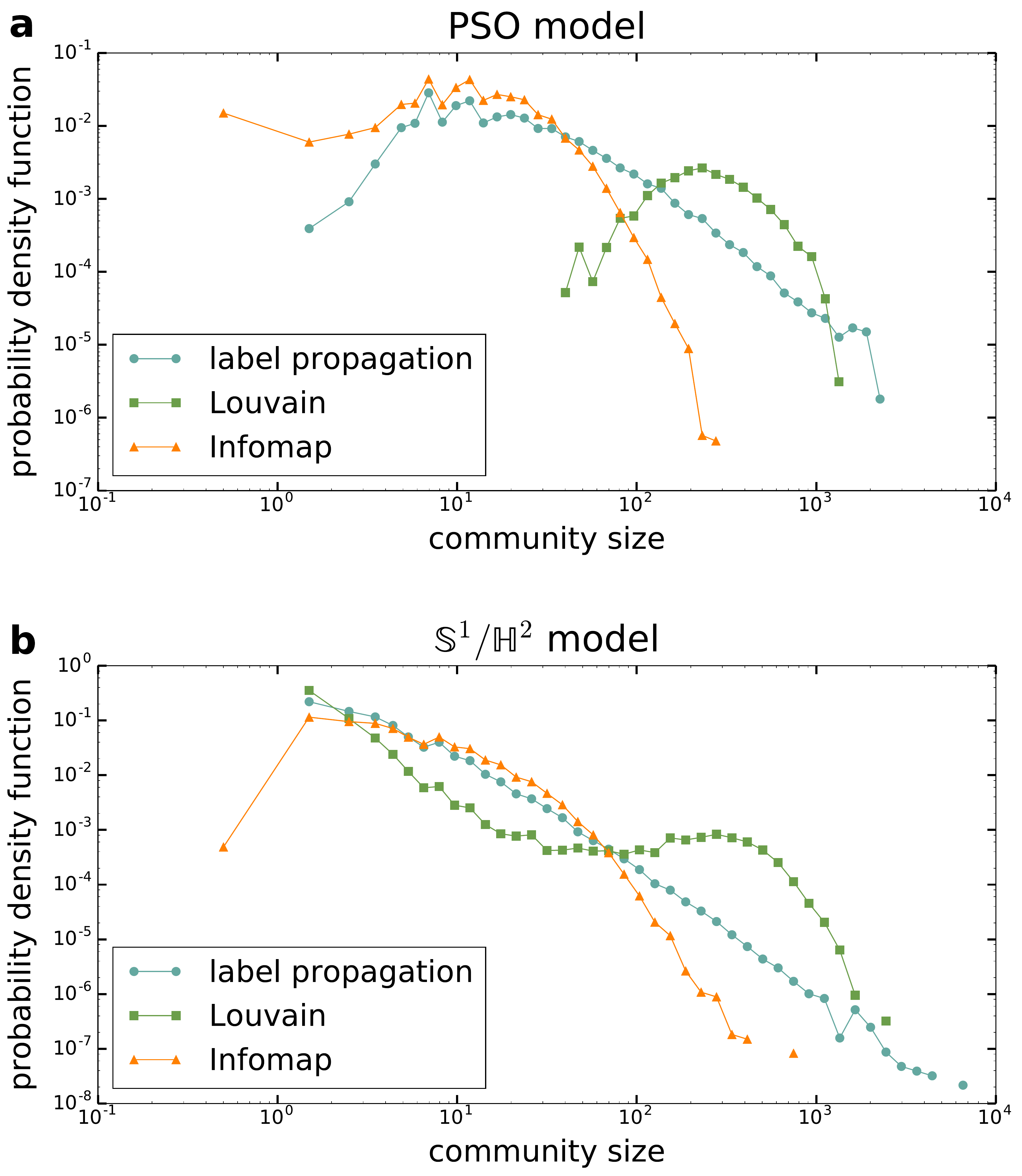}
    \caption{ {\bf Community size distributions.} (a) The probability density function of the community size in the PSO model according to asynchronous label propagation (blue circles), Louvain (green squares) and Infomap (orange triangles) based on 100 networks of size $N=10,000$, expected average degree $\left< k\right>=10$, temperature $T=0.2$ and popularity fading parameter $\beta=0.7$. (b) The probability density function of the community size in the $\mathbb{S}^1/\mathbb{H}^2$ model with the same symbol and colour coding as in panel (a), based on 100 networks of size $N=10,000$, expected average degree $\left< k\right>=10$, $1/\alpha=0.2$ and $1/(\gamma-1)=0.7$.}
    %minden hisztogram a 100 adott parameteru halozat osszesitett adata alapjan keszult
    \label{fig:com_sizes}
\end{figure}

A basic statistic regarding the revealed community structures is given by the community size distribution, which is exemplified by figure~\ref{fig:com_sizes} for the three examined community finding methods. According to that, the size of the communities found by the asynchronous label propagation follows more or less a power law for both the PSO model (figure~\ref{fig:com_sizes}(a)) and the $\mathbb{S}^1/\mathbb{H}^2$ model (figure~\ref{fig:com_sizes}(b)).
%According to that, the size of the communities found by Louvain follows a relatively narrow, bell-shaped curve, similar to a normal distribution for both the PSO model (figure~\ref{fig:com_sizes}(a)) and the $\mathbb{S}^1/\mathbb{H}^2$ model (figure~\ref{fig:com_sizes}(b)).
In the regime of small and middle-sized communities, the curve corresponding to Infomap seems to be close to that; however, towards the larger sizes it decays faster. In contrast, the community size distribution yielded by Louvain is quite distinct from the curves obtained with both asynchronous label propagation and Infomap, mostly due to a peak at higher community sizes for both the PSO model and the $\mathbb{S}^1/\mathbb{H}^2$ model. This difference between the community size distributions is in correspondence with the results seen for the AMI, where the output of Infomap and asynchronous label propagation turned out to be more similar to each other than to Louvain. 
%This apparent discrepancy between the community size distributions originates in the  difference between the approaches 

An interesting question related to the visibly strong community structure obtained with the studied hyperbolic models is how does it relate to the community structure of such networks where the angular distribution of the nodes is non-uniform, as in the case of the hyperbolic network models proposed in Refs.\cite{nPSO,nPSO_2,GPA_PSOsoftComms,S1softComms}. To address this question, here we define a transition between PSO networks with uniform angular node distribution and PSO networks generated with clear angular separation between modules in a similar fashion to the nPSO model introduced in Refs.\cite{nPSO,nPSO_2}, but with uniform angular distribution within the supposed communities instead of Gaussian distribution. Our related framework begins with generating a PSO network as outlined in section~\ref{sect:PSO}, and then running a community finding algorithm on the resulting network for locating its modules (we used Louvain for this purpose). Based on the found communities, we can then generate PSO networks with equally-sized gaps between the supposed modules by dividing the $[0,2\pi)$ interval into subintervals having a width proportional to the size of the given community, where the aggregated width of the subintervals can be expressed as $2\pi(1-g)$ when the aggregated width of the gaps is $2\pi g$. The number of nodes placed in a given subinterval is equal to the number of members of the corresponding community, and the angular coordinate of these nodes is distributed uniformly at random within the subinterval. Otherwise, the network generation process is identical to that in the original PSO model.

\begin{figure}
    \centering
    \captionsetup{width=\textwidth}
    \includegraphics[width=\textwidth]{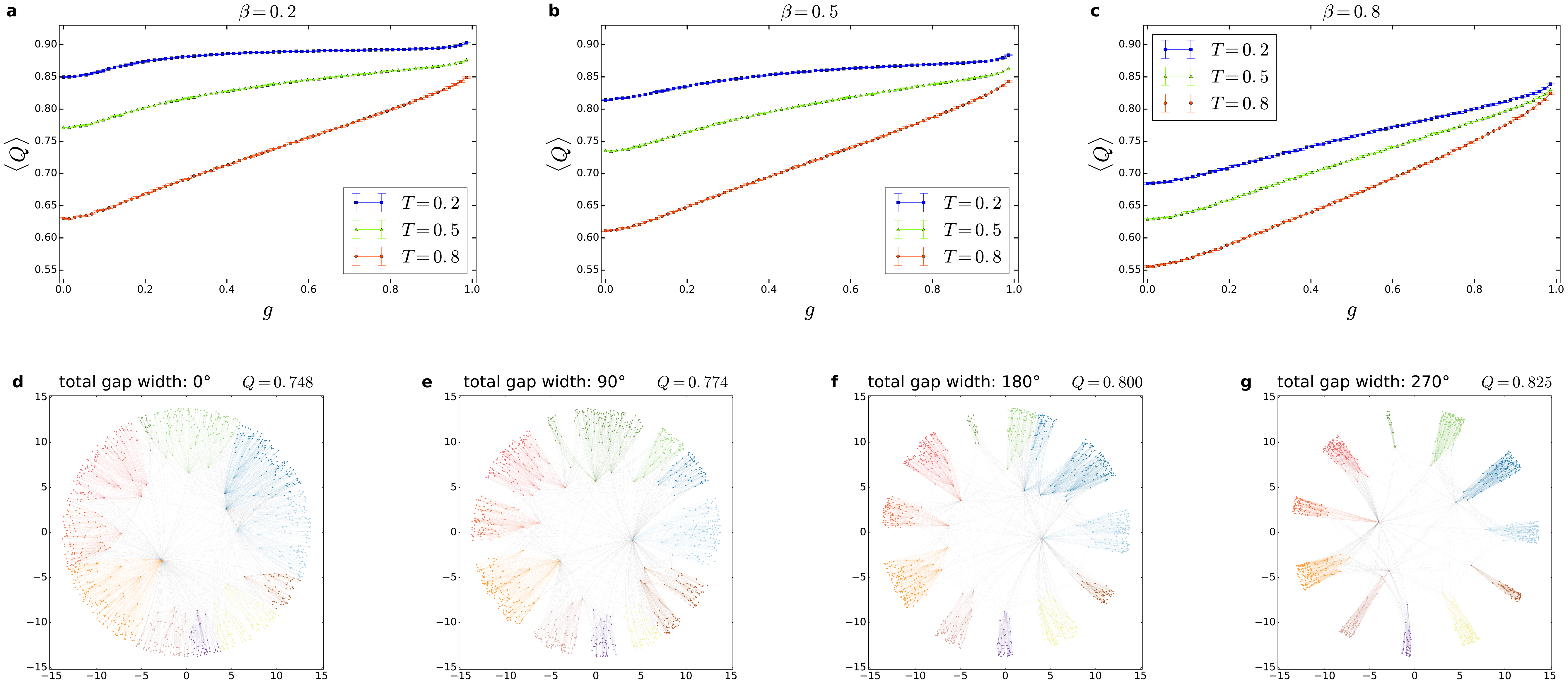}
    \caption{ {\bf Transition to non-uniform angular node distribution in the case of the PSO model.} The weighted modularity $Q$ averaged over 100 networks of size $N=1000$ and expected average degree $\left< k\right> = 10$ is shown as a function of the relative gap size $g$ between the modules for $\beta=0.2$ in panel (a), for $\beta=0.5$ in panel (b) and for $\beta=0.8$ in panel (c). (The error bars indicate the 95\% confidence intervals). In panels (d), (e), (f) and (g) we show a series of network layouts at increasing gap widths for $N=1000$, $\left< k\right> = 10$, $\beta=0.7$ and $T=0.2$, where the colours indicate the communities found by the Louvain algorithm.
    }
    \label{fig:uniform_nonuniform_transition}
\end{figure}

In figure~\ref{fig:uniform_nonuniform_transition} we show results obtained from this framework, where the top panels depict the modularity for communities found by the Louvain algorithm as a function of the relative gap size $g$, and the bottom panels provide layout examples at different values of $g$. According to the figure, although $Q$ increases as a function of the relative gap size $g$ as expected, this increase is rather mild, except for large $\beta$ or $T$ parameters. In other words, the modularity in the uniform PSO model can be quite close to the $Q$ that we obtain for modules with high angular separation, and therefore, the communities we observe in the uniform PSO model can be viewed also as a meaningful limit for the modular structure of systems where the angular distribution of the nodes is non-uniform.

As a closing of this section, we draw the attention to the supplementary materials, listing further results on the communities found in PSO and $\mathbb{S}^1/\mathbb{H}^2$ networks at different system sizes and average degree values (see sections~\ref{sect:detailedModularity}–\ref{sect:detailedAMI}). In addition, in the supplementary materials our analysis is repeated on an extension of the PSO model known as the E-PSO model \cite{EPSO_HyperMap} (described in section~\ref{sect:EPSO}), yielding results that are very similar to what we have detailed here. In section~\ref{sect:eqAngDist} we also examine what happens in the PSO model if the angular distribution of the nodes is strictly equidistant instead of homogeneous random. The qualitative behaviour of the communities found during these investigations is basically the same as seen here. Finally, in section~\ref{sect:unweightedCase} we show the results obtained for the examined PSO, E-PSO and $\mathbb{S}^1/\mathbb{H}^2$ networks when setting all the link weights to 1 instead of using the link weights given in equation~\ref{eq:linkWeights}.

\section{Discussion and conclusions}
\label{sect:Discuss}

Motivated by interesting signs of modules in hyperbolic networks with homogeneous angular node distribution reported in Refs.\cite{commSector_hypEmbBasedOnComms_2016,commSector_linkPred,commSector_commDetMethod,commSector_hypEmbBasedOnComms_2019,analogyBetweenHypEmbAndComms}, here we revisited the question of community structure in the PSO and $\mathbb{S}^1/\mathbb{H}^2$ models in a detailed in-depth study. Although for both of these models the model construction itself lacks any intentionally built-in community structure, the networks generated in these approaches still possess apparently strong communities for a wide range of the model parameters, as indicated by the high modularity values measured on the results of three independent community finding algorithms, namely asynchronous label propagation, Louvain and Infomap. The significance of the found communities is supported by the fact that only 1 out of the 3 applied methods is based on modularity optimisation, and that the comparison between the different partitions yielded reasonably high AMI values, indicating a considerable consistency between the results. Furthermore, the modularity values that can be achieved in Erd{\H o}s--R{\'e}nyi random graphs or Barab{\'a}si--Albert scale-free networks at the same average degree are way lower compared to the $Q$ values we observed in the hyperbolic networks. In addition, the ASI (corresponding to a quality measure independent of the modularity) was also very high for the major part of the parameter space.

%In the meantime, we measured reasonably high AMI values when comparing the partitions yielded by the three different community detection methods, indicating considerable consistency between them. 

The parameter plane in which we examined the behaviour of the modularity corresponded to the $(T,\beta)\in [0,1)\times(0,1]$ plane in the PSO model and the analogous $(\frac{1}{\alpha},\frac{1}{\gamma-1}) \in (0,1)\times(0,1)$ plane in the $\mathbb{S}^1/\mathbb{H}^2$ model. The intuitive meaning of these parameters can be summarised as follows: the average clustering coefficient of the generated networks is regulated by the temperature $T$ and its counterpart $1/\alpha$ (lower values result in higher average clustering coefficients), while the power-law decay exponent $\gamma$ of the degree distribution is controlled by the popularity fading parameter $\beta$ in the case of the PSO model according to the formula $\gamma=1+1/\beta$ and is itself a parameter of the $\mathbb{S}^1/\mathbb{H}^2$ model. According to our results, when changing these parameters, the behaviour of the modularity follows a similar pattern for both hyperbolic models and all three community finding algorithms, except for the PSO model combined with asynchronous label propagation. 

Putting aside the above-mentioned exception, for increasing $T$ (or $1/\alpha)$, together with a decrease in the average clustering coefficient the modularity also decreases (which is absolutely natural), and when $\beta$ (or equivalently, $1/(\gamma-1)$) is increased, resulting in a more fat-tailed degree distribution, $Q$ decreases again. However, the dependence of the modularity on the model parameters is not at all linear, instead we can observe a high, slightly decreasing plateau in the parameter plane with the maximum values in the origin and a relatively narrow belt of lower $Q$ values at the feet of the plateau, placed far from the origin. For the communities found by asynchronous label propagation in the networks generated by the PSO model, the behaviour is slightly different: although $Q$ is high close to the origin, for increasing $\beta$ it shows a slow increasing tendency, reaching its maximum in the medium $\beta$ range, followed by a drop for high $\beta$ values, similarly to the results seen for the other combinations between network generation models and community finding methods. 
%When taking a look at the dependency of $Q$ on the model parameters, we can see that $Q$ is increasing when $T$ or the corresponding $1/\alpha$ are decreased, which is quite natural, as these parameters control the average clustering coefficient of the models, and lowering their value is resulting in an increasing clustering coefficient. A similar tendency can be observed when examining the dependency of $Q$ on $\beta$ or the analogous $1/(\gamma-1)$, and since these parameters control the decay of the degree distribution, from this we can conclude that $Q$ is higher for the networks having a narrower degree distribution in the studied models.

%According to the above, the highest values in the modularity are achieved when $T\rightarrow 0,\beta\rightarrow 0$ in the PSO model and when $1/\alpha\rightarrow 0,1/(\gamma-1)\rightarrow 0$ in the $\mathbb{S}^1/\mathbb{H}^2$ model. Although this may be interesting from a theoretic point of view, 

When considering the parameter settings close to the origin ($T\rightarrow 0,\,\beta\rightarrow 0$ in the PSO model and $1/\alpha\rightarrow 0,\,1/(\gamma-1)\rightarrow 0$ in the $\mathbb{S}^1/\mathbb{H}^2$ model), which yield the largest modularity values in most of the cases, it is important to note that the corresponding networks are homogeneous in terms of the degree (the degree decay exponent $\gamma$ is large) and do not resemble scale-free real networks. However, when $\beta$ is increased (or equivalently, $\gamma$ is decreased), the modularity decreases only by a small magnitude for quite some range. E.g., at $\beta=0.6$, corresponding to $\gamma\simeq 2.67$, the modularity averaged over 100 networks can still reach up to $\left< Q\right>=0.929$ in the PSO model and $\left< Q\right>=0.898$ in the $\mathbb{S}^1/\mathbb{H}^2$ model. In other words, when setting the degree decay exponent to moderate values often seen in real systems with the help of $\beta$ or by directly tuning $\gamma$, the networks obtained with the studied models can still possess a strong community structure if the other parameter ($T$ or $1/\alpha$, controlling the clustering coefficient) is not pushed to extremely high values, meaning that the clustering coefficient is not reduced to extremely low values. 

The regime where $Q$ drops to lower values is on the one hand where $\beta\rightarrow 1$ (or equivalently $\gamma\rightarrow 2$ from above), corresponding to extremely fat-tailed degree distributions, and where $T\rightarrow 1$ (or equivalently $\alpha \rightarrow 1$ from above), corresponding to networks with clustering coefficients close to zero. Thus, if one would like to generate scale-free hyperbolic networks having communities and a degree decay exponent close to $\gamma=2$, it might be a better option to choose the models in Refs.\cite{GPA_PSOsoftComms,nPSO,nPSO_2,S1softComms}, where the community formation is helped by the non-uniform angular distribution of the nodes. Nevertheless, except the mentioned extreme regimes, the studied "traditional" hyperbolic models seem to produce a strong enough community structure that can be taken as a simple model for the apparent modular structures often observed in real systems. %??Esetleg kiemelném azt is, hogy a szög szerint egyenletes modelleknél nem kell önkényesen megszabni semmit a csoportok kapcsán (nPSO-nál csoportszámot és méretet is meg kell adni, GPA-nál egy $\Lambda$ initial attractiveness-t (the smaller the value of $\Lambda$, the more heterogeneous the distribution of angular coordinates.), ami "controls certain properties of the community size distribution" (ami nem tudom már mit jelent, de a lényeg, hogy van extra, önkényes paraméter a csoportstruktúra kapcsán a fokszámeloszlást és a klaszterezettséget meghatározni hivatott paramétereken felül)) ::: Ha gondolod fogalmazd ide. Viszont nem annyira szeretnék az nPSO és GPA szerzőkkel összeveszni, nem lenne jó, ha ez úgy tűnne fel, hogy a sima PSO még csoportokból is jobbakat generál mint azok a módszerek amik direkt csoportokat akarnak generálni. Az önkényes megadásnak az egyébként egy előnye tud lenni, hogy te kontrollálod, hogy hány csoportot akarsz, mert éppen olyan rendszerhez akarsz fittelni ahol ezt valamiért lehet tudni, és nem csak rácsodálkozol, hogy nahát hány csoport lett? -- Végülis ha valaki ránéz a non-uniform modellekre, rá fog jönni, hogy ott előre meg kell adni, milyen csoportokat akarunk, meg a non-uniformról eleve érezhető, hogy bonyolultabb mint a uniform, úgyhogy oké, ne erőltessük.

A remaining interesting question is why do the observed communities arise despite the absence of any explicit community formation mechanisms built into the construction of the studied models? In short, the same model properties that allow the development of a large clustering coefficient in the generated random graphs on the level of nodes also make the emergence of communities possible on a slightly larger scale. Communities are local structures in the sense that members connect to each other with a larger link density than to the rest of the system. %(if not in sheer numbers, at least when correcting with the size of the community and the size of the whole network (??igazából nem értem ezt a zárójeles megjegyzést, hogy mit jelent a korrekció)). 
As mentioned in the Introduction and as it can be seen in figures~\ref{fig:coms_illustrate}(a) and \ref{fig:coms_illustrate}(c), in hyperbolic networks such units correspond to well-defined angular regions \cite{commSector_hypEmbBasedOnComms_2016,commSector_commDetMethod,commSector_linkPred,commSector_hypEmbBasedOnComms_2019,Boguna_hyp_embed_coms,Cannistraci_ASI}, with a relatively low number of links across them. Thus, as noted in Ref.\cite{analogyBetweenHypEmbAndComms}, the community structure of a network can be also viewed as a coarse version of its layout in the hyperbolic space.

In our view, the key element in the formation of communities in the studied models is that due to the hyperbolicity of the native disk, for a node newly appearing at the periphery it is much easier to connect radially than "sideways" (i.e. to nodes with similarly large radial coordinate), as indicated by e.g. the distance formula in equation~(\ref{eq:hypDistApprox}). If the angular separation between the previously arrived nodes that are placed at smaller radii is large enough, they can become distinct attractive community cores to which the new nodes can connect with only a small interference (cross-links) between the different angular regions. In the PSO model, the condition for a large enough separation between the inner nodes is that they are pushed outwards (according to the popularity fading) relatively fast, i.e. $\beta$ is not large. In parallel, the cutoff in the connection probability as a function of the hyperbolic distance must also be sharp enough for localised connections; thus, $T$ must not be set large either to support community formation. A similar line of arguments holds also for the $\mathbb{S}^1/\mathbb{H}^2$ model. When $\gamma$ is large, then due to the relatively rapid decay in the degree distribution, the hidden variables $\kappa_i$ take low values that are mapped to relatively high radial coordinates even for the inner nodes, helping the formation of community cores. In parallel, a large $\alpha$ parameter in the $\mathbb{S}^1/\mathbb{H}^2$ model has a similar effect to a low $T$ value in the PSO model, sharpening the cutoff in the connection probability as a function of the metric distance.   

We also compared the community structure in the PSO model to the communities in networks with a non-uniform angular distribution of the nodes in a simple framework, motivated by the fact that the embedding of real networks is often non-homogeneous in terms of the angular coordinates, similarly to the hyperbolic models with built-in community formation introduced in Refs.\cite{nPSO,nPSO_2,GPA_PSOsoftComms,S1softComms}. Our framework enables a continuous transition between the homogeneous angular node distribution of the PSO model and an angular distribution with empty gaps between the supposed modules, where the angular coordinates are distributed uniformly at random inside the allowed angular regions. According to our results, the modularity shows only a mild increase as a function of the relative gap size for the majority of the parameter settings. Thus, the modules in the original PSO model can be quite close in strength to modules occurring in hyperbolic networks with a non-uniform angular node distribution, and the modular structure of the PSO model as a whole can be treated as a limiting case for those hyperbolic systems where the community structure is accompanied with a non-uniform distribution in the angular coordinates of the nodes.

%According to our results, the parameter regime where $Q$ eventually drops to lower values is corresponding to a relatively narrow belt at the feet of the plateau, far from the origin in the parameter plane. 

Our findings are also closely related to the community structures observed in networks grown with the help of simplicial complexes \cite{Bianconi_simplex_emergent_hyp,Mulder_geom_complexity} that were also shown to be hyperbolic. Explicit community formation is not built in these models either; however, the simplicial complexes form complete subgraphs (cliques), and when aggregating such dense structures, the appearance of communities seems to be more natural compared to the models studied here, where links are introduced one by one. Nevertheless, the formation of communities observed here deepens further the connection between hyperbolic networks and the models introduced in Refs.\cite{Bianconi_simplex_emergent_hyp,Mulder_geom_complexity}, that are known to possess a strong community structure. %Ezt azért csekkolni kéne a két cikkben.

In conclusion, our study draws the attention to the important but less known fact that the PSO and $\mathbb{S}^1/\mathbb{H}^2$ models are capable of generating random graphs that are not just small-world, highly clustered and scale-free, but in addition contain communities as well. Although the advantageous properties of hyperbolic models were already appreciated in the literature, this recognition makes them even more suitable for modelling real systems than thought before. In real systems, communities provide very important units at an intermediate level of the structural organisation of the network. Our detailed study of the behaviour of the community structure as a function of the model parameters show that modules are formed also in hyperbolic networks in an "automatic" way, simply as a consequence of the connection rules and the nature of the underlying hyperbolic geometry. These findings add a novel perspective and motivation for the studies and applications of hyperbolic network models.  
%In the meantime it is important to note that according to our measurements, in a relatively narrow parameter regime the modularity (and hence, the community structure) was not so strong, corresponding to the $\beta\rightarrow 1$ and $1/(\gamma-1)\rightarrow 1$ cases, where the degree distribution of the generated networks becomes extremely fat-tailed. 

\section*{Acknowledgments}
The authors are grateful for the enlightening discussions with Carlo Vittorio Cannistraci. The research was partially supported by the Hungarian National Research, Development and Innovation Office (grant no. K 128780, NVKP\_16-1-2016-0004) and by the Research Excellence Programme of the Ministry for Innovation and Technology in Hungary, within the framework of the Digital Biomarker thematic programme of the Semmelweis University.

\section*{References}
\bibliography{references}

\begin{thebibliography}{10}
\bibitem{modularity_code}
We calculated the modularity values with the Python function `modularity'
  available in the `networkx.algorithms.community.quality' package.;.
\bibitem{AMI_code}
We calculated the adjusted mutual information values with the Python function
  `adjusted\_mutual\_info\_score' available in the `sklearn.metrics.cluster'
  package.;.
\bibitem{alabprop_code}
We used the Python function `asyn\_lpa\_communities', an implementation of the
  asynchronous label propagation algorithm available in the
  `networkx.algorithms.community.label\_propagation' package.;.
\end{thebibliography}

\clearpage

\title{SUPPORTING INFORMATION}

\setcounter{section}{0}
\renewcommand{\thesection}{S\arabic{section}}

\section{The E-PSO model}
\label{sect:EPSO}
\setcounter{figure}{0}
\setcounter{table}{0}
\setcounter{equation}{0}
\renewcommand{\thefigure}{S1.\arabic{figure}}
\renewcommand{\thetable}{S1.\arabic{table}}
\renewcommand{\theequation}{S1.\arabic{equation}}

In our studies of the community structure of hyperbolic networks, besides the PSO model and the $\mathbb{S}^1/\mathbb{H}^2$ model, we also used the E-PSO model for random graph generation. The results on the communities found in this model are presented in sections~\ref{sect:detailedModularity}–\ref{sect:detailedAMI} and section~\ref{sect:unweightedCase}, whereas in the present section we provide a brief introduction to the model itself. 

The popularity-similarity optimisation (PSO) model~\cite{PSO} was generalised in the Supplementary Notes of Ref.~\cite{PSO} by the introduction of so-called \textit{internal} links: in this \textit{generalised PSO model}, in addition to the $m$ number of \textit{external} links connecting the new node to the already existing nodes, at each time step an $L$ number of further \textit{internal}
connections are created between disconnected pairs of previously appeared nodes, where the formation of all types of links is determined by the usual distance-dependent probabilities. It is straightforward to extend this model to negative values of the parameter $L$ \cite{our_embedding}, in which case after connecting the new node to $m$ number of the already existing nodes, the total number of links between the previously appeared nodes is changed by $L<0$, i.e. $|L|$ number of internal links are removed at each time step. To maintain the trend that mostly hyperbolically close nodes are connected to each other, it is natural to use the same probability formula for retaining a link as for the link creation and, accordingly, the complementary probability of link creation as the probability of link removal.

The E-PSO model~\cite{EPSO_HyperMap} is an equivalent of the generalised PSO model using solely external links, i.e. all edges of the network are established by the actual new node. Contrary to the original and the generalised PSO models, in the E-PSO model the number of links created at a time step is time-dependent. The expected number of links emerging at time $i\in[1,N]$ can be given as
\begin{equation}
    \hspace*{-1cm}
    \bar{m}_i = m+\bar{L}_i \simeq m+L\cdot\frac{2(1-\beta)}{(1-N^{-(1-\beta)})^2(2\beta -1)}\left[\left(\frac{N}{i}\right)^{2\beta-1}-1\right]\left(1-i^{-(1-\beta)}\right),
    \label{eq:expNumOfLinksATi}
\end{equation}
where $\bar{L}_i$ is the expected total number of internal links from previous nodes on the node appearing at iteration $i$ at the end of the network generation process in the generalised PSO model parametrised by the number of nodes $N$, the number $m$ of external links created at each time step, the change $L$ in the number of internal links at each time step and the popularity fading parameter $\beta$. Compared to the generalised PSO model, with the E-PSO model one can generate even large networks relatively fast, since in the generalised PSO model the hyperbolic distance needs to be calculated for all node pairs at each time step to update the probabilities of internal link creation or retainment in accordance with the updated node positions, whereas in the E-PSO model always only the distances from the newly appeared node need to be determined.

Not only the original PSO model, but the generalised PSO model and the analogous E-PSO model are capable of producing scale-free networks (with a degree decay exponent $\gamma=1+1/\beta$) that are highly clustered (in the case of small temperature $T$) and have the small-world property. Moreover, in the generalised PSO model and the E-PSO model, with the introduction of the parameter $L$ even it becomes adjustable how the average internal degree of the subgraphs spanning between nodes having a degree larger than a certain threshold depends on the degree threshold \cite{our_embedding}: for $0<L$ the average internal degree increases with the degree threshold and for $L=0$ (corresponding to the original PSO model) the average internal degree does not depend on the degree threshold until the degree threshold remains below a value at which the subgraphs become extremely small, while for $L<0$ the average internal degree gradually decreases as the degree threshold increases. Note that in these generalised versions of the PSO model the expected average degree of the resulted network can be calculated as $\left< k\right>=2(m+L)$ instead of $\left< k\right>=2m$.

\clearpage

\section{Quality of the detected communities}
\label{sect:detailedModularity}

\captionsetup[figure]{font=footnotesize,justification=justified,labelsep=period,labelfont=bf}

\setcounter{figure}{0}
\setcounter{table}{0}
\setcounter{equation}{0}
\renewcommand{\thefigure}{S2.\arabic{figure}}
\renewcommand{\thetable}{S2.\arabic{table}}
\renewcommand{\theequation}{S2.\arabic{equation}}

\begin{figure}[hbt]
    \centering
    \makebox[\textwidth][c]{\includegraphics[width=1.15\textwidth]{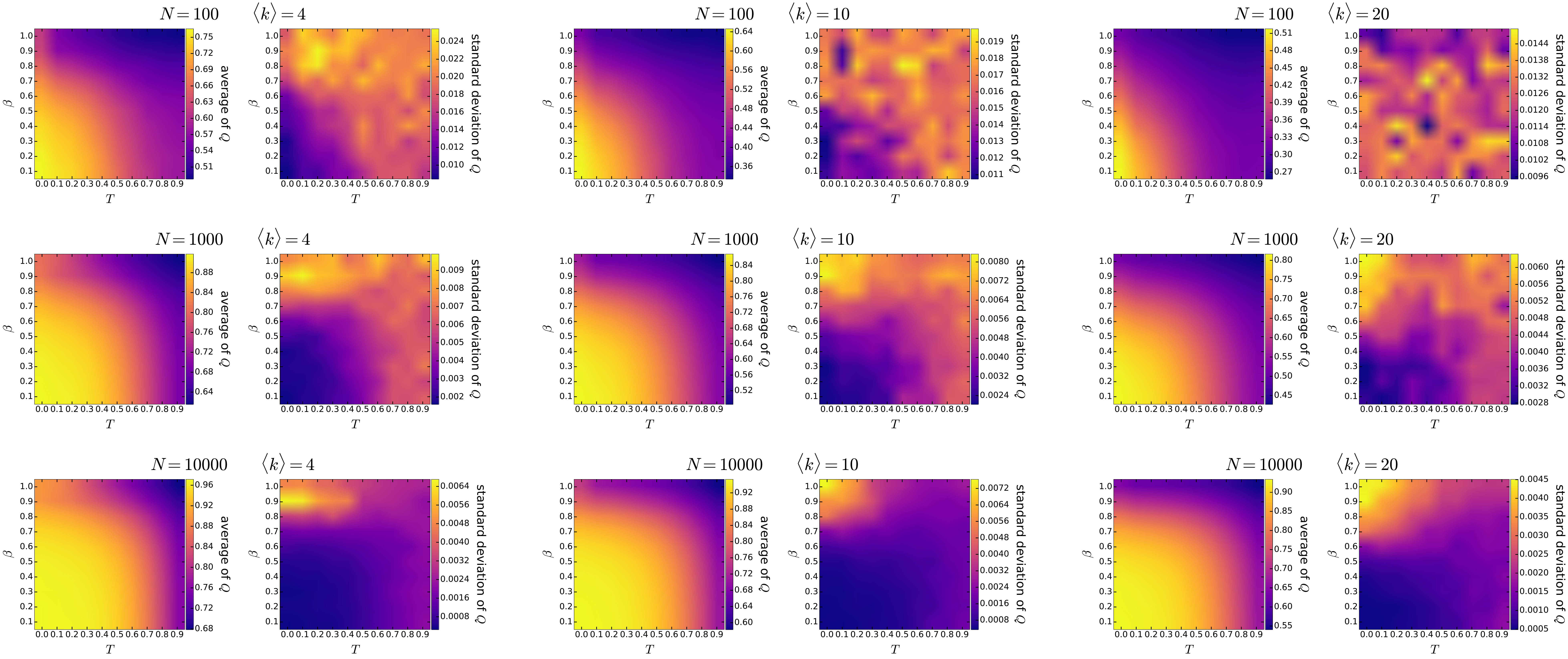}}
    \caption{{\bf The mean and the standard deviation of the highest weighted modularity $Q$ achieved among the \textit{asynchronous label propagation}, the \textit{Louvain} and the \textit{Infomap} algorithms in 100 \textit{PSO} networks of different parametrisations.} Each pair of subplots depicts the effect of changing the popularity fading parameter $\beta$ and the temperature $T$, with the number of nodes $N$ and the expected average degree $\langle k\rangle=2m$ given in the title of the subplot pair. The curvature of the hyperbolic plane $K$ was always set to $-1$, i.e. we used $\zeta=1$.}
    \label{fig:PSO_bestMod}
\end{figure}

\begin{figure}[hbt]
    \centering
    \makebox[\textwidth][c]{\includegraphics[width=1.15\textwidth]{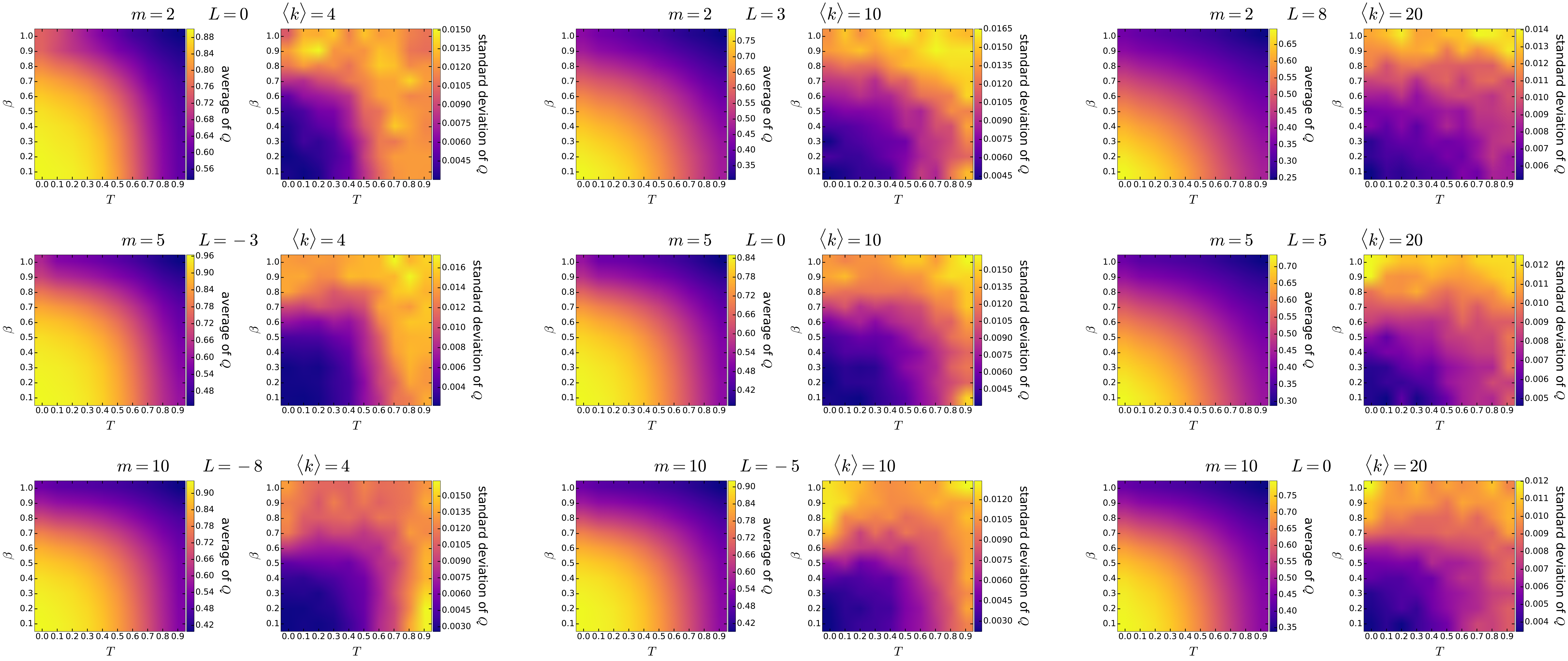}}
    \caption{{\bf The mean and the standard deviation of the highest weighted modularity $Q$ achieved among the \textit{asynchronous label propagation}, the \textit{Louvain} and the \textit{Infomap} algorithms in 100 \textit{E-PSO} networks of different parametrisations.} Each pair of subplots depicts the effect of changing the popularity fading parameter $\beta$ and the temperature $T$, with the parameters $m$ and $L$ given in the title of the subplot pair together with the corresponding expected average degree $\langle k\rangle=2(m+L)$. The number of nodes $N$ was 1000 in each case. The curvature of the hyperbolic plane $K$ was always set to $-1$, i.e. we used $\zeta=1$.}
    \label{fig:EPSO_bestMod}
\end{figure}

\begin{figure}[hbt]
    \centering
    \makebox[\textwidth][c]{\includegraphics[width=1.15\textwidth]{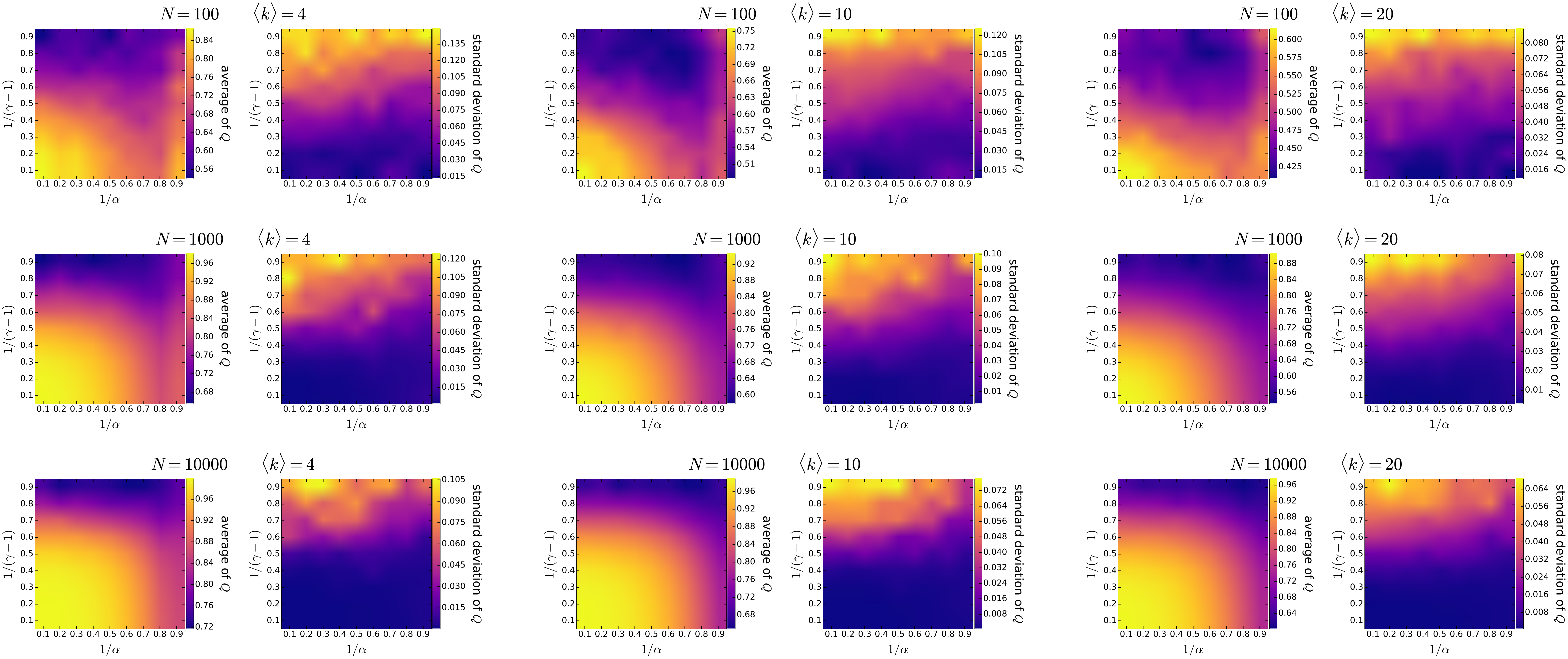}}
    \caption{{\bf The mean and the standard deviation of the highest weighted modularity $Q$ achieved among the \textit{asynchronous label propagation}, the \textit{Louvain} and the \textit{Infomap} algorithms in 100 $\mathbb{S}^1/\mathbb{H}^2$ networks of different parametrisations.} Each pair of subplots depicts the effect of changing $1/(\gamma-1)$ (equivalent to the popularity fading parameter $\beta$ in the E-PSO model) and $1/\alpha$ (analogous to the temperature $T$ in the E-PSO model), with the number of nodes $N$ and the expected average degree $\langle k\rangle$ given in the title of the subplot pair. We used $K=-1$ as the curvature of the hyperbolic plane in each case.}
    \label{fig:S1_bestMod}
\end{figure}

\begin{figure}[hbt]
    \centering
    \makebox[\textwidth][c]{\includegraphics[width=1.15\textwidth]{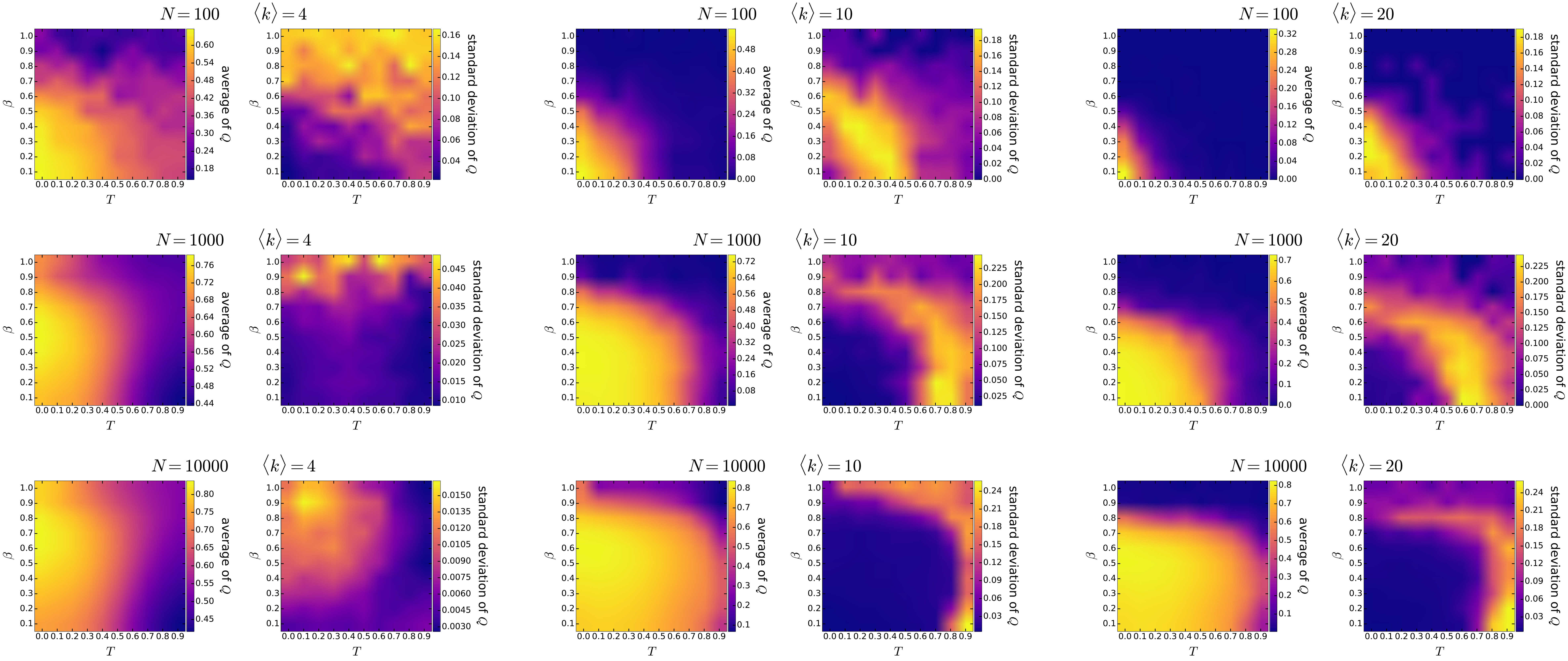}}
    \caption{{\bf The mean and the standard deviation of the weighted modularity $Q$ of the community structure detected by the \textit{asynchronous label propagation} algorithm in 100 \textit{PSO} networks of different parametrisations.} Each pair of subplots depicts the effect of changing the popularity fading parameter $\beta$ and the temperature $T$, with the number of nodes $N$ and the expected average degree $\langle k\rangle=2m$ given in the title of the subplot pair. The curvature of the hyperbolic plane $K$ was always set to $-1$, i.e. we used $\zeta=1$.}
    \label{fig:PSO_alabpropMod}
\end{figure}

\begin{figure}[hbt]
    \centering
    \makebox[\textwidth][c]{\includegraphics[width=1.15\textwidth]{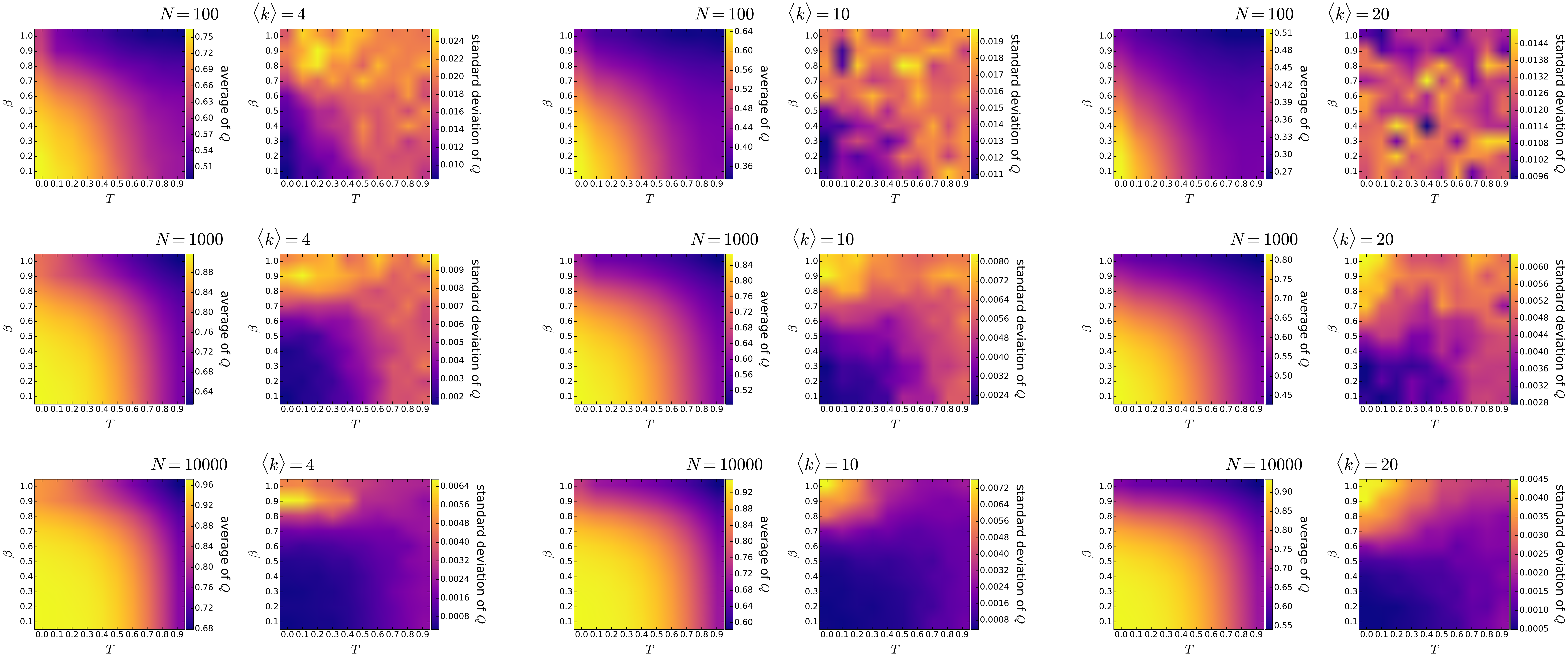}}
    \caption{{\bf The mean and the standard deviation of the weighted modularity $Q$ of the community structure detected by the \textit{Louvain} algorithm in 100 \textit{PSO} networks of different parametrisations.} Each pair of subplots depicts the effect of changing the popularity fading parameter $\beta$ and the temperature $T$, with the number of nodes $N$ and the expected average degree $\langle k\rangle=2m$ given in the title of the subplot pair. The curvature of the hyperbolic plane $K$ was always set to $-1$, i.e. we used $\zeta=1$.}
    \label{fig:PSO_LouvainMod}
\end{figure}

\begin{figure}[hbt]
    \centering
    \makebox[\textwidth][c]{\includegraphics[width=1.15\textwidth]{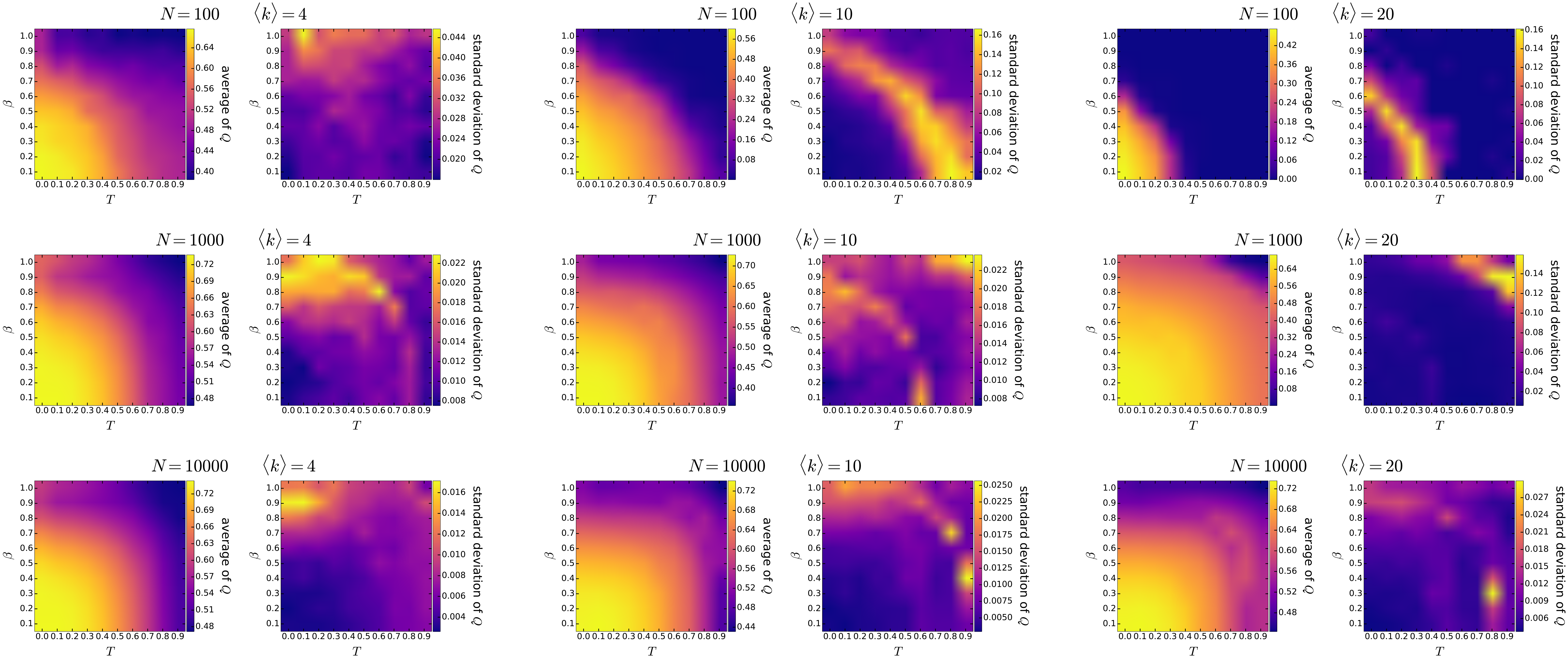}}
    \caption{{\bf The mean and the standard deviation of the weighted modularity $Q$ of the community structure detected by the \textit{Infomap} algorithm in 100 \textit{PSO} networks of different parametrisations.} Each pair of subplots depicts the effect of changing the popularity fading parameter $\beta$ and the temperature $T$, with the number of nodes $N$ and the expected average degree $\langle k\rangle=2m$ given in the title of the subplot pair. The curvature of the hyperbolic plane $K$ was always set to $-1$, i.e. we used $\zeta=1$.}
    \label{fig:PSO_InfomapMod}
\end{figure}

\begin{figure}[hbt]
    \centering
    \makebox[\textwidth][c]{\includegraphics[width=1.15\textwidth]{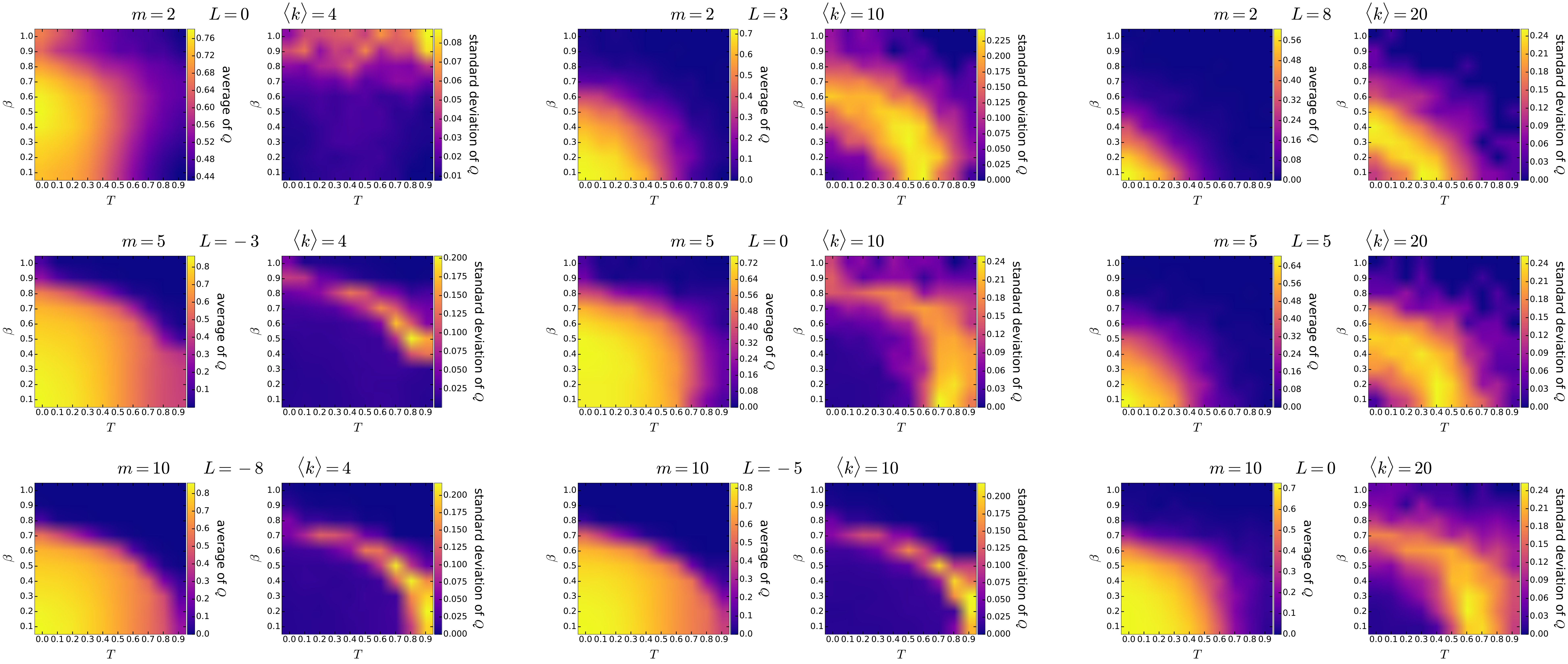}}
    \caption{{\bf The mean and the standard deviation of the weighted modularity $Q$ of the community structure detected by the \textit{asynchronous label propagation} algorithm in 100 \textit{E-PSO} networks of different parametrisations.} Each pair of subplots depicts the effect of changing the popularity fading parameter $\beta$ and the temperature $T$, with the parameters $m$ and $L$ given in the title of the subplot pair together with the corresponding expected average degree $\langle k\rangle=2(m+L)$. The number of nodes $N$ was 1000 in each case. The curvature of the hyperbolic plane $K$ was always set to $-1$, i.e. we used $\zeta=1$.}
    \label{fig:EPSO_alabpropMod}
\end{figure}

\begin{figure}[hbt]
    \centering
    \makebox[\textwidth][c]{\includegraphics[width=1.15\textwidth]{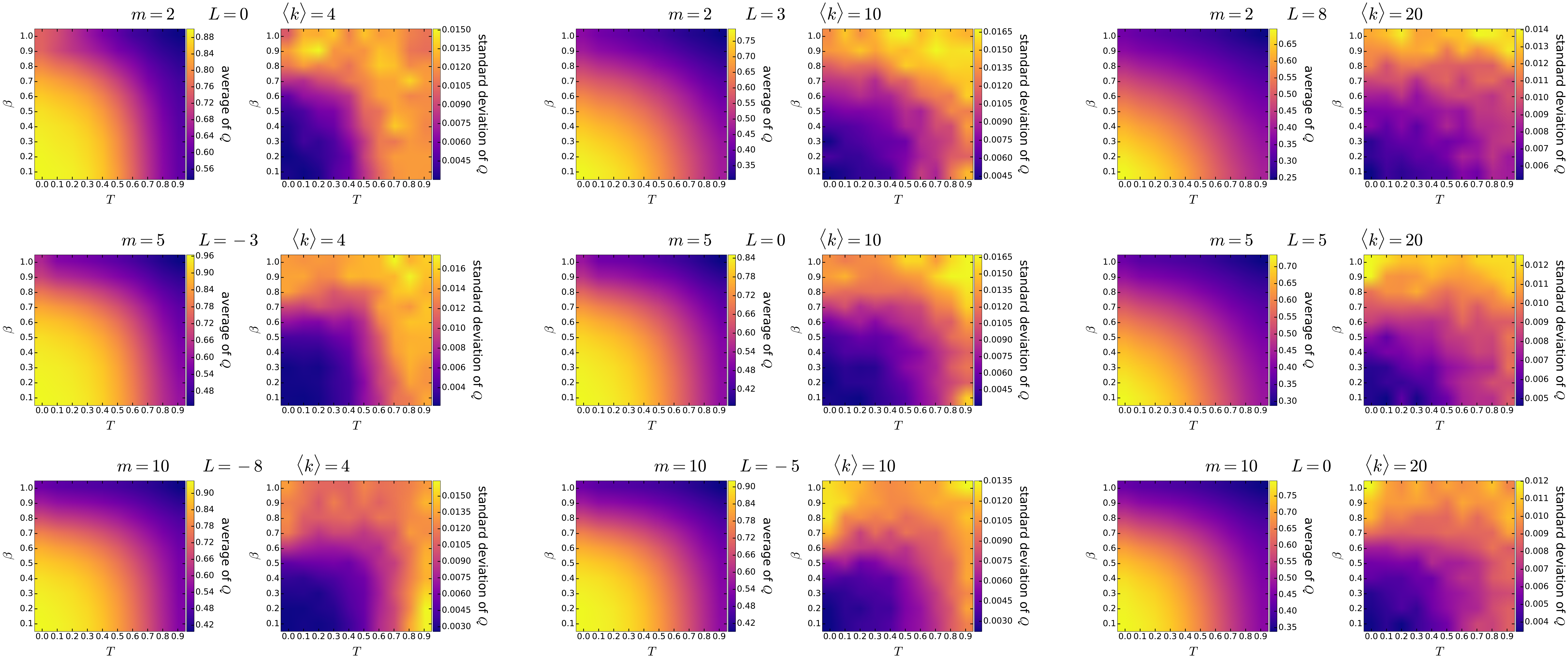}}
    \caption{{\bf The mean and the standard deviation of the weighted modularity $Q$ of the community structure detected by the \textit{Louvain} algorithm in 100 \textit{E-PSO} networks of different parametrisations.} Each pair of subplots depicts the effect of changing the popularity fading parameter $\beta$ and the temperature $T$, with the parameters $m$ and $L$ given in the title of the subplot pair together with the corresponding expected average degree $\langle k\rangle=2(m+L)$. The number of nodes $N$ was 1000 in each case. The curvature of the hyperbolic plane $K$ was always set to $-1$, i.e. we used $\zeta=1$.}
    \label{fig:EPSO_LouvainMod}
\end{figure}

\begin{figure}[hbt]
    \centering
    \makebox[\textwidth][c]{\includegraphics[width=1.15\textwidth]{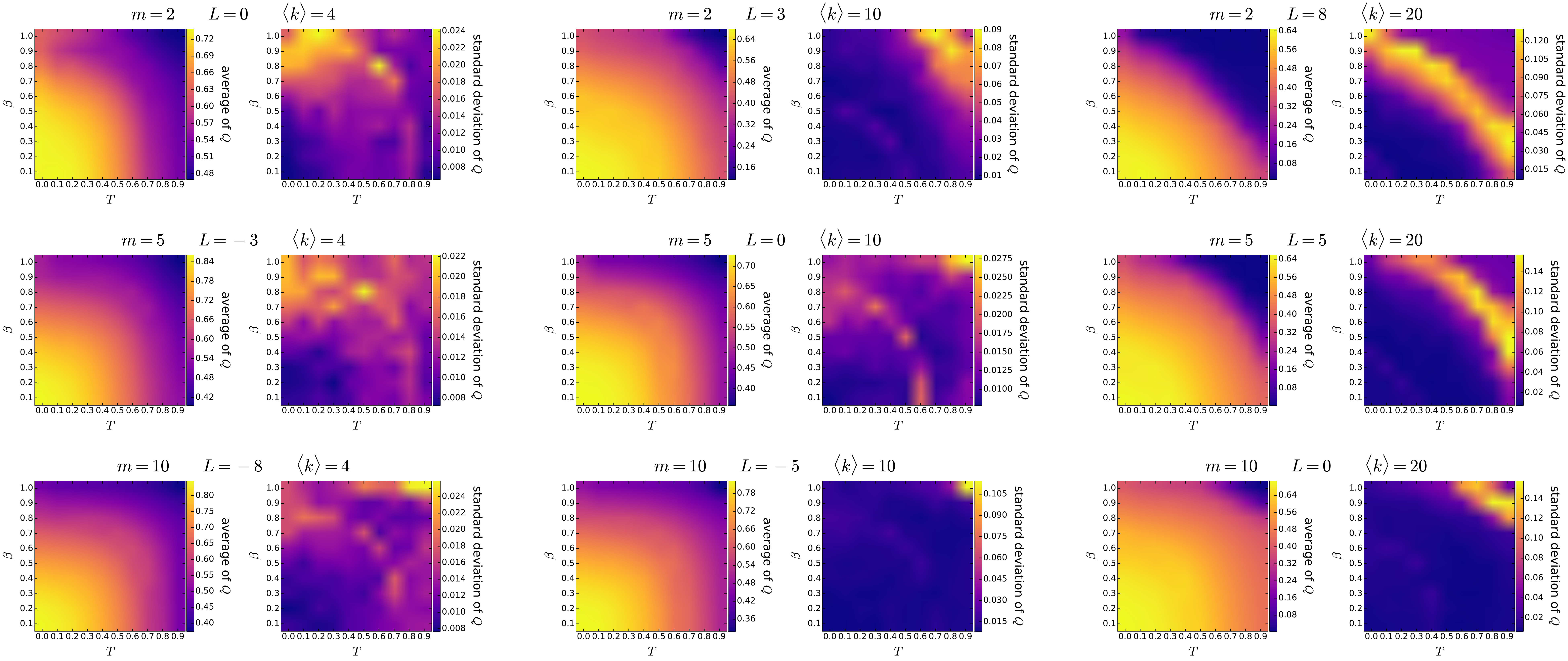}}
    \caption{{\bf The mean and the standard deviation of the weighted modularity $Q$ of the community structure detected by the \textit{Infomap} algorithm in 100 \textit{E-PSO} networks of different parametrisations.} Each pair of subplots depicts the effect of changing the popularity fading parameter $\beta$ and the temperature $T$, with the parameters $m$ and $L$ given in the title of the subplot pair together with the corresponding expected average degree $\langle k\rangle=2(m+L)$. The number of nodes $N$ was 1000 in each case. The curvature of the hyperbolic plane $K$ was always set to $-1$, i.e. we used $\zeta=1$.}
    \label{fig:EPSO_InfomapMod}
\end{figure}

\begin{figure}[hbt]
    \centering
    \makebox[\textwidth][c]{\includegraphics[width=1.15\textwidth]{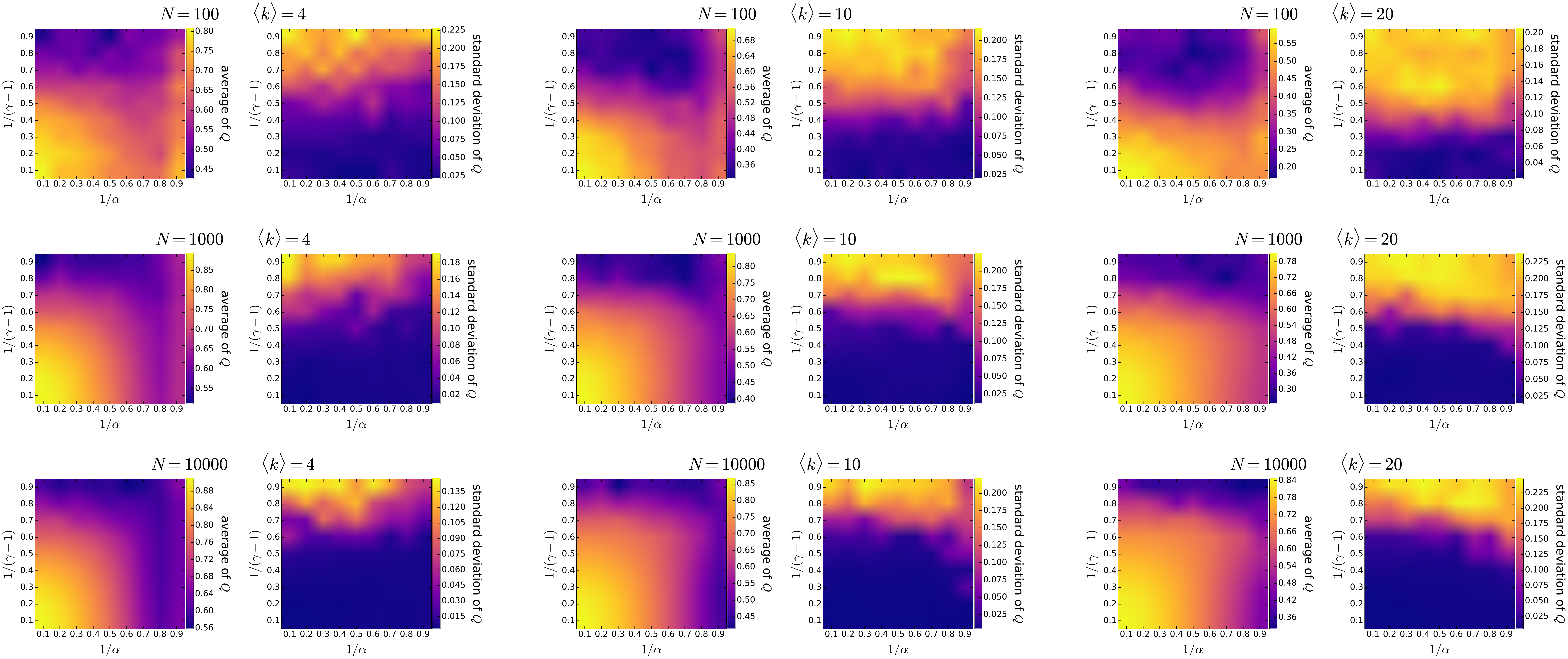}}
    \caption{{\bf The mean and the standard deviation of the weighted modularity $Q$ of the community structure detected by the \textit{asynchronous label propagation} algorithm in 100 $\mathbb{S}^1/\mathbb{H}^2$ networks of different parametrisations.} Each pair of subplots depicts the effect of changing $1/(\gamma-1)$ (equivalent to the popularity fading parameter $\beta$ in the E-PSO model) and $1/\alpha$ (analogous to the temperature $T$ in the E-PSO model), with the number of nodes $N$ and the expected average degree $\langle k\rangle$ given in the title of the subplot pair. We used $K=-1$ as the curvature of the hyperbolic plane in each case.}
    \label{fig:S1_alabpropMod}
\end{figure}

\begin{figure}[hbt]
    \centering
    \makebox[\textwidth][c]{\includegraphics[width=1.15\textwidth]{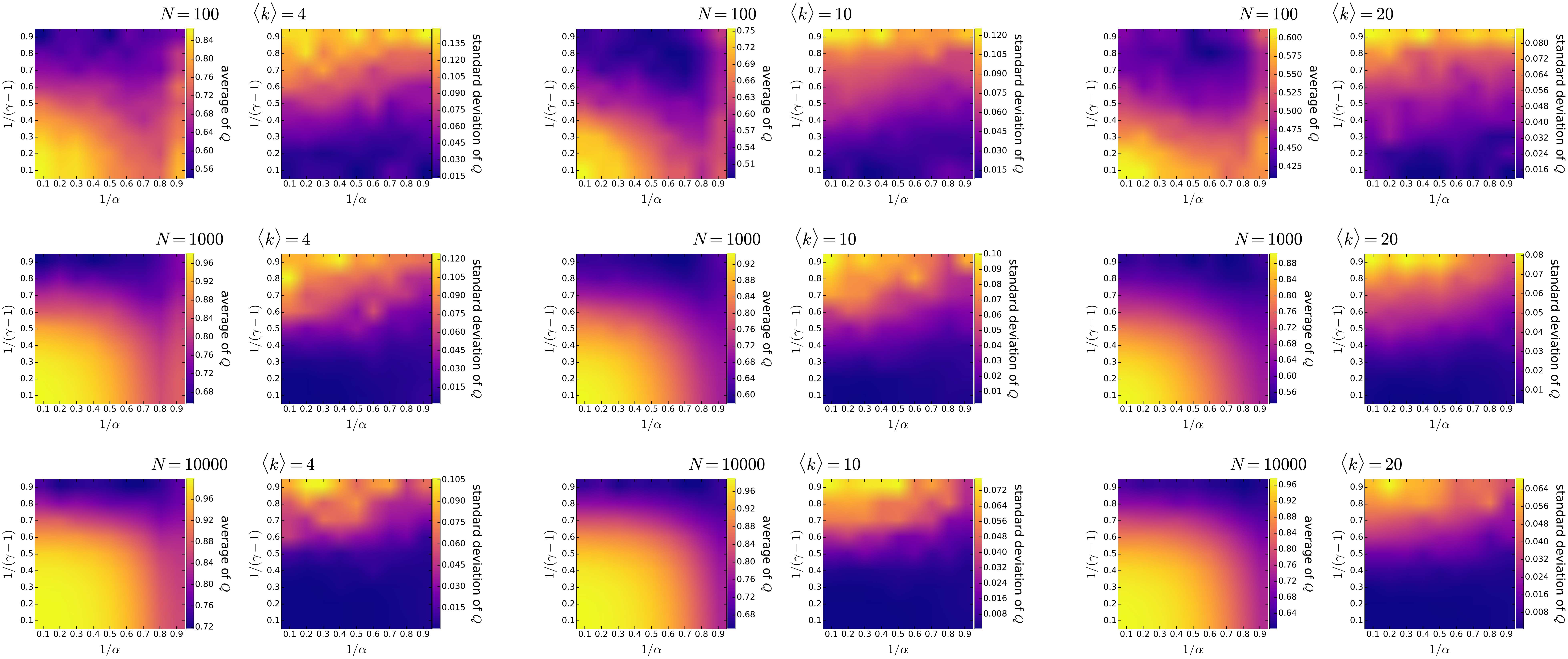}}
    \caption{{\bf The mean and the standard deviation of the weighted modularity $Q$ of the community structure detected by the \textit{Louvain} algorithm in 100 $\mathbb{S}^1/\mathbb{H}^2$ networks of different parametrisations.} Each pair of subplots depicts the effect of changing $1/(\gamma-1)$ (equivalent to the popularity fading parameter $\beta$ in the E-PSO model) and $1/\alpha$ (analogous to the temperature $T$ in the E-PSO model), with the number of nodes $N$ and the expected average degree $\langle k\rangle$ given in the title of the subplot pair. We used $K=-1$ as the curvature of the hyperbolic plane in each case.}
    \label{fig:S1_LouvainMod}
\end{figure}

\begin{figure}[hbt]
    \centering
    \makebox[\textwidth][c]{\includegraphics[width=1.15\textwidth]{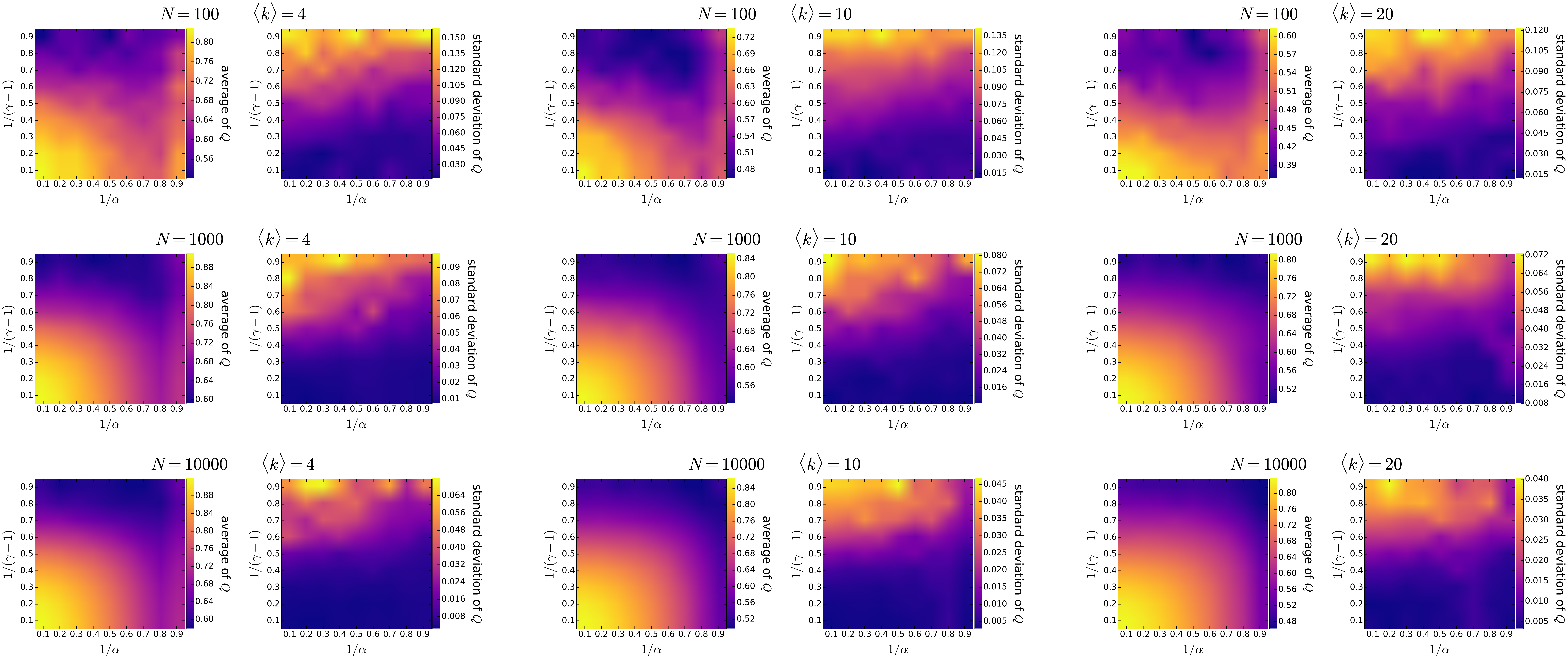}}
    \caption{{\bf The mean and the standard deviation of the weighted modularity $Q$ of the community structure detected by the \textit{Infomap} algorithm in 100 $\mathbb{S}^1/\mathbb{H}^2$ networks of different parametrisations.} Each pair of subplots depicts the effect of changing $1/(\gamma-1)$ (equivalent to the popularity fading parameter $\beta$ in the E-PSO model) and $1/\alpha$ (analogous to the temperature $T$ in the E-PSO model), with the number of nodes $N$ and the expected average degree $\langle k\rangle$ given in the title of the subplot pair. We used $K=-1$ as the curvature of the hyperbolic plane in each case.}
    \label{fig:S1_InfomapMod}
\end{figure}

We studied the quality of the community structures detected by the asynchronous label propagation~\cite{alabprop}, the Louvain~\cite{Louvain} and the Infomap~\cite{Infomap} algorithms in PSO \cite{PSO}, E-PSO \cite{EPSO_HyperMap,our_embedding} and $\mathbb{S}^1/\mathbb{H}^2$~\cite{S1,S1H2_Mercator} networks of various parameter combinations. The isolated nodes emerging in the case of the $\mathbb{S}^1/\mathbb{H}^2$ model and occasionally also in the networks generated by the E-PSO model of $L<0$ were removed before the community detection, meaning that the actual size of the examined networks does not necessarily reach the number of nodes $N$ inputted in these models. Each community detection algorithm was executed once for each network. Figures \ref{fig:PSO_bestMod}–\ref{fig:S1_bestMod} show the achieved highest weighted modularity averaged over 100 networks of each parameter setting together with the corresponding standard deviations. Figures \ref{fig:PSO_alabpropMod}–\ref{fig:S1_InfomapMod} present how the performance of the three different community detection algorithms depends on the parameters of the examined network generation models.

Figure \ref{fig:PSO_bestMod} depicts the effect on the achieved highest weighted modularity of changing the number of nodes $N$, the expected average degree $\langle k\rangle$, the popularity fading parameter $\beta$ and the temperature $T$ in the PSO model, which corresponds to the E-PSO model with $L=0$. For large $N$ and small $\langle k\rangle$ most of the nodes have the possibility to create connections only with hyperbolically close nodes, %(the number of possible neighbours of the outer nodes increases with $N$ and for small $\langle k\rangle$ the number of nodes to actually choose from all the possibilities is small), 
while for small $N/\langle k\rangle$ ratios the nodes are forced more often to connect even with farther nodes to create all the expected number of links. For this reason, a larger $N/\langle k\rangle$ ratio leads to connections that are more strongly determined by the hyperbolic distances, and thus to a more clear separation between the angular regions of the hyperbolic disk, i.e. a community structure with higher modularity. Besides, by sharpening the cutoff in the connection probability, small values of $T$ also facilitate the localisation of the node-node connections; thus, with the decrease of the temperature $T$ the modularity of the detected community structures increases. Furthermore, for smaller values of $\beta$ the inner nodes drift faster away from each other during the network growth, forming thereby more separated attraction centres for the outer nodes, due to which most of the network nodes can make a more definite choice between the community centres, which leads to communities with less external connections, i.e. larger modularity. 
%Ahhoz, hogy a belső pontok egymástól eléggé távol legyenek a hálópontok többségének érkezésekor, az kell, hogy a béta kicsi legyen, azaz gyorsan toljuk kifelé a korán megjelenő pontokat, ezáltal növelve köztük a távolságot. Ahhoz pedig, hogy a kötéseket viszonylag egyértelműen meghatározzák a hiperbolikus távolságok, az kell, hogy ne kényszerüljön arra egy pont, hogy tőle távoliakhoz kössön, azaz legyen elegendő pont tőle nem túl messze, akik közül választhat, hogy kikhez köt (ez akkor teljesül, ha N nagy (ekkor a pontok nagy része már viszonylag sok pont közül válogathat) és az átlagfokszám kicsi (ekkor nem kell olyan sok élt behúzni, hogy kifogyjunk a rövid élekből)), illetve az, hogy a kötési valószínűségben tényleg erősen számítson a hiperbolikus távolság, azaz a T legyen kicsi.

Figure \ref{fig:EPSO_bestMod} display how the parameters $m$ and $L$ affect the achieved highest weighted modularity in E-PSO networks. Based on these, not only the expected average degree $\langle k\rangle=2(m+L)$, but even $m$ and $L$ itself has an effect on the community structure of the generated networks. According to equation (S1.1), for $L<0$ the number of links created by a new node at its appearance is an increasing function of the node's appearance time, meaning that the early-appearing, inner nodes connect to few nodes only, whereas the outer nodes, for which the number of realisable connections (i.e. the number of previously appeared nodes) is not limited as much, form more connections. This way, nodes of any appearance time have the possibility to create connections only with hyperbolically close nodes. On the other hand, for $0<L$ the inner nodes create at their appearance a relatively large number of connections with the previously appeared nodes, which -- in the absence of enough hyperbolically close candidates -- leads to the emergence of connections between not so close nodes too. Taking into consideration the concept that the more the connections are restricted to hyperbolically close node pairs, the stronger the arising community structure, we can conclude that for a given expected average degree $\langle k\rangle$ the modularity can be increased by decreasing the parameter $L$ and, at the same time, increasing the parameter $m$ accordingly. %[DE Louvain, Infomap: <k>=4-nél L=-3 modularitása nagyobb, mint L=-8-é!..]
%The E-PSO model is analogous to the generalised PSO model, in which $L$ number of internal links are removed ($L<0$) or created ($0<L$) in each step of the network growth. 

According to figure \ref{fig:S1_bestMod}, the strength of the community structure in $\mathbb{S}^1/\mathbb{H}^2$ networks depends the same way on the model parameters as in PSO networks: the achieved highest weighted modularity is higher for larger number of nodes $N$, smaller average degree $\langle k\rangle$, larger degree decay exponent $\gamma$ (corresponding to smaller popularity fading parameter $\beta$ in the E-PSO model) and larger $\alpha$ (which is analogous to lower temperature $T$ in the E-PSO model).

%best

%PSO (E-PSO L=0-val)

%E-PSO (Lnem0)

%S1

\clearpage

\section{Community size distributions}
\label{sect:detailedCommSizeDist}

\setcounter{figure}{0}
\setcounter{table}{0}
\setcounter{equation}{0}

\renewcommand{\thefigure}{S3.\arabic{figure}}
\renewcommand{\thetable}{S3.\arabic{table}}
\renewcommand{\theequation}{S3.\arabic{equation}}

\captionsetup[figure]{font=footnotesize,justification=justified,labelsep=period,labelfont=bf}

\begin{figure}[hbt]
    \centering
    \makebox[\textwidth][c]{\includegraphics[width=1.15\textwidth]{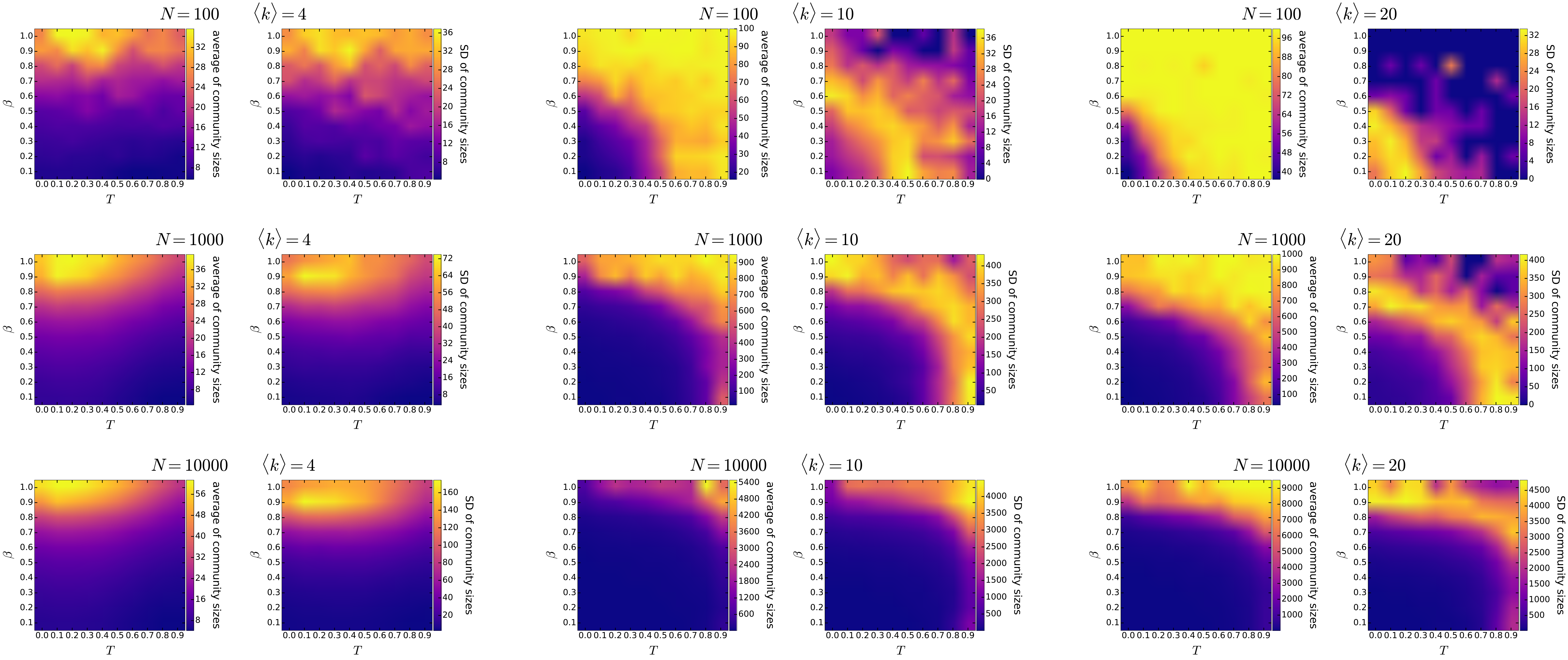}}
    \caption{{\bf The mean and the standard deviation of the size of communities detected by the \textit{asynchronous label propagation} algorithm in 100 \textit{PSO} networks of different parametrisations.} Each pair of subplots depicts the effect of changing the popularity fading parameter $\beta$ and the temperature $T$, with the number of nodes $N$ and the expected average degree $\langle k\rangle=2m$ given in the title of the subplot pair. The curvature of the hyperbolic plane $K$ was always set to $-1$, i.e. we used $\zeta=1$.}
    \label{fig:PSO_alabpropGSavstd}
\end{figure}

\begin{figure}[hbt]
    \centering
    \makebox[\textwidth][c]{\includegraphics[width=1.15\textwidth]{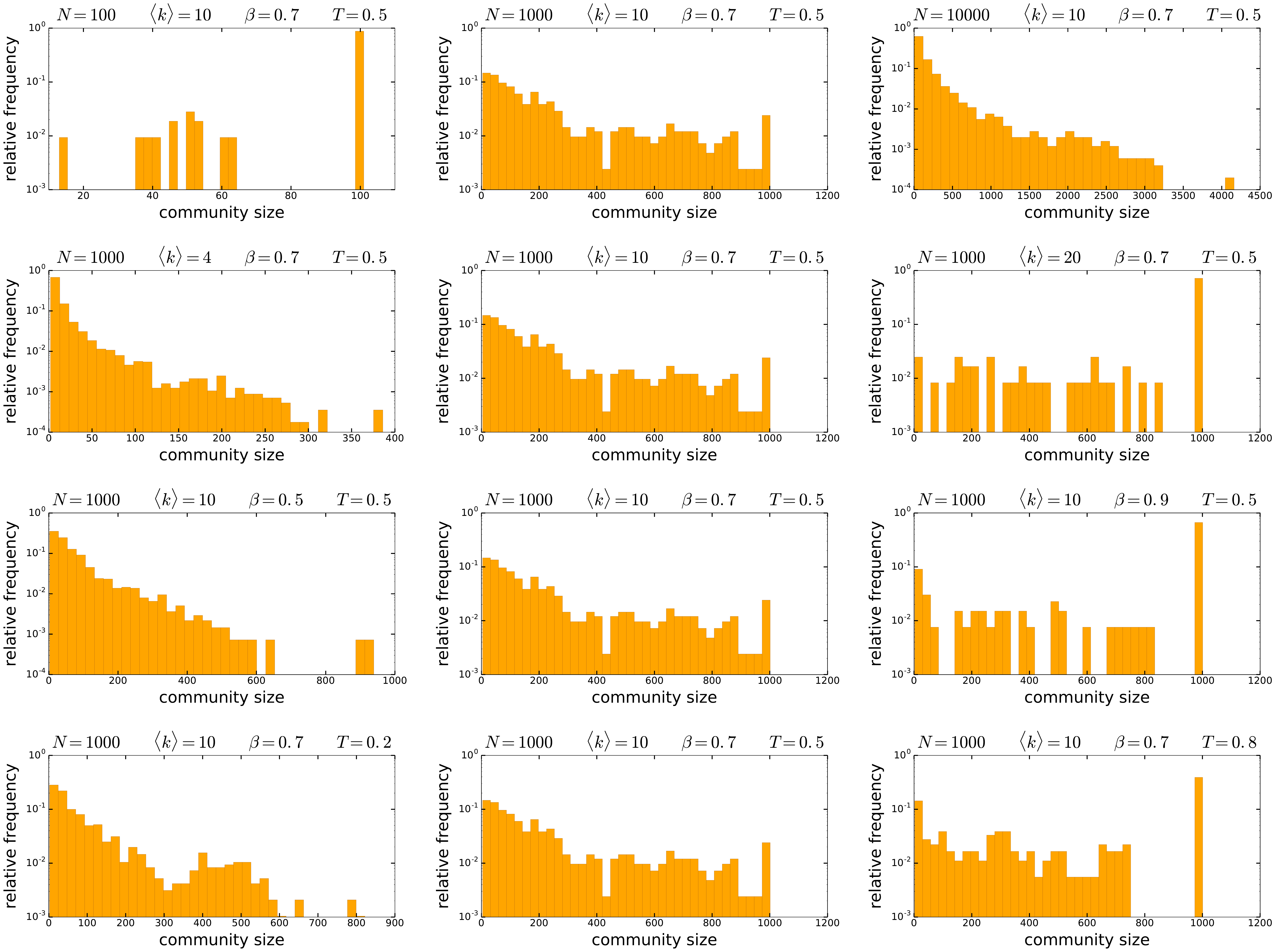}}
    \caption{{\bf The size distribution of the communities detected by the \textit{asynchronous label propagation} algorithm in 100 \textit{PSO} networks of different parametrisations.} The parameters of the network generation are listed in the title for each subplot. The curvature of the hyperbolic plane $K$ was always set to $-1$, i.e. we used $\zeta=1$. Each row of the figure demonstrates the effect of the change in a given network generation parameter: from top to bottom, the number of nodes $N$, the expected average degree $\langle k\rangle=2m$, the popularity fading parameter $\beta$ and the temperature $T$.}
    \label{fig:PSO_alabpropGShist}
\end{figure}

\begin{figure}[hbt]
    \centering
    \makebox[\textwidth][c]{\includegraphics[width=1.15\textwidth]{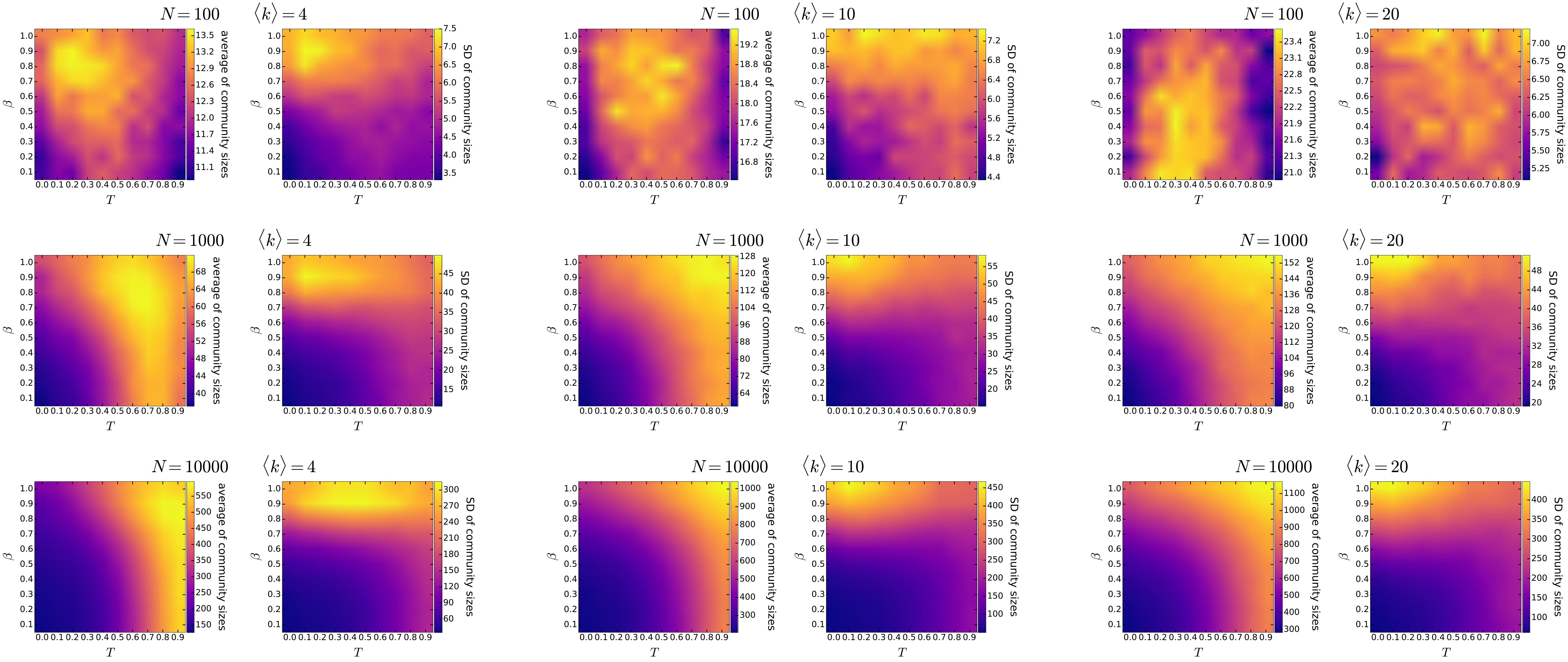}}
    \caption{{\bf The mean and the standard deviation of the size of communities detected by the \textit{Louvain} algorithm in 100 \textit{PSO} networks of different parametrisations.} Each pair of subplots depicts the effect of changing the popularity fading parameter $\beta$ and the temperature $T$, with the number of nodes $N$ and the expected average degree $\langle k\rangle=2m$ given in the title of the subplot pair. The curvature of the hyperbolic plane $K$ was always set to $-1$, i.e. we used $\zeta=1$.}
    \label{fig:PSO_LouvainGSavstd}
\end{figure}

\begin{figure}[hbt]
    \centering
    \makebox[\textwidth][c]{\includegraphics[width=1.15\textwidth]{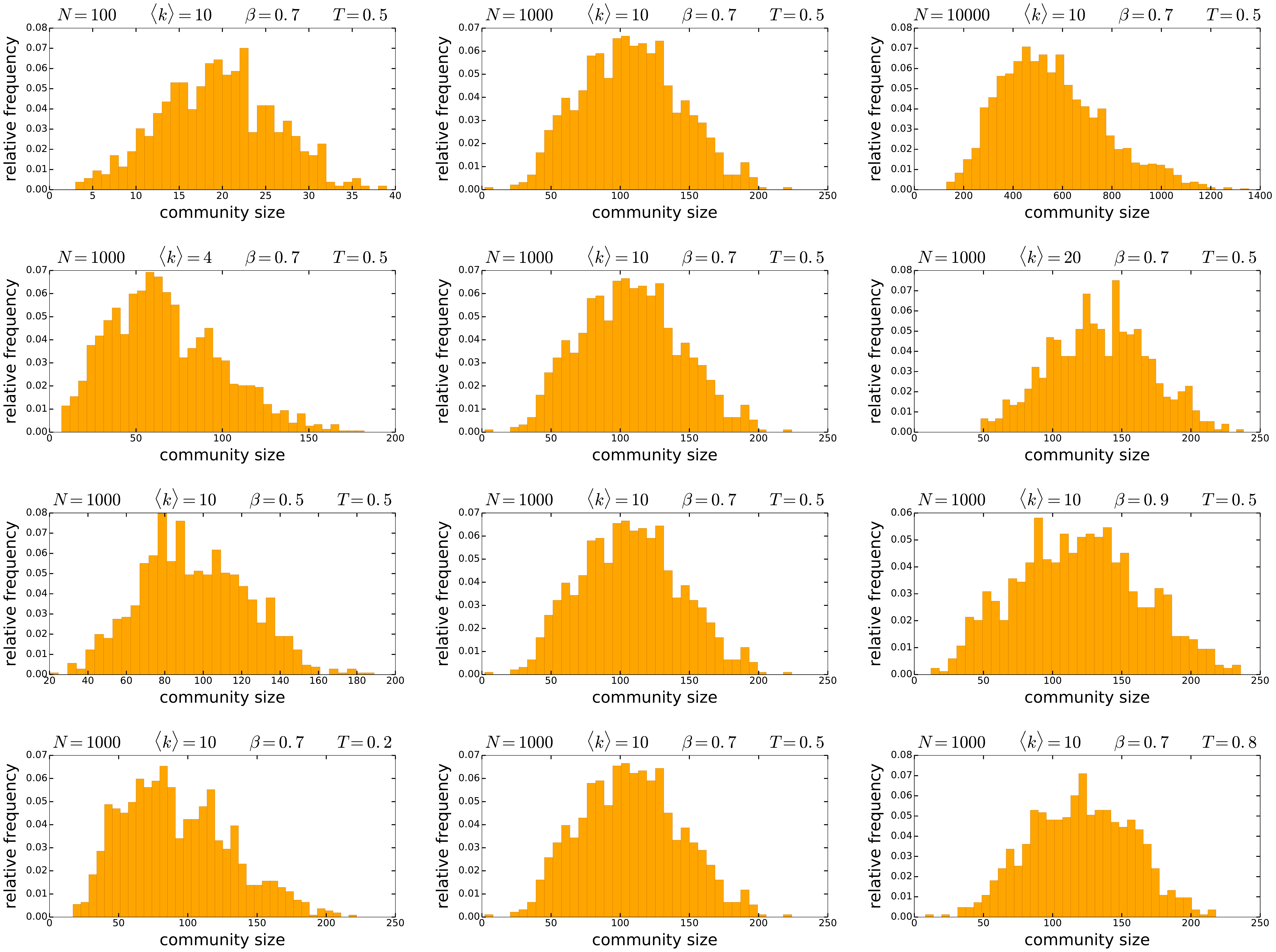}}
    \caption{{\bf The size distribution of the communities detected by the \textit{Louvain} algorithm in 100 \textit{PSO} networks of different parametrisations.} The parameters of the network generation are listed in the title for each subplot. The curvature of the hyperbolic plane $K$ was always set to $-1$, i.e. we used $\zeta=1$. Each row of the figure demonstrates the effect of the change in a given network generation parameter: from top to bottom, the number of nodes $N$, the expected average degree $\langle k\rangle=2m$, the popularity fading parameter $\beta$ and the temperature $T$.}
    \label{fig:PSO_LouvainGShist}
\end{figure}

\begin{figure}[hbt]
    \centering
    \makebox[\textwidth][c]{\includegraphics[width=1.15\textwidth]{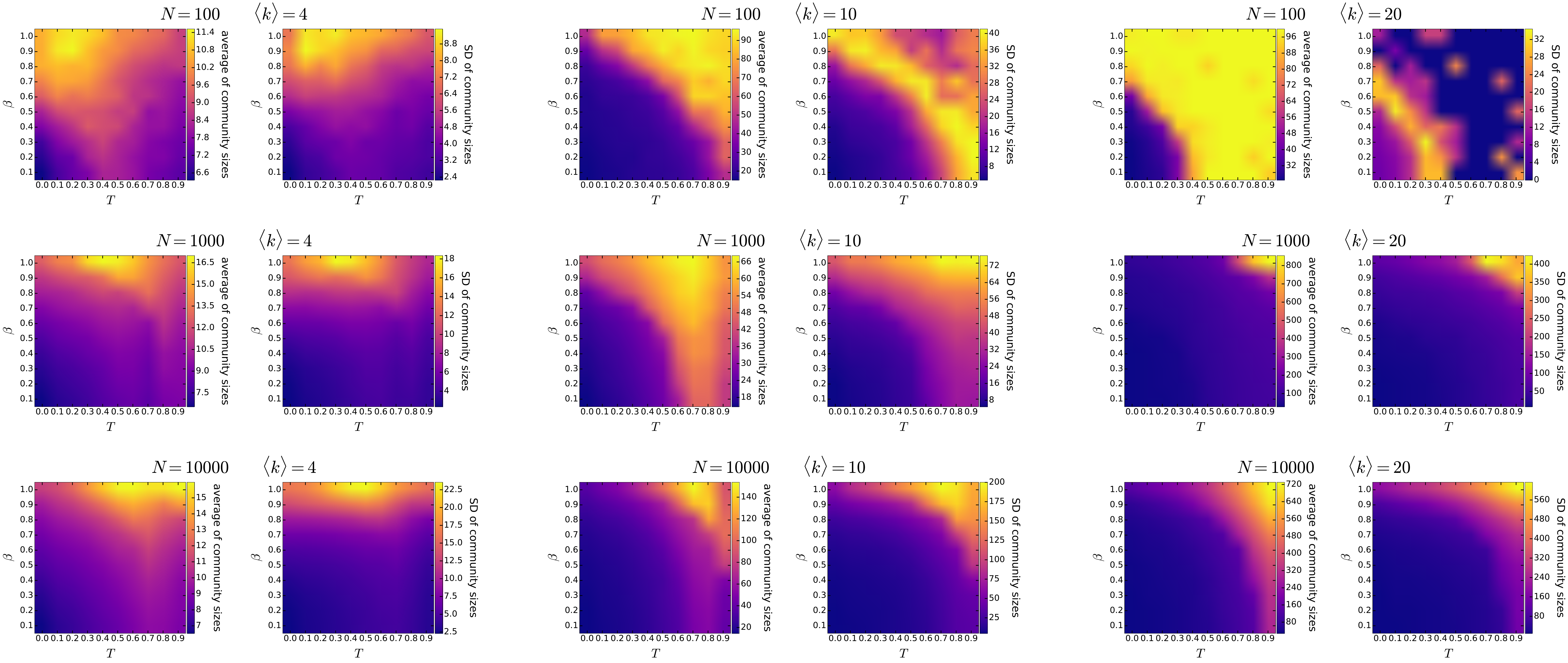}}
    \caption{{\bf The mean and the standard deviation of the size of communities detected by the \textit{Infomap} algorithm in 100 \textit{PSO} networks of different parametrisations.} Each pair of subplots depicts the effect of changing the popularity fading parameter $\beta$ and the temperature $T$, with the number of nodes $N$ and the expected average degree $\langle k\rangle=2m$ given in the title of the subplot pair. The curvature of the hyperbolic plane $K$ was always set to $-1$, i.e. we used $\zeta=1$.}
    \label{fig:PSO_InfomapGSavstd}
\end{figure}

\begin{figure}[hbt]
    \centering
    \makebox[\textwidth][c]{\includegraphics[width=1.15\textwidth]{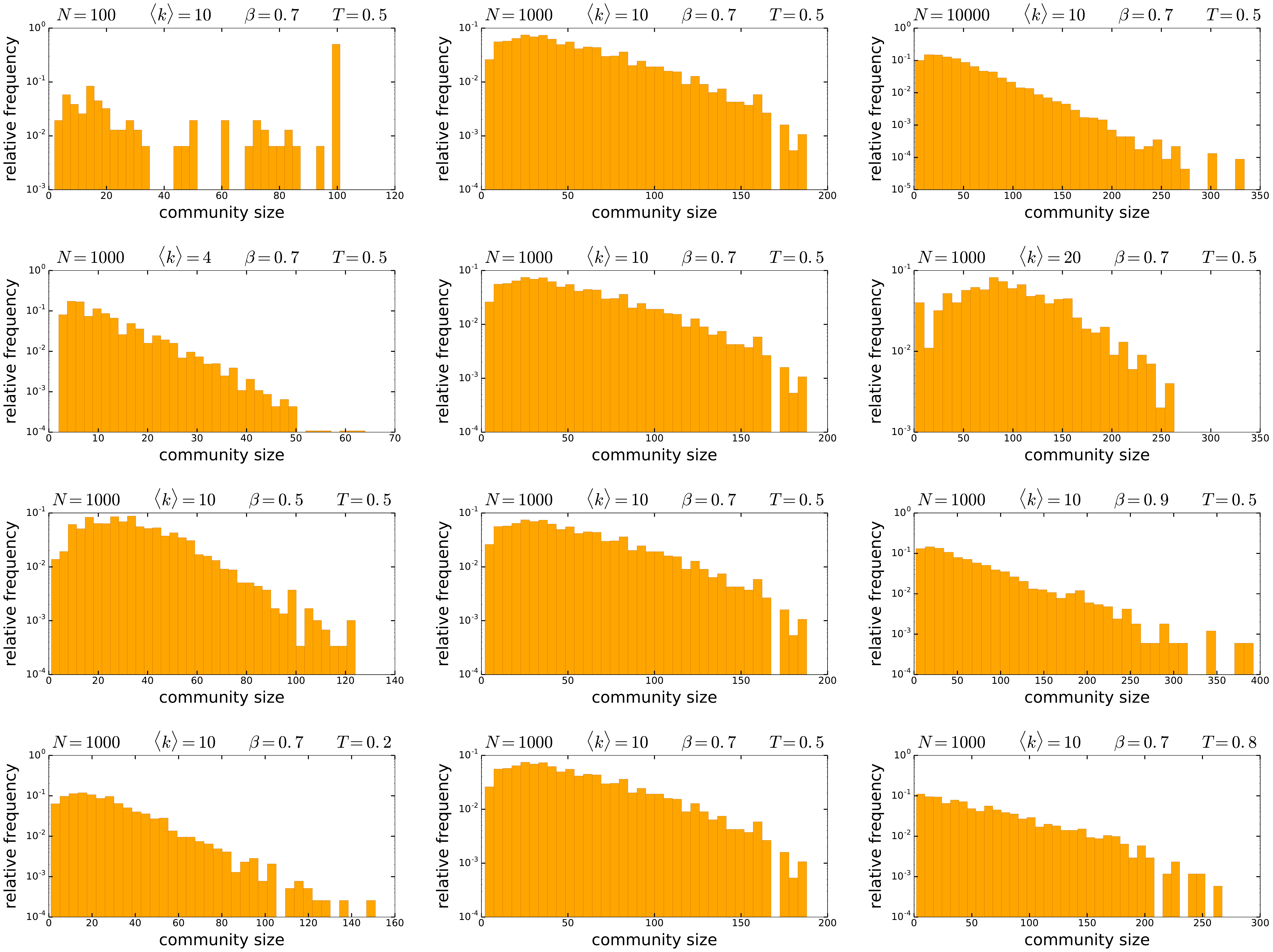}}
    \caption{{\bf The size distribution of the communities detected by the \textit{Infomap} algorithm in 100 \textit{PSO} networks of different parametrisations.} The parameters of the network generation are listed in the title for each subplot. The curvature of the hyperbolic plane $K$ was always set to $-1$, i.e. we used $\zeta=1$. Each row of the figure demonstrates the effect of the change in a given network generation parameter: from top to bottom, the number of nodes $N$, the expected average degree $\langle k\rangle=2m$, the popularity fading parameter $\beta$ and the temperature $T$.}
    \label{fig:PSO_InfomapGShist}
\end{figure}

\begin{figure}[hbt]
    \centering
    \makebox[\textwidth][c]{\includegraphics[width=1.15\textwidth]{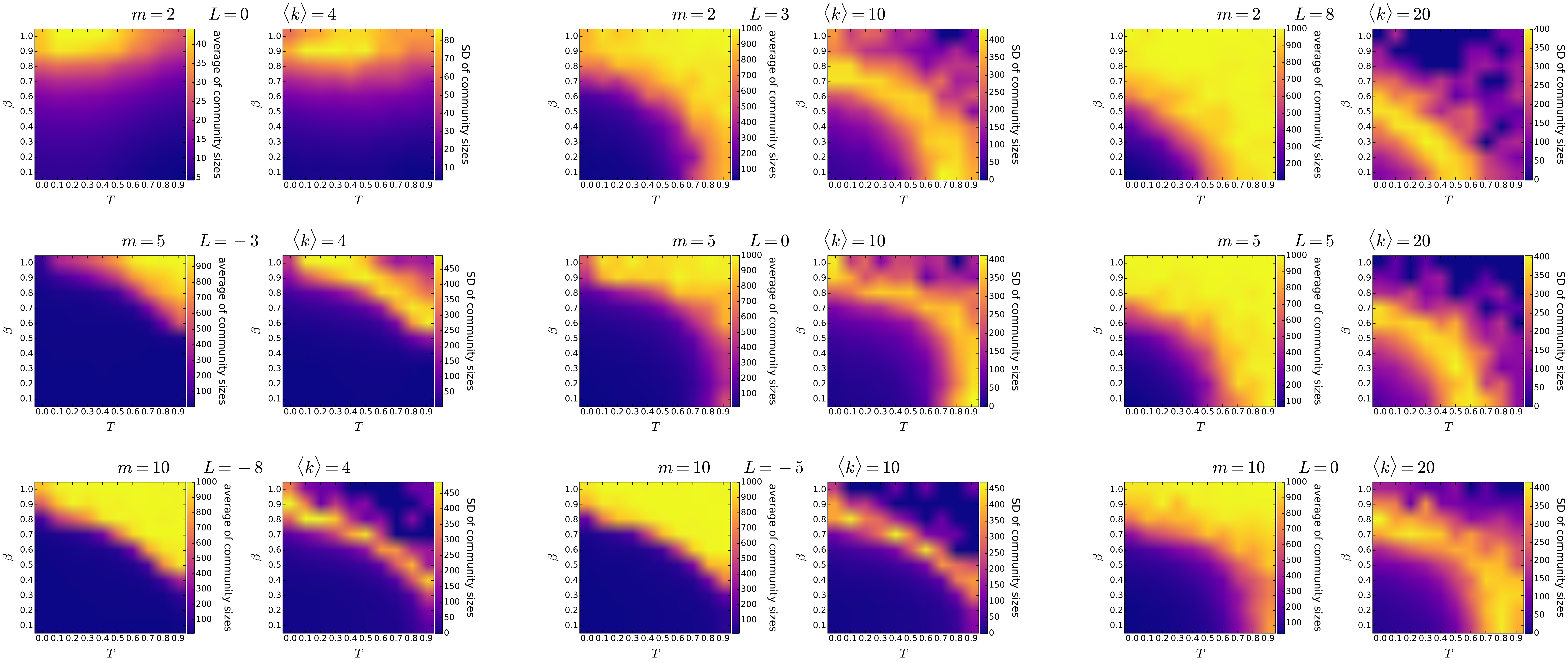}}
    \caption{{\bf The mean and the standard deviation of the size of communities detected by the \textit{asynchronous label propagation} algorithm in 100 \textit{E-PSO} networks of different parametrisations.} Each pair of subplots depicts the effect of changing the popularity fading parameter $\beta$ and the temperature $T$, with the parameters $m$ and $L$ given in the title of the subplot pair together with the corresponding expected average degree $\langle k\rangle=2(m+L)$. The number of nodes $N$ was 1000 in each case. The curvature of the hyperbolic plane $K$ was always set to $-1$, i.e. we used $\zeta=1$.}
    \label{fig:EPSO_alabpropGSavstd}
\end{figure}

\begin{figure}[hbt]
    \centering
    \makebox[\textwidth][c]{\includegraphics[width=1.15\textwidth]{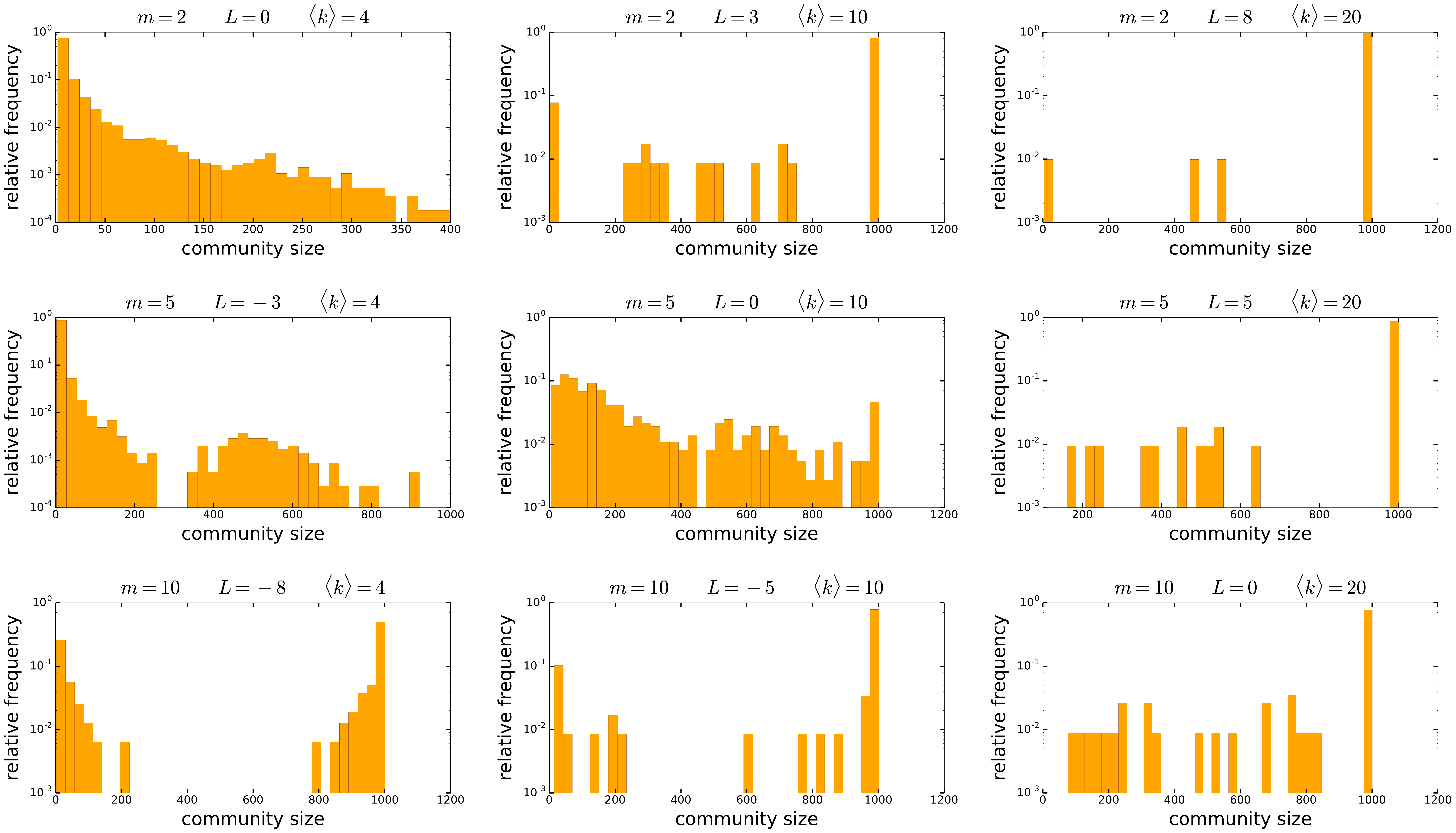}}
    \caption{{\bf The size distribution of the communities detected by the \textit{asynchronous label propagation} algorithm in 100 \textit{E-PSO} networks of different parametrisations.} We used $\zeta=1$, i.e. $K=-1$ as the curvature of the hyperbolic plane, the number of nodes $N$ was 1000, the popularity fading parameter $\beta$ was 0.7 and the temperature $T$ was 0.5 in each case. The parameters $m$ and $L$ are given in the title for each subplot together with the corresponding expected average degree $\langle k\rangle=2(m+L)$.}
    \label{fig:EPSO_alabpropGShist}
\end{figure}

\begin{figure}[hbt]
    \centering
    \makebox[\textwidth][c]{\includegraphics[width=1.15\textwidth]{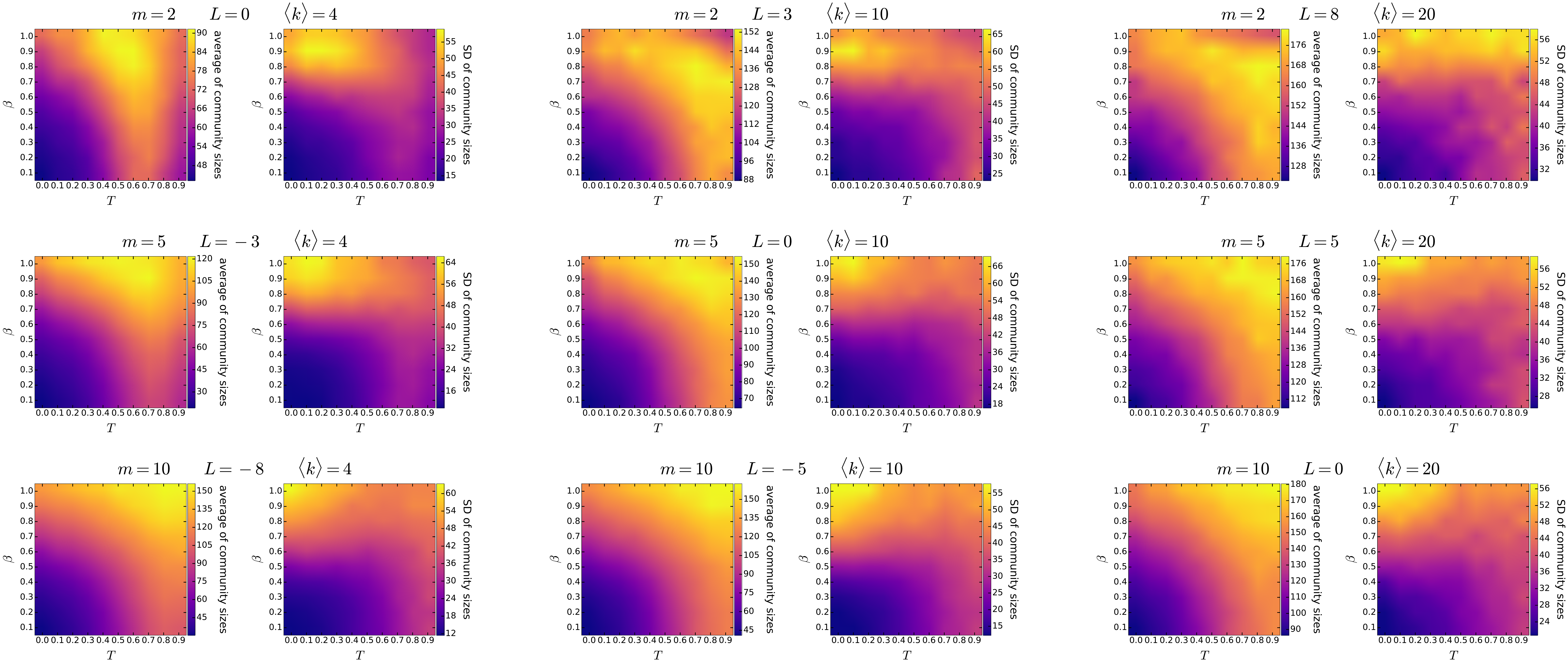}}
    \caption{{\bf The mean and the standard deviation of the size of communities detected by the \textit{Louvain} algorithm in 100 \textit{E-PSO} networks of different parametrisations.} Each pair of subplots depicts the effect of changing the popularity fading parameter $\beta$ and the temperature $T$, with the parameters $m$ and $L$ given in the title of the subplot pair together with the corresponding expected average degree $\langle k\rangle=2(m+L)$. The number of nodes $N$ was 1000 in each case. The curvature of the hyperbolic plane $K$ was always set to $-1$, i.e. we used $\zeta=1$.}
    \label{fig:EPSO_LouvainGSavstd}
\end{figure}

\begin{figure}[hbt]
    \centering
    \makebox[\textwidth][c]{\includegraphics[width=1.15\textwidth]{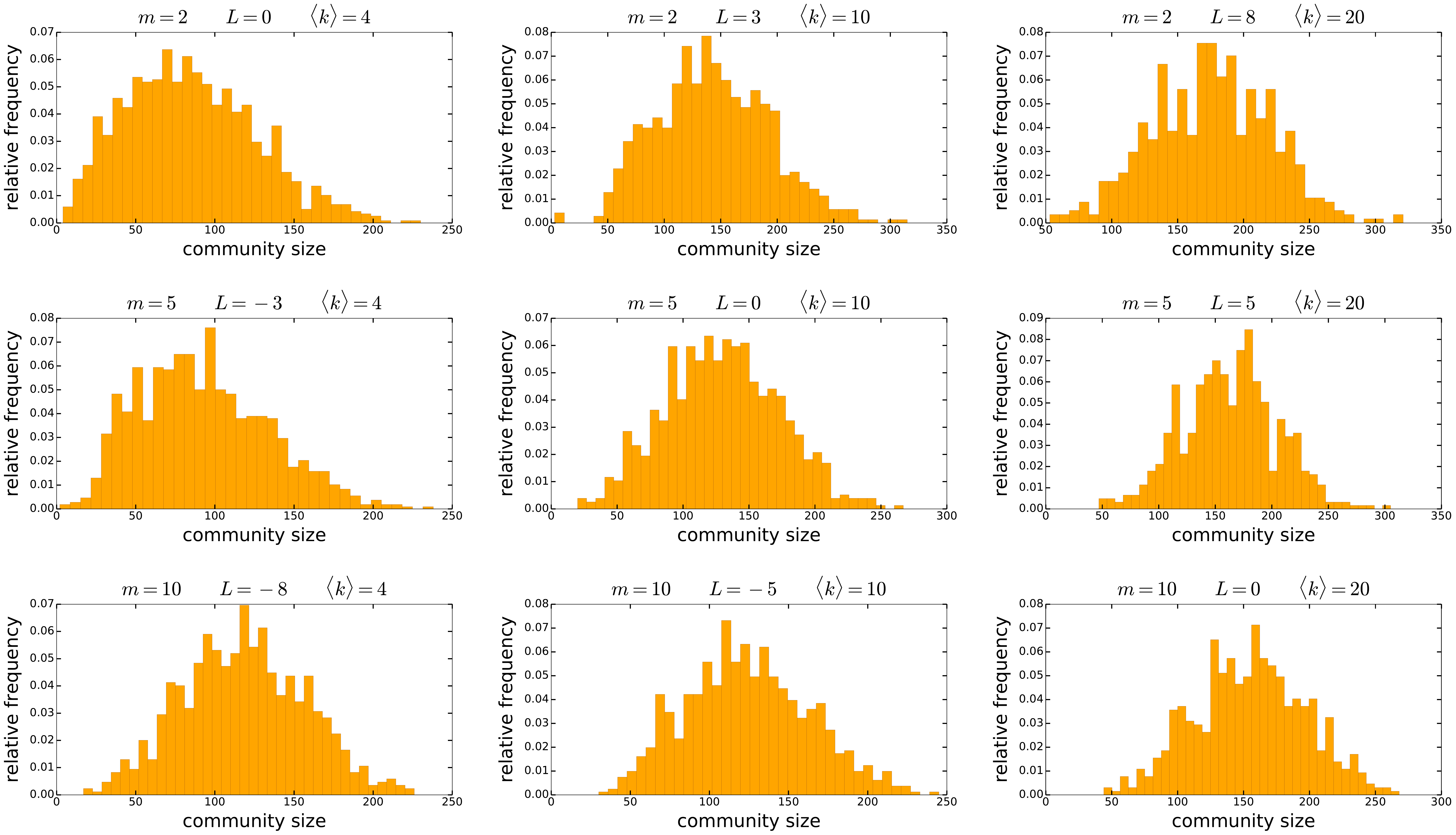}}
    \caption{{\bf The size distribution of the communities detected by the \textit{Louvain} algorithm in 100 \textit{E-PSO} networks of different parametrisations.} We used $\zeta=1$, i.e. $K=-1$ as the curvature of the hyperbolic plane, the number of nodes $N$ was 1000, the popularity fading parameter $\beta$ was 0.7 and the temperature $T$ was 0.5 in each case. The parameters $m$ and $L$ are given in the title for each subplot together with the corresponding expected average degree $\langle k\rangle=2(m+L)$.}
    \label{fig:EPSO_LouvainGShist}
\end{figure}

\begin{figure}[hbt]
    \centering
    \makebox[\textwidth][c]{\includegraphics[width=1.15\textwidth]{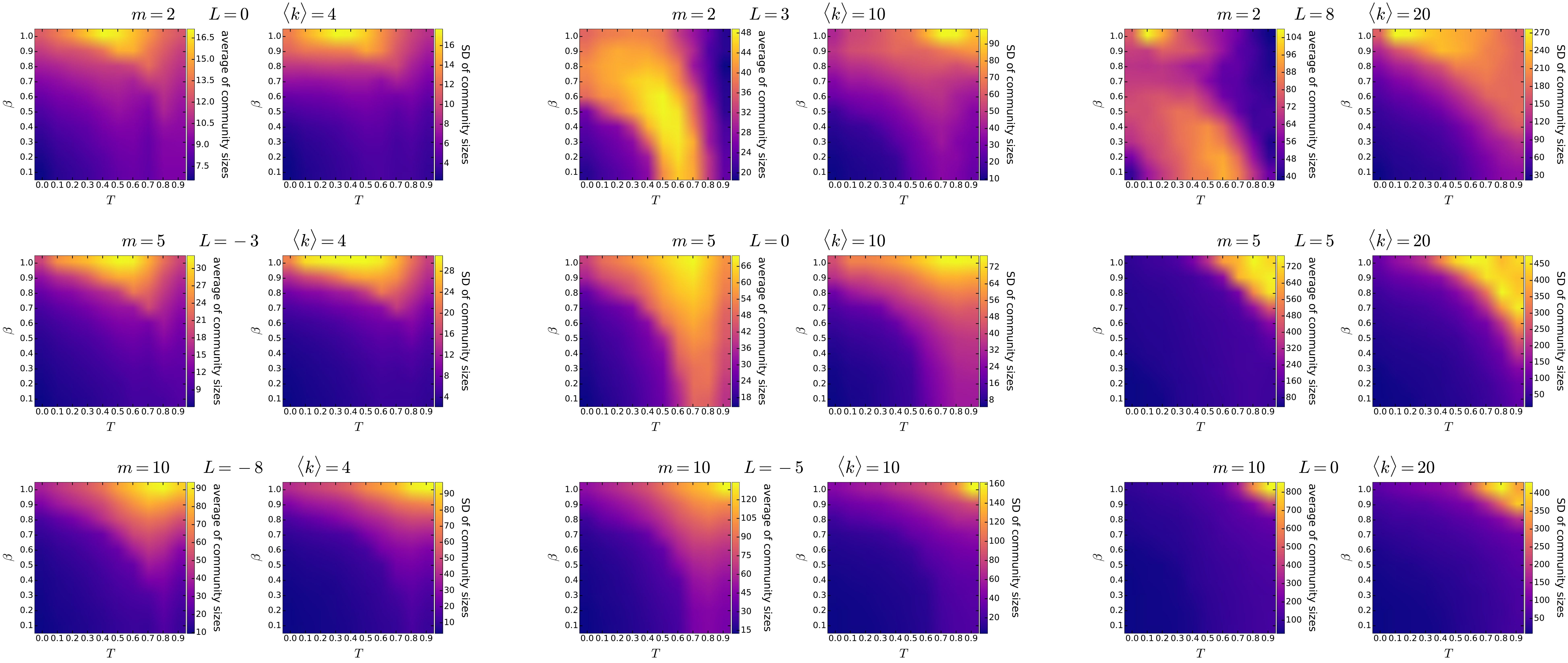}}
    \caption{{\bf The mean and the standard deviation of the size of communities detected by the \textit{Infomap} algorithm in 100 \textit{E-PSO} networks of different parametrisations.} Each pair of subplots depicts the effect of changing the popularity fading parameter $\beta$ and the temperature $T$, with the parameters $m$ and $L$ given in the title of the subplot pair together with the corresponding expected average degree $\langle k\rangle=2(m+L)$. The number of nodes $N$ was 1000 in each case. The curvature of the hyperbolic plane $K$ was always set to $-1$, i.e. we used $\zeta=1$.}
    \label{fig:EPSO_InfomapGSavstd}
\end{figure}

\begin{figure}[hbt]
    \centering
    \makebox[\textwidth][c]{\includegraphics[width=1.15\textwidth]{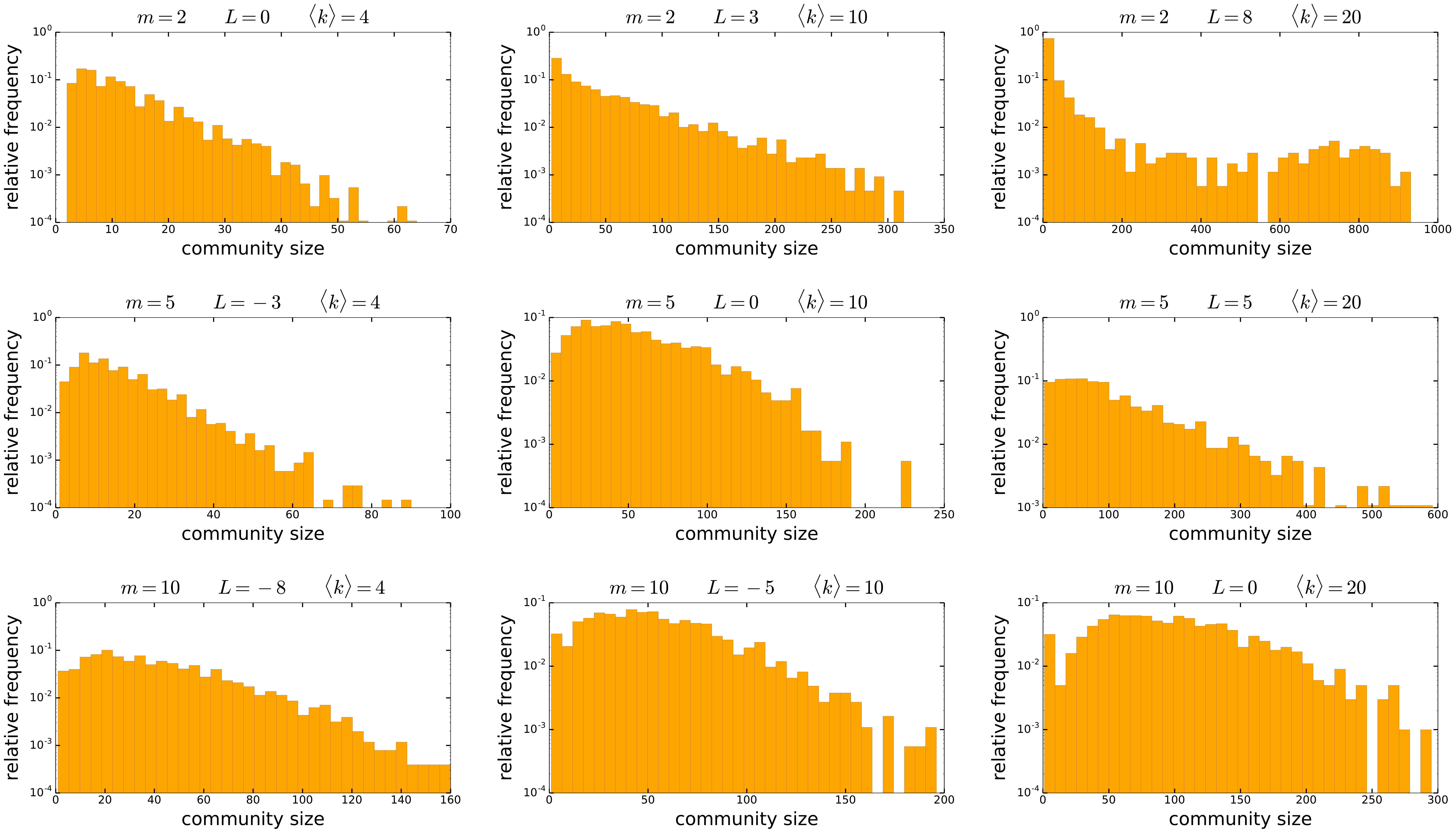}}
    \caption{{\bf The size distribution of the communities detected by the \textit{Infomap} algorithm in 100 \textit{E-PSO} networks of different parametrisations.} We used $\zeta=1$, i.e. $K=-1$ as the curvature of the hyperbolic plane, the number of nodes $N$ was 1000, the popularity fading parameter $\beta$ was 0.7 and the temperature $T$ was 0.5 in each case. The parameters $m$ and $L$ are given in the title for each subplot together with the corresponding expected average degree $\langle k\rangle=2(m+L)$.}
    \label{fig:EPSO_InfomapGShist}
\end{figure}

\begin{figure}[hbt]
    \centering
    \makebox[\textwidth][c]{\includegraphics[width=1.15\textwidth]{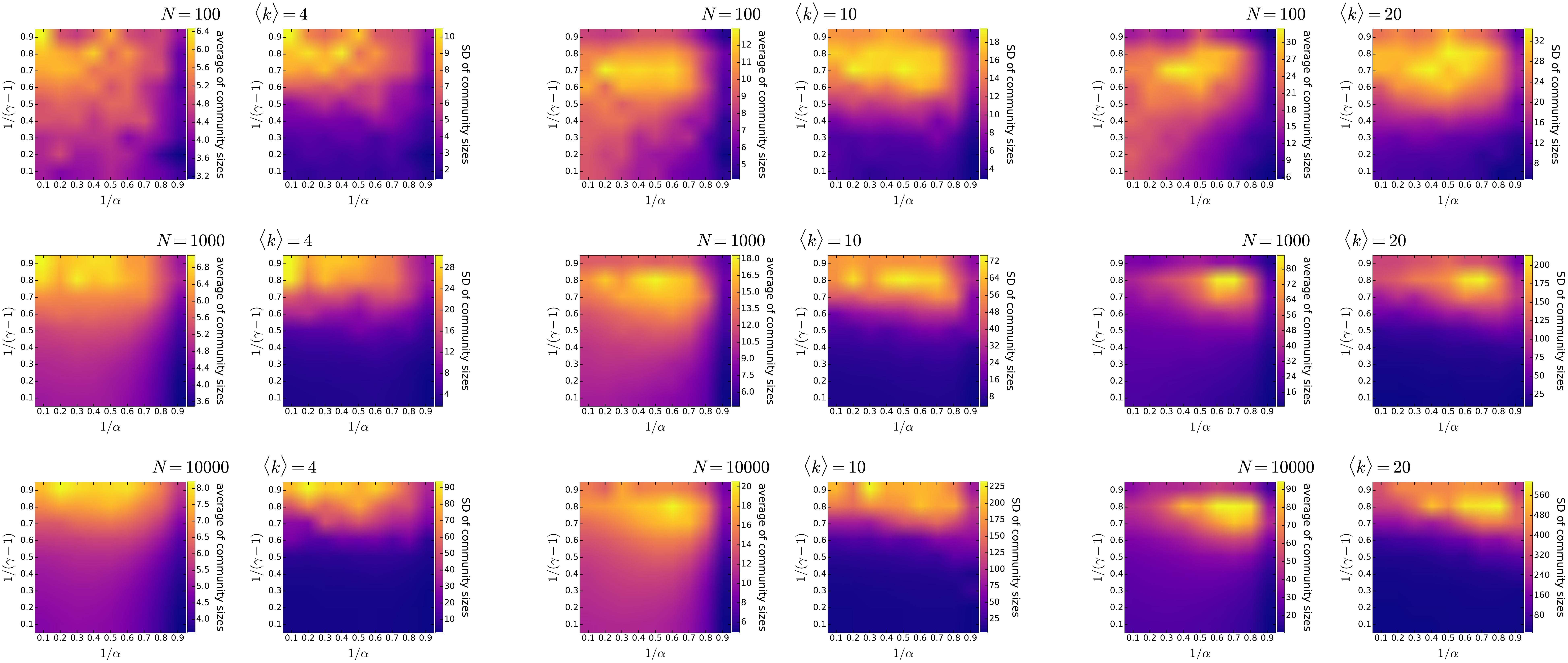}}
    \caption{{\bf The mean and the standard deviation of the size of communities detected by the \textit{asynchronous label propagation} algorithm in 100 $\mathbb{S}^1/\mathbb{H}^2$ networks of different parametrisations.} Each pair of subplots depicts the effect of changing $1/(\gamma-1)$ (equivalent to the popularity fading parameter $\beta$ in the E-PSO model) and $1/\alpha$ (analogous to the temperature $T$ in the E-PSO model), with the number of nodes $N$ and the expected average degree $\langle k\rangle$ given in the title of the subplot pair. We used $K=-1$ as the curvature of the hyperbolic plane in each case.}
    \label{fig:S1_alabpropGSavstd}
\end{figure}

\begin{figure}[hbt]
    \centering
    \makebox[\textwidth][c]{\includegraphics[width=1.15\textwidth]{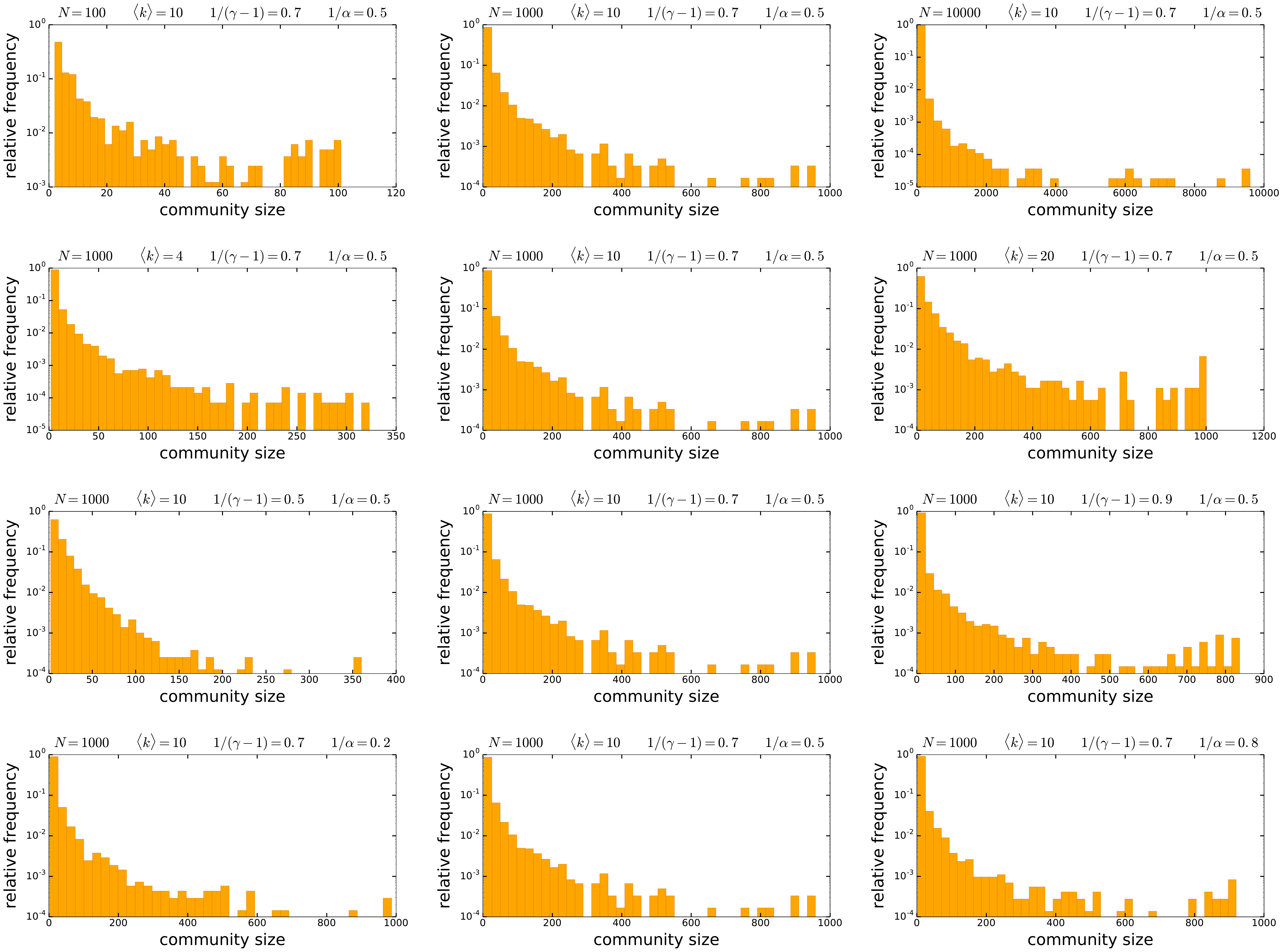}}
    \caption{{\bf The size distribution of the communities detected by the \textit{asynchronous label propagation} algorithm in 100 $\mathbb{S}^1/\mathbb{H}^2$ networks of different parametrisations.} The parameters of the network generation are listed in the title for each subplot. We used $K=-1$ as the curvature of the hyperbolic plane in each case. Each row of the figure demonstrates the effect of the change in a given network generation parameter: from top to bottom, the number of nodes $N$, the expected average degree $\langle k\rangle$, $1/(\gamma-1)$ (equivalent to the popularity fading parameter $\beta$ in the E-PSO model) and $1/\alpha$ (analogous to the temperature $T$ in the E-PSO model).}
    \label{fig:S1_alabpropGShist}
\end{figure}

\begin{figure}[hbt]
    \centering
    \makebox[\textwidth][c]{\includegraphics[width=1.15\textwidth]{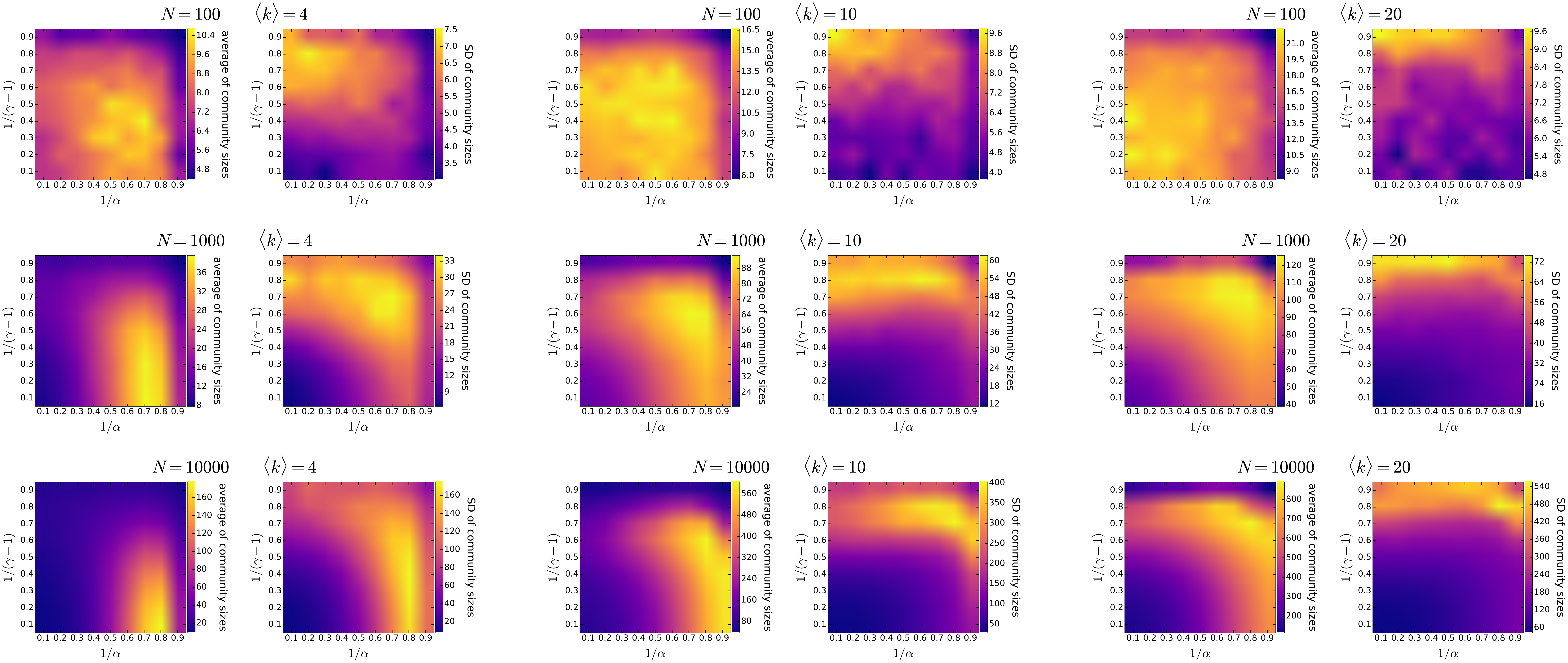}}
    \caption{{\bf The mean and the standard deviation of the size of communities detected by the \textit{Louvain} algorithm in 100 $\mathbb{S}^1/\mathbb{H}^2$ networks of different parametrisations.} Each pair of subplots depicts the effect of changing $1/(\gamma-1)$ (equivalent to the popularity fading parameter $\beta$ in the E-PSO model) and $1/\alpha$ (analogous to the temperature $T$ in the E-PSO model), with the number of nodes $N$ and the expected average degree $\langle k\rangle$ given in the title of the subplot pair. We used $K=-1$ as the curvature of the hyperbolic plane in each case.}
    \label{fig:S1_LouvainGSavstd}
\end{figure}

\begin{figure}[hbt]
    \centering
    \makebox[\textwidth][c]{\includegraphics[width=1.15\textwidth]{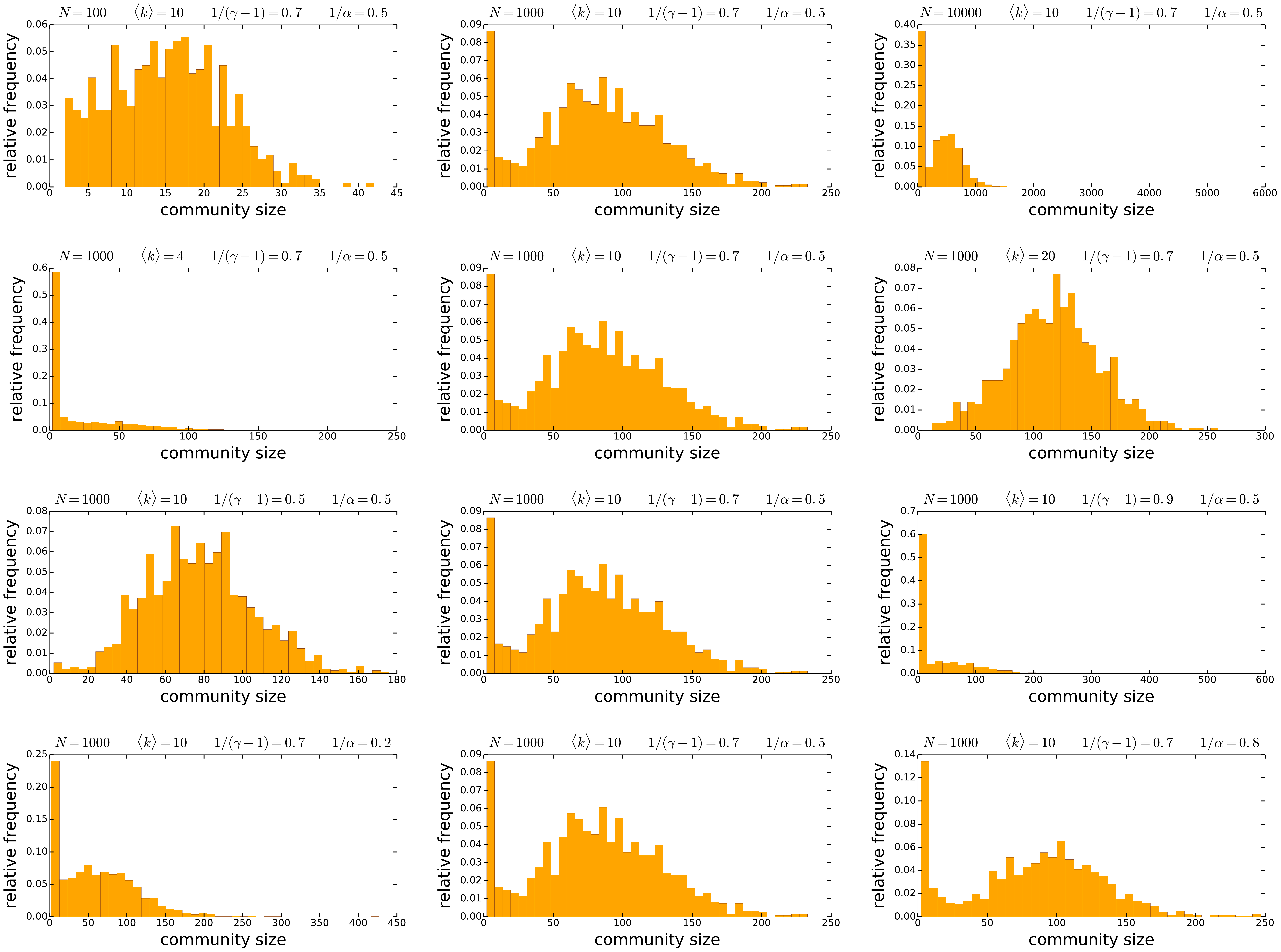}}
    \caption{{\bf The size distribution of the communities detected by the \textit{Louvain} algorithm in 100 $\mathbb{S}^1/\mathbb{H}^2$ networks of different parametrisations.} The parameters of the network generation are listed in the title for each subplot. We used $K=-1$ as the curvature of the hyperbolic plane in each case. Each row of the figure demonstrates the effect of the change in a given network generation parameter: from top to bottom, the number of nodes $N$, the expected average degree $\langle k\rangle$, $1/(\gamma-1)$ (equivalent to the popularity fading parameter $\beta$ in the E-PSO model) and $1/\alpha$ (analogous to the temperature $T$ in the E-PSO model).}
    \label{fig:S1_LouvainGShist}
\end{figure}

\begin{figure}[hbt]
    \centering
    \makebox[\textwidth][c]{\includegraphics[width=1.15\textwidth]{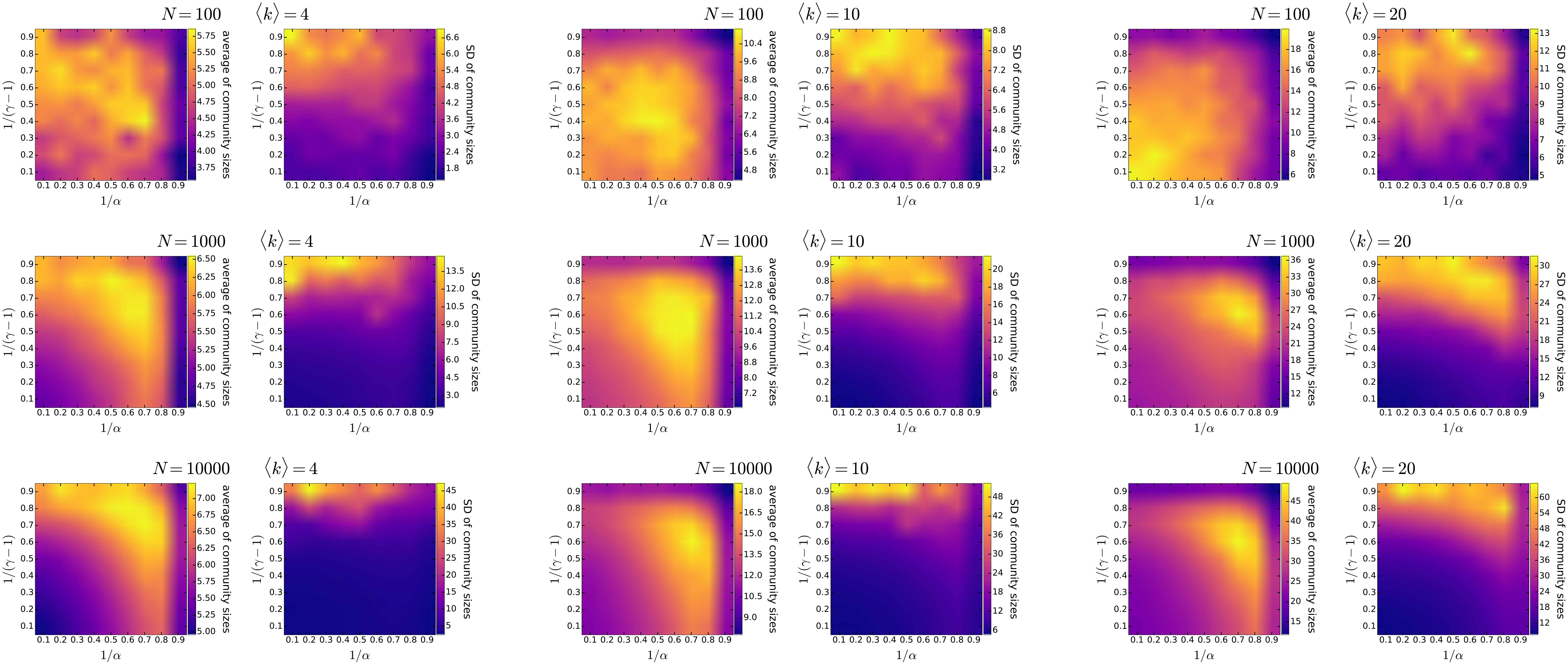}}
    \caption{{\bf The mean and the standard deviation of the size of communities detected by the \textit{Infomap} algorithm in 100 $\mathbb{S}^1/\mathbb{H}^2$ networks of different parametrisations.} Each pair of subplots depicts the effect of changing $1/(\gamma-1)$ (equivalent to the popularity fading parameter $\beta$ in the E-PSO model) and $1/\alpha$ (analogous to the temperature $T$ in the E-PSO model), with the number of nodes $N$ and the expected average degree $\langle k\rangle$ given in the title of the subplot pair. We used $K=-1$ as the curvature of the hyperbolic plane in each case.}
    \label{fig:S1_InfomapGSavstd}
\end{figure}

\begin{figure}[hbt]
    \centering
    \makebox[\textwidth][c]{\includegraphics[width=1.15\textwidth]{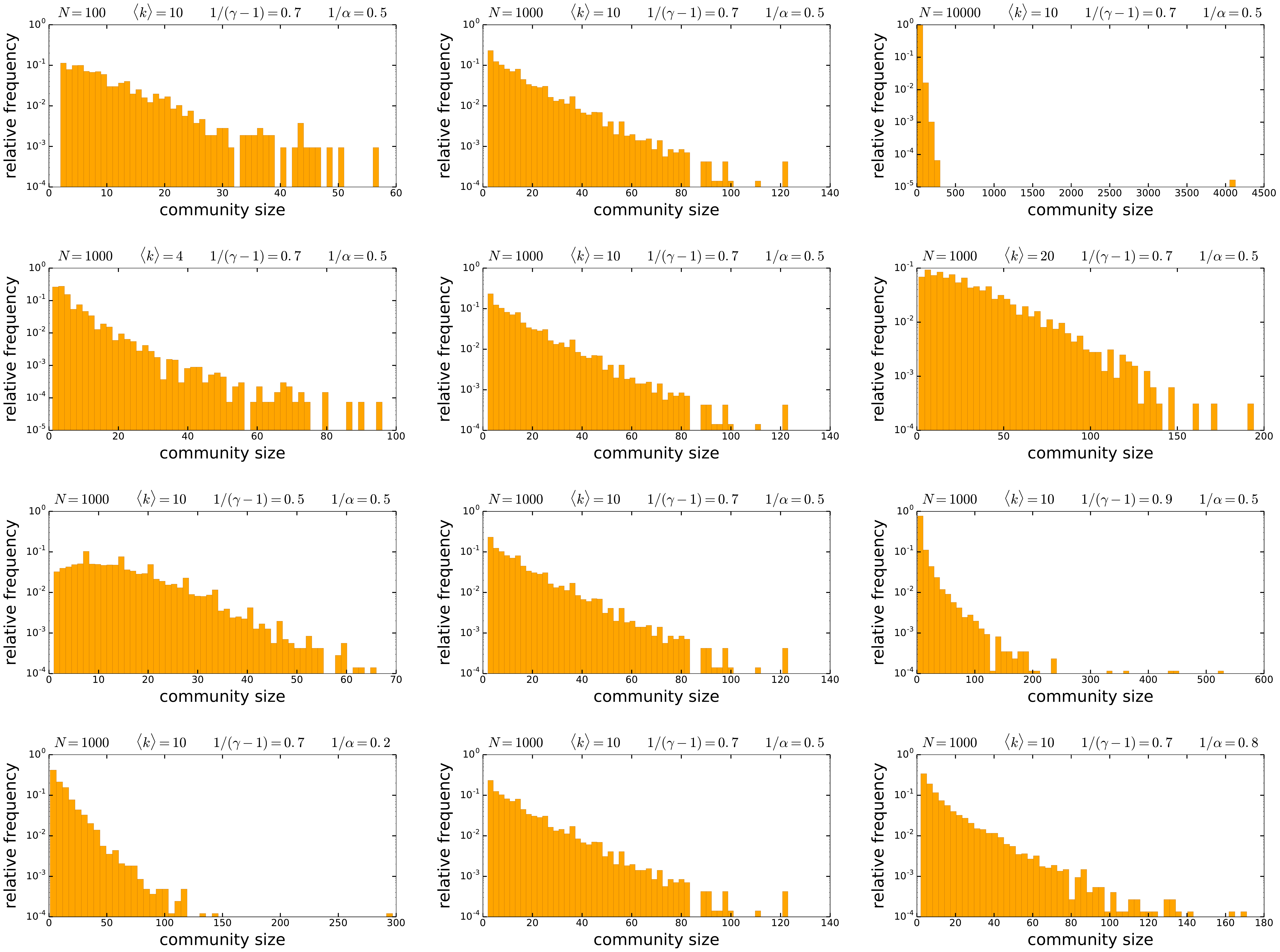}}
    \caption{{\bf The size distribution of the communities detected by the \textit{Infomap} algorithm in 100 $\mathbb{S}^1/\mathbb{H}^2$ networks of different parametrisations.} The parameters of the network generation are listed in the title for each subplot. We used $K=-1$ as the curvature of the hyperbolic plane in each case. Each row of the figure demonstrates the effect of the change in a given network generation parameter: from top to bottom, the number of nodes $N$, the expected average degree $\langle k\rangle$, $1/(\gamma-1)$ (equivalent to the popularity fading parameter $\beta$ in the E-PSO model) and $1/\alpha$ (analogous to the temperature $T$ in the E-PSO model).}
    \label{fig:S1_InfomapGShist}
\end{figure}

Here we present the characteristics of the community size distributions obtained with the asynchronous label propagation~\cite{alabprop}, the Louvain~\cite{Louvain} and the Infomap~\cite{Infomap} algorithms for the PSO \cite{PSO}, E-PSO \cite{EPSO_HyperMap,our_embedding} and $\mathbb{S}^1/\mathbb{H}^2$~\cite{S1,S1H2_Mercator} networks of various parameter combinations. Each community detection algorithm was executed once for each network. The isolated nodes emerging in the case of the $\mathbb{S}^1/\mathbb{H}^2$ model and occasionally also in the networks generated by the E-PSO model of $L<0$ were removed before the community detection, meaning that the actual size of the examined networks does not necessarily reach the number of nodes $N$ inputted in these models. We generated 100 networks with each parametrisation and investigated the sample created by assembling the occurring community sizes from all the 100 networks. Figures \ref{fig:PSO_alabpropGSavstd}–\ref{fig:S1_InfomapGShist} display how the mean and the standard deviation of this sample depend on the network generation parameters, as well as some examples for the corresponding community size distributions. Figures \ref{fig:PSO_alabpropGSavstd}–\ref{fig:PSO_InfomapGShist} deal with the E-PSO model with $L=0$ (i.e. the PSO model), figures \ref{fig:EPSO_alabpropGSavstd}–\ref{fig:EPSO_InfomapGShist} show the effect of changing the parameter $L$, and figures \ref{fig:S1_alabpropGSavstd}–\ref{fig:S1_InfomapGShist} refers to the $\mathbb{S}^1/\mathbb{H}^2$ model. The community size distribution is typically bell-shaped according to the Louvain algorithm, whereas rather skewed according to the Infomap and the asynchronous label propagation algorithms. In the parameter regime where we observed low $Q$ values, the community finding methods tend to merge the nodes into large communities of sizes comparable with $N$.
%When comparing the size distributions with each other, the presence of The asynchronous label propagation algorithms tends to merge a relatively large proportion of the network nodes into one group. %(EZ IGAZ?) Van valami általánosan jellemző paraméterfüggéáses tendencia, amit ide kéne írni?

%PSO (E-PSO L=0-val)

%E-PSO (Lnem0)

%S1

\clearpage

\section{Adjusted mutual information of the community structures found by different community detection algorithms}
\label{sect:detailedAMI}

\setcounter{figure}{0}
\setcounter{table}{0}
\setcounter{equation}{0}
\renewcommand{\thefigure}{S4.\arabic{figure}}
\renewcommand{\thetable}{S4.\arabic{table}}
\renewcommand{\theequation}{S4.\arabic{equation}}

\captionsetup[figure]{font=footnotesize,justification=justified,labelsep=period,labelfont=bf}

\begin{figure}[hbt]
    \centering
    \makebox[\textwidth][c]{\includegraphics[width=1.15\textwidth]{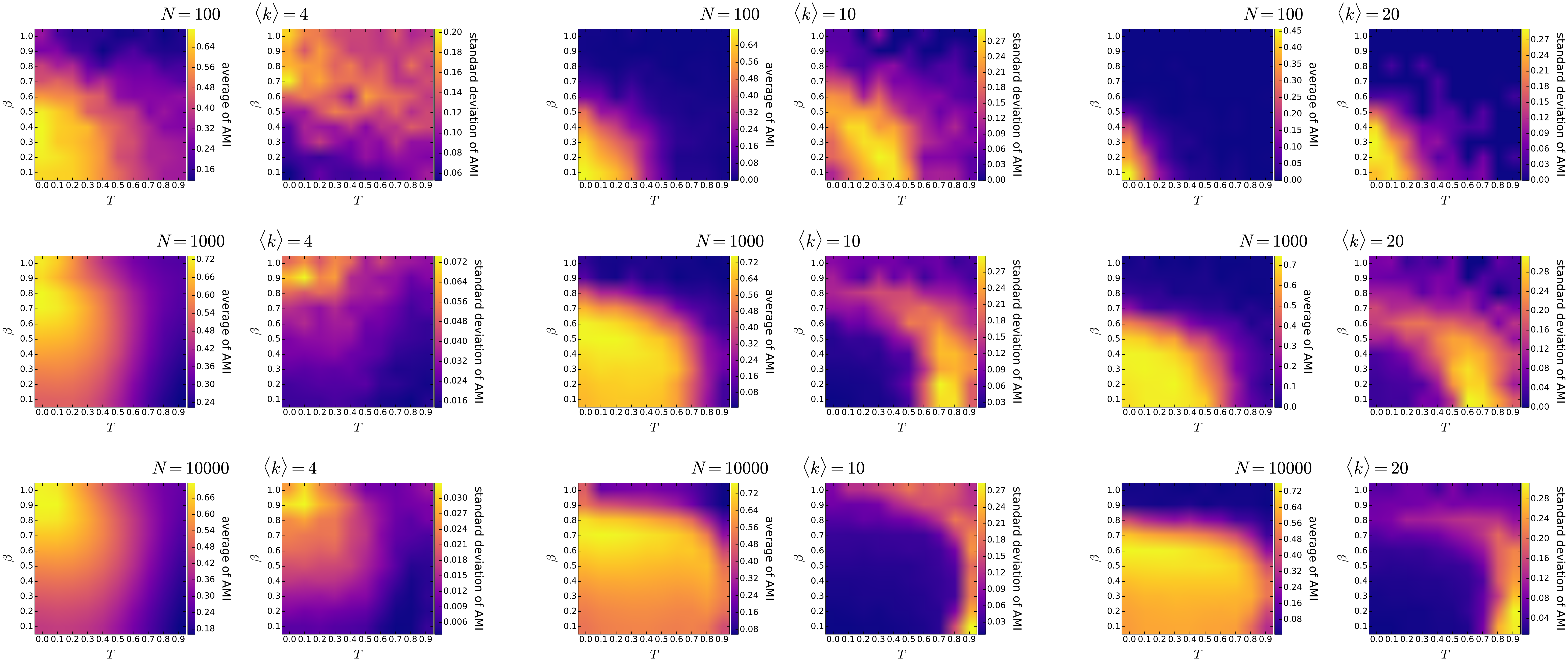}}
    \caption{{\bf The mean and the standard deviation of the adjusted mutual information of the two community structures detected by the \textit{asynchronous label propagation} and the \textit{Louvain} algorithms in 100 \textit{PSO} networks of different parametrisations.} Each pair of subplots depicts the effect of changing the popularity fading parameter $\beta$ and the temperature $T$, with the number of nodes $N$ and the expected average degree $\langle k\rangle=2m$ given in the title of the subplot pair. The curvature of the hyperbolic plane $K$ was always set to $-1$, i.e. we used $\zeta=1$.}
    \label{fig:AMI_alabprop_Louv_PSO}
\end{figure}

\begin{figure}[hbt]
    \centering
    \makebox[\textwidth][c]{\includegraphics[width=1.15\textwidth]{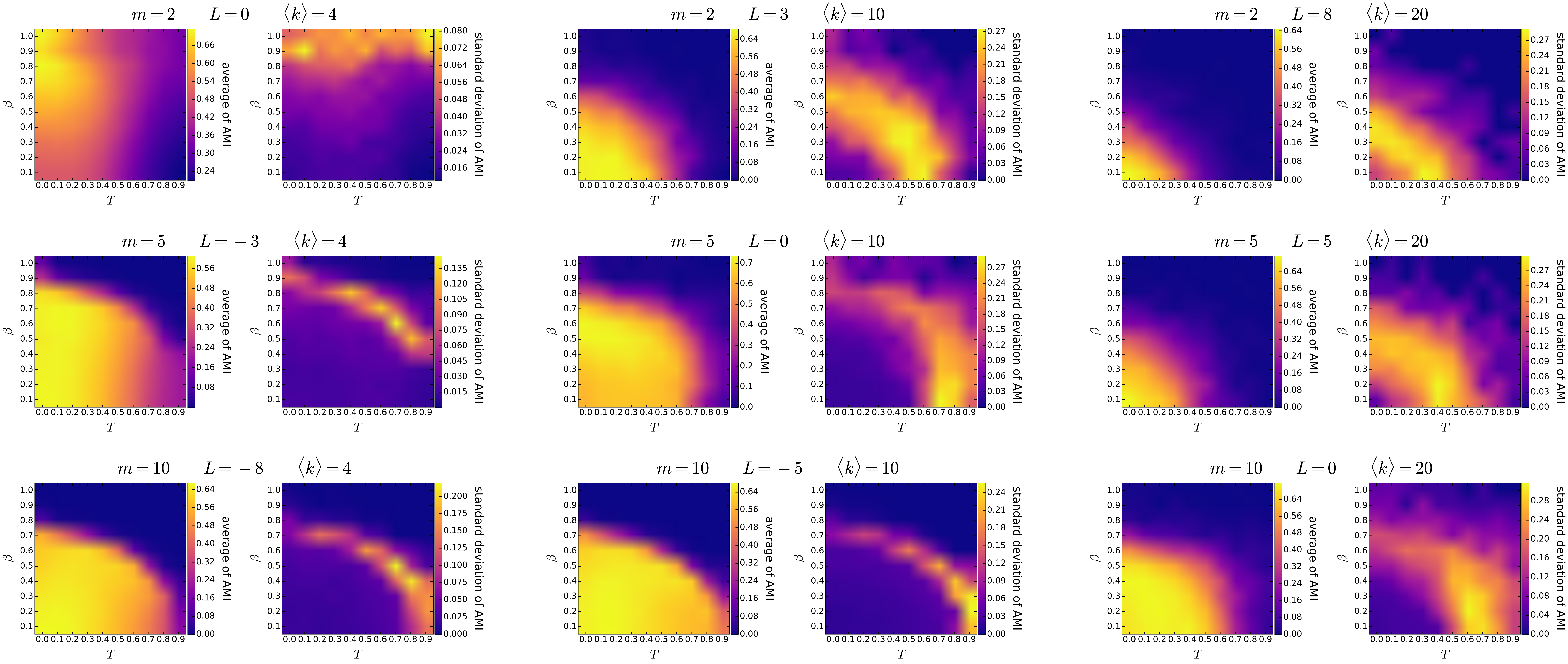}}
    \caption{{\bf The mean and the standard deviation of the adjusted mutual information of the two community structures detected by the \textit{asynchronous label propagation} and the \textit{Louvain} algorithms in 100 \textit{E-PSO} networks of different parametrisations.} Each pair of subplots depicts the effect of changing the popularity fading parameter $\beta$ and the temperature $T$, with the parameters $m$ and $L$ given in the title of the subplot pair together with the corresponding expected average degree $\langle k\rangle=2(m+L)$. The number of nodes $N$ was 1000 in each case. The curvature of the hyperbolic plane $K$ was always set to $-1$, i.e. we used $\zeta=1$.}
    \label{fig:AMI_alabprop_Louv_EPSO}
\end{figure}

\begin{figure}[hbt]
    \centering
    \makebox[\textwidth][c]{\includegraphics[width=1.15\textwidth]{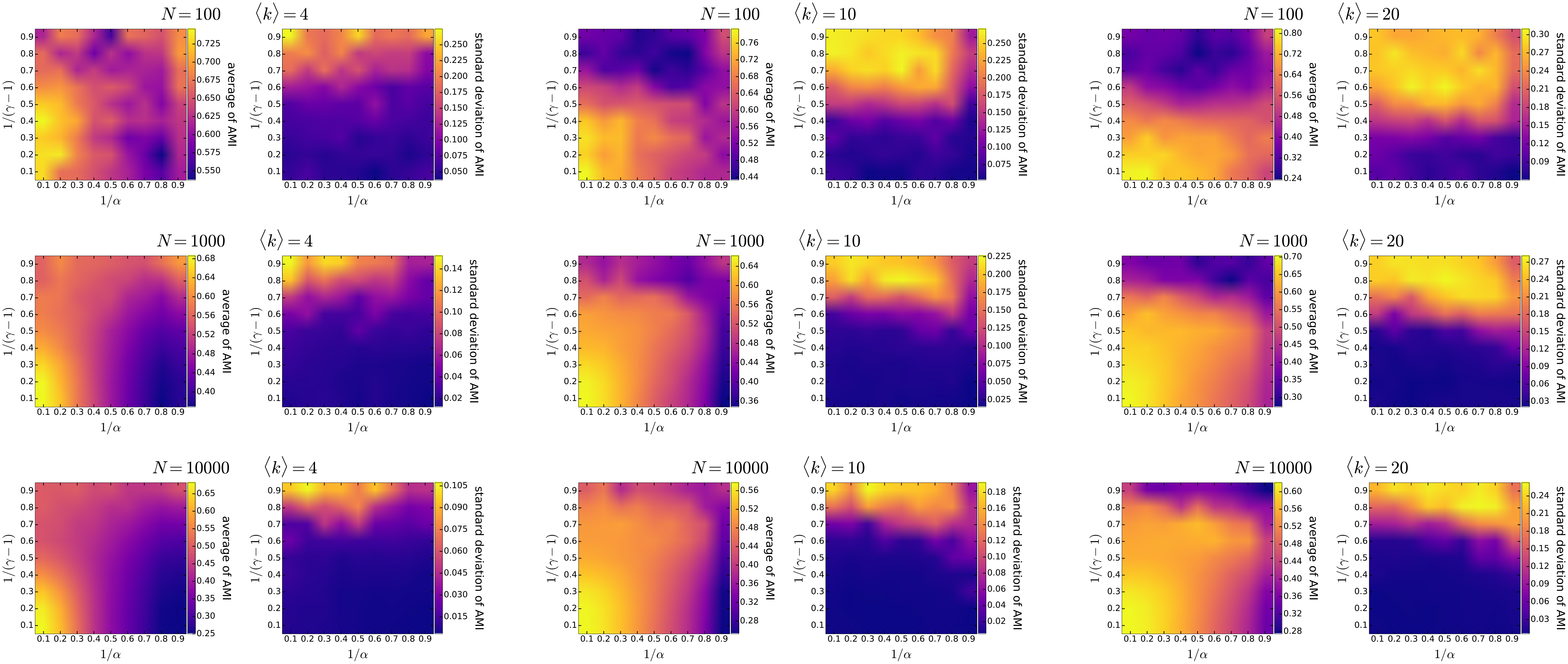}}
    \caption{{\bf The mean and the standard deviation of the adjusted mutual information of the two community structures detected by the \textit{asynchronous label propagation} and the \textit{Louvain} algorithms in 100 $\mathbb{S}^1/\mathbb{H}^2$ networks of different parametrisations.} Each pair of subplots depicts the effect of changing $1/(\gamma-1)$ (equivalent to the popularity fading parameter $\beta$ in the E-PSO model) and $1/\alpha$ (analogous to the temperature $T$ in the E-PSO model), with the number of nodes $N$ and the expected average degree $\langle k\rangle$ given in the title of the subplot pair. We used $K=-1$ as the curvature of the hyperbolic plane in each case.}
    \label{fig:AMI_alabprop_Louv_S1}
\end{figure}

\begin{figure}[hbt]
    \centering
    \makebox[\textwidth][c]{\includegraphics[width=1.15\textwidth]{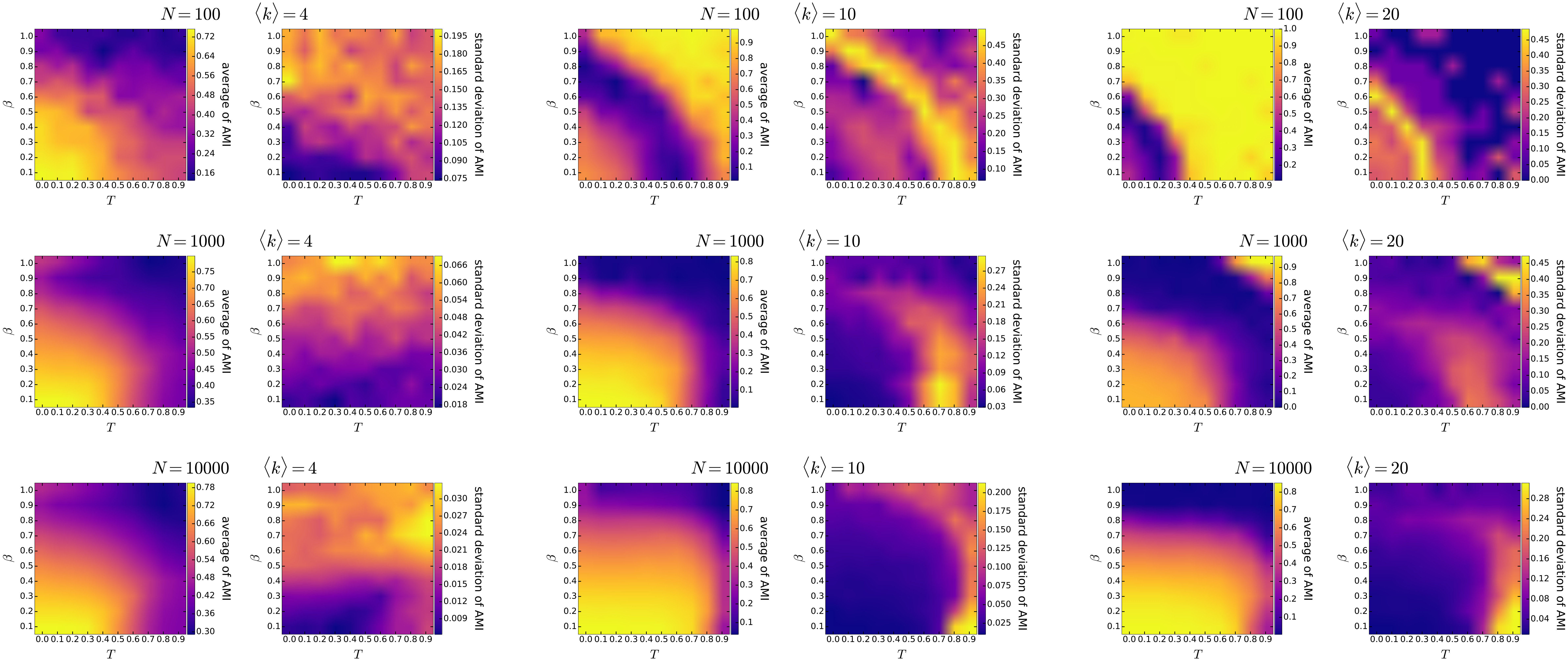}}
    \caption{{\bf The mean and the standard deviation of the adjusted mutual information of the two community structures detected by the \textit{asynchronous label propagation} and the \textit{Infomap} algorithms in 100 \textit{PSO} networks of different parametrisations.} Each pair of subplots depicts the effect of changing the popularity fading parameter $\beta$ and the temperature $T$, with the number of nodes $N$ and the expected average degree $\langle k\rangle=2m$ given in the title of the subplot pair. The curvature of the hyperbolic plane $K$ was always set to $-1$, i.e. we used $\zeta=1$.}
    \label{fig:AMI_alabprop_Inf_PSO}
\end{figure}

\begin{figure}[hbt]
    \centering
    \makebox[\textwidth][c]{\includegraphics[width=1.15\textwidth]{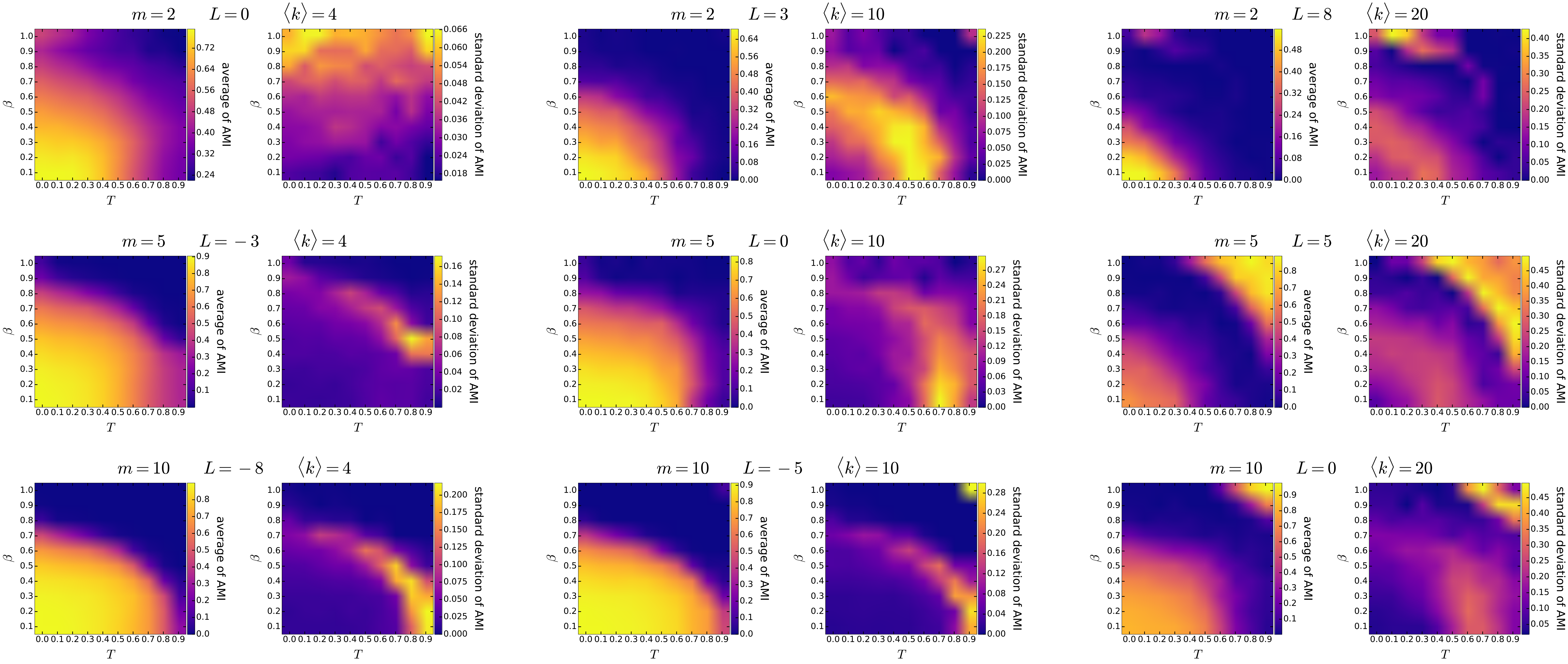}}
    \caption{{\bf The mean and the standard deviation of the adjusted mutual information of the two community structures detected by the \textit{asynchronous label propagation} and the \textit{Infomap} algorithms in 100 \textit{E-PSO} networks of different parametrisations.} Each pair of subplots depicts the effect of changing the popularity fading parameter $\beta$ and the temperature $T$, with the parameters $m$ and $L$ given in the title of the subplot pair together with the corresponding expected average degree $\langle k\rangle=2(m+L)$. The number of nodes $N$ was 1000 in each case. The curvature of the hyperbolic plane $K$ was always set to $-1$, i.e. we used $\zeta=1$.}
    \label{fig:AMI_alabprop_Inf_EPSO}
\end{figure}

\begin{figure}[hbt]
    \centering
    \makebox[\textwidth][c]{\includegraphics[width=1.15\textwidth]{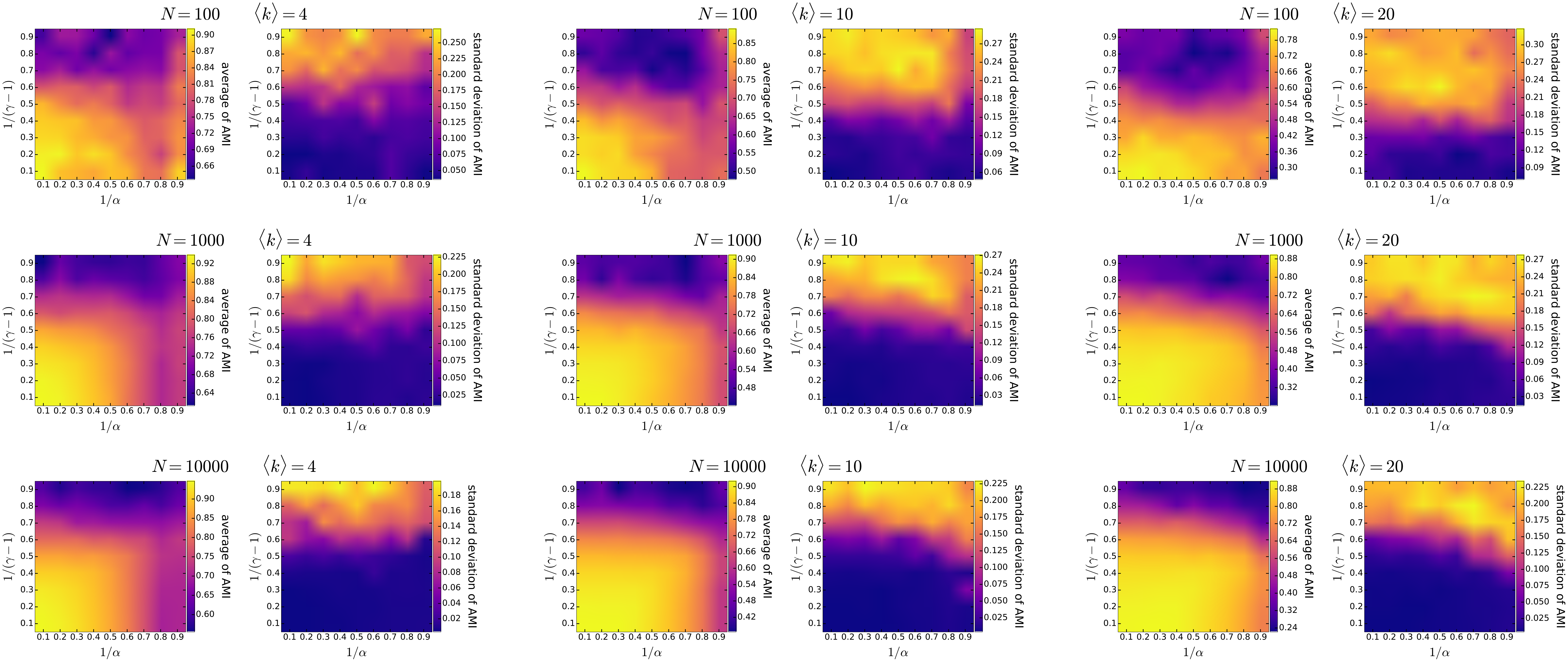}}
    \caption{{\bf The mean and the standard deviation of the adjusted mutual information of the two community structures detected by the \textit{asynchronous label propagation} and the \textit{Infomap} algorithms in 100 $\mathbb{S}^1/\mathbb{H}^2$ networks of different parametrisations.} Each pair of subplots depicts the effect of changing $1/(\gamma-1)$ (equivalent to the popularity fading parameter $\beta$ in the E-PSO model) and $1/\alpha$ (analogous to the temperature $T$ in the E-PSO model), with the number of nodes $N$ and the expected average degree $\langle k\rangle$ given in the title of the subplot pair. We used $K=-1$ as the curvature of the hyperbolic plane in each case.}
    \label{fig:AMI_alabprop_Inf_S1}
\end{figure}

\begin{figure}[hbt]
    \centering
    \makebox[\textwidth][c]{\includegraphics[width=1.15\textwidth]{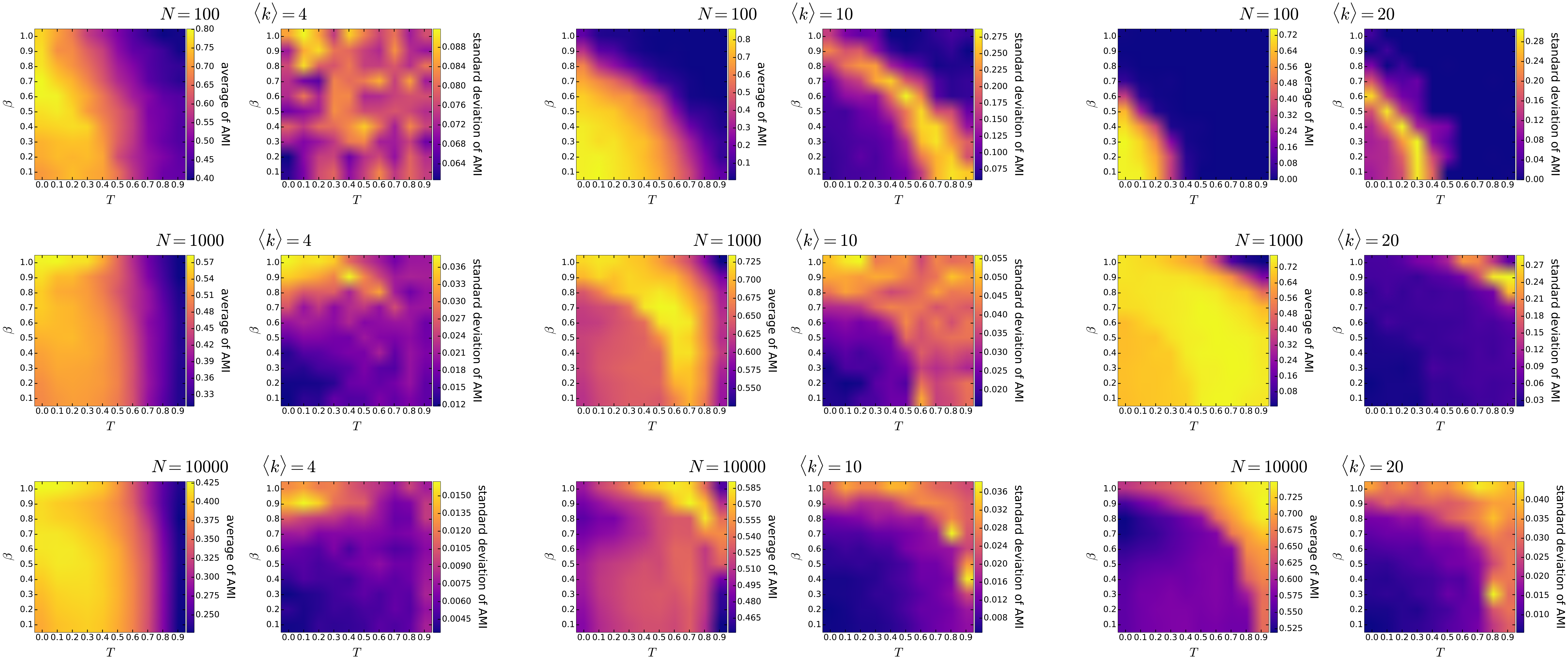}}
    \caption{{\bf The mean and the standard deviation of the adjusted mutual information of the two community structures detected by the \textit{Louvain} and the \textit{Infomap} algorithms in 100 \textit{PSO} networks of different parametrisations.} Each pair of subplots depicts the effect of changing the popularity fading parameter $\beta$ and the temperature $T$, with the number of nodes $N$ and the expected average degree $\langle k\rangle=2m$ given in the title of the subplot pair. The curvature of the hyperbolic plane $K$ was always set to $-1$, i.e. we used $\zeta=1$.}
    \label{fig:AMI_Louv_Inf_PSO}
\end{figure}

\begin{figure}[hbt]
    \centering
    \makebox[\textwidth][c]{\includegraphics[width=1.15\textwidth]{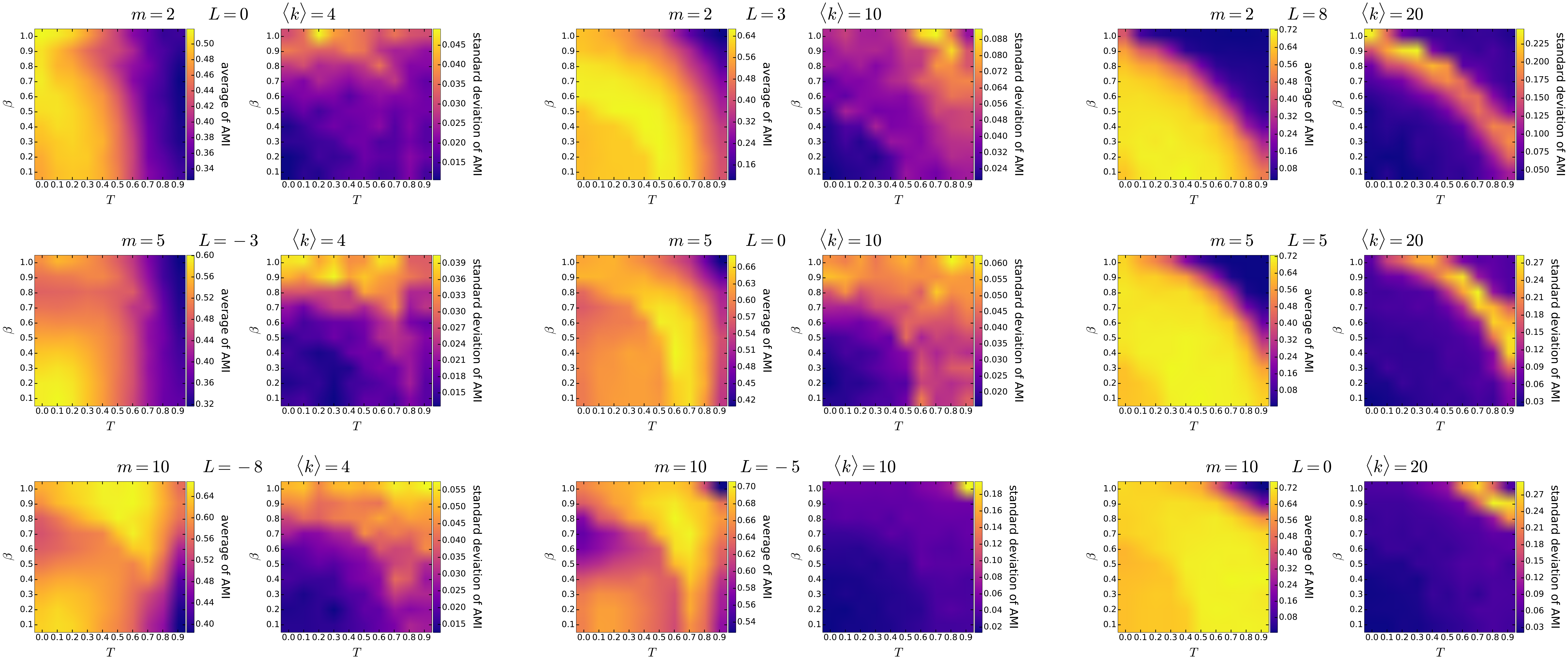}}
    \caption{{\bf The mean and the standard deviation of the adjusted mutual information of the two community structures detected by the \textit{Louvain} and the \textit{Infomap} algorithms in 100 \textit{E-PSO} networks of different parametrisations.} Each pair of subplots depicts the effect of changing the popularity fading parameter $\beta$ and the temperature $T$, with the parameters $m$ and $L$ given in the title of the subplot pair together with the corresponding expected average degree $\langle k\rangle=2(m+L)$. The number of nodes $N$ was 1000 in each case. The curvature of the hyperbolic plane $K$ was always set to $-1$, i.e. we used $\zeta=1$.}
    \label{fig:AMI_Louv_Inf_EPSO}
\end{figure}

\begin{figure}[hbt]
    \centering
    \makebox[\textwidth][c]{\includegraphics[width=1.15\textwidth]{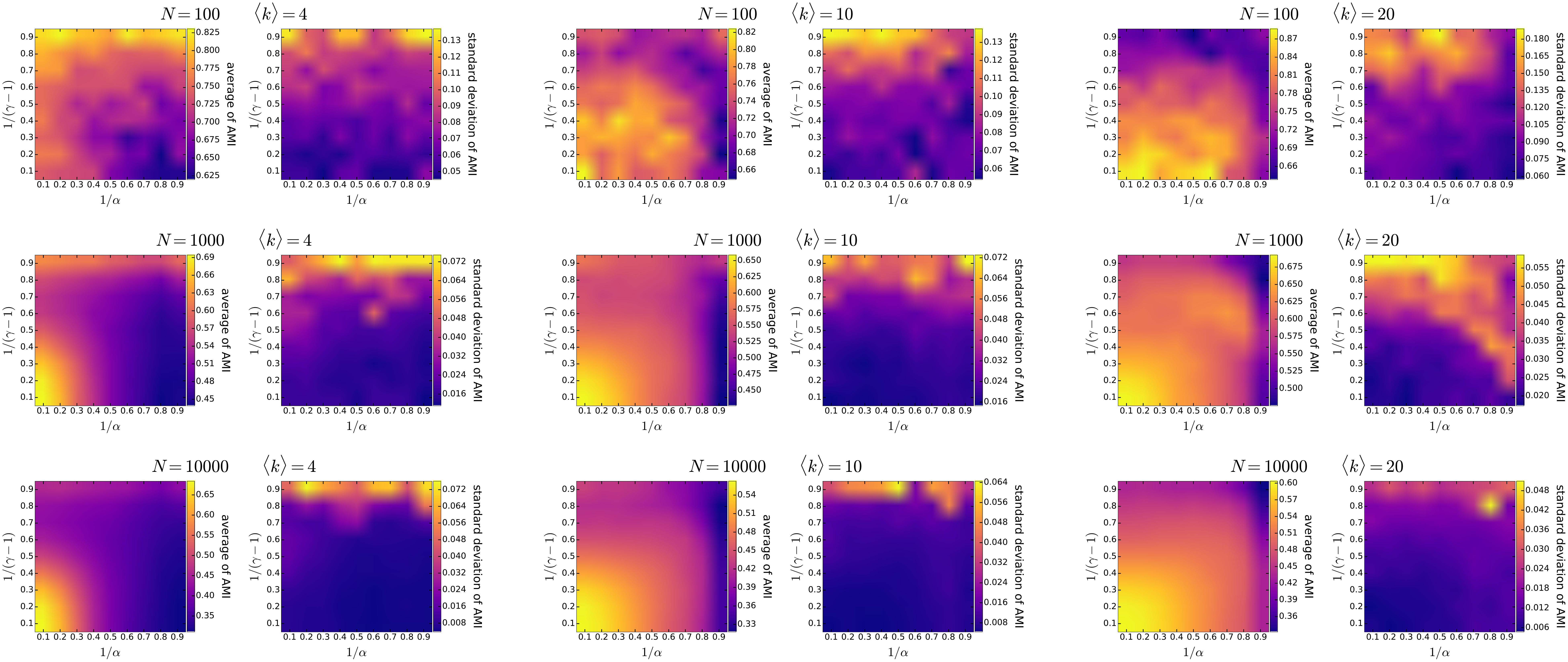}}
    \caption{{\bf The mean and the standard deviation of the adjusted mutual information of the two community structures detected by the \textit{Louvain} and the \textit{Infomap} algorithms in 100 $\mathbb{S}^1/\mathbb{H}^2$ networks of different parametrisations.} Each pair of subplots depicts the effect of changing $1/(\gamma-1)$ (equivalent to the popularity fading parameter $\beta$ in the E-PSO model) and $1/\alpha$ (analogous to the temperature $T$ in the E-PSO model), with the number of nodes $N$ and the expected average degree $\langle k\rangle$ given in the title of the subplot pair. We used $K=-1$ as the curvature of the hyperbolic plane in each case.}
    \label{fig:AMI_Louv_Inf_S1}
\end{figure}

We compared the community structures found by the asynchronous label propagation~\cite{alabprop}, the Louvain~\cite{Louvain} and the Infomap~\cite{Infomap} algorithms in the PSO \cite{PSO}, E-PSO \cite{EPSO_HyperMap,our_embedding} and $\mathbb{S}^1/\mathbb{H}^2$~\cite{S1,S1H2_Mercator} networks of various parameter combinations. Each community detection algorithm was executed once for each network. The isolated nodes emerging in the case of the $\mathbb{S}^1/\mathbb{H}^2$ model and occasionally also in the networks generated by the E-PSO model of $L<0$ were removed before the community detection, meaning that the actual size of the examined networks does not necessarily reach the number of nodes $N$ inputted in these models. We generated 100 networks with each parameter setting and calculated the adjusted mutual information (AMI)~\cite{AMI,McCarthy_AMI} of the resulted 3 partitions for each network. Figures \ref{fig:AMI_alabprop_Louv_PSO}–\ref{fig:AMI_alabprop_Louv_S1} display the average and the standard deviation of the AMI between the community structures obtained with asynchronous label propagation and Louvain, figures \ref{fig:AMI_alabprop_Inf_PSO}–\ref{fig:AMI_alabprop_Inf_S1} compare the result of asynchronous label propagation with the result of Infomap, while the consistency between the community structures detected by Louvain and Infomap is examined in figures \ref{fig:AMI_Louv_Inf_PSO}–\ref{fig:AMI_Louv_Inf_S1}.

According to these figures, asynchronous label propagation and Infomap produce the most similar partitions, while the most different is the result of Louvain and Infomap. In most of the cases, the AMI depends similarly on the network generation parameters as the weighted modularity $Q$. For the high modularity regions of the parameter space (large number of nodes $N$, small average degree $\langle k\rangle$, small popularity fading parameter $\beta$ or large degree decay exponent $\gamma$ and low temperature $T$ or large $\alpha$) the AMI is relatively large for each pair of community detection methods, indicating in these parameter regions the emergence of really apparent communities that are detectable for all the 3 investigated algorithms alike. 
%(EZEK MINDEN ÁBRA ALAPJÁN IGAZAK?)

%alabprop-Louvain

%alabprop-Infomap

%Louvain-Infomap

\clearpage

\section{Equidistant angular node arrangement}
\label{sect:eqAngDist}

\setcounter{figure}{0}
\setcounter{table}{0}
\setcounter{equation}{0}
\renewcommand{\thefigure}{S5.\arabic{figure}}
\renewcommand{\thetable}{S5.\arabic{table}}
\renewcommand{\theequation}{S5.\arabic{equation}}

\captionsetup[figure]{font=footnotesize,justification=justified,labelsep=period,labelfont=bf}

\begin{figure}[hbt]
    \centering
    \makebox[\textwidth][c]{\includegraphics[width=1.15\textwidth]{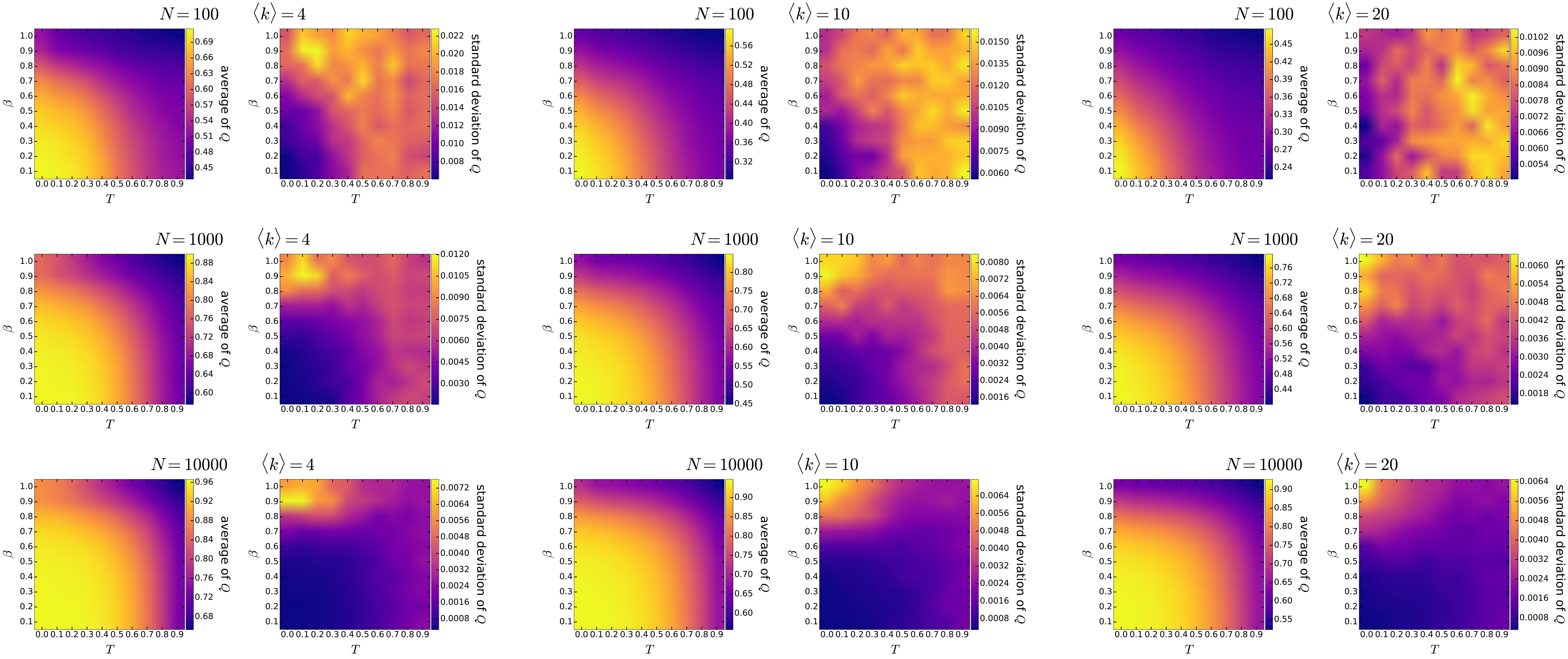}}
    \caption{{\bf The mean and the standard deviation of the highest weighted modularity $Q$ achieved among the \textit{asynchronous label propagation}, the \textit{Louvain} and the \textit{Infomap} algorithms in 100 \textit{PSO} networks of different parametrisations \textit{with strictly equidistant angular arrangement}.} Each pair of subplots depicts the effect of changing the popularity fading parameter $\beta$ and the temperature $T$, with the number of nodes $N$ and the expected average degree $\langle k\rangle=2m$ given in the title of the subplot pair. The curvature of the hyperbolic plane $K$ was always set to $-1$, i.e. we used $\zeta=1$.}
    \label{fig:eqAng_bestMod}
\end{figure}

\begin{figure}[hbt]
    \centering
    \makebox[\textwidth][c]{\includegraphics[width=1.15\textwidth]{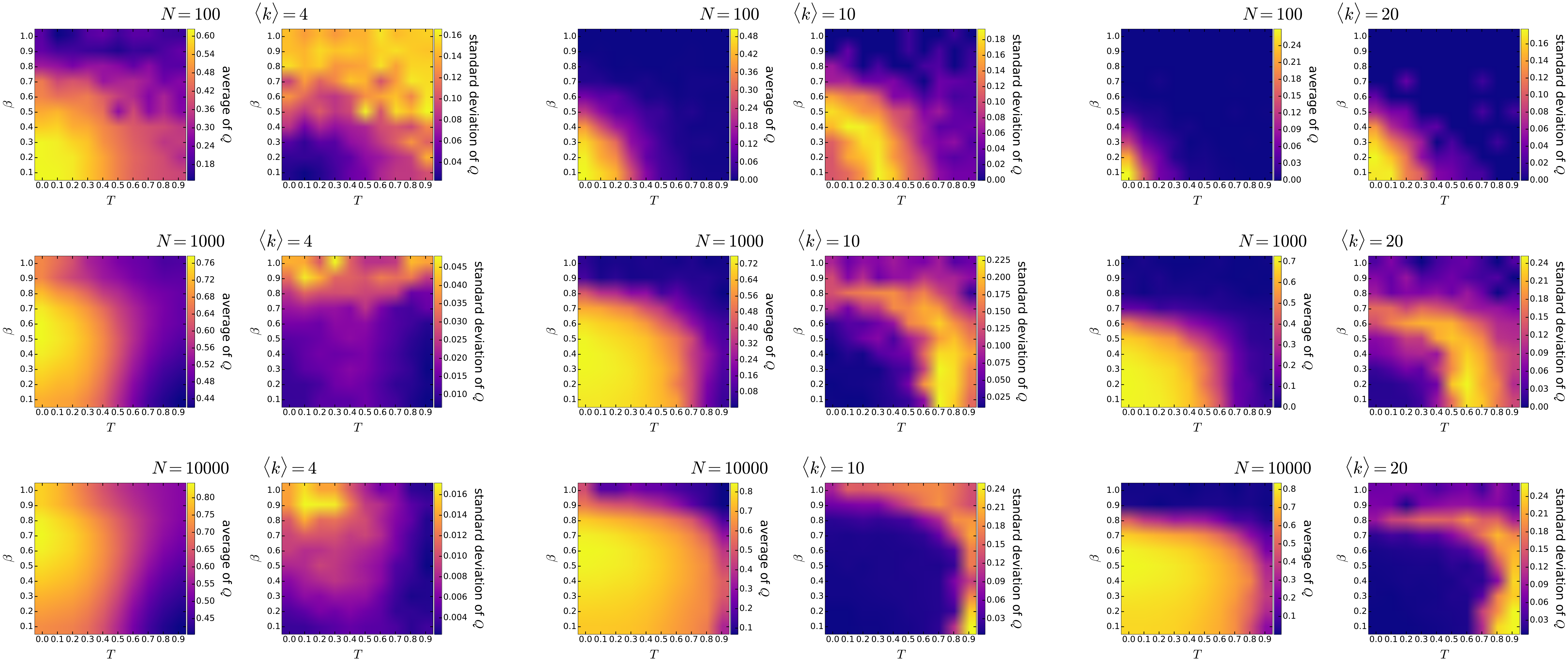}}
    \caption{{\bf The mean and the standard deviation of the weighted modularity $Q$ of the community structure detected by the \textit{asynchronous label propagation} algorithm in 100 \textit{PSO} networks of different parametrisations \textit{with strictly equidistant angular arrangement}.} Each pair of subplots depicts the effect of changing the popularity fading parameter $\beta$ and the temperature $T$, with the number of nodes $N$ and the expected average degree $\langle k\rangle=2m$ given in the title of the subplot pair. The curvature of the hyperbolic plane $K$ was always set to $-1$, i.e. we used $\zeta=1$.}
    \label{fig:eqAng_alabpropMod}
\end{figure}

\begin{figure}[hbt]
    \centering
    \makebox[\textwidth][c]{\includegraphics[width=1.15\textwidth]{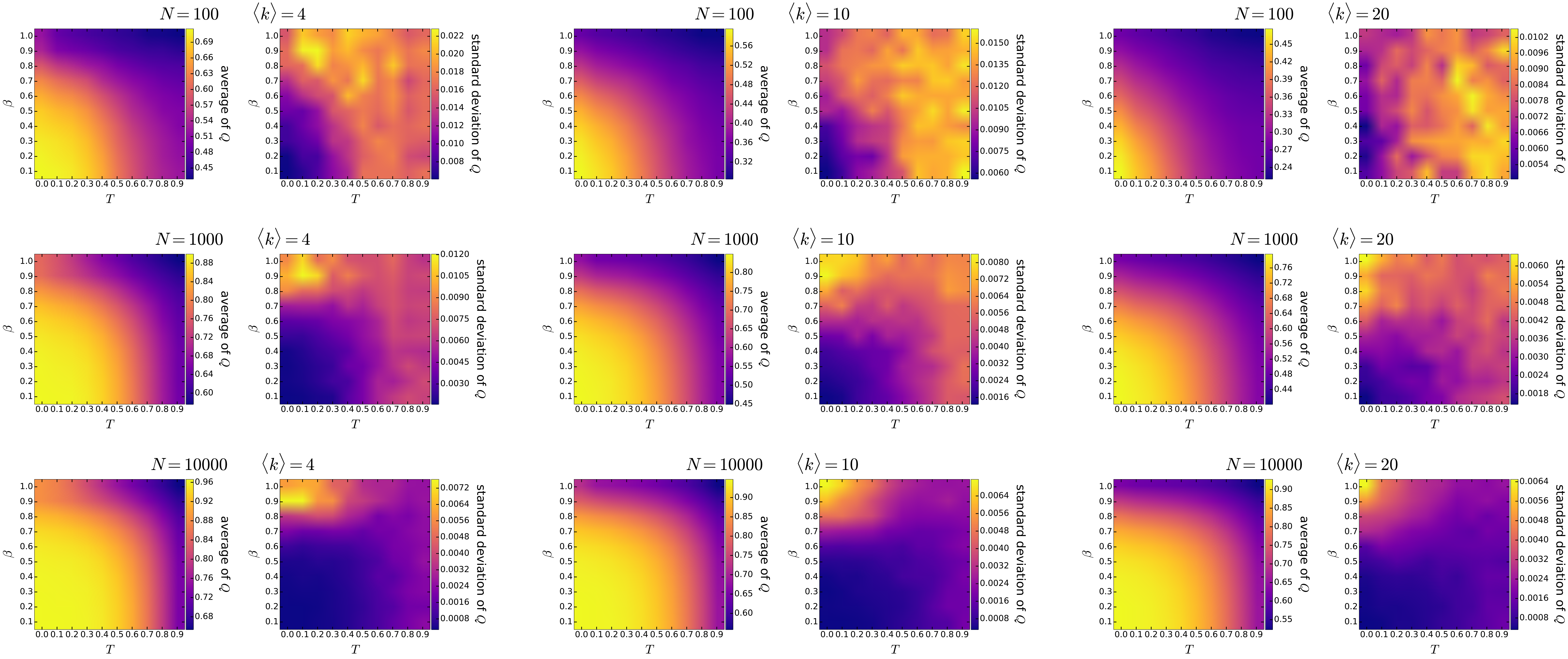}}
    \caption{{\bf The mean and the standard deviation of the weighted modularity $Q$ of the community structure detected by the \textit{Louvain} algorithm in 100 \textit{PSO} networks of different parametrisations \textit{with strictly equidistant angular arrangement}.} Each pair of subplots depicts the effect of changing the popularity fading parameter $\beta$ and the temperature $T$, with the number of nodes $N$ and the expected average degree $\langle k\rangle=2m$ given in the title of the subplot pair. The curvature of the hyperbolic plane $K$ was always set to $-1$, i.e. we used $\zeta=1$.}
    \label{fig:eqAng_LouvainMod}
\end{figure}

\begin{figure}[hbt]
    \centering
    \makebox[\textwidth][c]{\includegraphics[width=1.15\textwidth]{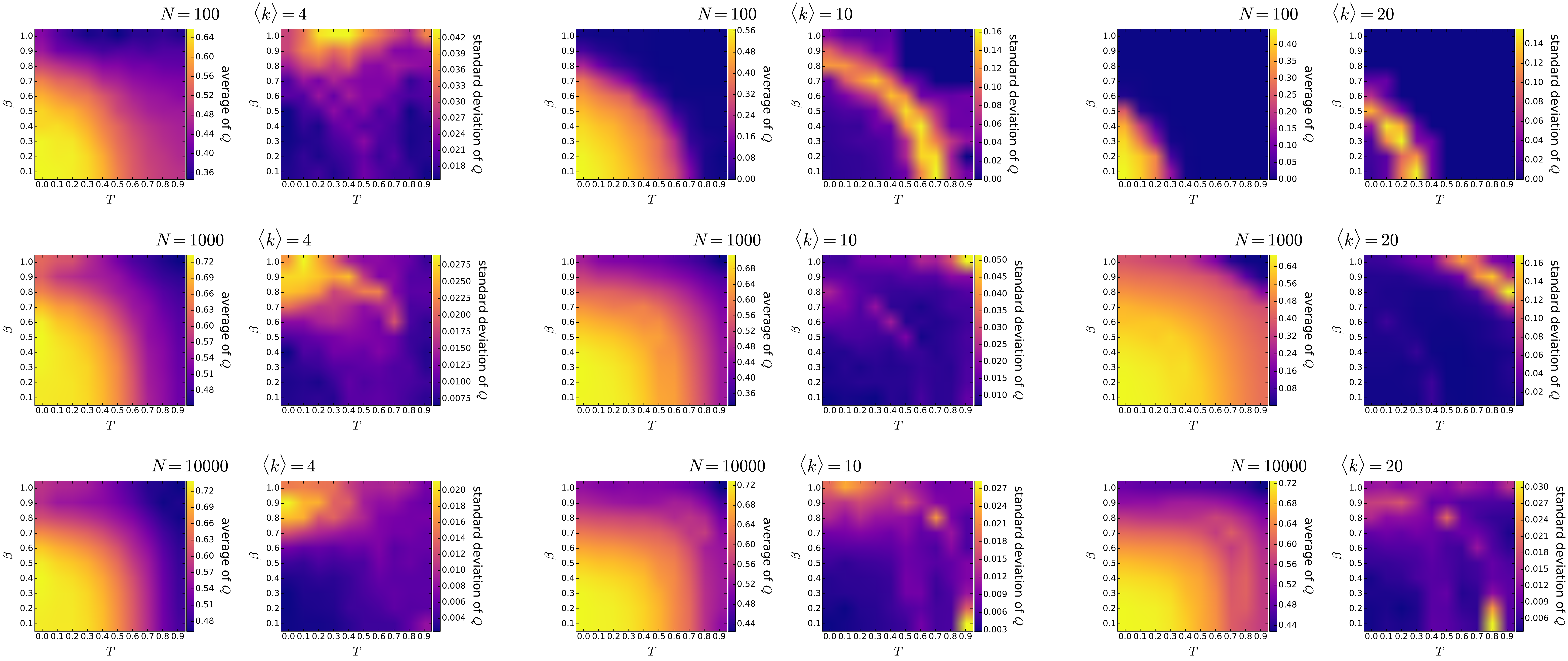}}
    \caption{{\bf The mean and the standard deviation of the weighted modularity $Q$ of the community structure detected by the \textit{Infomap} algorithm in 100 \textit{PSO} networks of different parametrisations \textit{with strictly equidistant angular arrangement}.} Each pair of subplots depicts the effect of changing the popularity fading parameter $\beta$ and the temperature $T$, with the number of nodes $N$ and the expected average degree $\langle k\rangle=2m$ given in the title of the subplot pair. The curvature of the hyperbolic plane $K$ was always set to $-1$, i.e. we used $\zeta=1$.}
    \label{fig:eqAng_InfomapMod}
\end{figure}

\begin{figure}[hbt]
    \centering
    \makebox[\textwidth][c]{\includegraphics[width=1.15\textwidth]{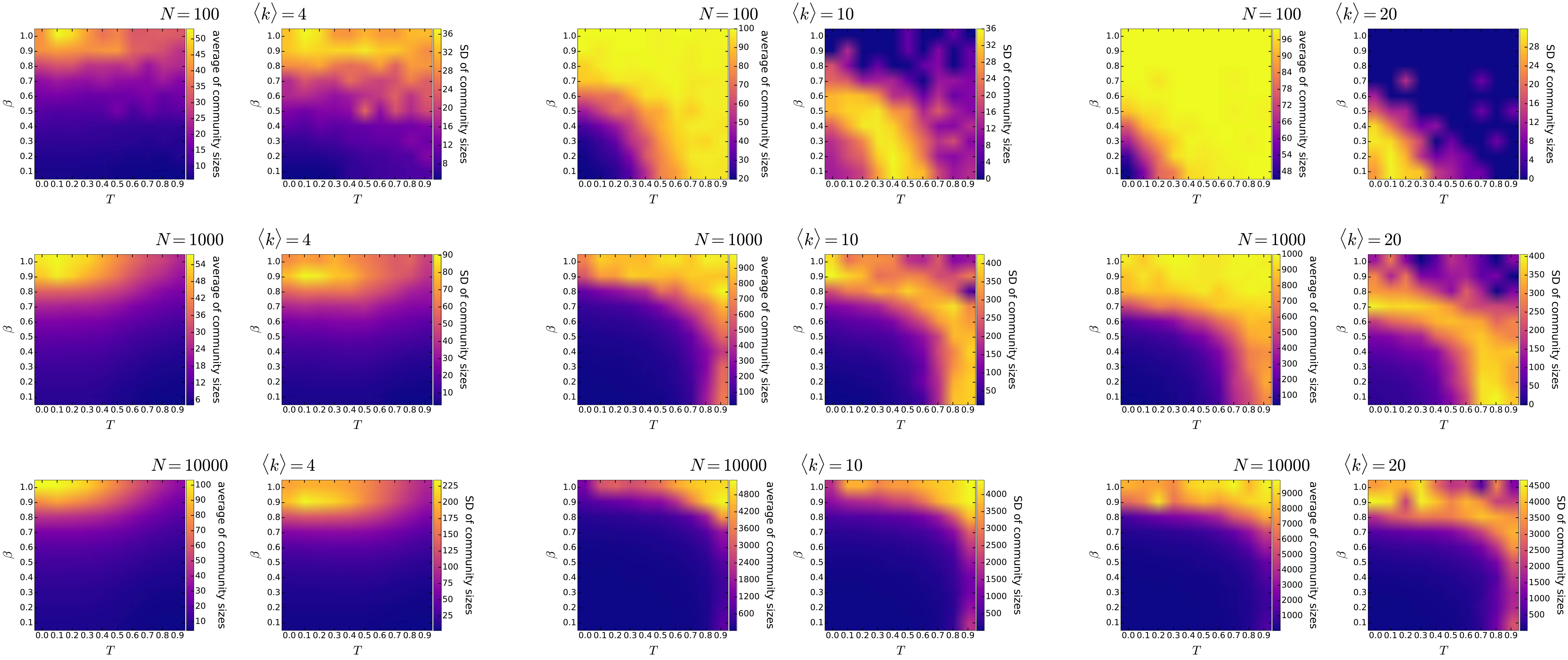}}
    \caption{{\bf The mean and the standard deviation of the size of communities detected by the \textit{asynchronous label propagation} algorithm in 100 \textit{PSO} networks of different parametrisations \textit{with strictly equidistant angular arrangement}.} Each pair of subplots depicts the effect of changing the popularity fading parameter $\beta$ and the temperature $T$, with the number of nodes $N$ and the expected average degree $\langle k\rangle=2m$ given in the title of the subplot pair. The curvature of the hyperbolic plane $K$ was always set to $-1$, i.e. we used $\zeta=1$.}
    \label{fig:eqAng_alabpropGSavstd}
\end{figure}

\begin{figure}[hbt]
    \centering
    \makebox[\textwidth][c]{\includegraphics[width=1.15\textwidth]{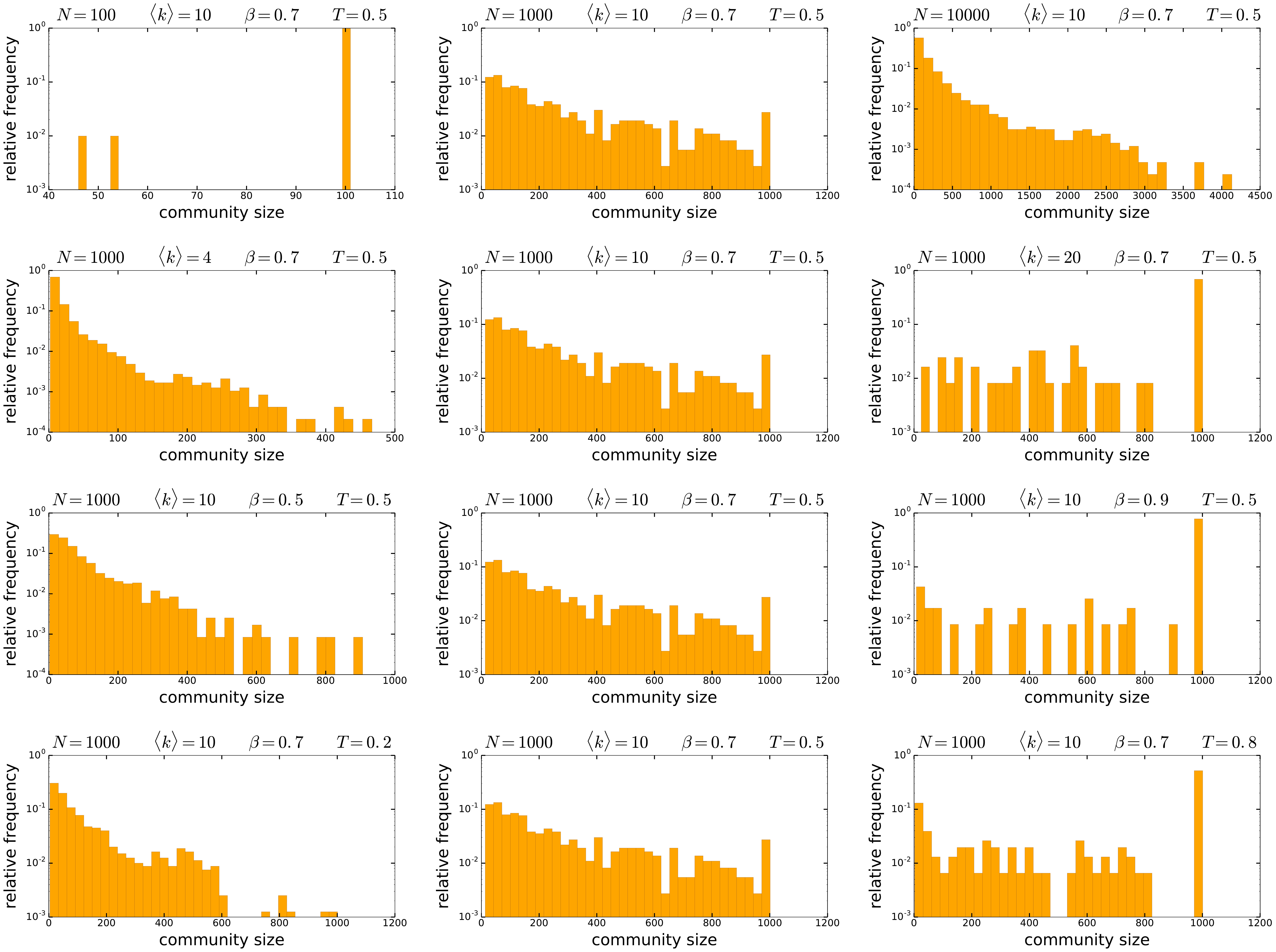}}
    \caption{{\bf The size distribution of the communities detected by the \textit{asynchronous label propagation} algorithm in 100 \textit{PSO} networks of different parametrisations \textit{with strictly equidistant angular arrangement}.} The parameters of the network generation are listed in the title for each subplot. The curvature of the hyperbolic plane $K$ was always set to $-1$, i.e. we used $\zeta=1$. Each row of the figure demonstrates the effect of the change in a given network generation parameter: from top to bottom, the number of nodes $N$, the expected average degree $\langle k\rangle=2m$, the popularity fading parameter $\beta$ and the temperature $T$.}
    \label{fig:eqAng_alabpropGShist}
\end{figure}

\begin{figure}[hbt]
    \centering
    \makebox[\textwidth][c]{\includegraphics[width=1.15\textwidth]{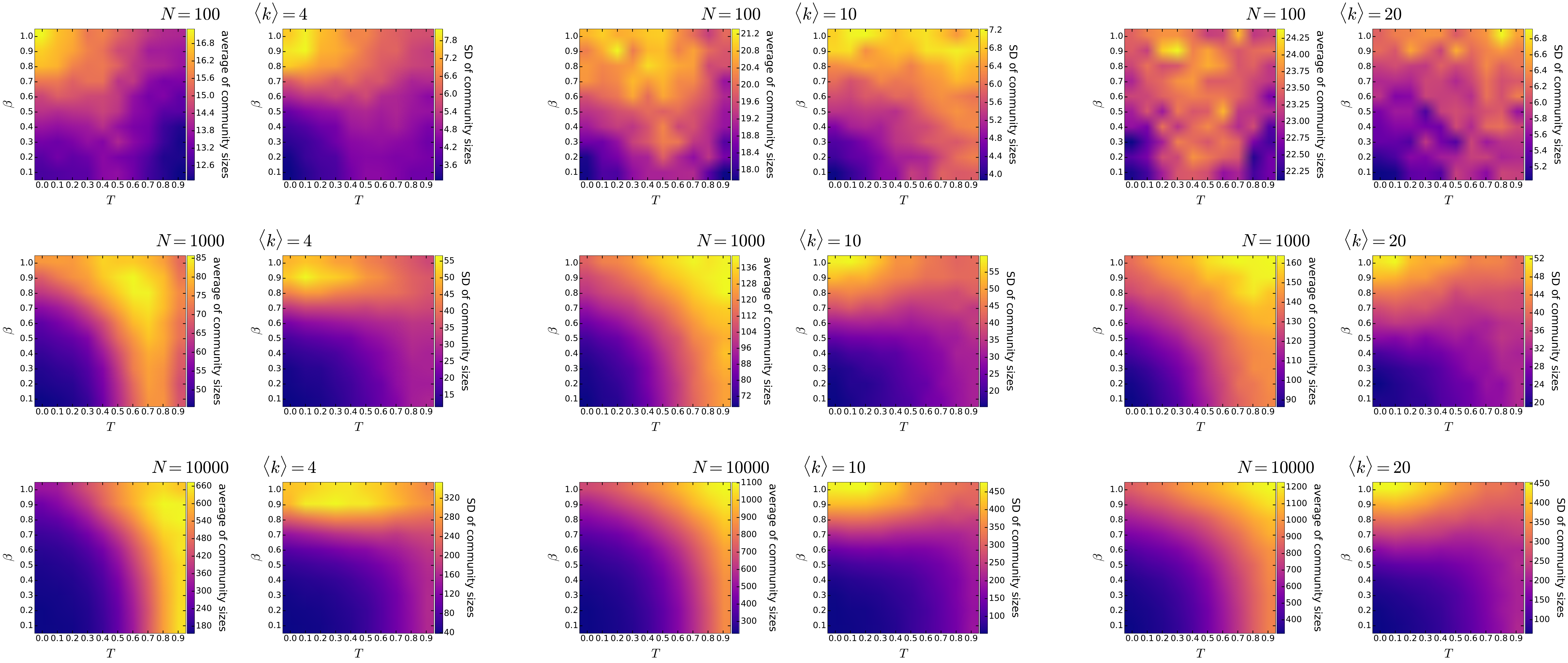}}
    \caption{{\bf The mean and the standard deviation of the size of communities detected by the \textit{Louvain} algorithm in 100 \textit{PSO} networks of different parametrisations \textit{with strictly equidistant angular arrangement}.} Each pair of subplots depicts the effect of changing the popularity fading parameter $\beta$ and the temperature $T$, with the number of nodes $N$ and the expected average degree $\langle k\rangle=2m$ given in the title of the subplot pair. The curvature of the hyperbolic plane $K$ was always set to $-1$, i.e. we used $\zeta=1$.}
    \label{fig:eqAng_LouvainGSavstd}
\end{figure}

\begin{figure}[hbt]
    \centering
    \makebox[\textwidth][c]{\includegraphics[width=1.15\textwidth]{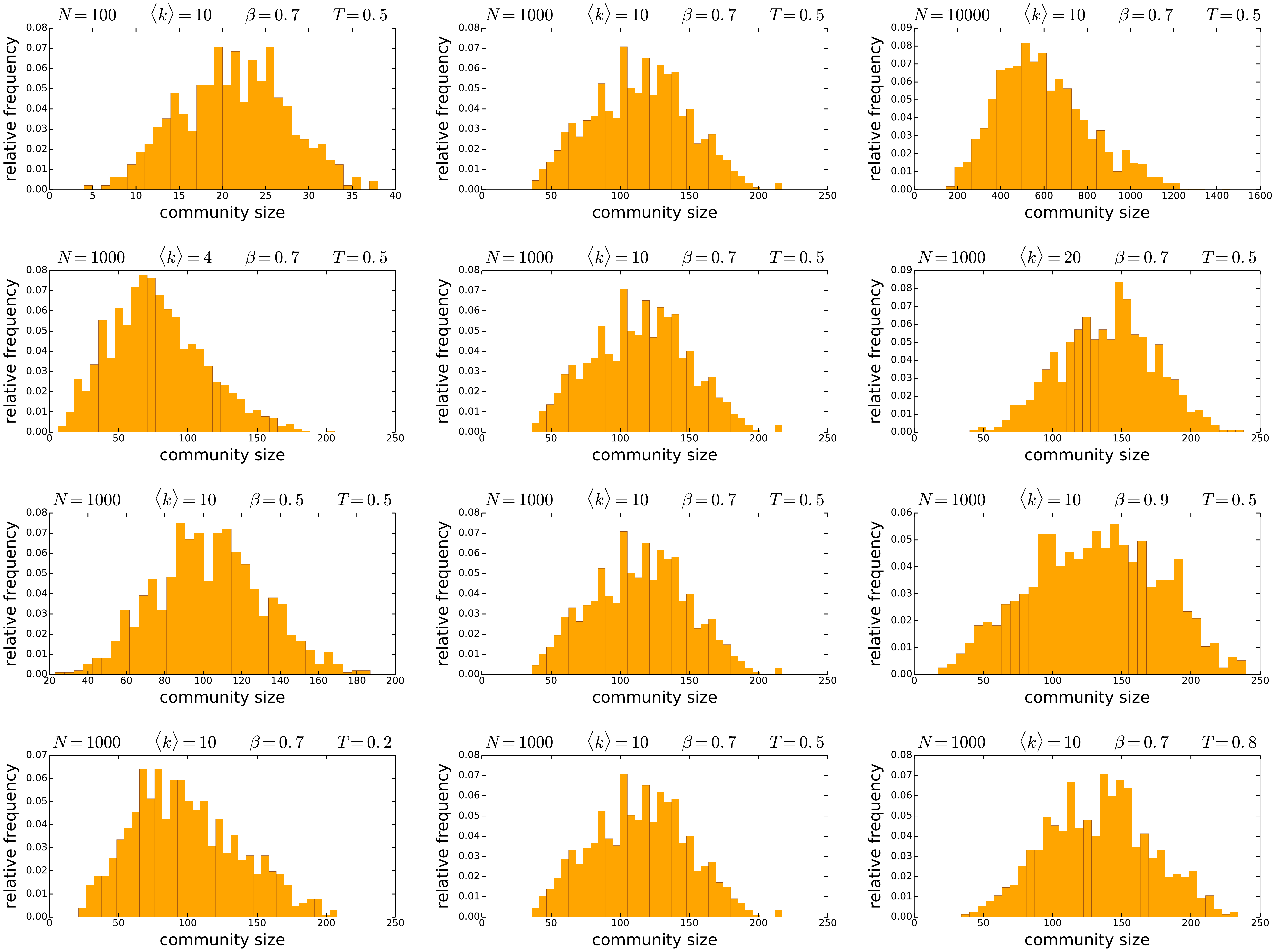}}
    \caption{{\bf The size distribution of the communities detected by the \textit{Louvain} algorithm in 100 \textit{PSO} networks of different parametrisations \textit{with strictly equidistant angular arrangement}.} The parameters of the network generation are listed in the title for each subplot. The curvature of the hyperbolic plane $K$ was always set to $-1$, i.e. we used $\zeta=1$. Each row of the figure demonstrates the effect of the change in a given network generation parameter: from top to bottom, the number of nodes $N$, the expected average degree $\langle k\rangle=2m$, the popularity fading parameter $\beta$ and the temperature $T$.}
    \label{fig:eqAng_LouvainGShist}
\end{figure}

\begin{figure}[hbt]
    \centering
    \makebox[\textwidth][c]{\includegraphics[width=1.15\textwidth]{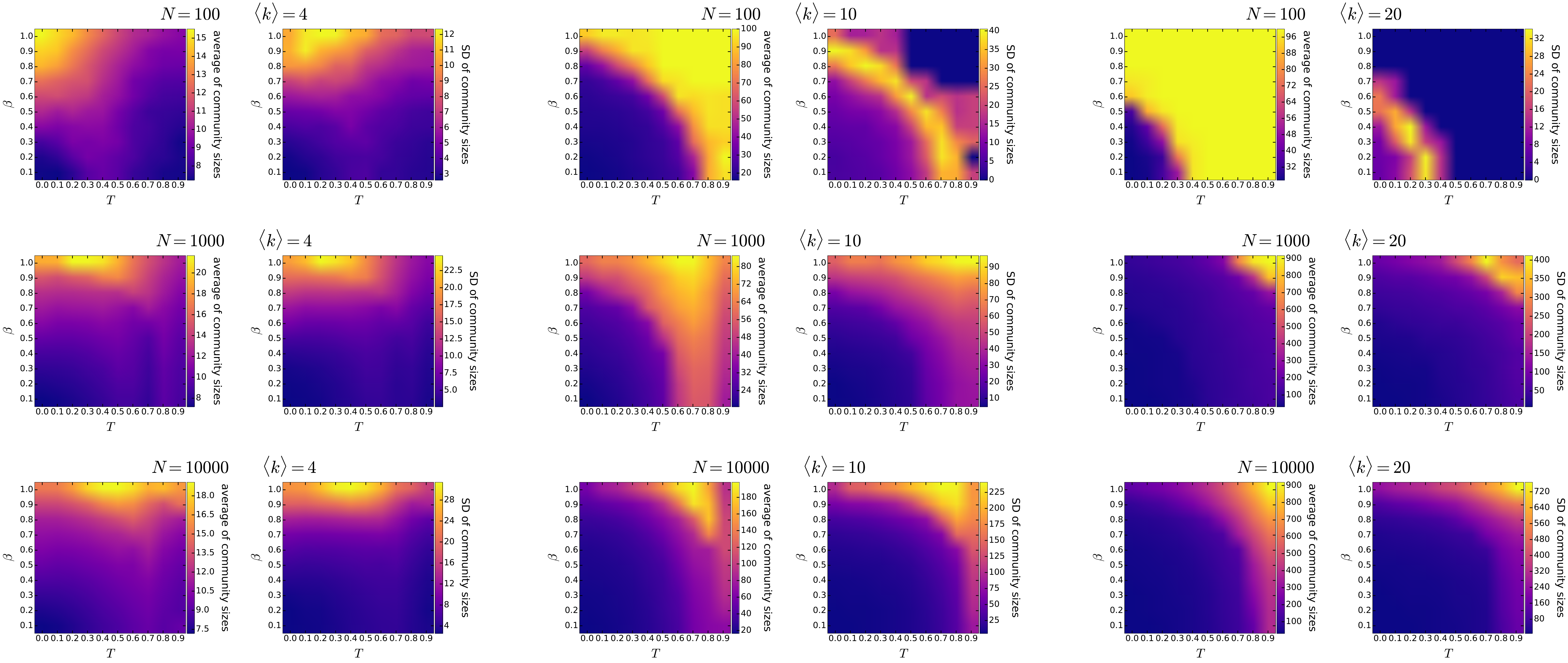}}
    \caption{{\bf The mean and the standard deviation of the size of communities detected by the \textit{Infomap} algorithm in 100 \textit{PSO} networks of different parametrisations \textit{with strictly equidistant angular arrangement}.} Each pair of subplots depicts the effect of changing the popularity fading parameter $\beta$ and the temperature $T$, with the number of nodes $N$ and the expected average degree $\langle k\rangle=2m$ given in the title of the subplot pair. The curvature of the hyperbolic plane $K$ was always set to $-1$, i.e. we used $\zeta=1$.}
    \label{fig:eqAng_InfomapGSavstd}
\end{figure}

\begin{figure}[hbt]
    \centering
    \makebox[\textwidth][c]{\includegraphics[width=1.15\textwidth]{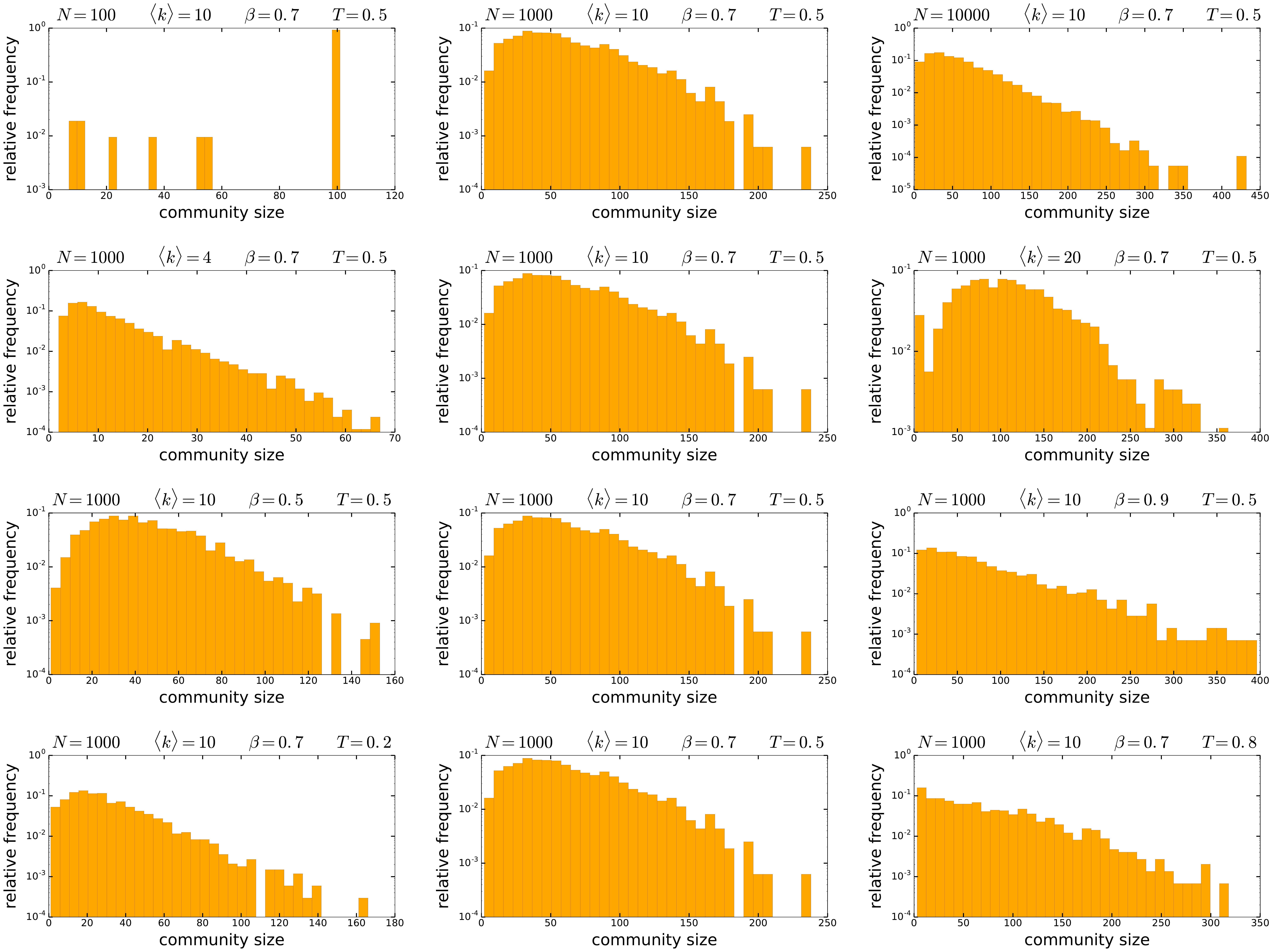}}
    \caption{{\bf The size distribution of the communities detected by the \textit{Infomap} algorithm in 100 \textit{PSO} networks of different parametrisations \textit{with strictly equidistant angular arrangement}.} The parameters of the network generation are listed in the title for each subplot. The curvature of the hyperbolic plane $K$ was always set to $-1$, i.e. we used $\zeta=1$. Each row of the figure demonstrates the effect of the change in a given network generation parameter: from top to bottom, the number of nodes $N$, the expected average degree $\langle k\rangle=2m$, the popularity fading parameter $\beta$ and the temperature $T$.}
    \label{fig:eqAng_InfomapGShist}
\end{figure}

\begin{figure}[hbt]
    \centering
    \makebox[\textwidth][c]{\includegraphics[width=1.15\textwidth]{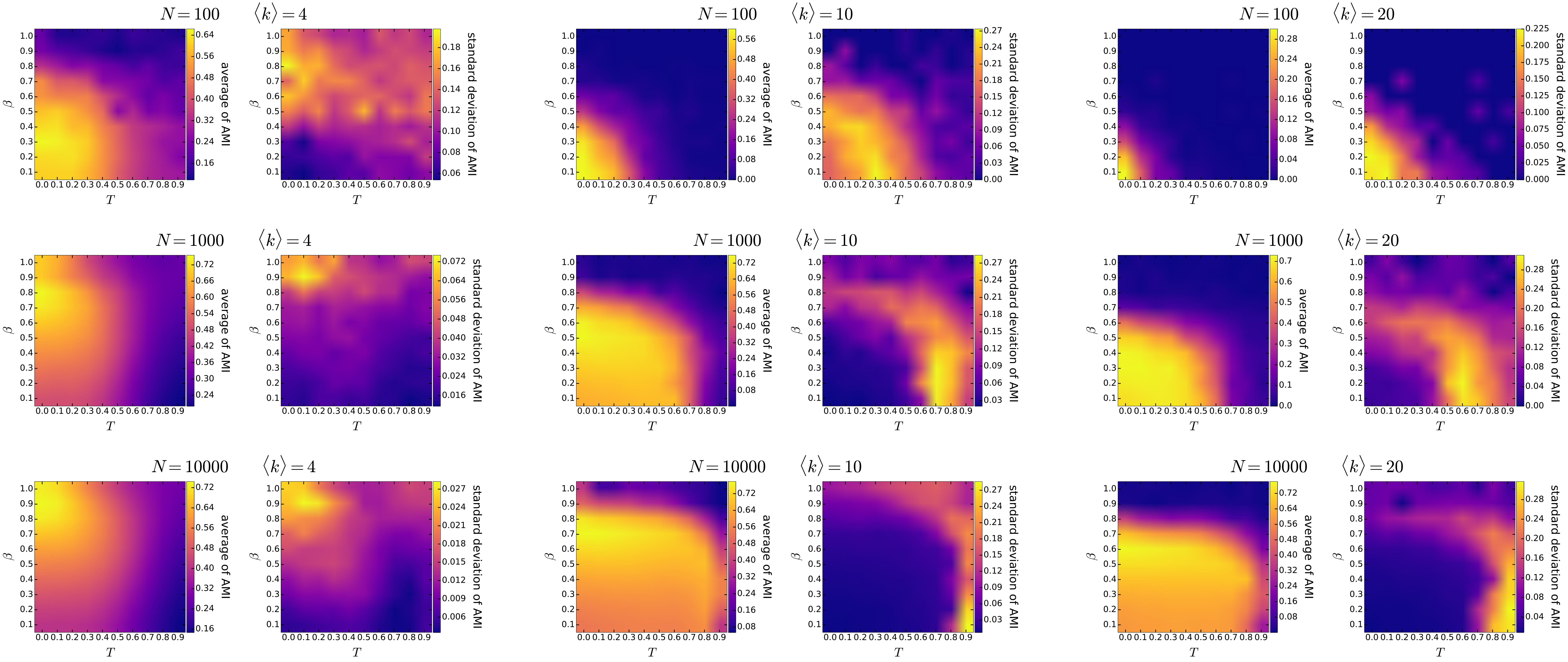}}
    \caption{{\bf The mean and the standard deviation of the adjusted mutual information of the two community structures detected by the \textit{asynchronous label propagation} and the \textit{Louvain} algorithms in 100 \textit{PSO} networks of different parametrisations \textit{with strictly equidistant angular arrangement}.} Each pair of subplots depicts the effect of changing the popularity fading parameter $\beta$ and the temperature $T$, with the number of nodes $N$ and the expected average degree $\langle k\rangle=2m$ given in the title of the subplot pair. The curvature of the hyperbolic plane $K$ was always set to $-1$, i.e. we used $\zeta=1$.}
    \label{fig:AMI_alabprop_Louv_eqAng}
\end{figure}

\begin{figure}[hbt]
    \centering
    \makebox[\textwidth][c]{\includegraphics[width=1.15\textwidth]{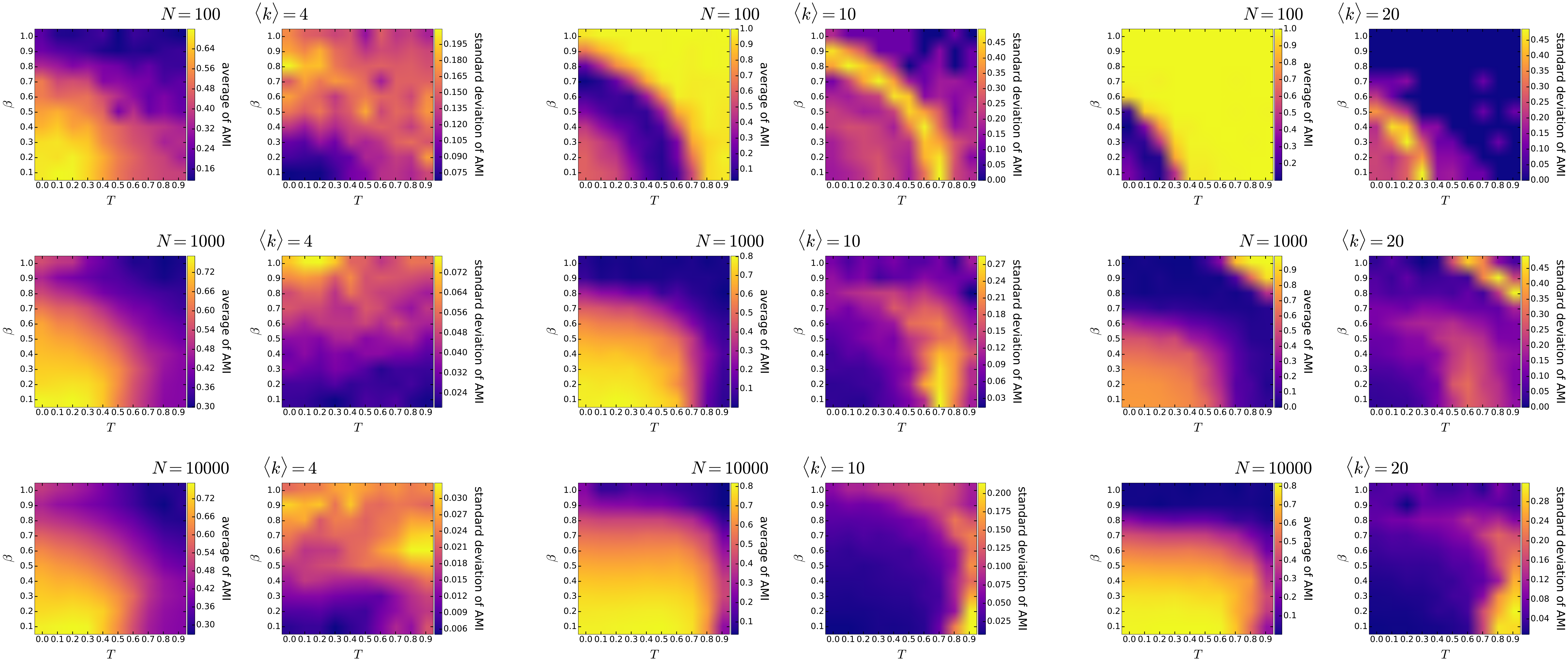}}
    \caption{{\bf The mean and the standard deviation of the adjusted mutual information of the two community structures detected by the \textit{asynchronous label propagation} and the \textit{Infomap} algorithms in 100 \textit{PSO} networks of different parametrisations \textit{with strictly equidistant angular arrangement}.} Each pair of subplots depicts the effect of changing the popularity fading parameter $\beta$ and the temperature $T$, with the number of nodes $N$ and the expected average degree $\langle k\rangle=2m$ given in the title of the subplot pair. The curvature of the hyperbolic plane $K$ was always set to $-1$, i.e. we used $\zeta=1$.}
    \label{fig:AMI_alabprop_Inf_eqAng}
\end{figure}

\begin{figure}[hbt]
    \centering
    \makebox[\textwidth][c]{\includegraphics[width=1.15\textwidth]{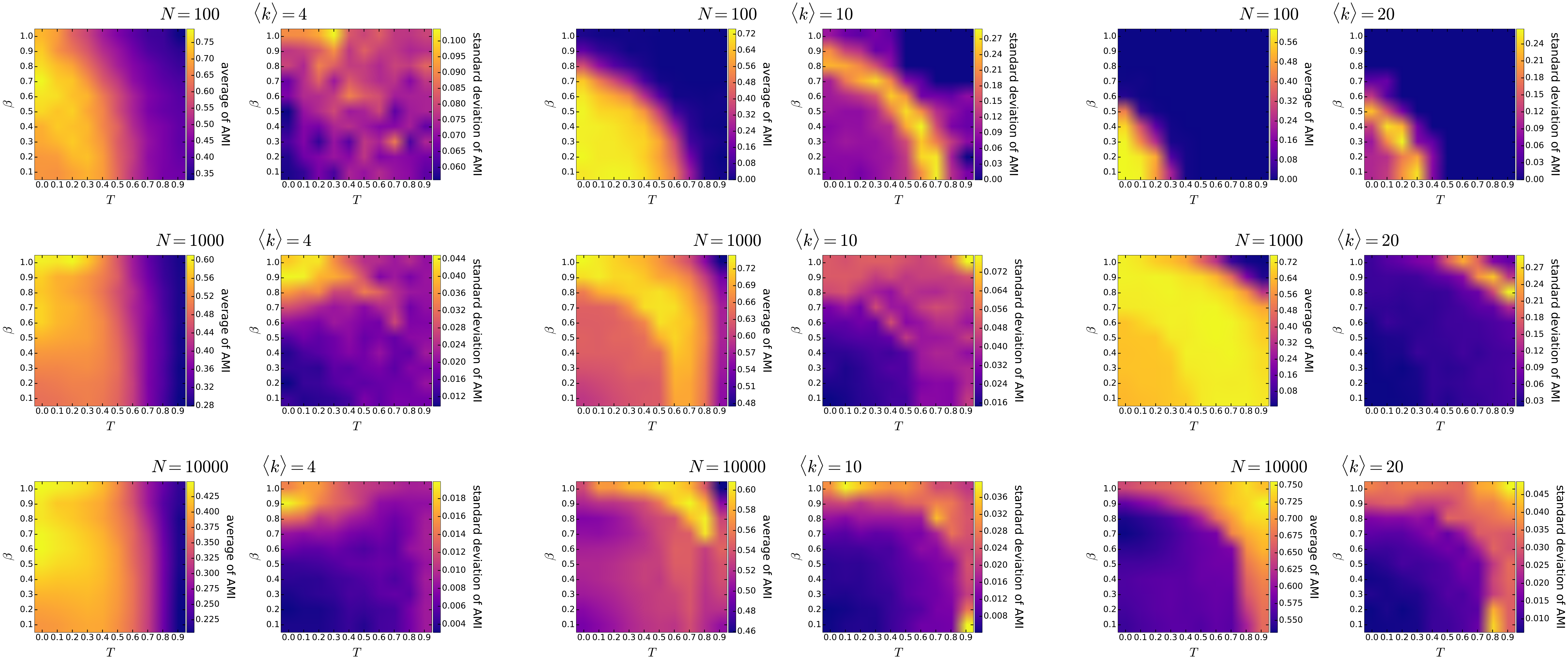}}
    \caption{{\bf The mean and the standard deviation of the adjusted mutual information of the two community structures detected by the \textit{Louvain} and the \textit{Infomap} algorithms in 100 \textit{PSO} networks of different parametrisations \textit{with strictly equidistant angular arrangement}.} Each pair of subplots depicts the effect of changing the popularity fading parameter $\beta$ and the temperature $T$, with the number of nodes $N$ and the expected average degree $\langle k\rangle=2m$ given in the title of the subplot pair. The curvature of the hyperbolic plane $K$ was always set to $-1$, i.e. we used $\zeta=1$.}
    \label{fig:AMI_Louv_Inf_eqAng}
\end{figure}

In order to exclude the possibility that the emergence of communities is a result of the inhomogeneities in the angular node arrangement arising inevitably when a finite number of angular coordinates is sampled from a uniform distribution in $[0,2\pi)$, we modified the PSO model~\cite{PSO} to use strictly equidistant angular arrangement, i.e. instead of sampling the angular coordinates uniformly randomly, assign the coordinates $\theta_i=(i-1)\cdot\frac{2\pi}{N},\,i=1,2,...,N$ to the network nodes in a randomly chosen order. %(ez nem úgy néz ki, mintha az i. pont koordinátája lenne az i-1-es szorzójú?). 
According to figures \ref{fig:eqAng_bestMod}–\ref{fig:eqAng_InfomapMod}, the networks generated by this modified PSO model possess a community structure with similarly high weighted modularity as the usual PSO networks (see figures \ref{fig:PSO_bestMod} and \ref{fig:PSO_alabpropMod}-\ref{fig:PSO_InfomapMod}). Hence, we can conclude that the emergence of communities is not just an effect of the finite network size, but the inherent property of the studied %angularly uniform(?) 
hyperbolic network models. %(Bár az S1-re nem néztünk ilyet, mert annak macerásabb lett volna átírni a kódját.) 
Figures \ref{fig:eqAng_alabpropGSavstd}–\ref{fig:eqAng_InfomapGShist} show similar results for the networks generated by the modified PSO model with regard to the community size distributions as figures \ref{fig:PSO_alabpropGSavstd}-\ref{fig:PSO_InfomapGShist} in the case of the original PSO model. Lastly, the adjusted mutual information (AMI)~\cite{AMI,McCarthy_AMI} between the community structures detected by asynchronous label propagation~\cite{alabprop}, Louvain~\cite{Louvain} and Infomap~\cite{Infomap} behaves the same way for the modified PSO model (figures \ref{fig:AMI_alabprop_Louv_eqAng}–\ref{fig:AMI_Louv_Inf_eqAng}) as for the original PSO model (figures \ref{fig:AMI_alabprop_Louv_PSO}, \ref{fig:AMI_alabprop_Inf_PSO} and \ref{fig:AMI_Louv_Inf_PSO}). %(EZEK MIND IGAZAK MINDENFÉLE PARAMÉTERNÉL?..)

%modularitás

%csoportméret - átlag és szórás

%eqAng-os AMI

\clearpage
\section{Community detection on unweighted hyperbolic networks}
\label{sect:unweightedCase}

\setcounter{figure}{0}
\setcounter{table}{0}
\setcounter{equation}{0}
\renewcommand{\thefigure}{S6.\arabic{figure}}
\renewcommand{\thetable}{S6.\arabic{table}}
\renewcommand{\theequation}{S6.\arabic{equation}}

\captionsetup[figure]{font=footnotesize,justification=justified,labelsep=period,labelfont=bf}

\begin{figure}[hbt]
    \centering
    \makebox[\textwidth][c]{\includegraphics[width=1.15\textwidth]{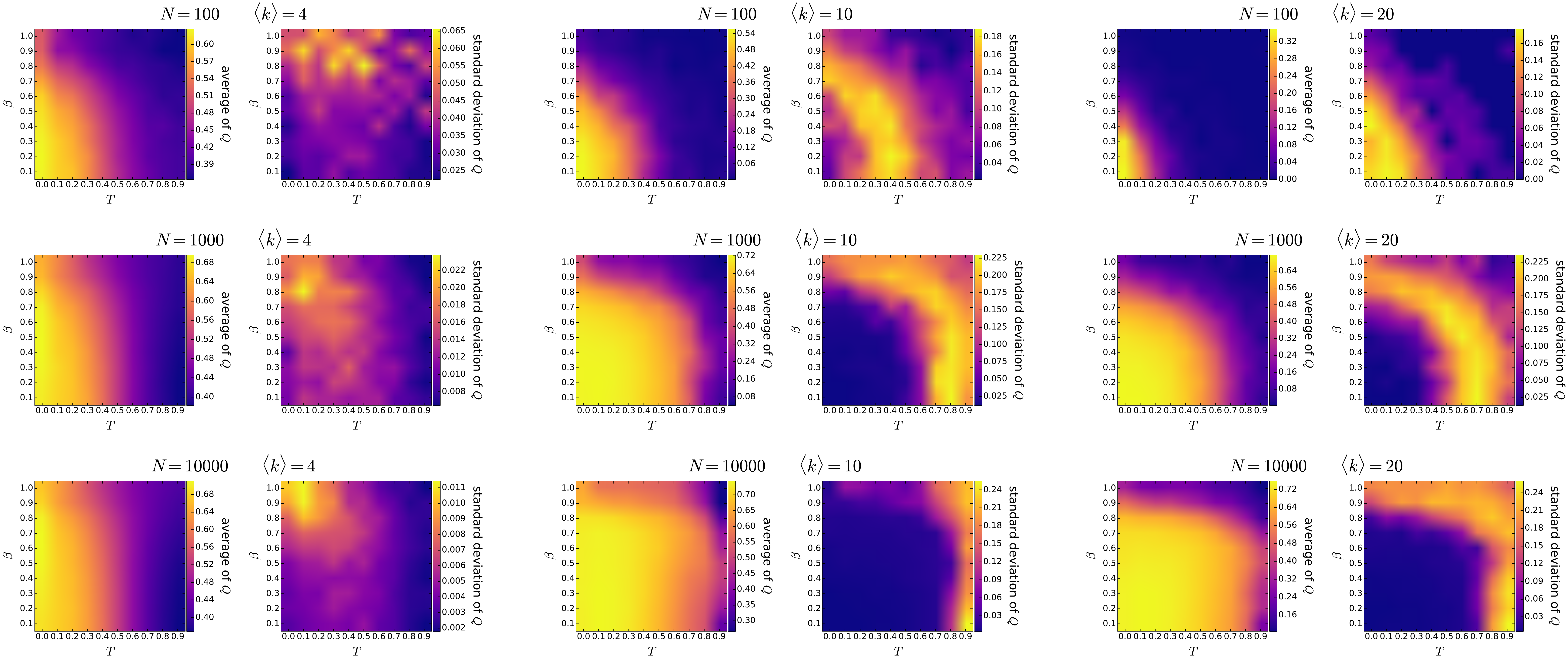}}
    \caption{{\bf The mean and the standard deviation of the unweighted modularity $Q$ of the community structure detected by the \textit{asynchronous label propagation} algorithm in 100 \textit{unweighted} \textit{PSO} networks of different parametrisations.} Each pair of subplots depicts the effect of changing the popularity fading parameter $\beta$ and the temperature $T$, with the number of nodes $N$ and the expected average degree $\langle k\rangle=2m$ given in the title of the subplot pair. The curvature of the hyperbolic plane $K$ was always set to $-1$, i.e. we used $\zeta=1$.}
    \label{fig:PSO_alabpropMod_uw}
\end{figure}

\begin{figure}[hbt]
    \centering
    \makebox[\textwidth][c]{\includegraphics[width=1.15\textwidth]{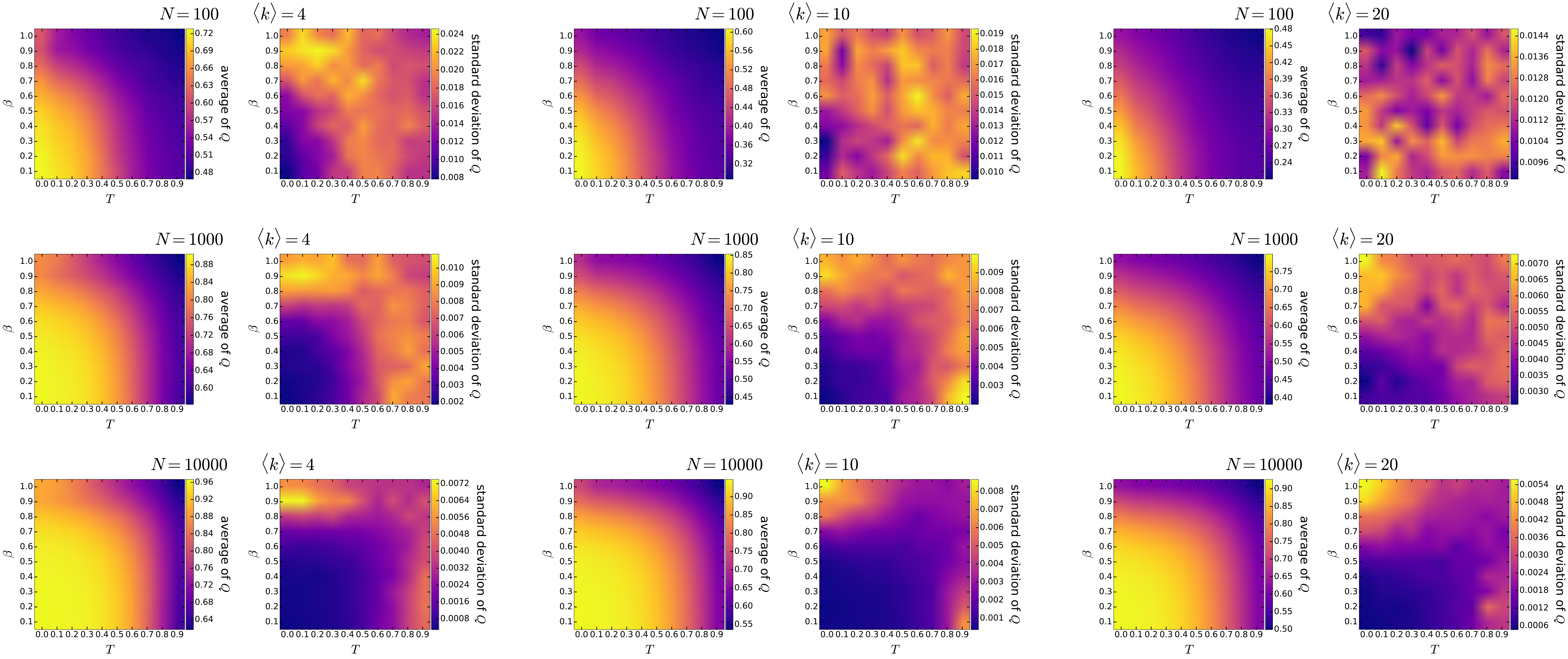}}
    \caption{{\bf The mean and the standard deviation of the unweighted modularity $Q$ of the community structure detected by the \textit{Louvain} algorithm in 100 \textit{unweighted} \textit{PSO} networks of different parametrisations.} Each pair of subplots depicts the effect of changing the popularity fading parameter $\beta$ and the temperature $T$, with the number of nodes $N$ and the expected average degree $\langle k\rangle=2m$ given in the title of the subplot pair. The curvature of the hyperbolic plane $K$ was always set to $-1$, i.e. we used $\zeta=1$.}
    \label{fig:PSO_LouvainMod_uw}
\end{figure}

\begin{figure}[hbt]
    \centering
    \makebox[\textwidth][c]{\includegraphics[width=1.15\textwidth]{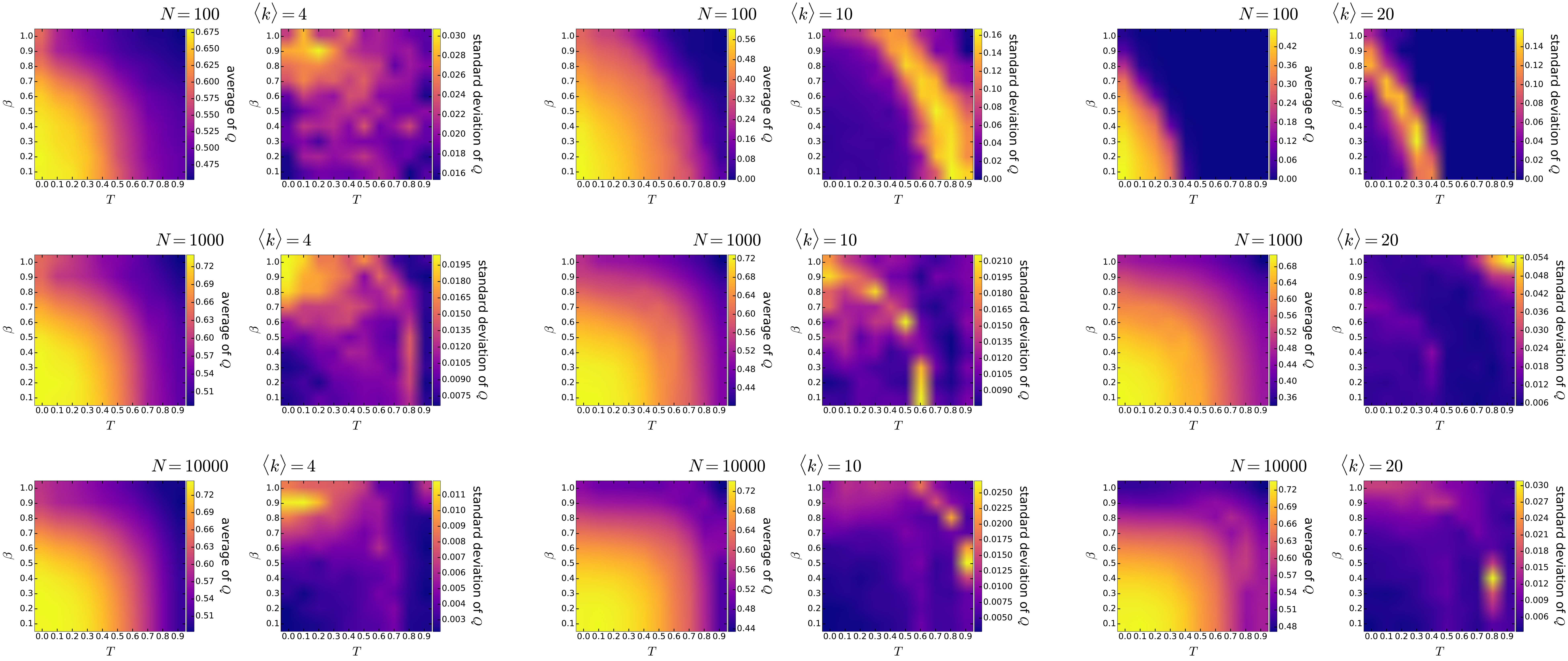}}
    \caption{{\bf The mean and the standard deviation of the unweighted modularity $Q$ of the community structure detected by the \textit{Infomap} algorithm in 100 \textit{unweighted} \textit{PSO} networks of different parametrisations.} Each pair of subplots depicts the effect of changing the popularity fading parameter $\beta$ and the temperature $T$, with the number of nodes $N$ and the expected average degree $\langle k\rangle=2m$ given in the title of the subplot pair. The curvature of the hyperbolic plane $K$ was always set to $-1$, i.e. we used $\zeta=1$.}
    \label{fig:PSO_InfomapMod_uw}
\end{figure}

\begin{figure}[hbt]
    \centering
    \makebox[\textwidth][c]{\includegraphics[width=1.15\textwidth]{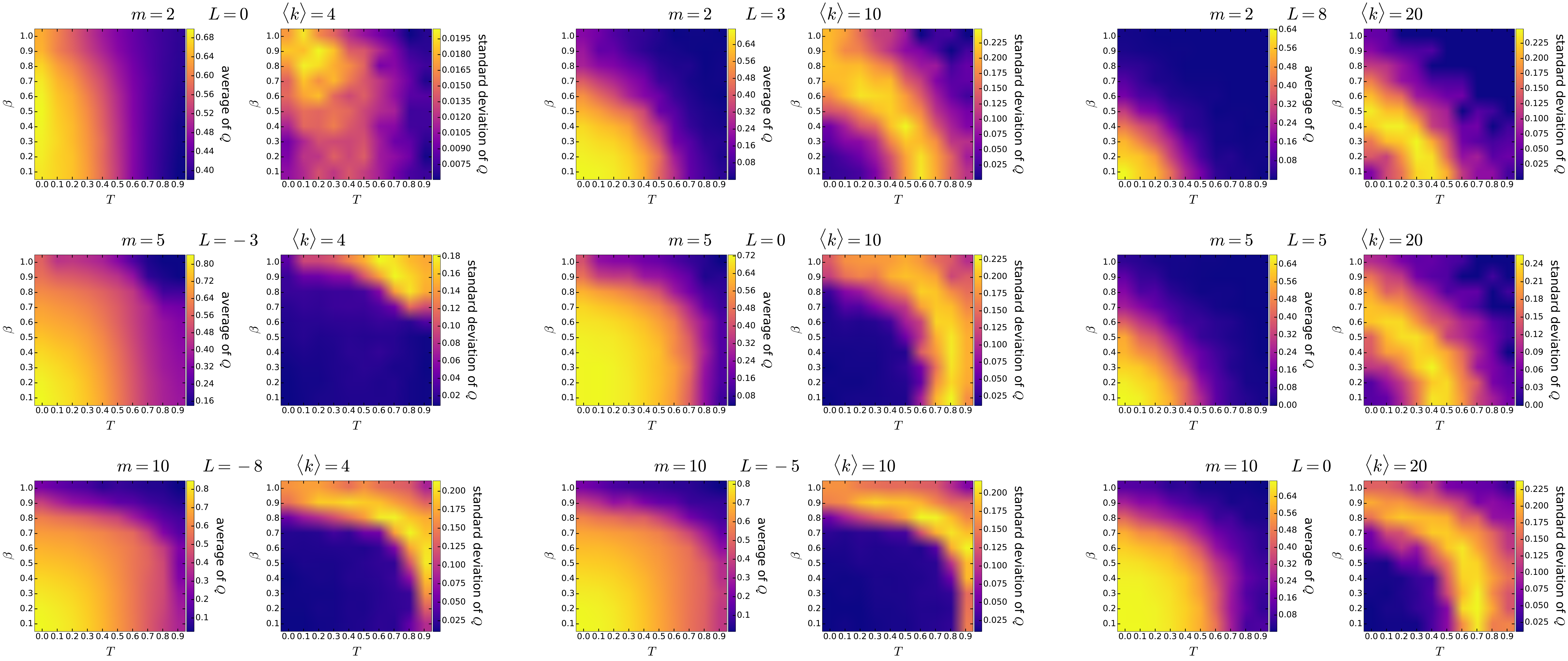}}
    \caption{{\bf The mean and the standard deviation of the unweighted modularity $Q$ of the community structure detected by the \textit{asynchronous label propagation} algorithm in 100 \textit{unweighted} \textit{E-PSO} networks of different parametrisations.} Each pair of subplots depicts the effect of changing the popularity fading parameter $\beta$ and the temperature $T$, with the parameters $m$ and $L$ given in the title of the subplot pair together with the corresponding expected average degree $\langle k\rangle=2(m+L)$. The number of nodes $N$ was 1000 in each case. The curvature of the hyperbolic plane $K$ was always set to $-1$, i.e. we used $\zeta=1$.}
    \label{fig:EPSO_alabpropMod_uw}
\end{figure}

\begin{figure}[hbt]
    \centering
    \makebox[\textwidth][c]{\includegraphics[width=1.15\textwidth]{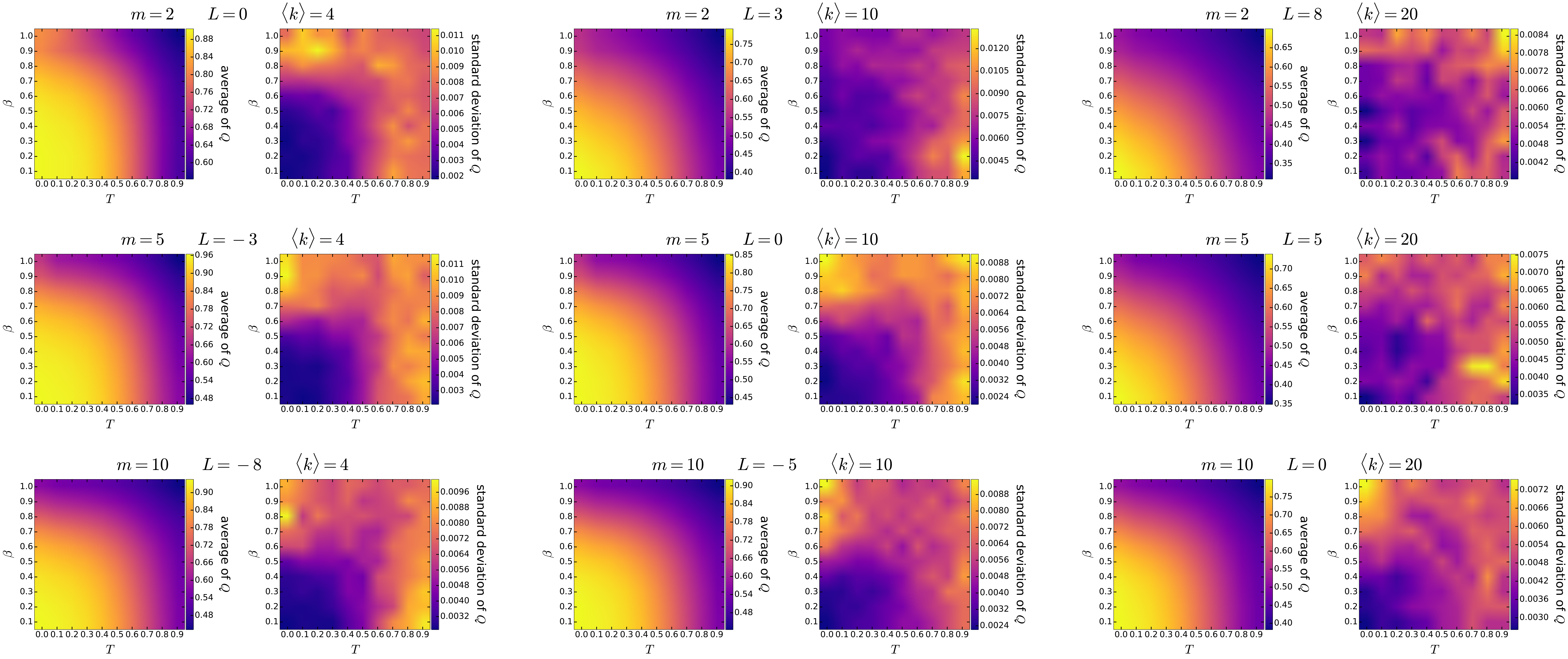}}
    \caption{{\bf The mean and the standard deviation of the unweighted modularity $Q$ of the community structure detected by the \textit{Louvain} algorithm in 100 \textit{unweighted} \textit{E-PSO} networks of different parametrisations.} Each pair of subplots depicts the effect of changing the popularity fading parameter $\beta$ and the temperature $T$, with the parameters $m$ and $L$ given in the title of the subplot pair together with the corresponding expected average degree $\langle k\rangle=2(m+L)$. The number of nodes $N$ was 1000 in each case. The curvature of the hyperbolic plane $K$ was always set to $-1$, i.e. we used $\zeta=1$.}
    \label{fig:EPSO_LouvainMod_uw}
\end{figure}

\begin{figure}[hbt]
    \centering
    \makebox[\textwidth][c]{\includegraphics[width=1.15\textwidth]{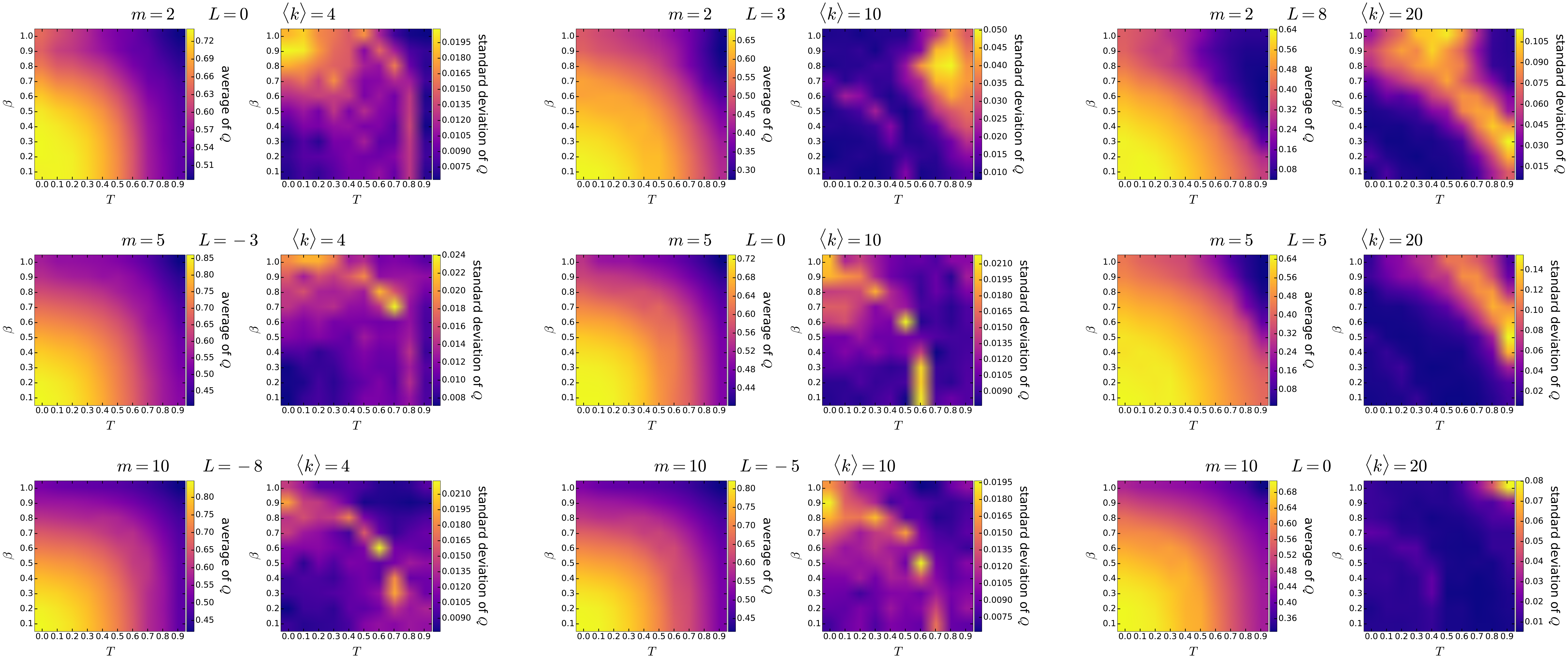}}
    \caption{{\bf The mean and the standard deviation of the unweighted modularity $Q$ of the community structure detected by the \textit{Infomap} algorithm in 100 \textit{unweighted} \textit{E-PSO} networks of different parametrisations.} Each pair of subplots depicts the effect of changing the popularity fading parameter $\beta$ and the temperature $T$, with the parameters $m$ and $L$ given in the title of the subplot pair together with the corresponding expected average degree $\langle k\rangle=2(m+L)$. The number of nodes $N$ was 1000 in each case. The curvature of the hyperbolic plane $K$ was always set to $-1$, i.e. we used $\zeta=1$.}
    \label{fig:EPSO_InfomapMod_uw}
\end{figure}

\begin{figure}[hbt]
    \centering
    \makebox[\textwidth][c]{\includegraphics[width=1.15\textwidth]{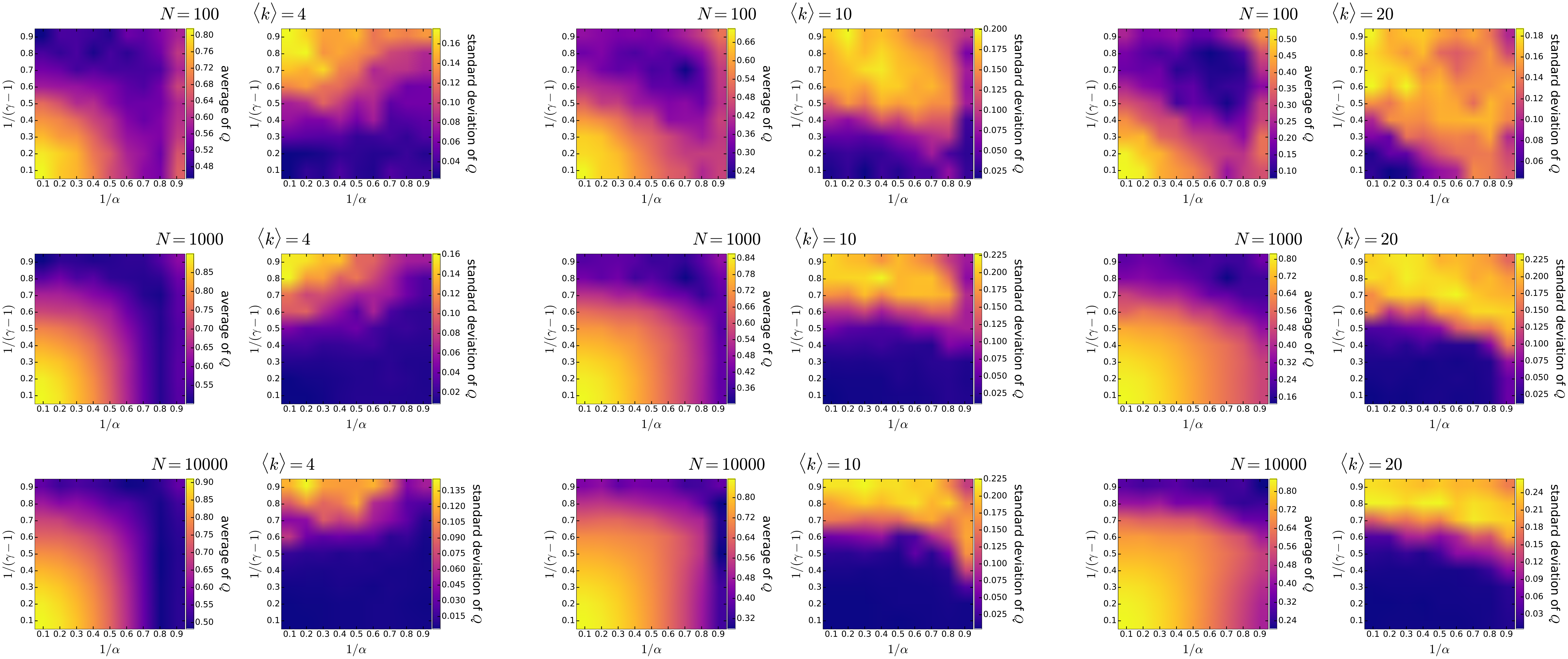}}
    \caption{{\bf The mean and the standard deviation of the unweighted modularity $Q$ of the community structure detected by the \textit{asynchronous label propagation} algorithm in 100 \textit{unweighted} $\mathbb{S}^1/\mathbb{H}^2$ networks of different parametrisations.} Each pair of subplots depicts the effect of changing $1/(\gamma-1)$ (equivalent to the popularity fading parameter $\beta$ in the E-PSO model) and $1/\alpha$ (analogous to the temperature $T$ in the E-PSO model), with the number of nodes $N$ and the expected average degree $\langle k\rangle$ given in the title of the subplot pair. We used $K=-1$ as the curvature of the hyperbolic plane in each case.}
    \label{fig:S1_alabpropMod_uw}
\end{figure}

\begin{figure}[hbt]
    \centering
    \makebox[\textwidth][c]{\includegraphics[width=1.15\textwidth]{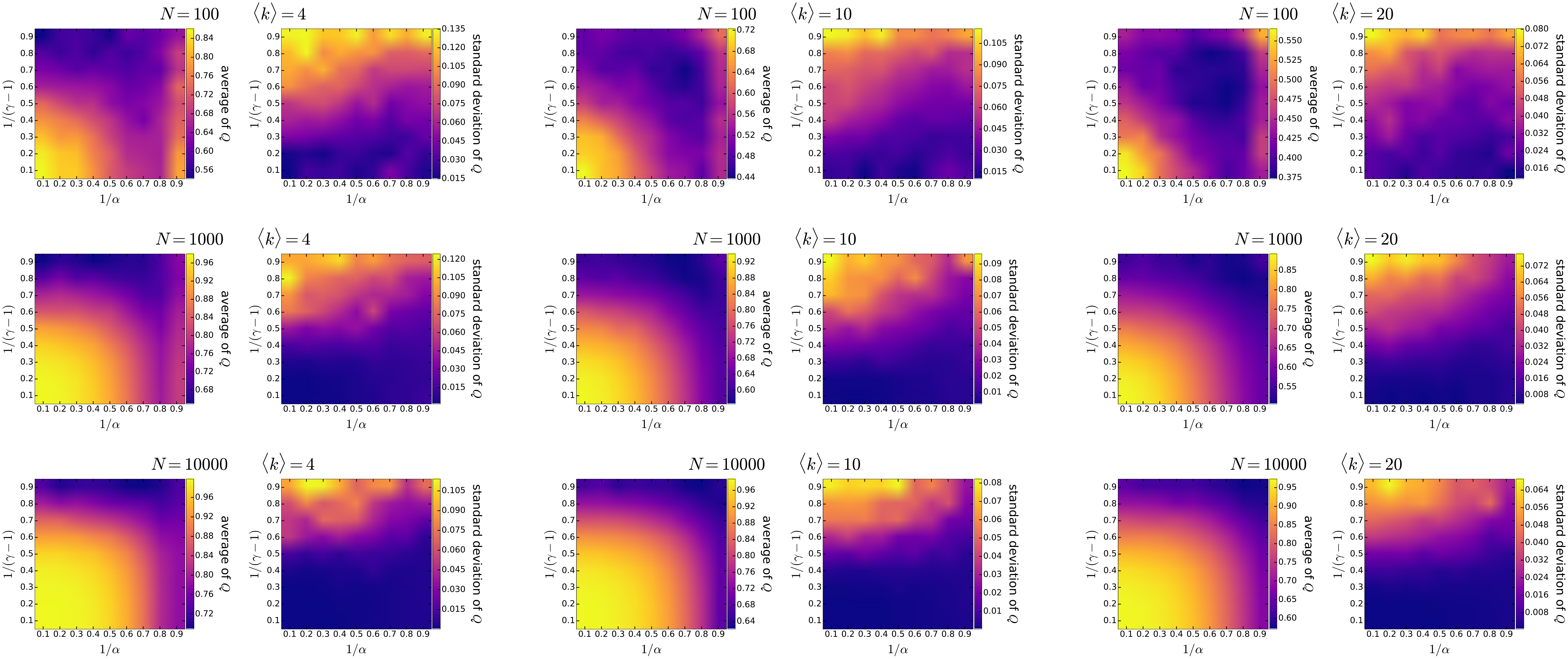}}
    \caption{{\bf The mean and the standard deviation of the unweighted modularity $Q$ of the community structure detected by the \textit{Louvain} algorithm in 100 \textit{unweighted} $\mathbb{S}^1/\mathbb{H}^2$ networks of different parametrisations.} Each pair of subplots depicts the effect of changing $1/(\gamma-1)$ (equivalent to the popularity fading parameter $\beta$ in the E-PSO model) and $1/\alpha$ (analogous to the temperature $T$ in the E-PSO model), with the number of nodes $N$ and the expected average degree $\langle k\rangle$ given in the title of the subplot pair. We used $K=-1$ as the curvature of the hyperbolic plane in each case.}
    \label{fig:S1_LouvainMod_uw}
\end{figure}

\begin{figure}[hbt]
    \centering
    \makebox[\textwidth][c]{\includegraphics[width=1.15\textwidth]{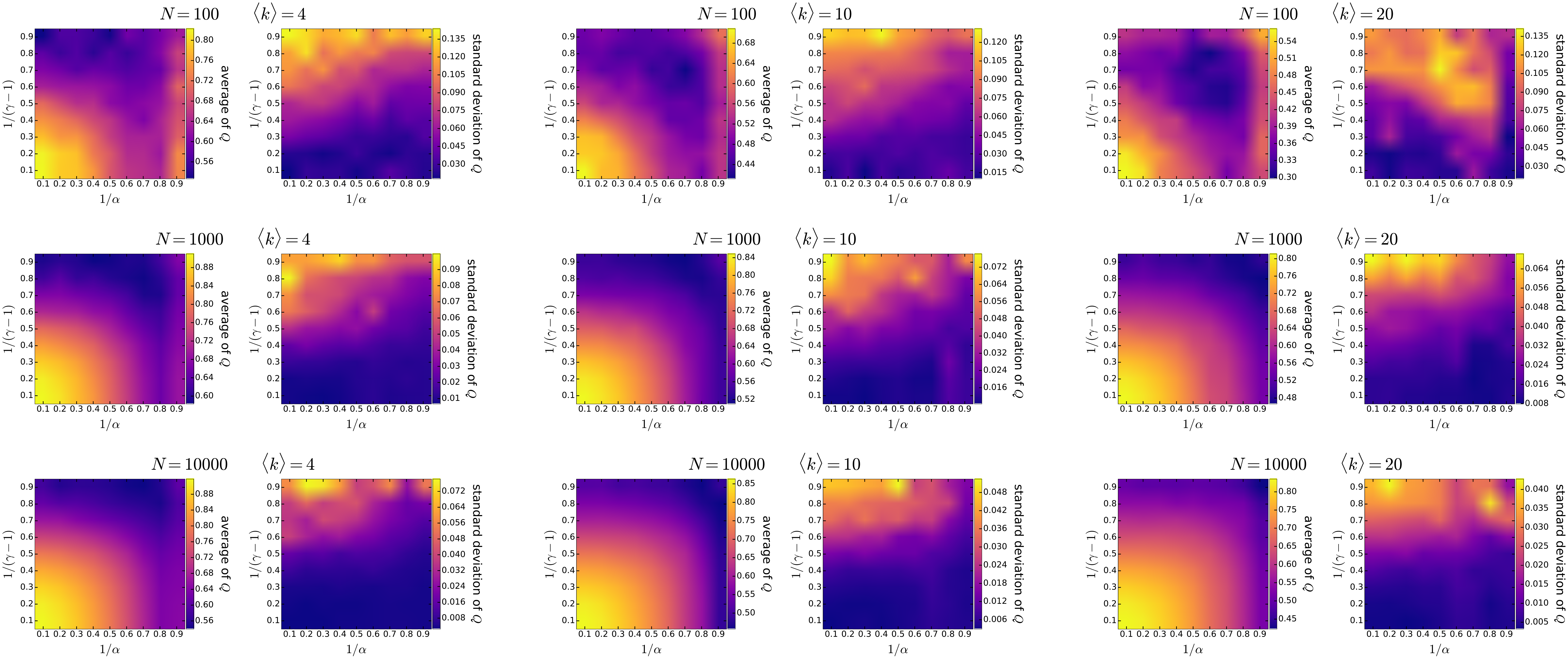}}
    \caption{{\bf The mean and the standard deviation of the unweighted modularity $Q$ of the community structure detected by the \textit{Infomap} algorithm in 100 \textit{unweighted} $\mathbb{S}^1/\mathbb{H}^2$ networks of different parametrisations.} Each pair of subplots depicts the effect of changing $1/(\gamma-1)$ (equivalent to the popularity fading parameter $\beta$ in the E-PSO model) and $1/\alpha$ (analogous to the temperature $T$ in the E-PSO model), with the number of nodes $N$ and the expected average degree $\langle k\rangle$ given in the title of the subplot pair. We used $K=-1$ as the curvature of the hyperbolic plane in each case.}
    \label{fig:S1_InfomapMod_uw}
\end{figure}

\begin{figure}[hbt]
    \centering
    \makebox[\textwidth][c]{\includegraphics[width=1.15\textwidth]{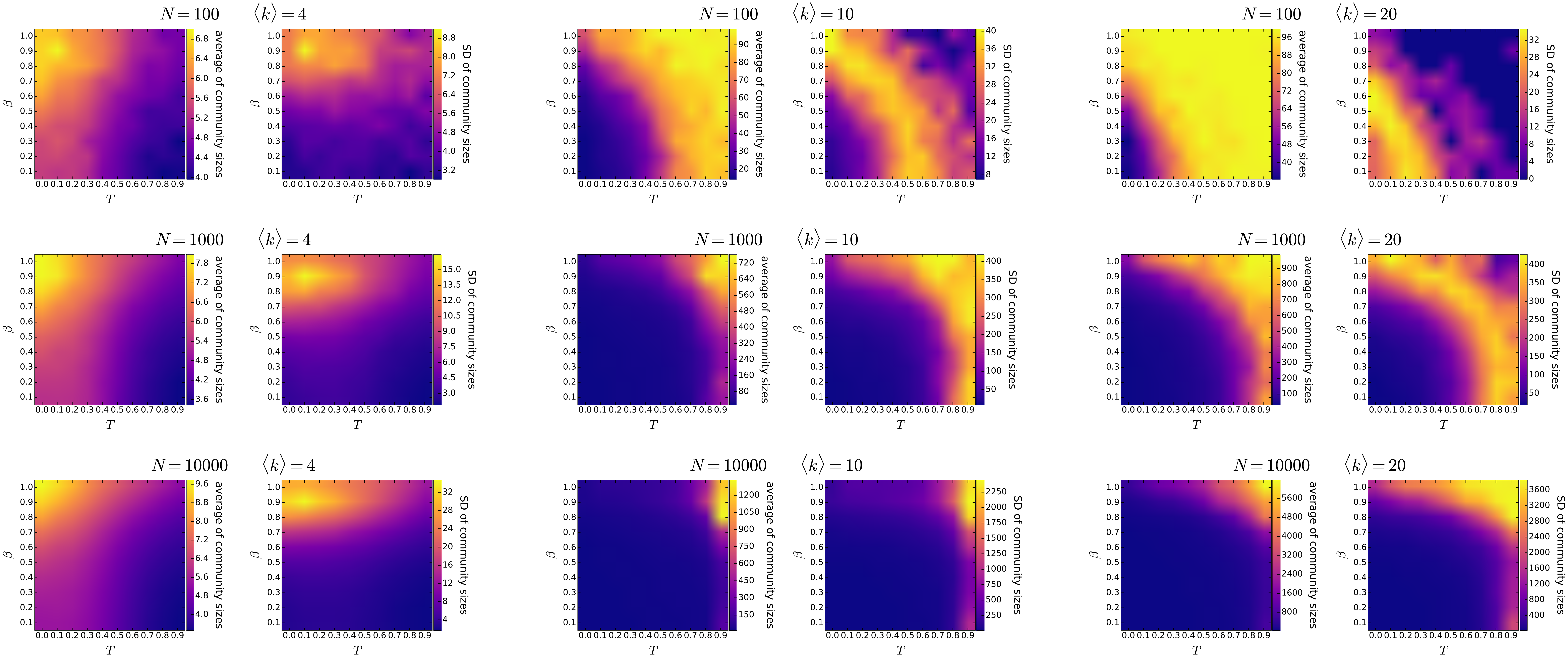}}
    \caption{{\bf The mean and the standard deviation of the size of communities detected by the \textit{asynchronous label propagation} algorithm in 100 \textit{unweighted} \textit{PSO} networks of different parametrisations.} Each pair of subplots depicts the effect of changing the popularity fading parameter $\beta$ and the temperature $T$, with the number of nodes $N$ and the expected average degree $\langle k\rangle=2m$ given in the title of the subplot pair. The curvature of the hyperbolic plane $K$ was always set to $-1$, i.e. we used $\zeta=1$.}
    \label{fig:PSO_alabpropGSavstd_uw}
\end{figure}

\begin{figure}[hbt]
    \centering
    \makebox[\textwidth][c]{\includegraphics[width=1.15\textwidth]{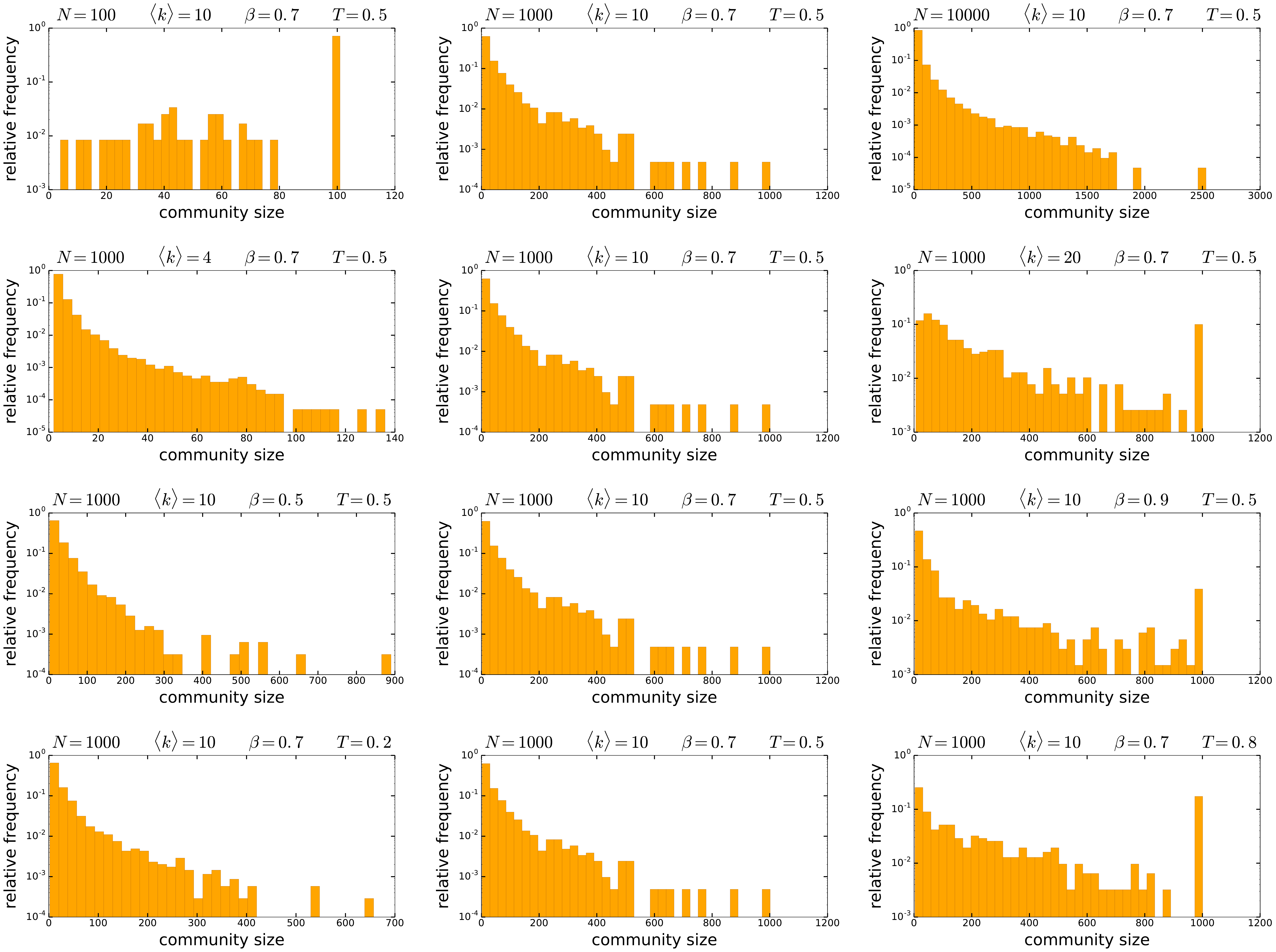}}
    \caption{{\bf The size distribution of the communities detected by the \textit{asynchronous label propagation} algorithm in 100 \textit{unweighted} \textit{PSO} networks of different parametrisations.} The parameters of the network generation are listed in the title for each subplot. The curvature of the hyperbolic plane $K$ was always set to $-1$, i.e. we used $\zeta=1$. Each row of the figure demonstrates the effect of the change in a given network generation parameter: from top to bottom, the number of nodes $N$, the expected average degree $\langle k\rangle=2m$, the popularity fading parameter $\beta$ and the temperature $T$.}
    \label{fig:PSO_alabpropGShist_uw}
\end{figure}

\begin{figure}[hbt]
    \centering
    \makebox[\textwidth][c]{\includegraphics[width=1.15\textwidth]{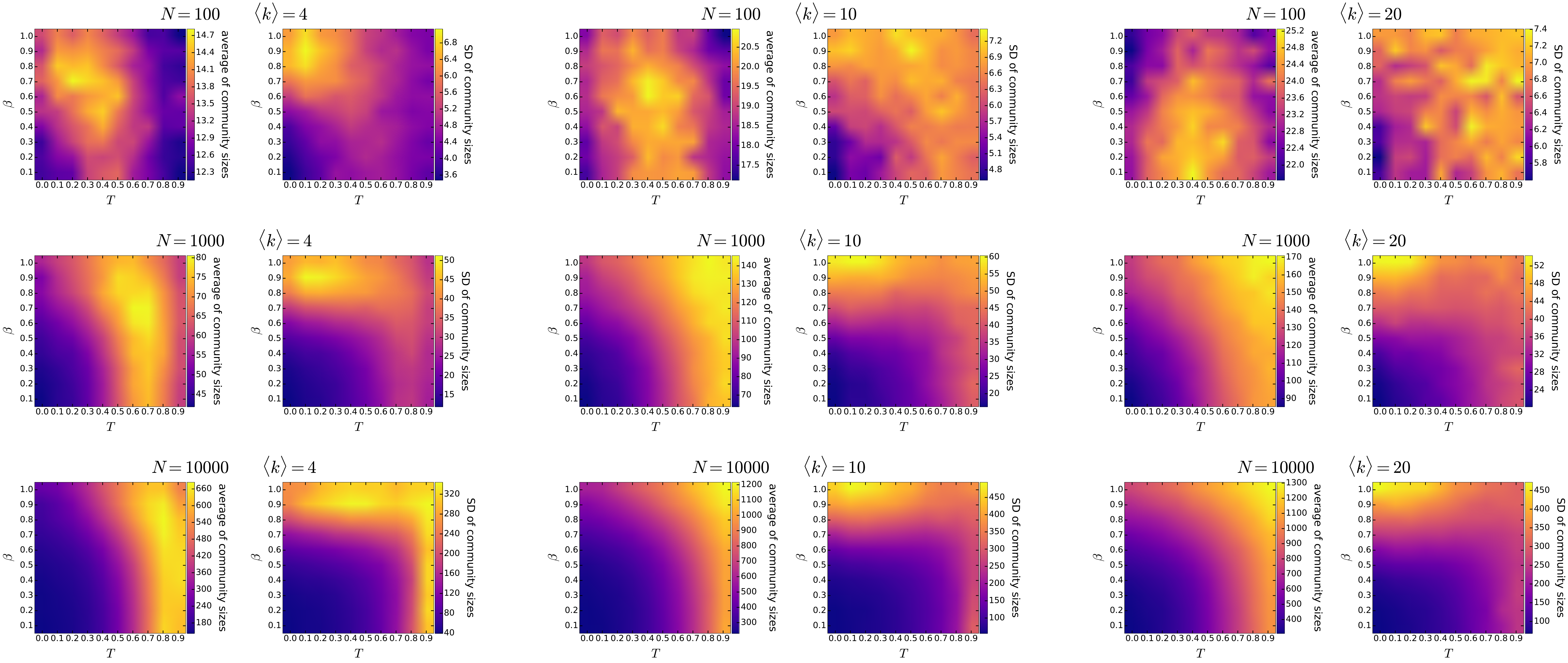}}
    \caption{{\bf The mean and the standard deviation of the size of communities detected by the \textit{Louvain} algorithm in 100 \textit{unweighted} \textit{PSO} networks of different parametrisations.} Each pair of subplots depicts the effect of changing the popularity fading parameter $\beta$ and the temperature $T$, with the number of nodes $N$ and the expected average degree $\langle k\rangle=2m$ given in the title of the subplot pair. The curvature of the hyperbolic plane $K$ was always set to $-1$, i.e. we used $\zeta=1$.}
    \label{fig:PSO_LouvainGSavstd_uw}
\end{figure}

\begin{figure}[hbt]
    \centering
    \makebox[\textwidth][c]{\includegraphics[width=1.15\textwidth]{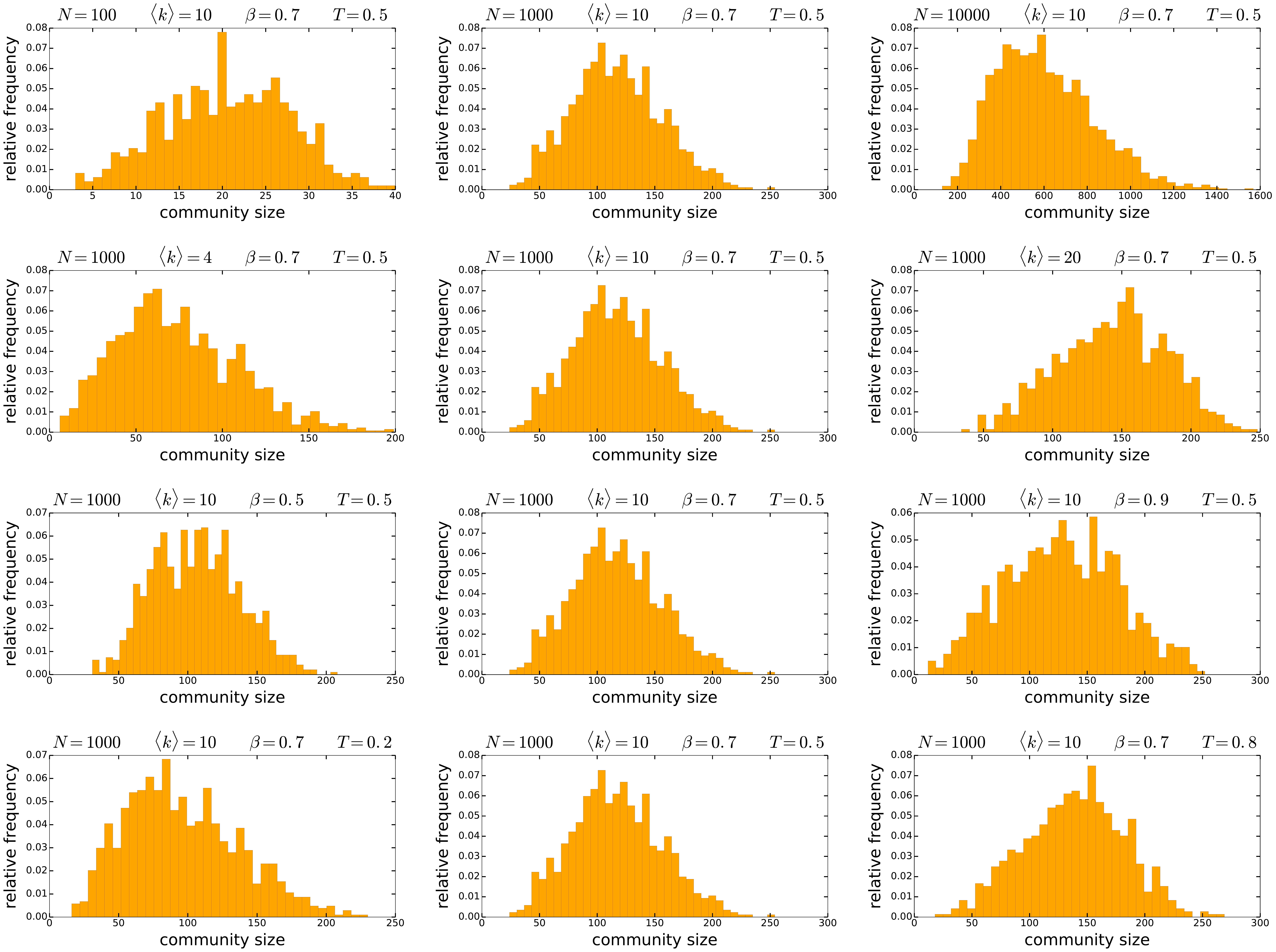}}
    \caption{{\bf The size distribution of the communities detected by the \textit{Louvain} algorithm in 100 \textit{unweighted} \textit{PSO} networks of different parametrisations.} The parameters of the network generation are listed in the title for each subplot. The curvature of the hyperbolic plane $K$ was always set to $-1$, i.e. we used $\zeta=1$. Each row of the figure demonstrates the effect of the change in a given network generation parameter: from top to bottom, the number of nodes $N$, the expected average degree $\langle k\rangle=2m$, the popularity fading parameter $\beta$ and the temperature $T$.}
    \label{fig:PSO_LouvainGShist_uw}
\end{figure}

\begin{figure}[hbt]
    \centering
    \makebox[\textwidth][c]{\includegraphics[width=1.15\textwidth]{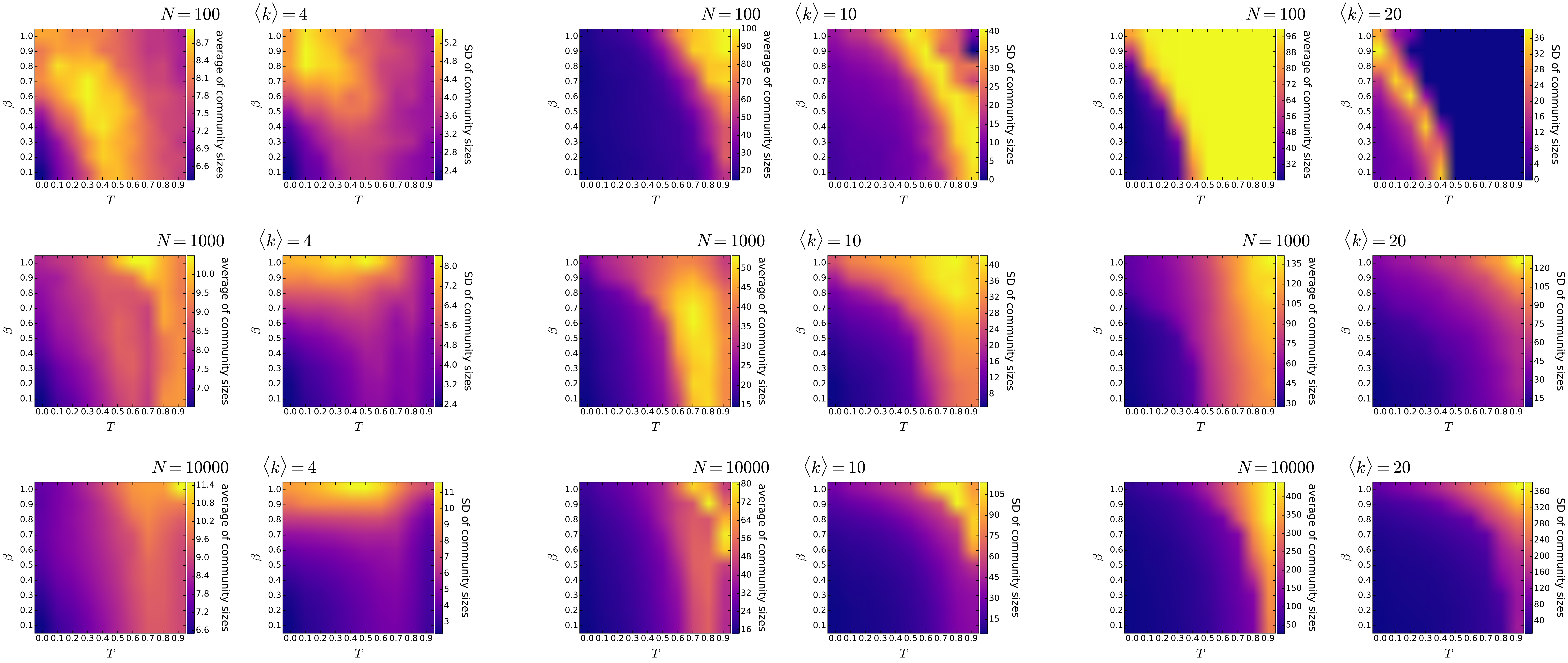}}
    \caption{{\bf The mean and the standard deviation of the size of communities detected by the \textit{Infomap} algorithm in 100 \textit{unweighted} \textit{PSO} networks of different parametrisations.} Each pair of subplots depicts the effect of changing the popularity fading parameter $\beta$ and the temperature $T$, with the number of nodes $N$ and the expected average degree $\langle k\rangle=2m$ given in the title of the subplot pair. The curvature of the hyperbolic plane $K$ was always set to $-1$, i.e. we used $\zeta=1$.}
    \label{fig:PSO_InfomapGSavstd_uw}
\end{figure}

\begin{figure}[hbt]
    \centering
    \makebox[\textwidth][c]{\includegraphics[width=1.15\textwidth]{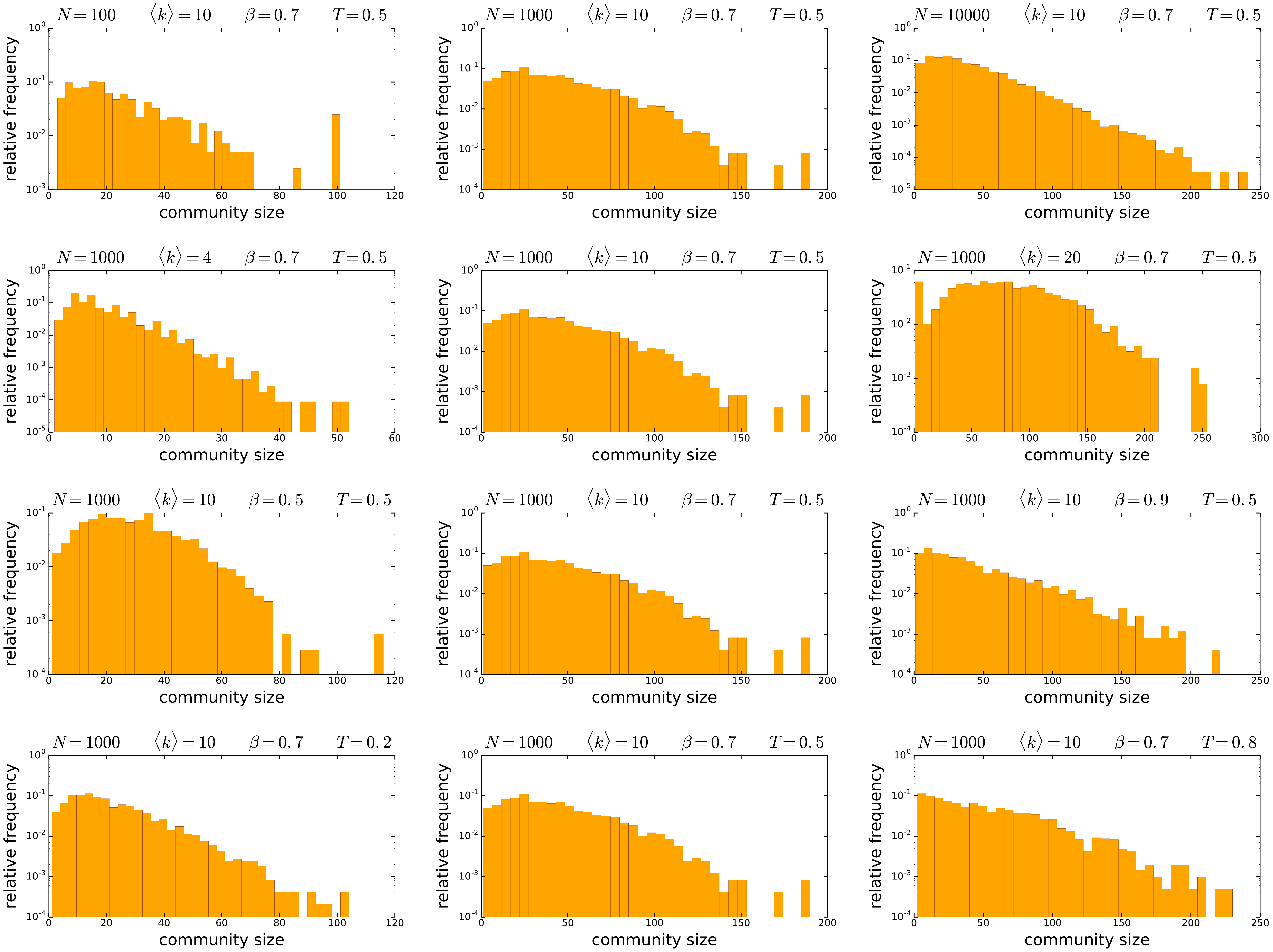}}
    \caption{{\bf The size distribution of the communities detected by the \textit{Infomap} algorithm in 100 \textit{unweighted} \textit{PSO} networks of different parametrisations.} The parameters of the network generation are listed in the title for each subplot. The curvature of the hyperbolic plane $K$ was always set to $-1$, i.e. we used $\zeta=1$. Each row of the figure demonstrates the effect of the change in a given network generation parameter: from top to bottom, the number of nodes $N$, the expected average degree $\langle k\rangle=2m$, the popularity fading parameter $\beta$ and the temperature $T$.}
    \label{fig:PSO_InfomapGShist_uw}
\end{figure}

\begin{figure}[hbt]
    \centering
    \makebox[\textwidth][c]{\includegraphics[width=1.15\textwidth]{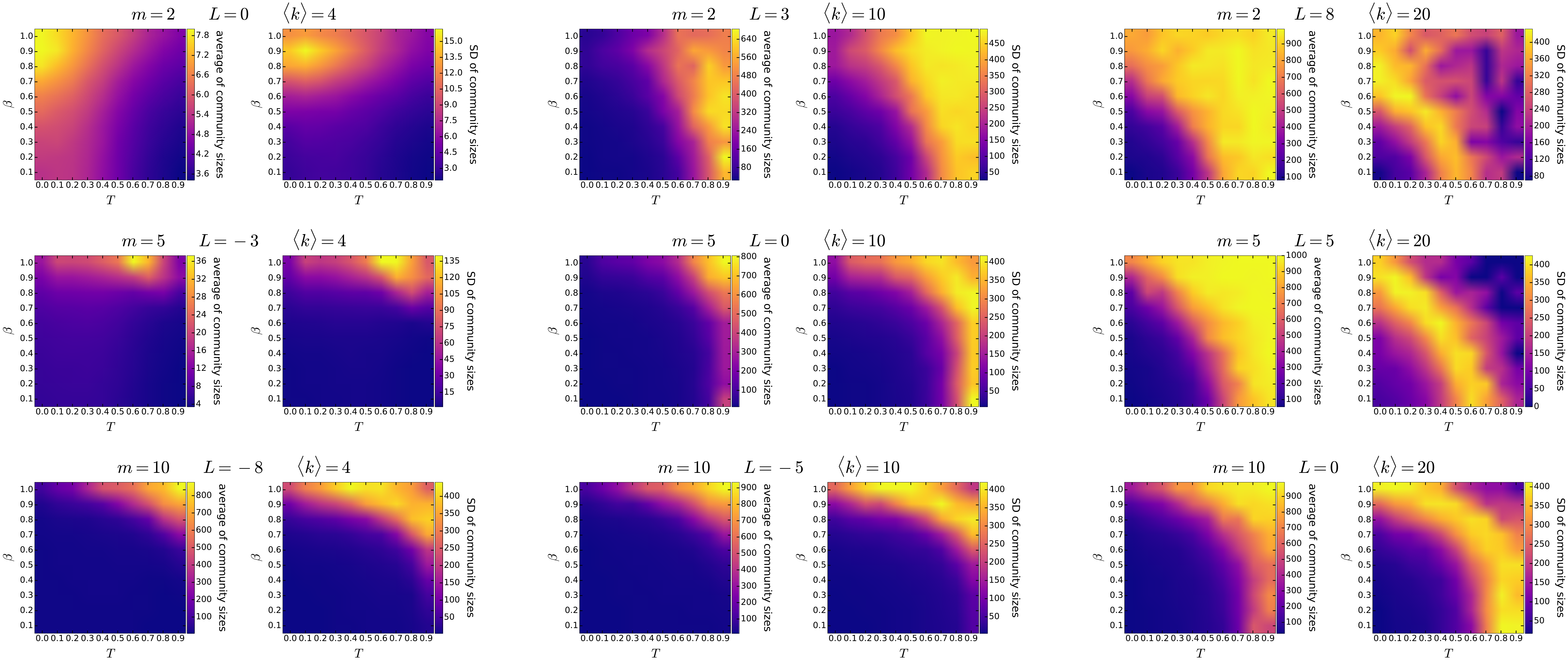}}
    \caption{{\bf The mean and the standard deviation of the size of communities detected by the \textit{asynchronous label propagation} algorithm in 100 \textit{unweighted} \textit{E-PSO} networks of different parametrisations.} Each pair of subplots depicts the effect of changing the popularity fading parameter $\beta$ and the temperature $T$, with the parameters $m$ and $L$ given in the title of the subplot pair together with the corresponding expected average degree $\langle k\rangle=2(m+L)$. The number of nodes $N$ was 1000 in each case. The curvature of the hyperbolic plane $K$ was always set to $-1$, i.e. we used $\zeta=1$.}
    \label{fig:EPSO_alabpropGSavstd_uw}
\end{figure}

\begin{figure}[hbt]
    \centering
    \makebox[\textwidth][c]{\includegraphics[width=1.15\textwidth]{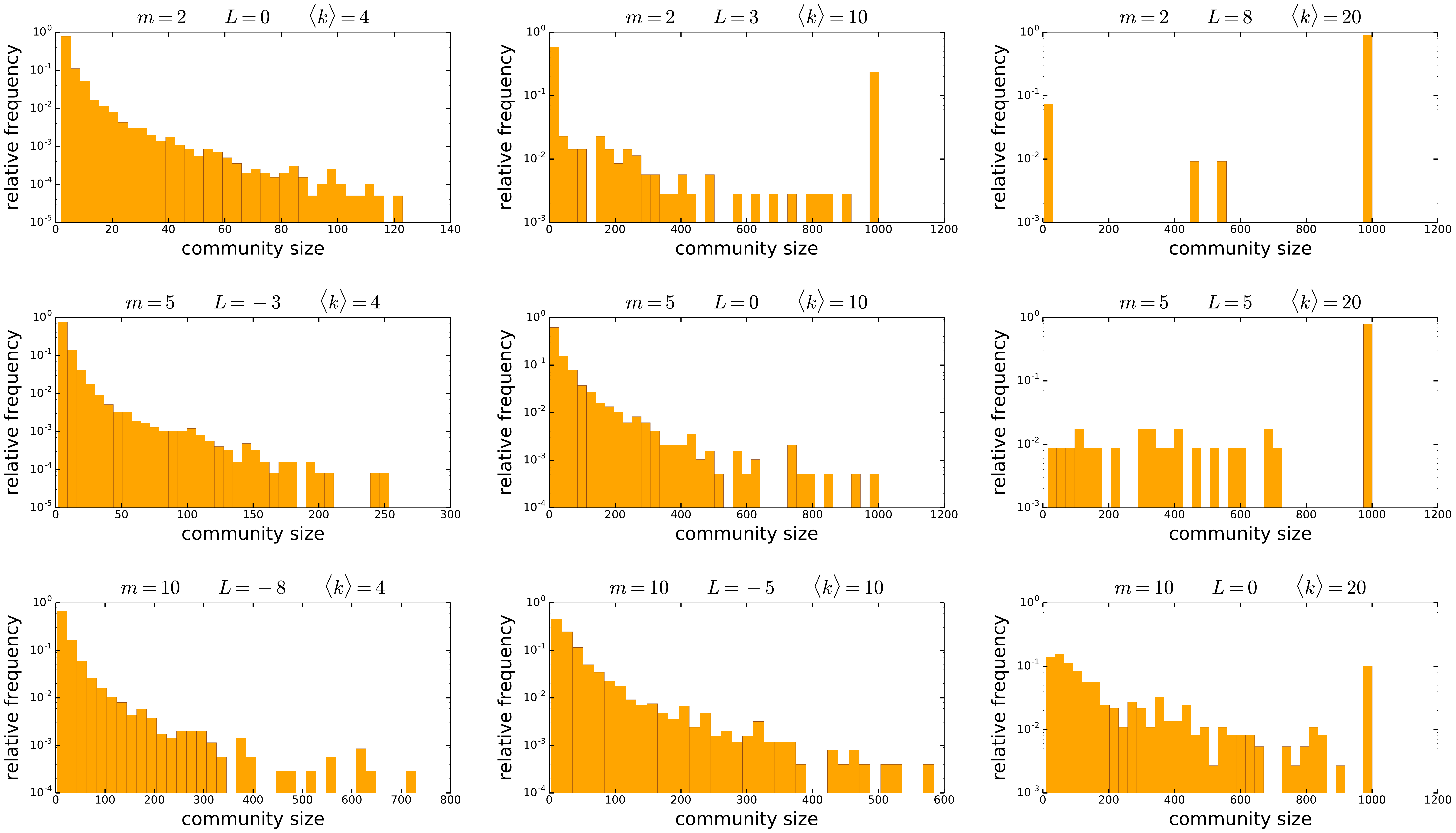}}
    \caption{{\bf The size distribution of the communities detected by the \textit{asynchronous label propagation} algorithm in 100 \textit{unweighted} \textit{E-PSO} networks of different parametrisations.} We used $\zeta=1$, i.e. $K=-1$ as the curvature of the hyperbolic plane, the number of nodes $N$ was 1000, the popularity fading parameter $\beta$ was 0.7 and the temperature $T$ was 0.5 in each case. The parameters $m$ and $L$ are given in the title for each subplot together with the corresponding expected average degree $\langle k\rangle=2(m+L)$.}
    \label{fig:EPSO_alabpropGShist_uw}
\end{figure}

\begin{figure}[hbt]
    \centering
    \makebox[\textwidth][c]{\includegraphics[width=1.15\textwidth]{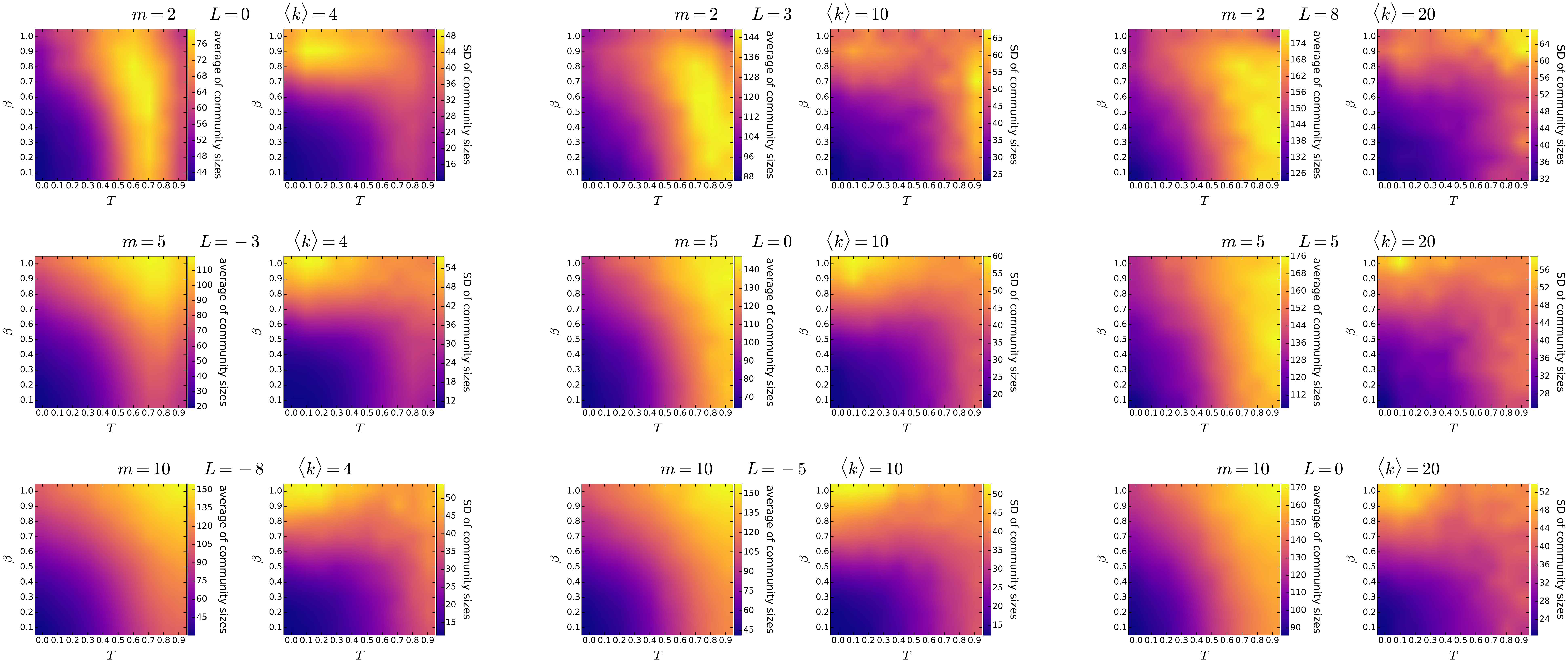}}
    \caption{{\bf The mean and the standard deviation of the size of communities detected by the \textit{Louvain} algorithm in 100 \textit{unweighted} \textit{E-PSO} networks of different parametrisations.} Each pair of subplots depicts the effect of changing the popularity fading parameter $\beta$ and the temperature $T$, with the parameters $m$ and $L$ given in the title of the subplot pair together with the corresponding expected average degree $\langle k\rangle=2(m+L)$. The number of nodes $N$ was 1000 in each case. The curvature of the hyperbolic plane $K$ was always set to $-1$, i.e. we used $\zeta=1$.}
    \label{fig:EPSO_LouvainGSavstd_uw}
\end{figure}

\begin{figure}[hbt]
    \centering
    \makebox[\textwidth][c]{\includegraphics[width=1.15\textwidth]{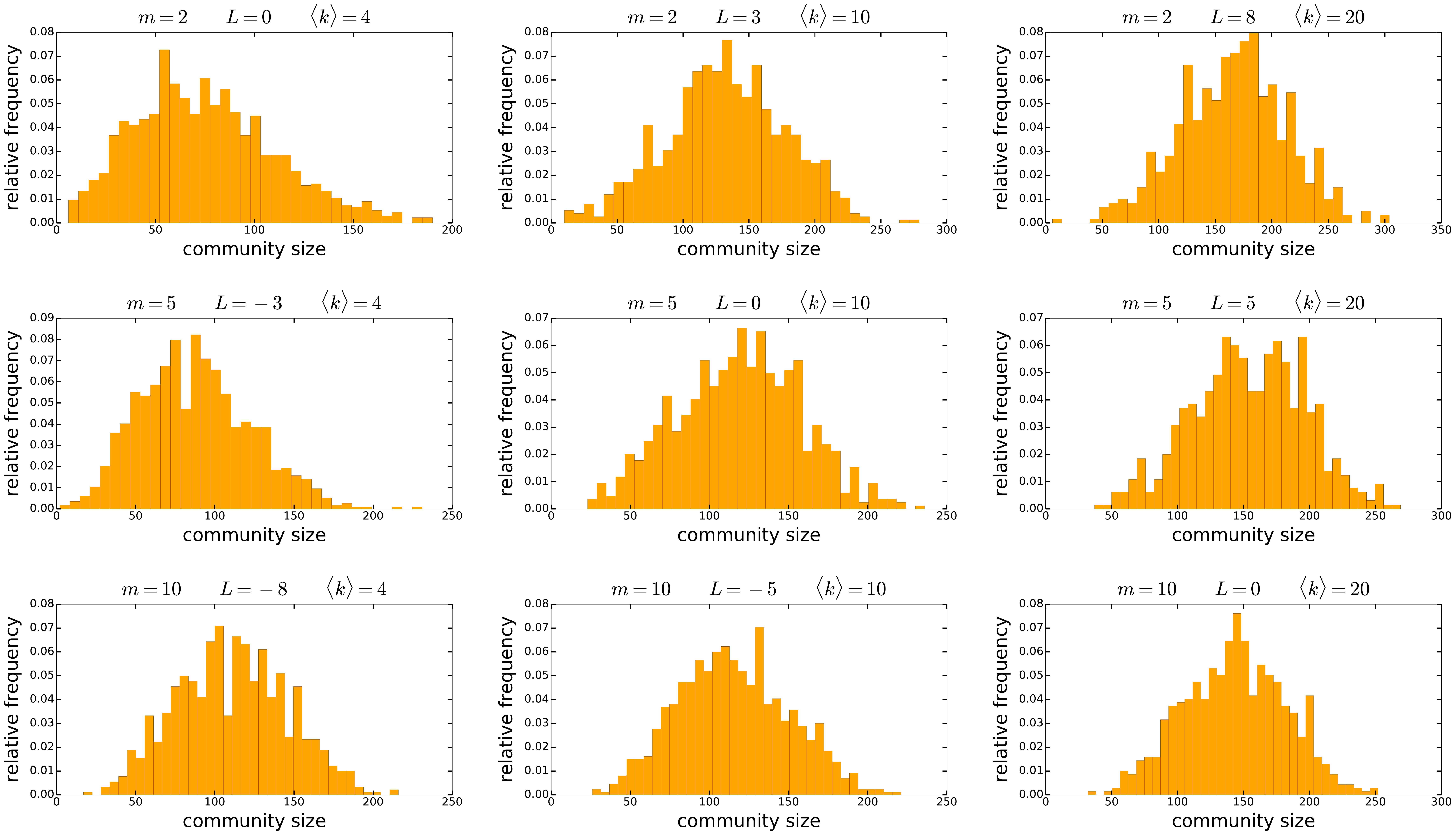}}
    \caption{{\bf The size distribution of the communities detected by the \textit{Louvain} algorithm in 100 \textit{unweighted} \textit{E-PSO} networks of different parametrisations.} We used $\zeta=1$, i.e. $K=-1$ as the curvature of the hyperbolic plane, the number of nodes $N$ was 1000, the popularity fading parameter $\beta$ was 0.7 and the temperature $T$ was 0.5 in each case. The parameters $m$ and $L$ are given in the title for each subplot together with the corresponding expected average degree $\langle k\rangle=2(m+L)$.}
    \label{fig:EPSO_LouvainGShist_uw}
\end{figure}

\begin{figure}[hbt]
    \centering
    \makebox[\textwidth][c]{\includegraphics[width=1.15\textwidth]{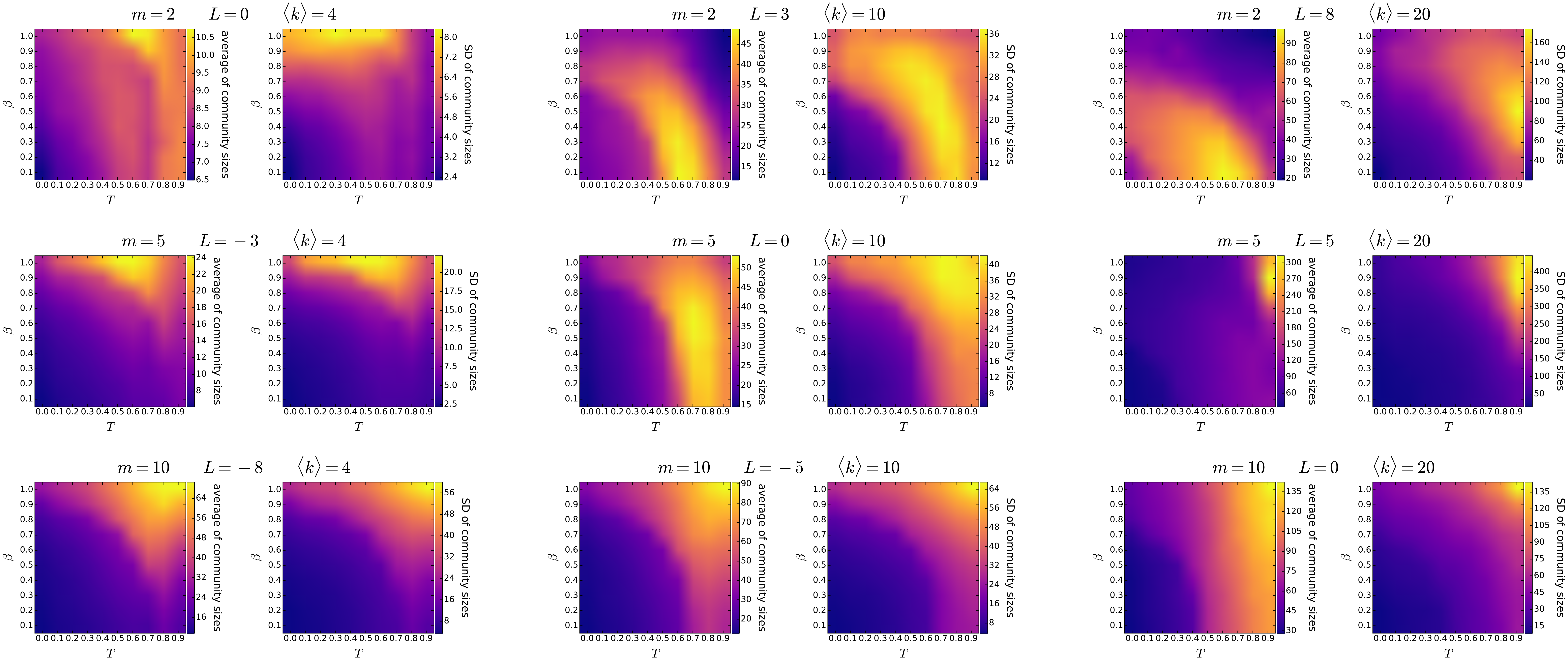}}
    \caption{{\bf The mean and the standard deviation of the size of communities detected by the \textit{Infomap} algorithm in 100 \textit{unweighted} \textit{E-PSO} networks of different parametrisations.} Each pair of subplots depicts the effect of changing the popularity fading parameter $\beta$ and the temperature $T$, with the parameters $m$ and $L$ given in the title of the subplot pair together with the corresponding expected average degree $\langle k\rangle=2(m+L)$. The number of nodes $N$ was 1000 in each case. The curvature of the hyperbolic plane $K$ was always set to $-1$, i.e. we used $\zeta=1$.}
    \label{fig:EPSO_InfomapGSavstd_uw}
\end{figure}

\begin{figure}[hbt]
    \centering
    \makebox[\textwidth][c]{\includegraphics[width=1.15\textwidth]{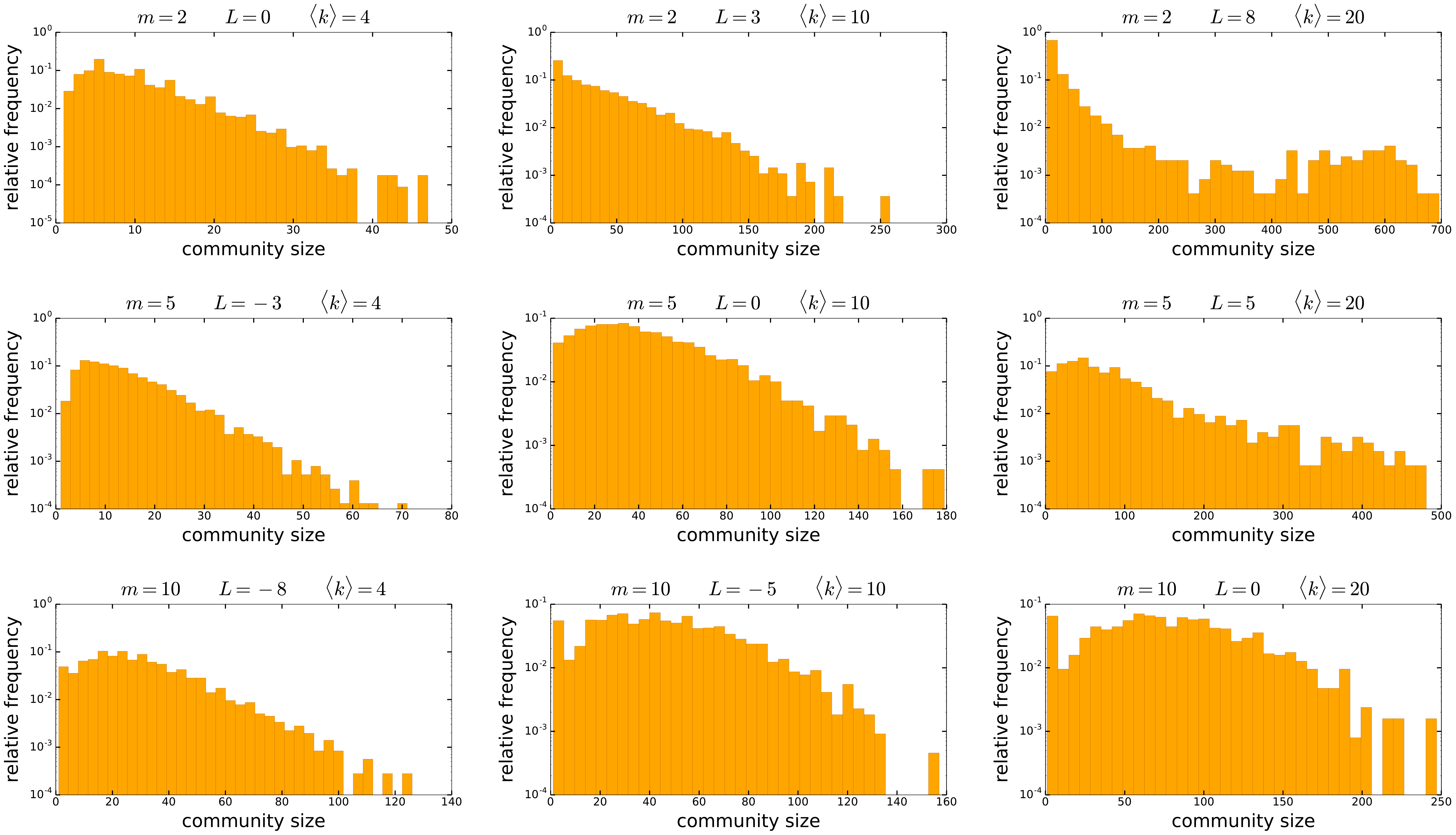}}
    \caption{{\bf The size distribution of the communities detected by the \textit{Infomap} algorithm in 100 \textit{unweighted} \textit{E-PSO} networks of different parametrisations.} We used $\zeta=1$, i.e. $K=-1$ as the curvature of the hyperbolic plane, the number of nodes $N$ was 1000, the popularity fading parameter $\beta$ was 0.7 and the temperature $T$ was 0.5 in each case. The parameters $m$ and $L$ are given in the title for each subplot together with the corresponding expected average degree $\langle k\rangle=2(m+L)$.}
    \label{fig:EPSO_InfomapGShist_uw}
\end{figure}

\begin{figure}[hbt]
    \centering
    \makebox[\textwidth][c]{\includegraphics[width=1.15\textwidth]{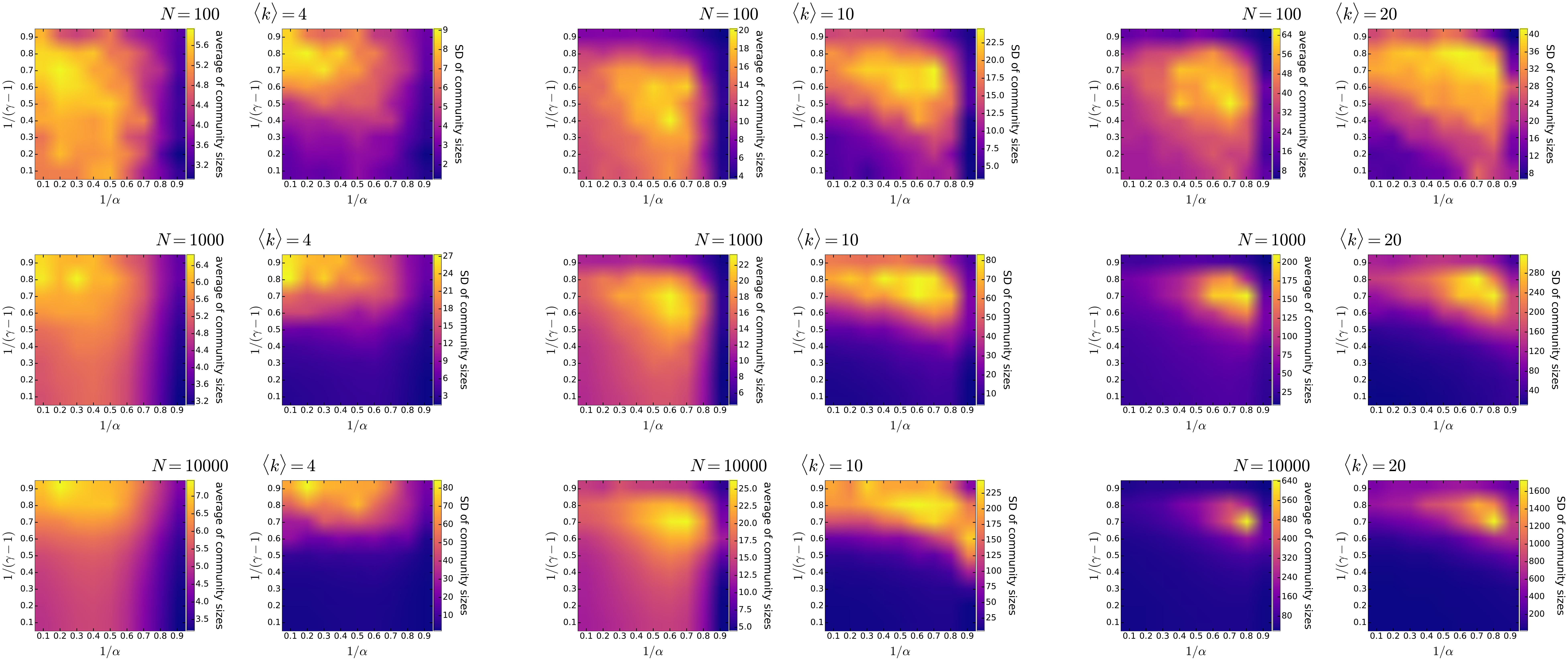}}
    \caption{{\bf The mean and the standard deviation of the size of communities detected by the \textit{asynchronous label propagation} algorithm in 100 \textit{unweighted} $\mathbb{S}^1/\mathbb{H}^2$ networks of different parametrisations.} Each pair of subplots depicts the effect of changing $1/(\gamma-1)$ (equivalent to the popularity fading parameter $\beta$ in the E-PSO model) and $1/\alpha$ (analogous to the temperature $T$ in the E-PSO model), with the number of nodes $N$ and the expected average degree $\langle k\rangle$ given in the title of the subplot pair. We used $K=-1$ as the curvature of the hyperbolic plane in each case.}
    \label{fig:S1_alabpropGSavstd_uw}
\end{figure}

\begin{figure}[hbt]
    \centering
    \makebox[\textwidth][c]{\includegraphics[width=1.15\textwidth]{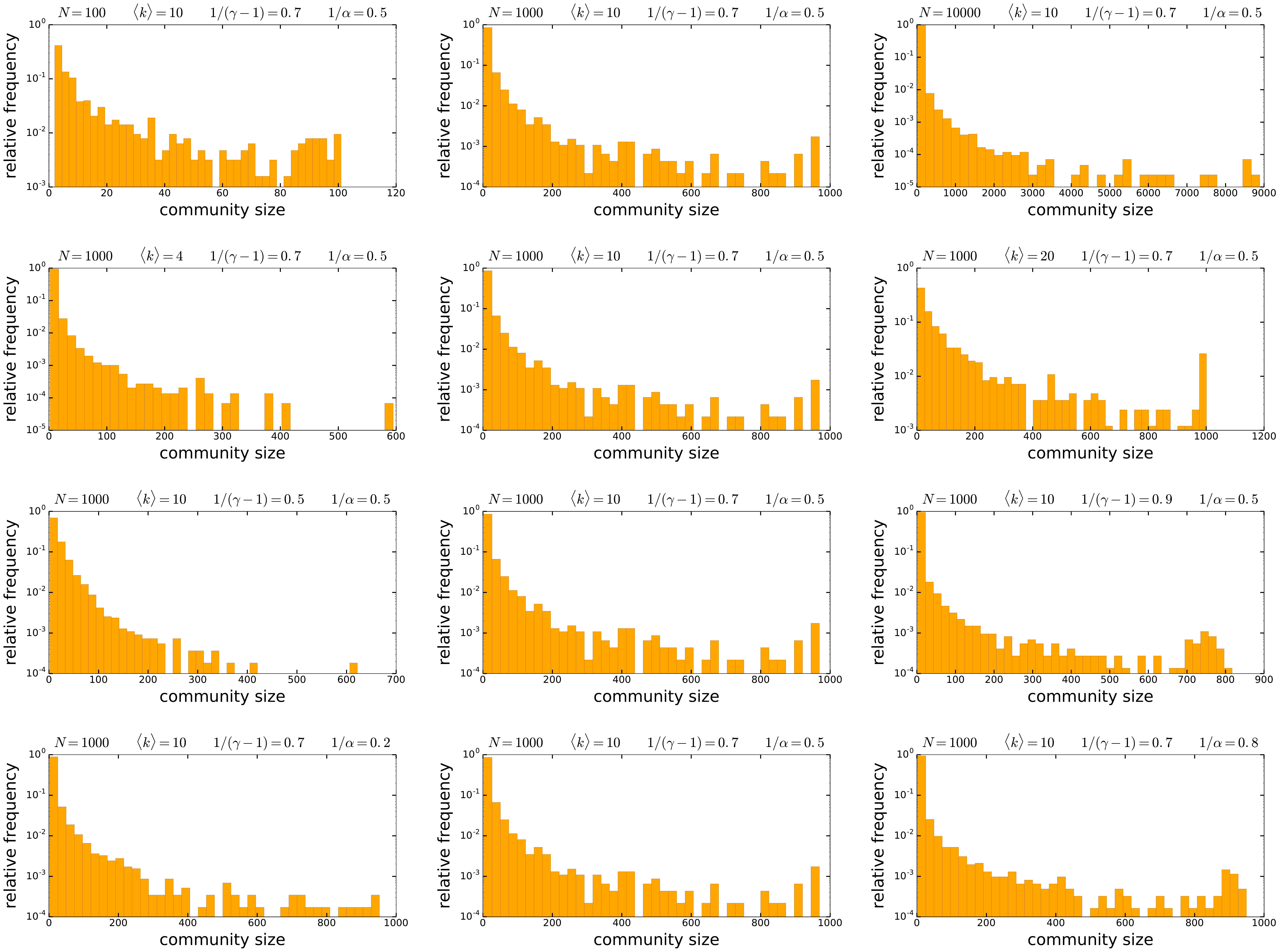}}
    \caption{{\bf The size distribution of the communities detected by the \textit{asynchronous label propagation} algorithm in 100 \textit{unweighted} $\mathbb{S}^1/\mathbb{H}^2$ networks of different parametrisations.} The parameters of the network generation are listed in the title for each subplot. We used $K=-1$ as the curvature of the hyperbolic plane in each case. Each row of the figure demonstrates the effect of the change in a given network generation parameter: from top to bottom, the number of nodes $N$, the expected average degree $\langle k\rangle$, $1/(\gamma-1)$ (equivalent to the popularity fading parameter $\beta$ in the E-PSO model) and $1/\alpha$ (analogous to the temperature $T$ in the E-PSO model).}
    \label{fig:S1_alabpropGShist_uw}
\end{figure}

\begin{figure}[hbt]
    \centering
    \makebox[\textwidth][c]{\includegraphics[width=1.15\textwidth]{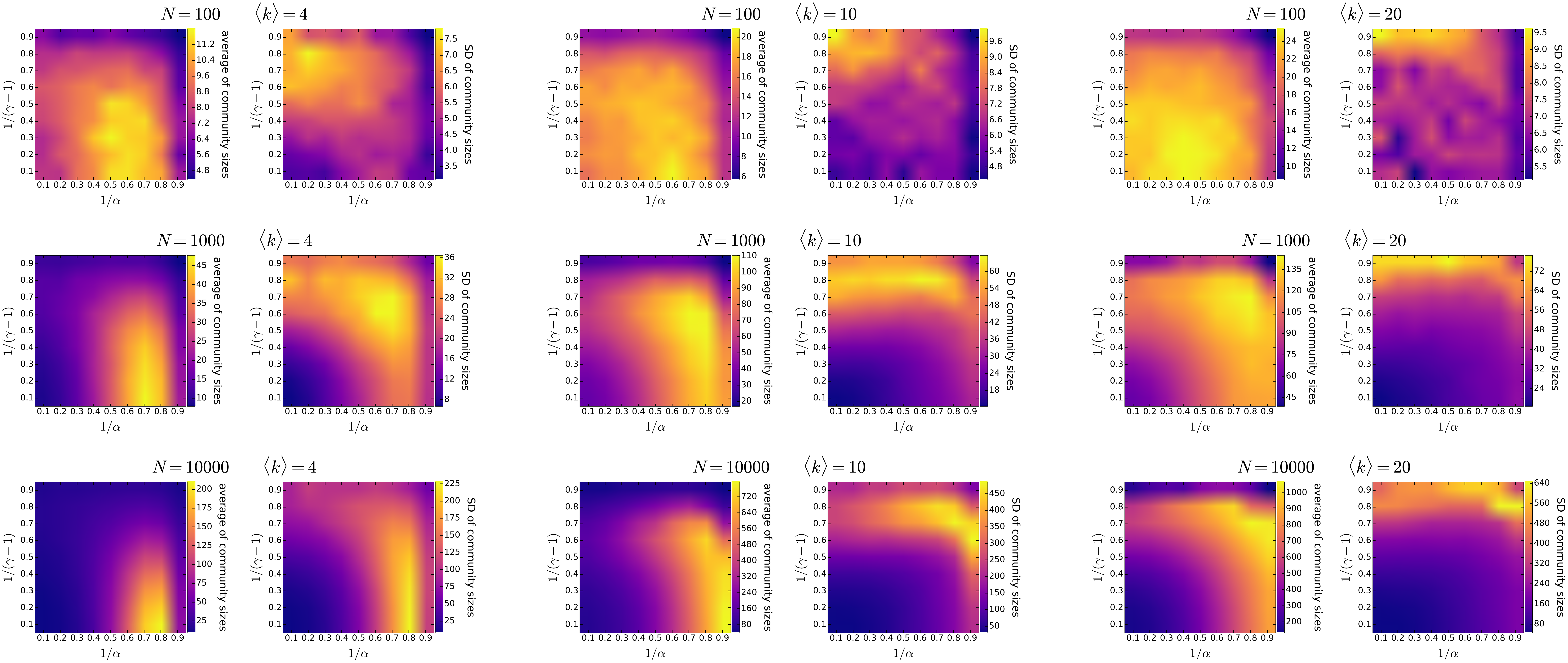}}
    \caption{{\bf The mean and the standard deviation of the size of communities detected by the \textit{Louvain} algorithm in 100 \textit{unweighted} $\mathbb{S}^1/\mathbb{H}^2$ networks of different parametrisations.} Each pair of subplots depicts the effect of changing $1/(\gamma-1)$ (equivalent to the popularity fading parameter $\beta$ in the E-PSO model) and $1/\alpha$ (analogous to the temperature $T$ in the E-PSO model), with the number of nodes $N$ and the expected average degree $\langle k\rangle$ given in the title of the subplot pair. We used $K=-1$ as the curvature of the hyperbolic plane in each case.}
    \label{fig:S1_LouvainGSavstd_uw}
\end{figure}

\begin{figure}[hbt]
    \centering
    \makebox[\textwidth][c]{\includegraphics[width=1.15\textwidth]{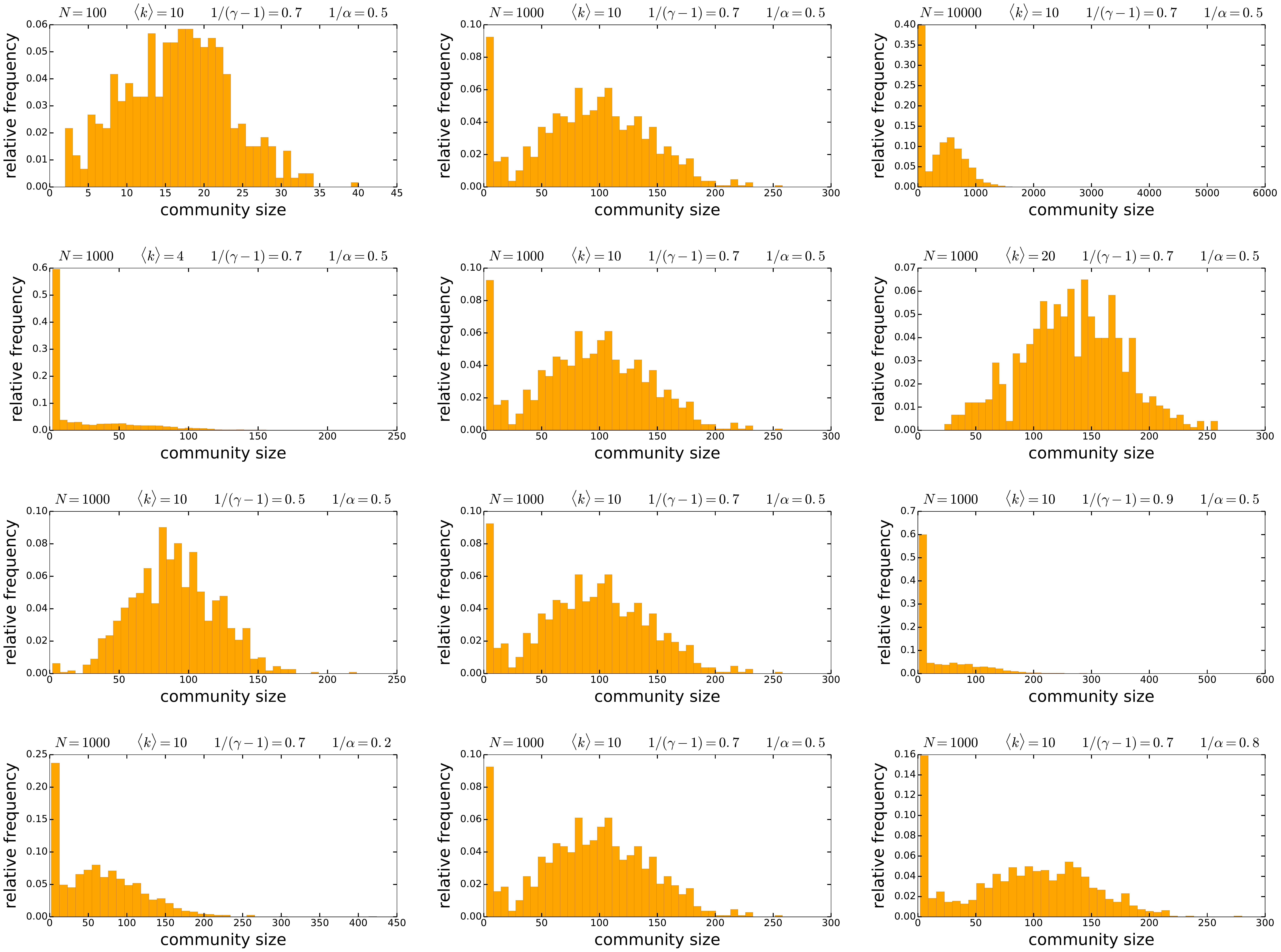}}
    \caption{{\bf The size distribution of the communities detected by the \textit{Louvain} algorithm in 100 \textit{unweighted} $\mathbb{S}^1/\mathbb{H}^2$ networks of different parametrisations.} The parameters of the network generation are listed in the title for each subplot. We used $K=-1$ as the curvature of the hyperbolic plane in each case. Each row of the figure demonstrates the effect of the change in a given network generation parameter: from top to bottom, the number of nodes $N$, the expected average degree $\langle k\rangle$, $1/(\gamma-1)$ (equivalent to the popularity fading parameter $\beta$ in the E-PSO model) and $1/\alpha$ (analogous to the temperature $T$ in the E-PSO model).}
    \label{fig:S1_LouvainGShist_uw}
\end{figure}

\begin{figure}[hbt]
    \centering
    \makebox[\textwidth][c]{\includegraphics[width=1.15\textwidth]{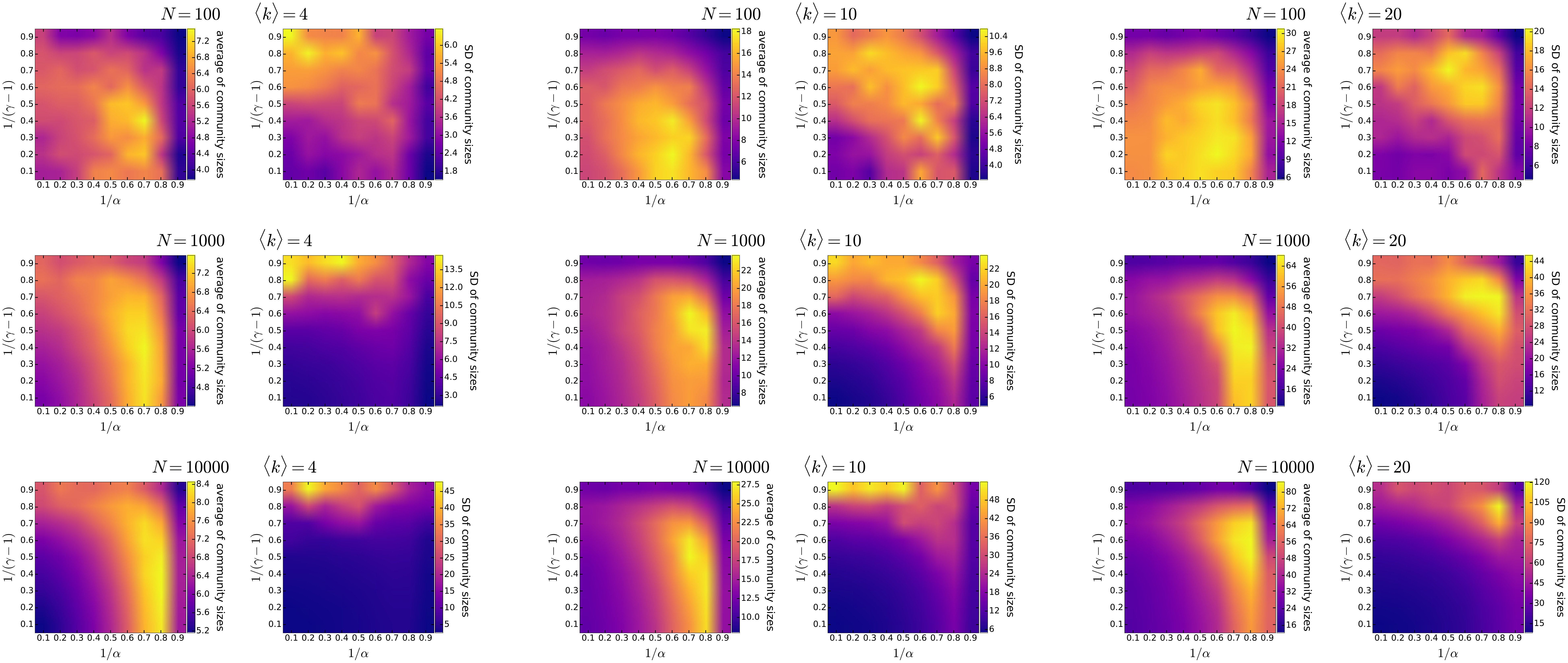}}
    \caption{{\bf The mean and the standard deviation of the size of communities detected by the \textit{Infomap} algorithm in 100 \textit{unweighted} $\mathbb{S}^1/\mathbb{H}^2$ networks of different parametrisations.} Each pair of subplots depicts the effect of changing $1/(\gamma-1)$ (equivalent to the popularity fading parameter $\beta$ in the E-PSO model) and $1/\alpha$ (analogous to the temperature $T$ in the E-PSO model), with the number of nodes $N$ and the expected average degree $\langle k\rangle$ given in the title of the subplot pair. We used $K=-1$ as the curvature of the hyperbolic plane in each case.}
    \label{fig:S1_InfomapGSavstd_uw}
\end{figure}

\begin{figure}[hbt]
    \centering
    \makebox[\textwidth][c]{\includegraphics[width=1.15\textwidth]{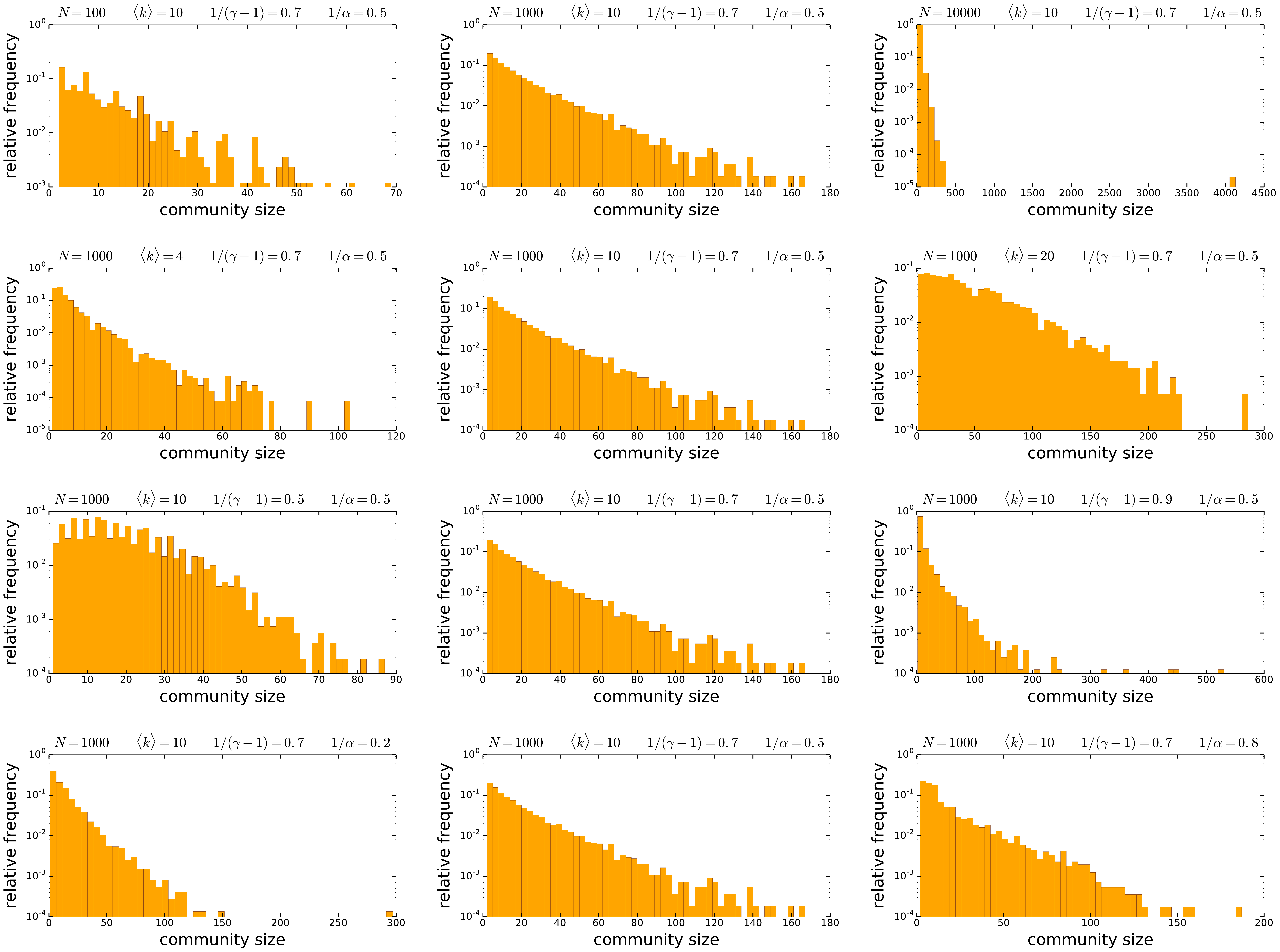}}
    \caption{{\bf The size distribution of the communities detected by the \textit{Infomap} algorithm in 100 \textit{unweighted} $\mathbb{S}^1/\mathbb{H}^2$ networks of different parametrisations.} The parameters of the network generation are listed in the title for each subplot. We used $K=-1$ as the curvature of the hyperbolic plane in each case. Each row of the figure demonstrates the effect of the change in a given network generation parameter: from top to bottom, the number of nodes $N$, the expected average degree $\langle k\rangle$, $1/(\gamma-1)$ (equivalent to the popularity fading parameter $\beta$ in the E-PSO model) and $1/\alpha$ (analogous to the temperature $T$ in the E-PSO model).}
    \label{fig:S1_InfomapGShist_uw}
\end{figure}

\begin{figure}[hbt]
    \centering
    \makebox[\textwidth][c]{\includegraphics[width=1.15\textwidth]{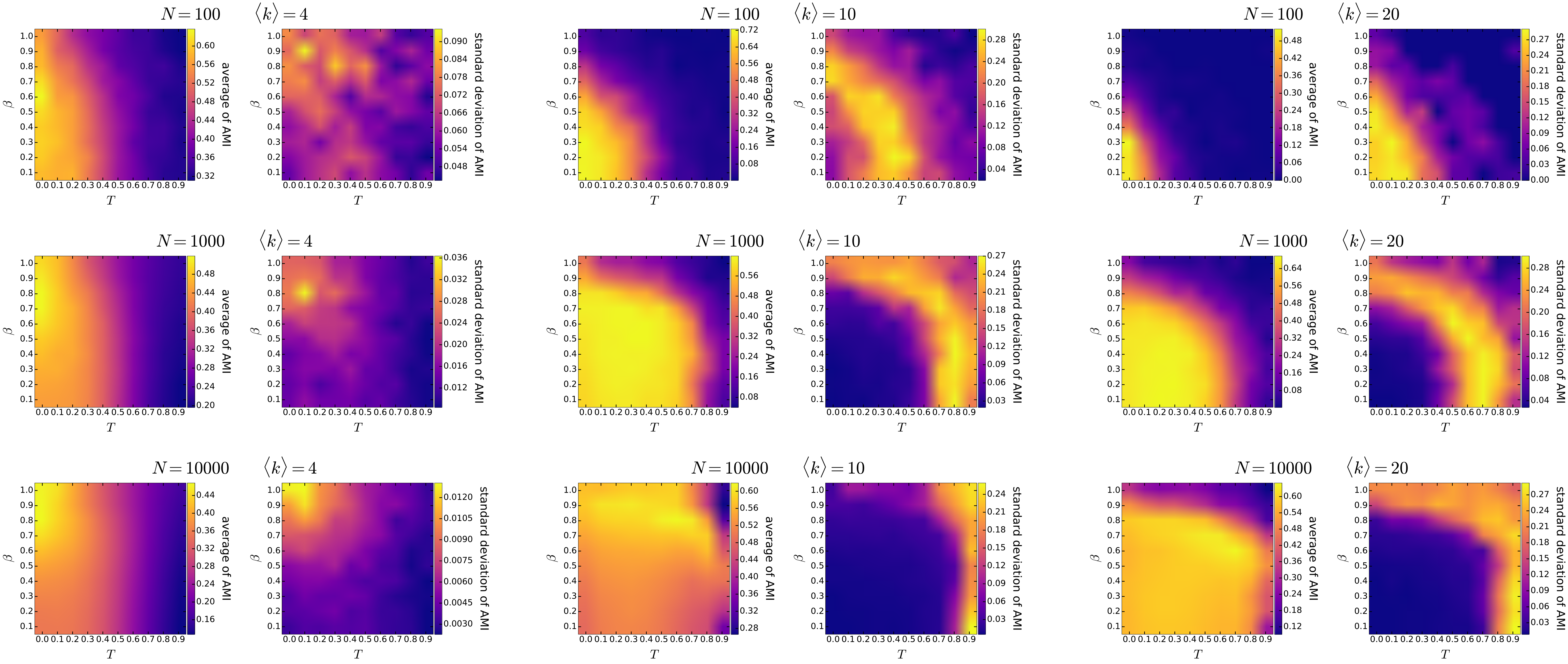}}
    \caption{{\bf The mean and the standard deviation of the adjusted mutual information of the two community structures detected by the \textit{asynchronous label propagation} and the \textit{Louvain} algorithms in 100 \textit{unweighted} \textit{PSO} networks of different parametrisations.} Each pair of subplots depicts the effect of changing the popularity fading parameter $\beta$ and the temperature $T$, with the number of nodes $N$ and the expected average degree $\langle k\rangle=2m$ given in the title of the subplot pair. The curvature of the hyperbolic plane $K$ was always set to $-1$, i.e. we used $\zeta=1$.}
    \label{fig:AMI_alabprop_Louv_PSO_uw}
\end{figure}

\begin{figure}[hbt]
    \centering
    \makebox[\textwidth][c]{\includegraphics[width=1.15\textwidth]{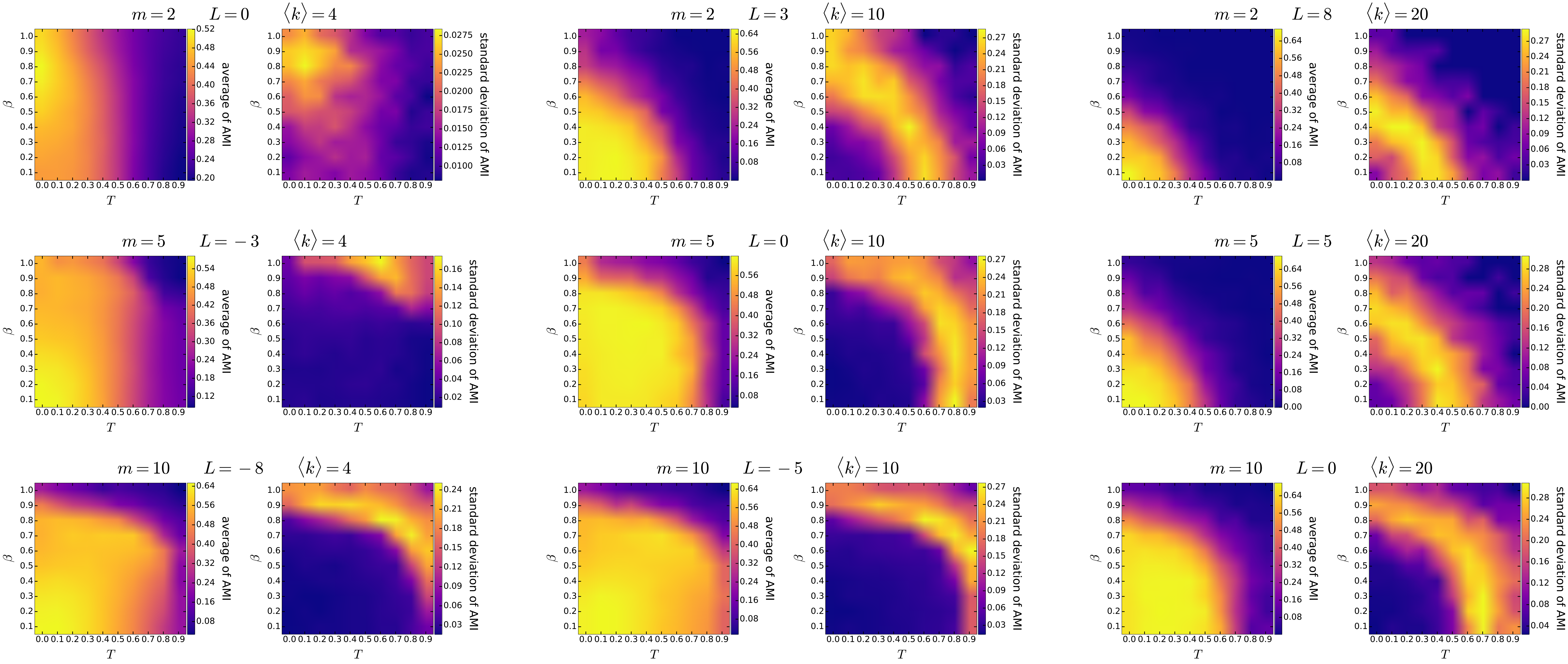}}
    \caption{{\bf The mean and the standard deviation of the adjusted mutual information of the two community structures detected by the \textit{asynchronous label propagation} and the \textit{Louvain} algorithms in 100 \textit{unweighted} \textit{E-PSO} networks of different parametrisations.} Each pair of subplots depicts the effect of changing the popularity fading parameter $\beta$ and the temperature $T$, with the parameters $m$ and $L$ given in the title of the subplot pair together with the corresponding expected average degree $\langle k\rangle=2(m+L)$. The number of nodes $N$ was 1000 in each case. The curvature of the hyperbolic plane $K$ was always set to $-1$, i.e. we used $\zeta=1$.}
    \label{fig:AMI_alabprop_Louv_EPSO_uw}
\end{figure}

\begin{figure}[hbt]
    \centering
    \makebox[\textwidth][c]{\includegraphics[width=1.15\textwidth]{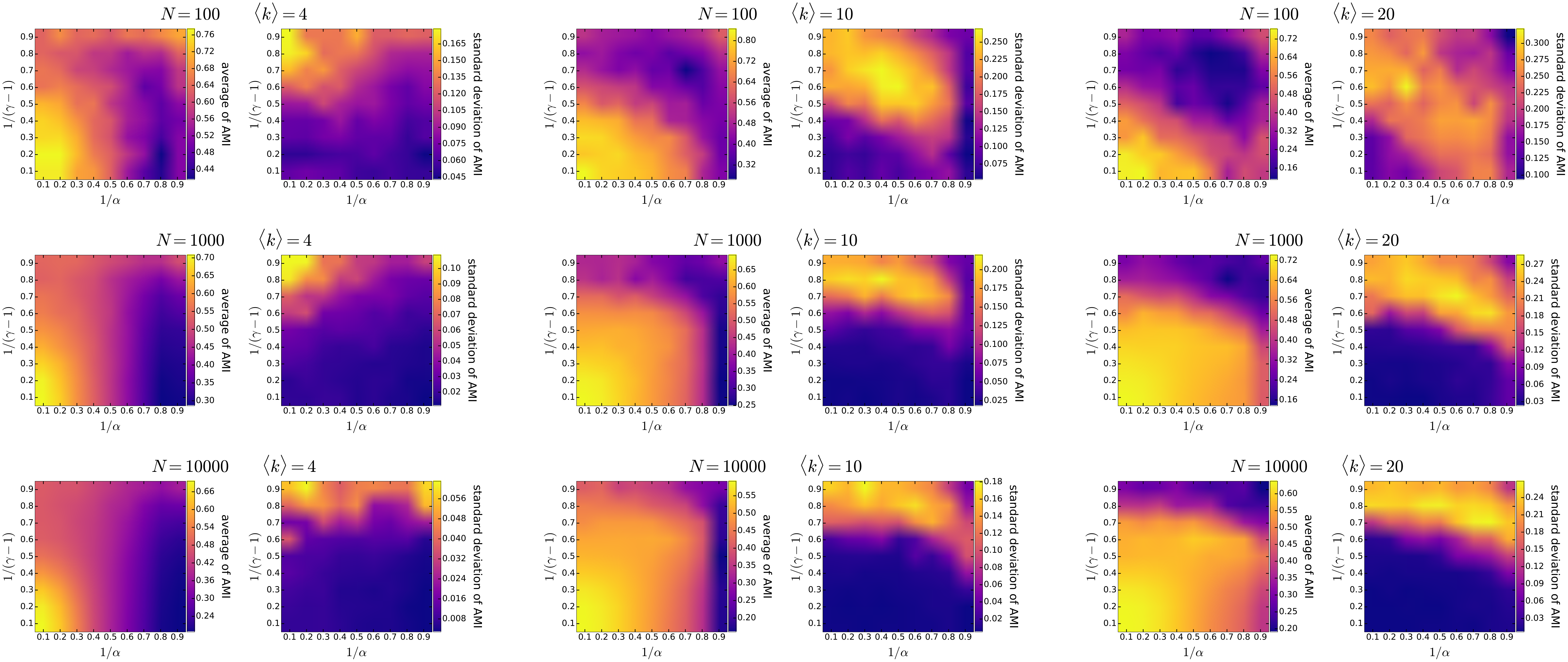}}
    \caption{{\bf The mean and the standard deviation of the adjusted mutual information of the two community structures detected by the \textit{asynchronous label propagation} and the \textit{Louvain} algorithms in 100 \textit{unweighted} $\mathbb{S}^1/\mathbb{H}^2$ networks of different parametrisations.} Each pair of subplots depicts the effect of changing $1/(\gamma-1)$ (equivalent to the popularity fading parameter $\beta$ in the E-PSO model) and $1/\alpha$ (analogous to the temperature $T$ in the E-PSO model), with the number of nodes $N$ and the expected average degree $\langle k\rangle$ given in the title of the subplot pair. We used $K=-1$ as the curvature of the hyperbolic plane in each case.}
    \label{fig:AMI_alabprop_Louv_S1_uw}
\end{figure}

\begin{figure}[hbt]
    \centering
    \makebox[\textwidth][c]{\includegraphics[width=1.15\textwidth]{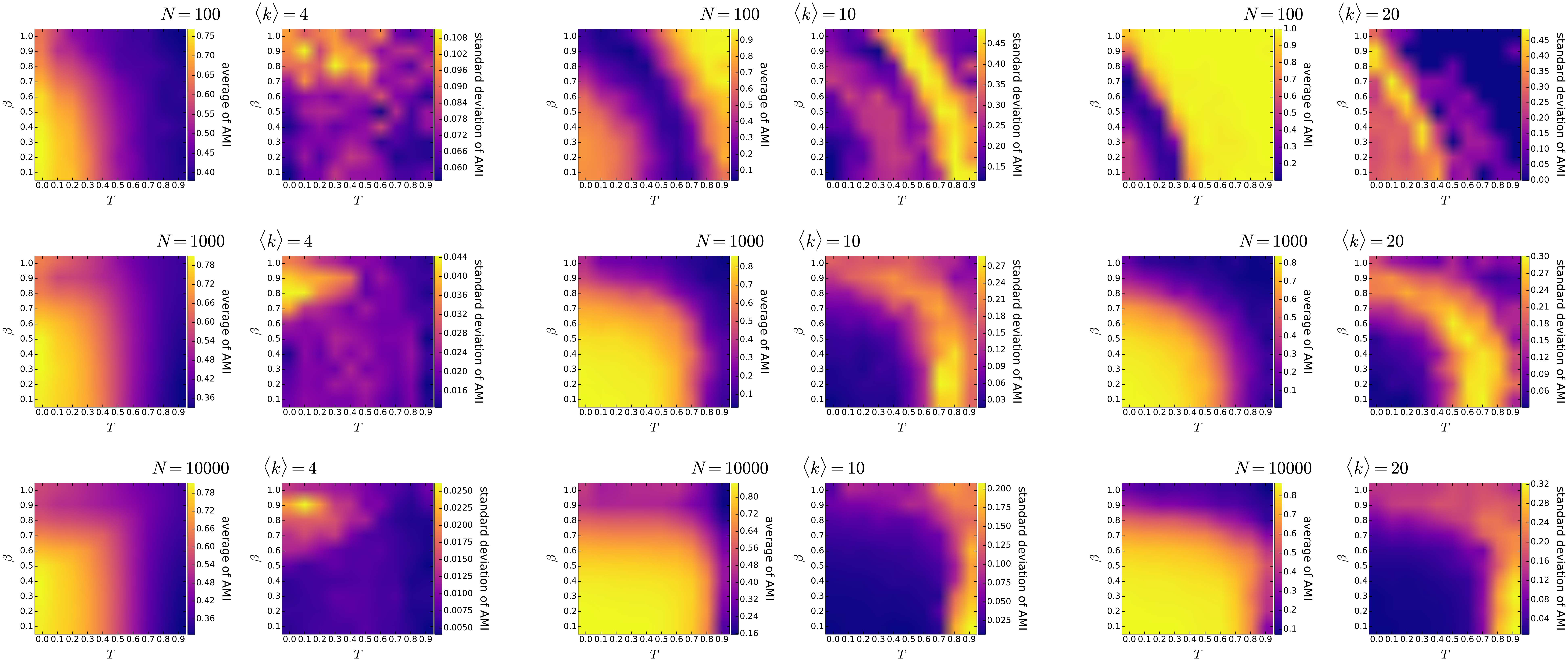}}
    \caption{{\bf The mean and the standard deviation of the adjusted mutual information of the two community structures detected by the \textit{asynchronous label propagation} and the \textit{Infomap} algorithms in 100 \textit{unweighted} \textit{PSO} networks of different parametrisations.} Each pair of subplots depicts the effect of changing the popularity fading parameter $\beta$ and the temperature $T$, with the number of nodes $N$ and the expected average degree $\langle k\rangle=2m$ given in the title of the subplot pair. The curvature of the hyperbolic plane $K$ was always set to $-1$, i.e. we used $\zeta=1$.}
    \label{fig:AMI_alabprop_Inf_PSO_uw}
\end{figure}

\begin{figure}[hbt]
    \centering
    \makebox[\textwidth][c]{\includegraphics[width=1.15\textwidth]{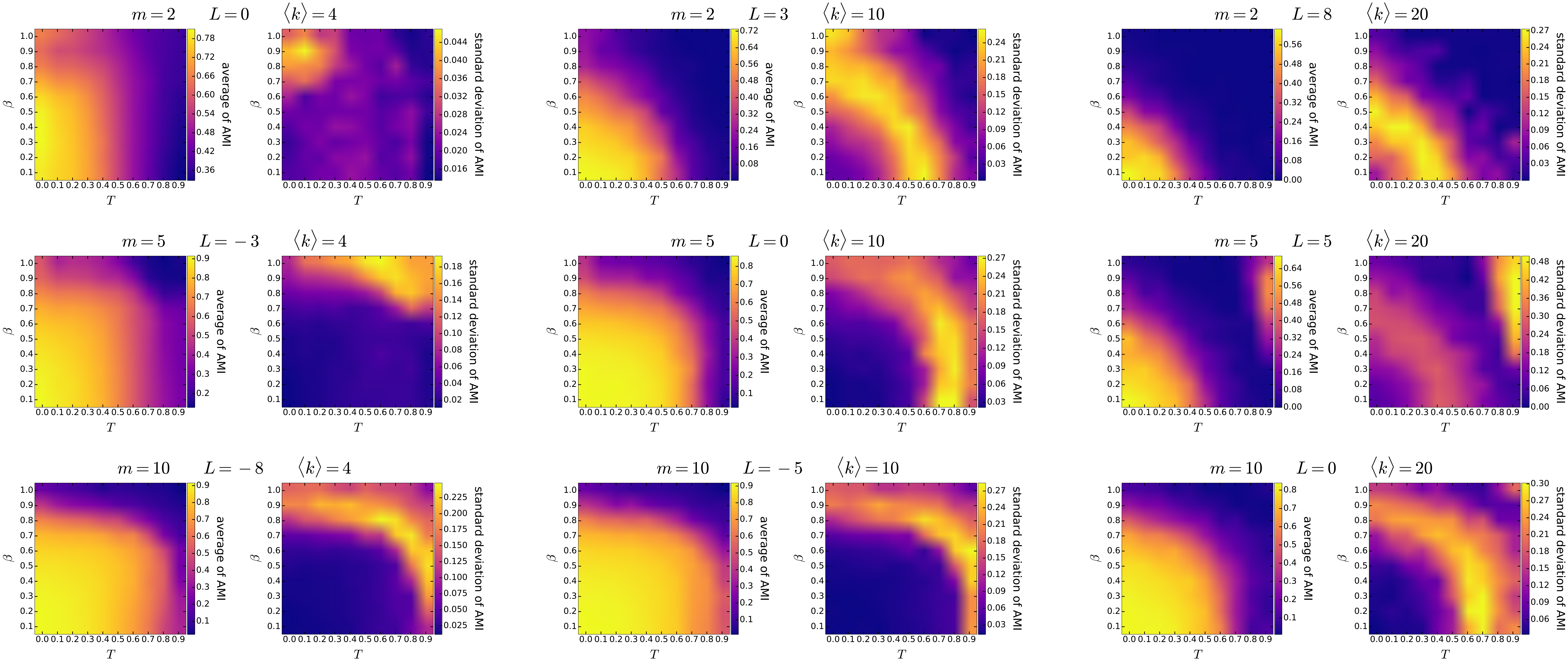}}
    \caption{{\bf The mean and the standard deviation of the adjusted mutual information of the two community structures detected by the \textit{asynchronous label propagation} and the \textit{Infomap} algorithms in 100 \textit{unweighted} \textit{E-PSO} networks of different parametrisations.} Each pair of subplots depicts the effect of changing the popularity fading parameter $\beta$ and the temperature $T$, with the parameters $m$ and $L$ given in the title of the subplot pair together with the corresponding expected average degree $\langle k\rangle=2(m+L)$. The number of nodes $N$ was 1000 in each case. The curvature of the hyperbolic plane $K$ was always set to $-1$, i.e. we used $\zeta=1$.}
    \label{fig:AMI_alabprop_Inf_EPSO_uw}
\end{figure}

\begin{figure}[hbt]
    \centering
    \makebox[\textwidth][c]{\includegraphics[width=1.15\textwidth]{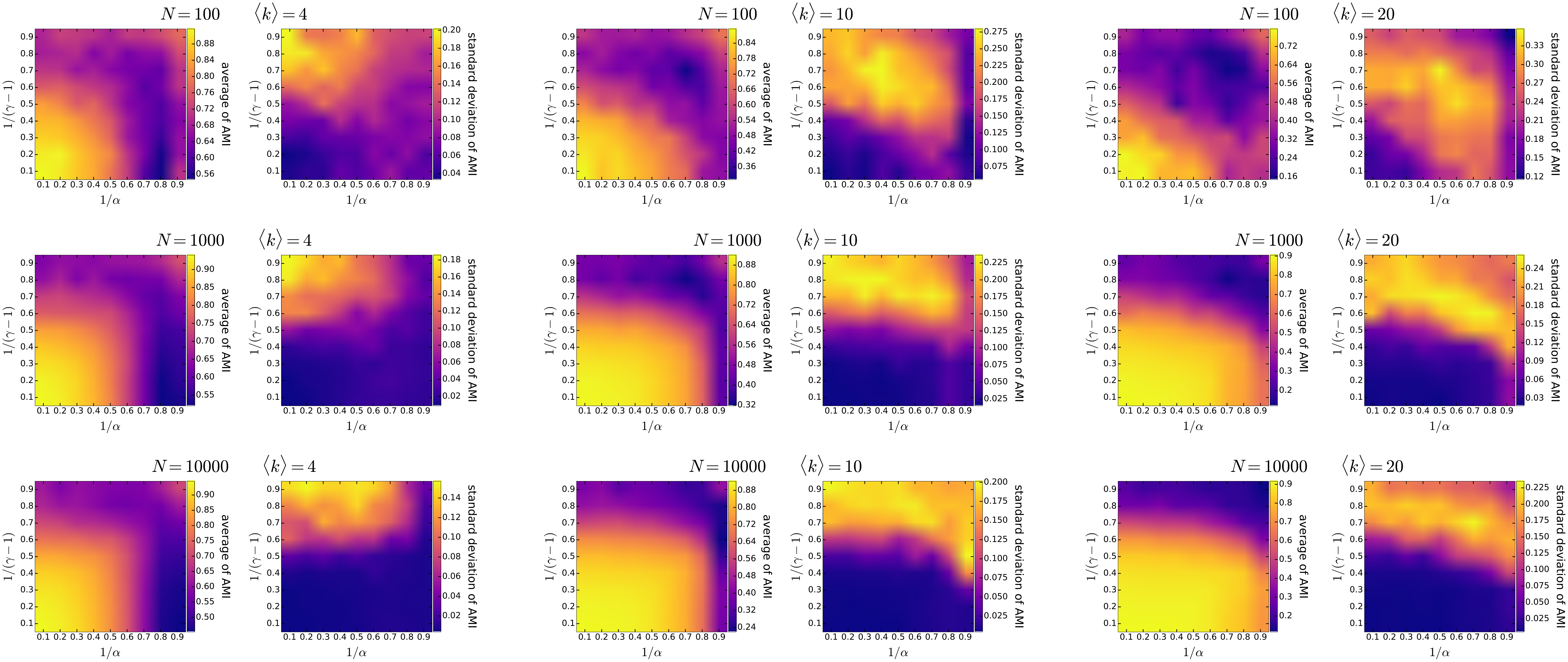}}
    \caption{{\bf The mean and the standard deviation of the adjusted mutual information of the two community structures detected by the \textit{asynchronous label propagation} and the \textit{Infomap} algorithms in 100 \textit{unweighted} $\mathbb{S}^1/\mathbb{H}^2$ networks of different parametrisations.} Each pair of subplots depicts the effect of changing $1/(\gamma-1)$ (equivalent to the popularity fading parameter $\beta$ in the E-PSO model) and $1/\alpha$ (analogous to the temperature $T$ in the E-PSO model), with the number of nodes $N$ and the expected average degree $\langle k\rangle$ given in the title of the subplot pair. We used $K=-1$ as the curvature of the hyperbolic plane in each case.}
    \label{fig:AMI_alabprop_Inf_S1_uw}
\end{figure}

\begin{figure}[hbt]
    \centering
    \makebox[\textwidth][c]{\includegraphics[width=1.15\textwidth]{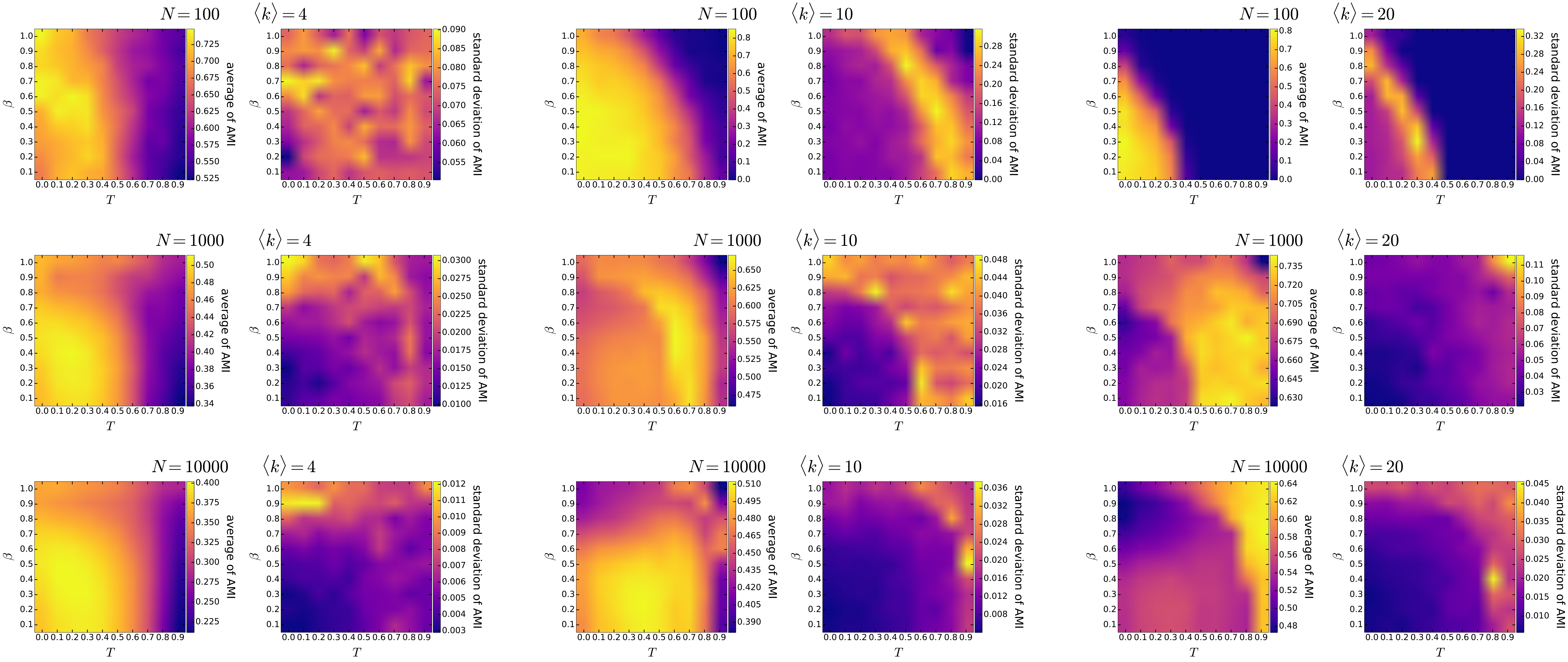}}
    \caption{{\bf The mean and the standard deviation of the adjusted mutual information of the two community structures detected by the \textit{Louvain} and the \textit{Infomap} algorithms in 100 \textit{unweighted} \textit{PSO} networks of different parametrisations.} Each pair of subplots depicts the effect of changing the popularity fading parameter $\beta$ and the temperature $T$, with the number of nodes $N$ and the expected average degree $\langle k\rangle=2m$ given in the title of the subplot pair. The curvature of the hyperbolic plane $K$ was always set to $-1$, i.e. we used $\zeta=1$.}
    \label{fig:AMI_Louv_Inf_PSO_uw}
\end{figure}

\begin{figure}[hbt]
    \centering
    \makebox[\textwidth][c]{\includegraphics[width=1.15\textwidth]{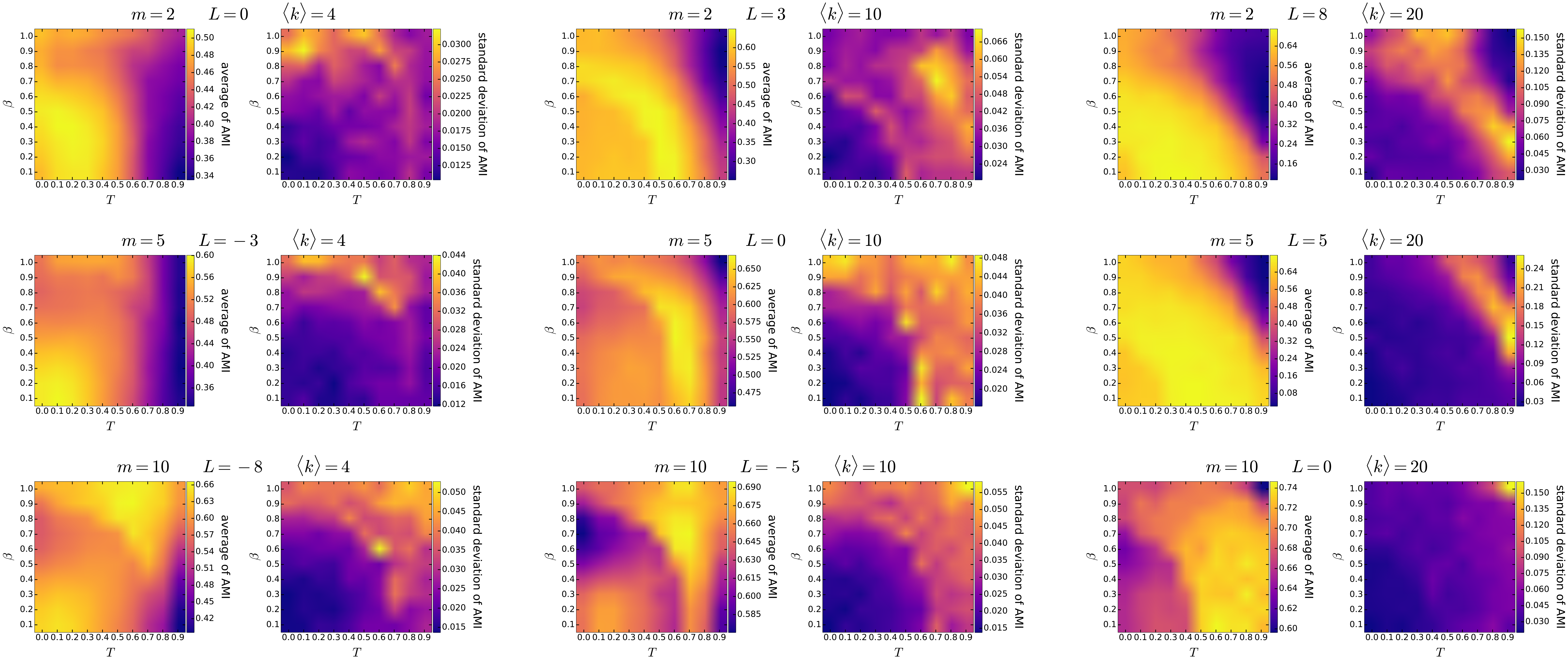}}
    \caption{{\bf The mean and the standard deviation of the adjusted mutual information of the two community structures detected by the \textit{Louvain} and the \textit{Infomap} algorithms in 100 \textit{unweighted} \textit{E-PSO} networks of different parametrisations.} Each pair of subplots depicts the effect of changing the popularity fading parameter $\beta$ and the temperature $T$, with the parameters $m$ and $L$ given in the title of the subplot pair together with the corresponding expected average degree $\langle k\rangle=2(m+L)$. The number of nodes $N$ was 1000 in each case. The curvature of the hyperbolic plane $K$ was always set to $-1$, i.e. we used $\zeta=1$.}
    \label{fig:AMI_Louv_Inf_EPSO_uw}
\end{figure}

\begin{figure}[hbt]
    \centering
    \makebox[\textwidth][c]{\includegraphics[width=1.15\textwidth]{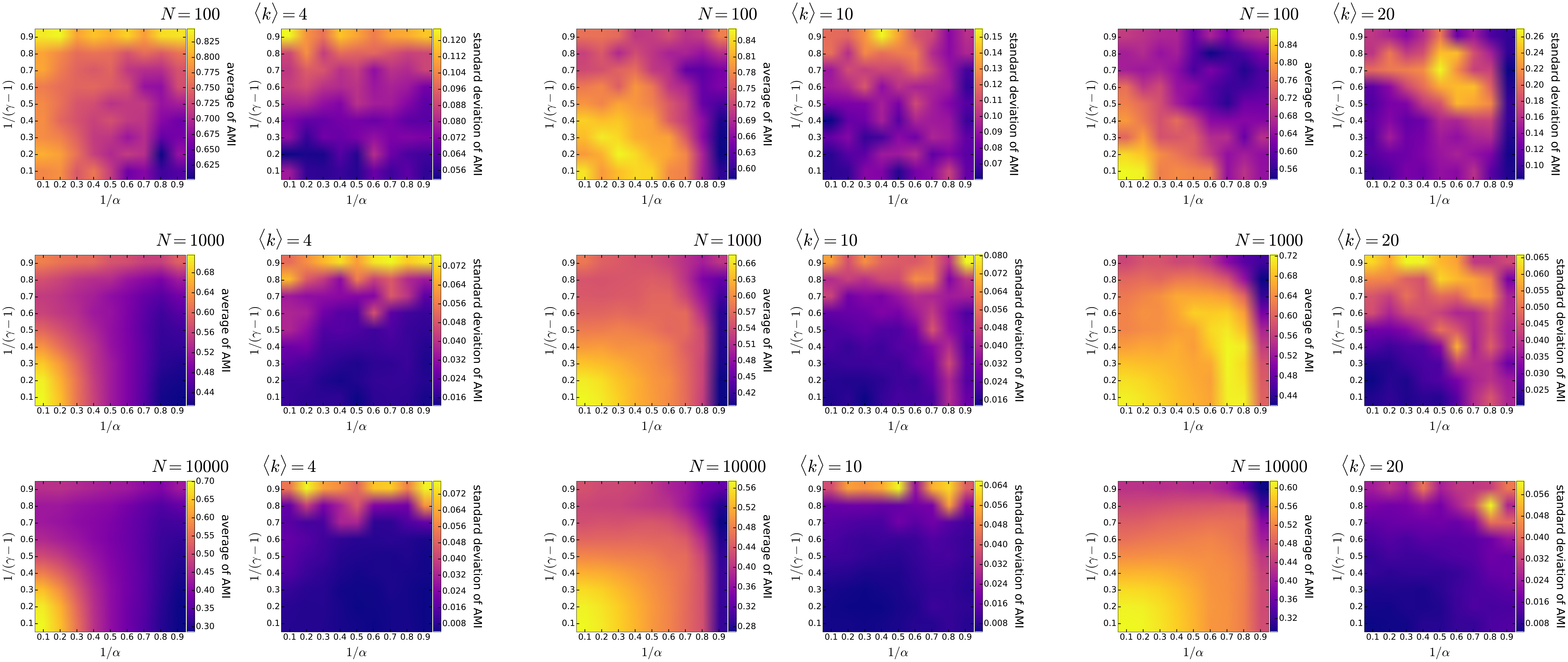}}
    \caption{{\bf The mean and the standard deviation of the adjusted mutual information of the two community structures detected by the \textit{Louvain} and the \textit{Infomap} algorithms in 100 \textit{unweighted} $\mathbb{S}^1/\mathbb{H}^2$ networks of different parametrisations.} Each pair of subplots depicts the effect of changing $1/(\gamma-1)$ (equivalent to the popularity fading parameter $\beta$ in the E-PSO model) and $1/\alpha$ (analogous to the temperature $T$ in the E-PSO model), with the number of nodes $N$ and the expected average degree $\langle k\rangle$ given in the title of the subplot pair. We used $K=-1$ as the curvature of the hyperbolic plane in each case.}
    \label{fig:AMI_Louv_Inf_S1_uw}
\end{figure}

\begin{figure}
    \centering
    \captionsetup{width=0.85\textwidth}
    \includegraphics[width=0.85\textwidth]{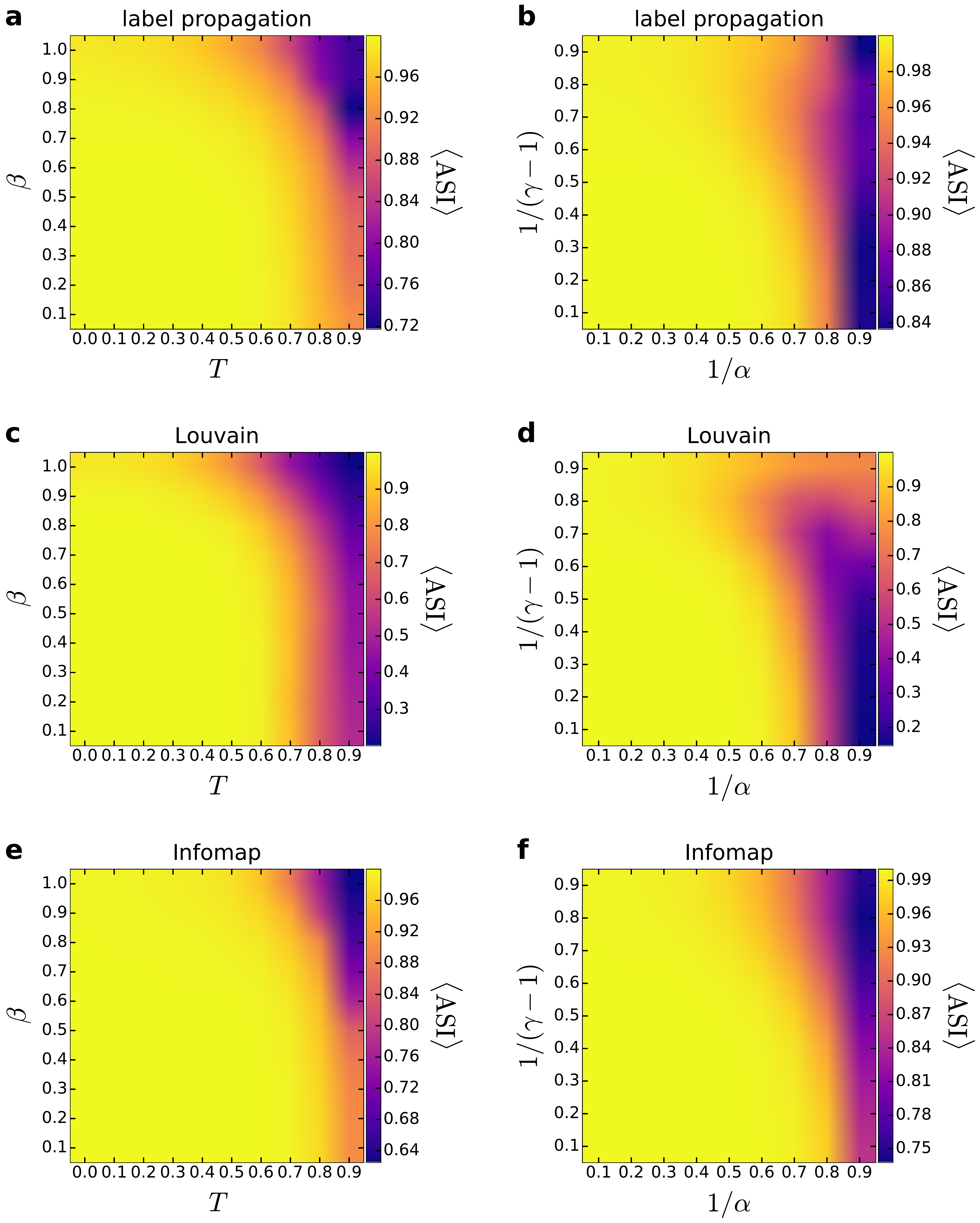}
    \caption[width=0.85\textwidth]{ {\bf Angular separation index in the \textit{unweighted} PSO and $\mathbb{S}^1/\mathbb{H}^2$ models.} The results for the PSO model are given in the left column (panels (a), (c) and (e)), whereas the ASI obtained for the $\mathbb{S}^1/\mathbb{H}^2$ model appears in the right column (panels (b), (d) and (f)). The ASI for the communities detected by asynchronous label propagation is given in the top row (panels (a) and (b)), the ASI regarding the results of Louvain is shown in the middle row (panels (c) and (d)) and the ASI for the partitions found by Infomap is presented in the bottom row (panels (e) and (f)). We show the measured ASI (indicated by the color, averaged over 100 samples) as a function of the model parameters $T$ and $\beta$, or $1/\alpha$ and $1/(\gamma -1)$ for networks of size $N=10,000$ and expected average degree $\left< k\right> =10$.}
    \label{fig:ASI_uw}
\end{figure}

We studied the quality of the community structures detected by the asynchronous label propagation~\cite{alabprop}, the Louvain~\cite{Louvain} and the Infomap~\cite{Infomap} algorithms in PSO \cite{PSO}, E-PSO \cite{EPSO_HyperMap,our_embedding} and $\mathbb{S}^1/\mathbb{H}^2$~\cite{S1,S1H2_Mercator} networks of various parameter combinations. The isolated nodes emerging in the case of the $\mathbb{S}^1/\mathbb{H}^2$ model and occasionally also in the networks generated by the E-PSO model of $L<0$ were removed before the community detection, meaning that the actual size of the examined networks does not necessarily reach the number of nodes $N$ inputted in these models. Each community detection algorithm was executed once for each network. This section presents the results obtained in the case of setting each link weight in the previously examined synthetic networks to 1, independently of the hyperbolic distance between the connected nodes. Similarly to figures~\ref{fig:PSO_alabpropMod}–\ref{fig:S1_InfomapMod}, figures \ref{fig:PSO_alabpropMod_uw}–\ref{fig:S1_InfomapMod_uw} show how the unweighted modularity \cite{Newman_modularity_original} achieved by the three different community detection algorithms depends on the network generation parameters. As in section~\ref{sect:detailedCommSizeDist}, figures \ref{fig:PSO_alabpropGSavstd_uw}–\ref{fig:S1_InfomapGShist_uw} display how the mean and the standard deviation of the community sizes depend on the network generation parameters, as well as some examples for the corresponding community size distributions. Figures \ref{fig:AMI_alabprop_Louv_PSO_uw}–\ref{fig:AMI_Louv_Inf_S1_uw} are analogous to the figures of section~\ref{sect:detailedAMI}, showing the similarity between the community structures detected by the different methods in the unweighted networks by means of the adjusted mutual information (AMI) \cite{AMI,McCarthy_AMI}. Lastly, figure~\ref{fig:ASI} of the main article depicting the angular separation index (ASI) of the detected communities \cite{Cannistraci_ASI} is repeated for the unweighted case in figure \ref{fig:ASI_uw}.

%%% modularity %%%%%%%%%%%%%%%%%%%%%%%%%%%%
%PSO (E-PSO L=0-val)

%E-PSO (Lnem0)

%S1

\clearpage
%%% group sizes %%%%%%%%%%%%%%%%%%%%%%%%%%%%%%%
%PSO (E-PSO L=0-val)

%E-PSO (Lnem0)

%S1

\clearpage
%%% AMI %%%%%%%%%%%%%%%%%%
%alabprop-Louvain

%alabprop-Infomap

%Louvain-Infomap

\clearpage

%%% ASI %%%%%%%%%%%

\end{document}